\documentclass[12pt]{article}
\usepackage{amsfonts,amssymb,graphicx,fullpage}
\usepackage[tight]{subfigure}
\usepackage{rotating}
\pdfoutput=1
\usepackage{ifpdf}
\ifpdf
  \usepackage{color}
  \definecolor{darkblue}{rgb}{0.3,0.3,0.6}
  \usepackage[pdftitle={Kaehler metrics},pdfauthor={Gabriele Honecker},pdfsubject={String theory},pdfkeywords={string theory, rigid branes},colorlinks=true,linkcolor=black,urlcolor=darkblue,filecolor=darkblue,citecolor=darkblue,menucolor=darkblue,breaklinks=true,bookmarks=true,anchorcolor=black,unicode=true]{hyperref}
\else
  
\fi
\usepackage[numbers,compress]{natbib}
\usepackage{a4wide}
\usepackage{longtable}
\usepackage{graphicx}
\usepackage{subfigure}
\usepackage[centertags]{amsmath}

\newcommand{\bCentering}{\centering}
\newcommand{\bCaption}{\caption}
\newcommand{\sgn}{{\rm sgn}}
\newcommand{\unity}{{\footnotesize\mbox{1\!\!I}}}

\def\muc{\multicolumn}

\def\Z{\mathbb{Z}}
\def\unity{1\!\!{\rm I}}
\def\ov{\overline}
\def\N{\mathbf{N}}
\def\Sym{\mathbf{Sym}}
\def\Anti{\mathbf{Anti}}
\def\Adj{\mathbf{Adj}}

\def\ov{\overline}
\def\1{{\bf 1}}
\def\2{{\bf 2}}
\def\3{{\bf 3}}
\def\4{{\bf 4}}
\def\6{{\bf 6}}
\def\OR{\Omega\mathcal{R}}
\def\half{\frac{1}{2}}
\def\pp{\uparrow\uparrow}
\def\ap{\uparrow\downarrow}
\def\targ#1#2{\genfrac{[}{]}{0pt}{}{#1}{#2}}
\def\half{{\textstyle\frac{1}{2}}}

\newcommand{\bCaptionfonts}{\small}
\makeatletter
\long\def\@makecaption#1#2{%
  \vskip\abovecaptionskip
  \sbox\@tempboxa{{\bCaptionfonts #1: #2}}%
  \ifdim \wd\@tempboxa >\hsize
    {\bCaptionfonts #1: #2\par}
  \else
    \hbox to\hsize{\hfil\box\@tempboxa\hfil}%
  \fi
  \vskip\belowcaptionskip}
\makeatother

\makeatletter
\let\ORIGINALlatex@openbib@code=\@openbib@code
\renewcommand{\@openbib@code}{\ORIGINALlatex@openbib@code\setlength{\itemsep}{1ex plus.5ex minus.5ex}\setlength{\parsep}{0pt}}
\makeatother

\def\mathtab#1#2#3{\begin{table}[th]\bCentering$#1$\bCaption{#3}\label{tab:#2}\end{table}}
\def\mathtabfix#1#2#3{\begin{table}[th]\bCentering\resizebox{\linewidth}{!}{$#1$}\bCaption{#3}\label{tab:#2}\end{table}}

\def\mathsidetabfix#1#2#3{\begin{sidewaystable}[H]\bCentering\resizebox{\linewidth}{!}{$#1$}\bCaption{#3}\label{tab:#2}\end{sidewaystable}}

\renewcommand{\arraystretch}{1.3}

\begin{document}
\begin{center}
\begin{flushright}
{\small MZ-TH/11-26\\September 2011}
\end{flushright}

\vspace{17mm}
{\Large\bf K\"ahler metrics and gauge kinetic functions for intersecting D6-branes on toroidal orbifolds}

\
\

{\Large\bf - The complete perturbative story -}

\vspace{5mm}
{\large Gabriele Honecker
}

\vspace{3mm}
{\it Institut f\"ur Physik  (WA THEP), Johannes-Gutenberg-Universit\"at, D - 55099 Mainz, Germany
\; {\tt Gabriele.Honecker@uni-mainz.de}}

\vspace{5mm}{\bf Abstract}\\[2ex]\parbox{140mm}{
We systematically derive the perturbatively exact holomorphic gauge kinetic function, the open string K\"ahler metrics and 
closed string K\"ahler potential on intersecting D6-branes by matching open string one-loop computations of gauge thresholds 
with field theoretical  gauge couplings in ${\cal N}=1$ supergravity. 
We consider all cases of bulk, fractional and rigid D6-branes on $T^6/\OR$
and the orbifolds $T^6/(\Z_N \times \OR)$ and $T^6/(\Z_2 \times \Z_{2M} \times \OR)$ without and 
with discrete torsion, which differ in the number of bulk complex structures and in the bulk K\"ahler potential. 
Our analysis includes all supersymmetric configurations of vanishing and 
non-vanishing angles among D6-branes and O6-planes, 
and all possible Wilson line and displacement
moduli are taken into account. 
The shape of the K\"ahler moduli turns out to be orbifold independent but angle dependent, whereas the holomorphic
gauge kinetic functions obtain three different kinds of one-loop corrections: a K\"ahler moduli dependent one for some vanishing angle independently 
of the orbifold background, another one depending on  complex structure moduli only for fractional and rigid D6-branes, and finally a constant term from intersections with O6-planes.
These results are of essential importance for the construction of the related effective 
field theory of phenomenologically appealing D-brane models.

As first examples, we compute the complete perturbative gauge kinetic functions and K\"ahler metrics for some
$T^6/\Z_2 \times \Z_2$ examples with rigid D-branes of~\cite{Angelantonj:2009yj}.
As a second class of examples, the K\"ahler metrics and gauge kinetic functions for the fractional QCD and leptonic D6-brane stacks of the Standard Model on
$T^6/\Z_6'$ from~\cite{Gmeiner:2008xq} are given.
}
\end{center}

\thispagestyle{empty}
\clearpage

\tableofcontents
\newpage
\setlength{\parskip}{1em plus1ex minus.5ex}
\section{Introduction}\label{S:intro}

Over recent years, considerable progress has been made in constructing supersymmetric
globally consistent string theory vacua with Standard Model gauge group and 
matter content, see e.g.~\cite{Cvetic:2001tj,Cvetic:2001nr,Honecker:2003vq,Honecker:2004kb,Honecker:2004np,Gmeiner:2005vz,Gmeiner:2008xq} 
and the review articles~\cite{Uranga:2003pz,Blumenhagen:2005mu,Blumenhagen:2006ci,Dudas:2006bj,Marchesano:2007de,Lust:2007kw}
for intersecting D6-branes,~\cite{Buchmuller:2005jr,Lebedev:2006kn,Buchmuller:2006ik,Lebedev:2008un} for 
heterotic orbifolds,~\cite{Braun:2005nv,Bouchard:2005ag} for heterotic Calabi-Yau compactifications with
$SU(N)$ bundles and~\cite{Blumenhagen:2005ga,Blumenhagen:2005pm,Blumenhagen:2005zg,Blumenhagen:2006ux} for $U(N)$ bundles, 
the review~\cite{Weigand:2010wm} and references therein for globally defined F-theory models 
and~\cite{Dijkstra:2004ym,Dijkstra:2004cc} for Gepner models.

Establishing the existence of Standard Model vacua, however, also requires 
the matching of the low-energy effective  action, in particular recovering
the perturbative gauge and Yukawa couplings of the Standard Model group and particles. 
Partial results at tree level can generically be obtained by dimensional reduction of the supergravity
and D-brane Chern-Simons and Born-Infeld actions in combination with charge selection 
rules, see e.g.~\cite{Aldazabal:2000cn} and~\cite{Camara:2011jg,Grimm:2011dx,Kerstan:2011dy} for very early and very recent works on D6-branes, respectively.
The exact dependence on moduli fields and numerical values at one loop, however, requires more powerful 
techniques of conformal field theory. The well-known methods from heterotic 
orbifolds, e.g.~\cite{Derendinger:1991hq,Kaplunovsky:1995jw}, have been translated to the case of {\it bulk} D6-branes in IIA
string theory on the six-torus and its T-dual variants of D-branes in the IIB theory, by identifying orbifold twists on the heterotic side with intersection angles in IIA 
and magnetic background fields in IIB, see e.g. the reviews~\cite{Angelantonj:2002ct,Blumenhagen:2006ci}.\footnote{Field theory results on K\"ahler metrics and gauge thresholds
at the orbifold point in the type IIB string, which are {\it not} T-dual to the intersecting D6-brane scenario include e.g. the  globally consistent models of~\cite{Antoniadis:1999ge,Billo:2007sw,Billo:2007py} and the local models of~\cite{Conlon:2009xf,Conlon:2009kt}.
}
But realistic string spectra require the use of {\it rigid} or at least {\it fractional} 
D-branes in order to project out adjoint moduli which would be responsible for arbitrary continuous
breakings of the gauge group along flat directions. These types of D6-branes on orbifolds of the type IIA string
with at least one $\Z_2$ subsymmetry, which leads to new non-trivial contributions to the one-string-loop gauge threshold 
computation as worked out in~\cite{Gmeiner:2009fb}, possess new chiral configurations
at some vanishing intersection angle, and the particle generations can emerge from various intersection sectors of orbifold-image D6-branes
for orbifolds other than $\Z_2 \times \Z_2$.  

The aim of the present article is to consistently and compactly formulate the perturbatively exact holomorphic gauge kinetic function, the bulk K\"ahler potential and 
open string K\"ahler metrics for (factorisable) toroidal orbifold backgrounds of type IIA orientifolds in the most general possible set-up, i.e. by including
all (untilted and tilted) background lattices and all (discrete or continuous) displacements and Wilson lines in such a way that it can be readily applied
to the existing Standard Model-like spectra on fractional D6-branes on $T^6/\Z_6'$~\cite{Gmeiner:2007zz,Gmeiner:2008xq,Gmeiner:2009fb} and 
$T^6/\Z_6$~\cite{Honecker:2004kb,Gmeiner:2007we} as well as expected new models on orbifolds with discrete torsion.
To this end, the previously computed gauge threshold corrections~\cite{Lust:2003ky,Akerblom:2007np,Blumenhagen:2007ip,Gmeiner:2009fb}
 are carefully regrouped for all backgrounds into lattice sums with beta function coefficients as prefactors plus constant terms from intersections with O6-planes 
 and complex structure moduli dependent contributions on fractional and rigid D6-branes only. While the first kind of correction has been used 
 before to derive K\"ahler metrics, e.g. on the six-torus in~\cite{Akerblom:2007np}, the two other kinds are to our knowledge fully appreciated here 
 for the first time.

\vspace{2mm}

The paper is organised as follows. In section~\ref{S:gauge-thres-revisit} we review the geometric set-up and 
computation of gauge thresholds at open string one-loop. The focus is on the comparison of bulk, fractional and rigid D6-branes,
and all gauge thresholds are reformulated such that the beta function coefficients appear as prefactors, wherever possible.
This is essential for the correct identification of the K\"ahler metrics in section~\ref{S:Kaehler_metrics+potential}.
In section~\ref{S:gauge-thres-revisit}, we furthermore focus, besides the unitary groups, on symplectic and orthogonal gauge 
factors as well as (anomalous) single $U(1)$s and anomaly-free massless linear combinations of Abelian gauge factors, 
all of which have to our knowledge not been discussed in detail before. \\
In section~\ref{S:Kaehler_metrics+potential}, the matching of the stringy gauge thresholds from the previous section
with the supergravity expressions is performed for each case, and the K\"ahler metrics and perturbatively exact holomorphic
gauge kinetic functions are extracted. The discussion includes all factorisable toroidal orbifolds with different numbers
$h_{21}^{\rm bulk}=3,1,0$ of bulk complex structures, possible one-loop field redefinitions, 
and all gauge groups $SU(N)$, $SO(2M)$, $Sp(2M)$ and $U(1)$.\\
\indent The use of the generic results is demonstrated in two classes of examples, first  in section~\ref{S:Compare-Example} on 
rigid D6-branes in the  $T^6/\Z_2 \times \Z_2$ background with discrete torsion dual to the magnetised D9/D5-brane 
set-up of~\cite{Angelantonj:2009yj},
and finally in section~\ref{S:Z6p-Example}, the generic expressions are applied to the Standard Model on fractional D6-branes
in the $T^6/\Z_6'$ orbifold background of~\cite{Gmeiner:2007zz,Gmeiner:2008xq,Gmeiner:2009fb}, which was the original motivation for studying the field theory
on various kinds of D6-branes, in particular including chiral matter at some vanishing angle, in detail. \\
Section~\ref{S:Conclusions} contains our conclusions, and technical details on the rewriting of the gauge threshold amplitudes
with M\"obius strip topology, the tree-level gauge couplings for
orbifolds with different numbers $h_{21}^{\rm bulk}=3,1,0$ of bulk complex structures and details on the 
three-cycles and intersection numbers of the $T^6/\Z_2 \times \Z_2$ examples dual to those in~\cite{Angelantonj:2009yj}
are collected in appendices~\ref{App:A} to~\ref{App:Magnetised}.

\section{The gauge thresholds revisited}\label{S:gauge-thres-revisit}

In this section, we review the computation of the  gauge thresholds for rigid D6-branes by means of the magnetic background field method on the 
least discussed orbifold background $T^6/\Z_2 \times \Z_{2M}$ with discrete torsion. Our discussion includes all possible supersymmetric 
D6-brane configurations at three or one vanishing angle or with all three angles non-vanishing. We comment on changes in the normalisation for all other known
bulk and fractional D6-branes on the factorisable six-torus $T^6$, orbifolds with one generator $T^6/\Z_N$ and with two generators $T^6/\Z_2 \times
\Z_{2M}$ without discrete torsion.

To this means, we discuss the background geometry and cycles in section~\ref{Ss:Geometry} and then briefly review the gauge threshold amplitudes in 
section~\ref{Ss:GeneralThresholds}.
Tables~\ref{tab:Z2Z2M-torsion-Bifundamentals-beta+thresholds},~\ref{tab:Six-torus-Bifund-beta+thresholds},~\ref{tab:Z2Z2M-no_torsion-Anti+Sym-beta+thresholds} and~\ref{tab:T6-Z2N-Bifund-beta+thresholds} contain the complete result for beta function coefficients and gauge thresholds from
bifundamental and adjoint matter of $SU(N_a)$ on all toroidal orbifold backgrounds
($T^6/\Z_2 \times \Z_{2M}$ with discrete torsion, $T^6$ and $T^6/\Z_3$, $T^6/\Z_2 \times \Z_{2M}$ without discrete torsion, $T^6/\Z_{2N}$, respectively)
 under considerations, and tables~\ref{tab:Z2Z2M-torsion-AntiSym-beta+thresholds},~\ref{tab:Six-torus-AntiSym-beta+thresholds},~\ref{tab:Z2Z2M-no_torsion-Anti+Sym-beta+thresholds} and~\ref{tab:T6-Z2N-Anti+Sym-beta+thresholds} give the analogous result for symmetric and antisymmetric matter. 
In section~\ref{Ss:Comments_on_U1}, we discuss the situation for Abelian gauge factors.
This complete presentation of all possible cases serves as preparation for determining the holomorphic gauge kinetic function and 
K\"ahler metrics for each case in section~\ref{S:Kaehler_metrics+potential}.

\subsection{Geometry, three-cycles and RR tadpole cancellation}\label{Ss:Geometry}

\subsubsection{Orbifolds, compactification lattices and one-cycles}\label{Sss:Lattices}

Throughout this paper we consider intersecting D6-branes on the factorisable six-torus, $T^6=T_{(1)}^2 \times T_{(2)}^2 \times T_{(3)}^2$,  and 
its orbifolds with all possibilities of one generator,
\begin{equation*}
T^6/\Z_N: \qquad \theta: z_i \rightarrow e^{2\pi i v_i}  \, z_i,
\end{equation*}
or all choices of two generators containing a $\Z_2 \times \Z_2$ subgroup,
\begin{equation*}
T^6/\Z_2 \times \Z_{2M}: \qquad \theta: z_i \rightarrow e^{2\pi i v_i}  \, z_i
\quad \text{ and } \quad 
\omega: z_i \rightarrow e^{2\pi i w_i}  \, z_i,
\end{equation*}
acting on the complex coordinates $z_i$ of the two-tori $T_{(i)}^2$.
The corresponding shift vectors for these orbifolds, which are singular limits of  Calabi-Yau threefolds, are listed in table~\ref{Tab:T6ZN+T6Z2Z2M-shifts}.
\begin{table}[ht]
\renewcommand{\arraystretch}{1.3}
  \begin{minipage}[b]{0.4\linewidth}\centering
   \begin{equation*}
\begin{array}{|c|c|}\hline
 T^6/& \vec{v} 
 \\\hline\hline
 \Z_3 & \frac{1}{3}(1,-2,1)
 \\
 \Z_4 & \frac{1}{4}(1,-2,1)
 \\
 \Z_6 & \frac{1}{6}(1,-2,1)
 \\
 \Z_6' &  \frac{1}{6}(1,2,-3)
\\\hline
\end{array}
    \end{equation*}
\end{minipage}
\hspace{0.5cm}
\begin{minipage}[b]{0.5\linewidth}
\centering
  \begin{equation*}
\begin{array}{|c|c|c|}\hline
 T^6/& \vec{v} & \vec{w}
 \\\hline\hline
\Z_2 \times \Z_2 & \frac{1}{2}(1,-1,0) & \frac{1}{2}(0,1,-1)
\\
\Z_2 \times \Z_4 & \frac{1}{2}(1,-1,0) & \frac{1}{4}(0,1,-1) 
\\
\Z_2 \times \Z_6 & \frac{1}{2}(1,-1,0) & \frac{1}{6}(0,1,-1)
\\
\Z_2 \times \Z_6'& \frac{1}{2}(1,-1,0) & \frac{1}{6}(-2,1,1) 
\\\hline
\end{array}
    \end{equation*}
\end{minipage}
\caption{Left: shift vectors for all toroidal orbifolds on factorisable tori with one generator. The two-tori are ordered 
such that the $\Z_2 \equiv \Z_2^{(2)}$ sub-symmetry leaves the second two-torus invariant.
Right: shift vectors for all toroidal orbifolds with two generators and $\Z_2 \times \Z_2$ sub-group. For each of these orbifolds, there 
exist two inequivalent choices of the phase, $\eta = \pm 1$, with which one $\Z_2$ sub-group acts on the twisted states of the other $\Z_2$
and preserves either the two- or three-cycles.
These are the orbifolds `without discrete torsion' $(\eta=1)$ and `with discrete torsion' $(\eta=-1)$. 
}
\label{Tab:T6ZN+T6Z2Z2M-shifts}
\end{table}

\begin{figure}[ht]
\begin{minipage}[b]{0.5\linewidth}
\begin{center}
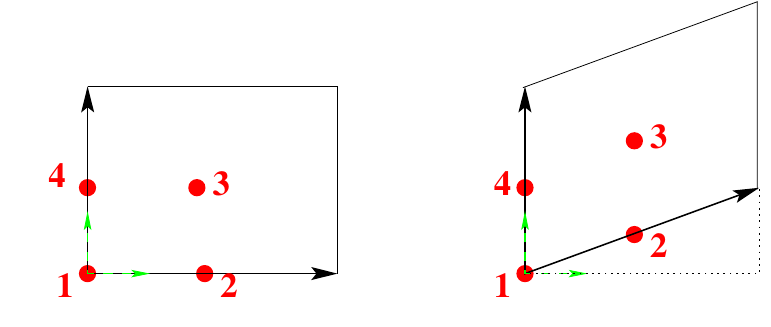
\end{center}
\caption{The $\Z_2$ invariant `untilted' {\bf a}-type (left) and `tilted' 
{\bf b}-type (right)  tori, which are parameterised by $b=0,\frac{1}{2}$, respectively.
The $\Z_2$ fixed points are depicted in blue. The points 1,4 are 
invariant under ${\cal R}$, whereas the other two points are on 
the tilted torus exchanged under ${\cal R}$,
$2 \stackrel{\cal R}{\leftrightarrow}
2+2b$ and $3  \stackrel{\cal R}{\leftrightarrow} 3-2b$. 
The untilted torus with $R_1=R_2=r$ corresponds to the {\bf A}-type
$\Z_4$ invariant lattice in figure~\protect\ref{Fig:Z4-Z6lattice}. 
The tilted torus for $R_2/R_1=2,2\sqrt{3},2/\sqrt{3}$ corresponds to the 
{\bf B}-type $\Z_4$ lattice and the {\bf A}- and {\bf B}-type $\Z_3$ 
(and $\Z_6$) invariant lattices in  figure~\protect\ref{Fig:Z4-Z6lattice} 
with radii $r=\sqrt{2} R_2, R_2/\sqrt{3},R_2$, respectively.
}
\label{Fig:Z2-lattice}
\end{minipage}
\hspace{0.5cm}
\begin{minipage}[b]{0.5\linewidth}
\begin{center}
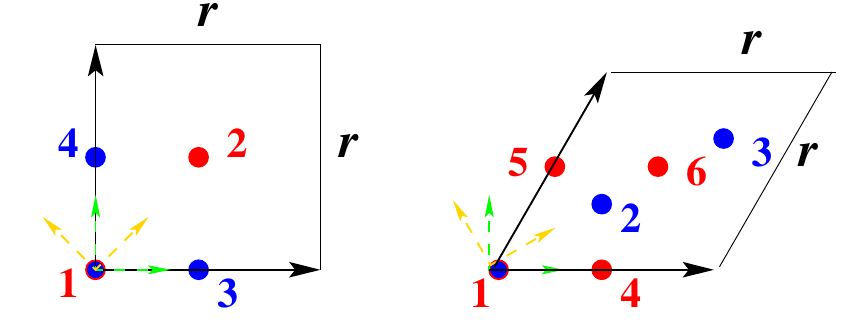
\end{center}
\caption{The $\Z_4$ (left) and $\Z_3$ (right) invariant lattices.
For the {\bf A} orientation (with green coordinate axes), $\pi_{2i-1}$ spans
the $\Re(z)$ axis, and on the {\bf B} lattice (axes in yellow), the 
$\Re(z)$ axis extends along $\pi_{2i-1} + \pi_{2i}$. The $\Z_4$ invariant points 
1,2 (left, in red) are ${\cal R}$ invariant, the additional $\Z_2$ fixed points
$3 \stackrel{\Z_4}{\leftrightarrow} 4$ (left, in blue) are ${\cal R}$ invariant
on the {\bf A}-lattice, but are permuted under ${\cal R}$ on the {\bf B}-lattice. 
The $\Z_3$ invariant points $2\stackrel{\Z_2}{\leftrightarrow} 3$ (right, in blue)
are exchanged under ${\cal R}$ on the {\bf A} orientation and are invariant on the 
{\bf B}-lattice. The $\Z_2$ fixed points 
$6 \stackrel{\Z_3}{\rightarrow} 5 \stackrel{\Z_3}{\rightarrow} 4$ (right, in red)
contain one point fixed under ${\cal R}$, the other two are exchanged under ${\cal R}$.
The origin 1 is fixed under the full ${\cal R}$ and $\Z_6$ symmetry.
}
\label{Fig:Z4-Z6lattice}
\end{minipage}
\end{figure}

The $\Z_2$ rotations are consistent with any choice of two-torus lattice, whereas a $\Z_4$ rotation requires a square torus and a $\Z_3$ (or $\Z_6$)
rotation requires a rhombus with acute angle $\pi/3$. The situation is depicted in figure~\ref{Fig:Z2-lattice} for $\Z_2$ and figure~\ref{Fig:Z4-Z6lattice} for $\Z_4$ and $\Z_3$ symmetries, respectively. 
The orientifold projection $\OR$ in Type IIA string theory contains an anti-holomorphic involution ${\cal R}$ 
on the compact space. On the factorisable torus, the involution is simply given by complex conjugation,
\begin{equation*}
{\cal R}: z_i \rightarrow \ov{z}_i.
\end{equation*}
This constrains the shape of the $\Z_2$ invariant tori to be `untilted' (rectangular) or `tilted' parameterised by the 
discrete choices $b=0,1/2$ of the real part of the complex structure, cf. figure~\ref{Fig:Z2-lattice}, which in the 
T-dual IIB language correspond to two discrete choices of the $B$-field. 
For the $\Z_4$ and $\Z_3$ invariant lattices, there exist two inequivalent orientations w.r.t. the $\Re(z)$ axis:
for the {\bf A}-lattice the basic one-cycle $\pi_{2i-1}$ spans the real axis, and for the {\bf B}-type lattice 
the real axis extends along $\pi_{2i-1} + \pi_{2i}$.
The untilted and titled torus differ in the number of parallel $\OR$ invariant O6-planes, $N^{(i)}_{\OR}=2(1-b_i)$.
The Hodge numbers $(h_{11},h_{21})$ for all toroidal orbifolds listed in table~\ref{Tab:T6ZN+T6Z2Z2M-shifts} 
can be found in the appendix of~\cite{Forste:2010gw} together with the decomposition $(h_{11}^+,h_{11}^-)$ into 
massless vectors and K\"ahler moduli in dependence of the lattice orientations. For those orbifolds with $\Z_2 \times \Z_2$
sub-symmetry, also all inequivalent choices of discrete torsion and some exotic O6-plane are taken into account in~\cite{Forste:2010gw}.

Any one-cycle on the two-torus $T^2_{(i)}$ can be expressed in terms of the coprime wrapping numbers $(n^i_a,m^i_a)$
along the basic lattice vectors $\pi_{2i-1}$ and $\pi_{2i}$. The angle of such an one-cycle w.r.t. the ${\cal R}$ invariant 
axis is given by 
\begin{equation}\label{Eq:tan-angles}
\tan \left( \pi \phi^{(i)}_{a}\right) = \left\{\begin{array}{cc}
\frac{m^i_a + b_i n^i_a}{n^i_a} \frac{R_2^{(i)}}{R_1^{(i)}} & \Z_2 ({\bf a},{\bf b})\\
\frac{m^i_a }{n^i_a} & \Z_4({\bf A})\\
 \frac{m^i_a-n^i_a}{m^i_a+n^i_a} & \Z_4 ({\bf B})\\
\sqrt{3} \frac{m^i_a}{2 \, n^i_a + m^i_a} & \Z_3({\bf A})\\
\frac{1}{\sqrt{3}} \frac{m^i_a-n^i_a}{m^i_a+n^i_a} & \Z_3({\bf B})
\end{array}\right.
,
\end{equation}
and relative angles between D$6_a$ and D$6_b$-branes as well as with the $\OR\theta^n\omega^m$ invariant 
O6-plane are given by 
\begin{equation*}
\phi^{(i)}_{ab} \equiv \phi^{(i)}_b - \phi^{(i)}_a 
\quad \text{ and } \quad
\phi^{(i)}_{a,\OR\theta^{n}\omega^m} \equiv \phi^{(i)}_{\OR\theta^{n}\omega^m} - \phi^{(i)}_a,
\end{equation*}
where our notation is chosen to fit with the $T^6/\Z_2 \times \Z_{2M}$ orbifold backgrounds in~\cite{Forste:2010gw}.
For the O6-plane, we used the notation that the action of the orbifold generator produces a new one-cycle 
with the following torus wrapping numbers,
\begin{equation}\label{Eq:Torus-wrapping-adjoint}
\begin{aligned}
(n^i_a,m^i_a)  \quad \stackrel{\text{rotation by } 2\pi v_i}{\longrightarrow} \quad 
(n^i_{(\omega a)},m^i_{(\omega a)}) =
\left\{\begin{array}{cc}
(-n^i_a,-m^i_a) & w_i=\frac{1}{2}
\\
(-m^i_a,n^i_a)  & w_i=\frac{1}{4}
\\
(-m^i_a,n^i_a + m^i_a) & w_i=\frac{1}{6}
\end{array}\right. ,
\end{aligned}
\end{equation}
where the $\Z_2$ rotation applies to all allowed lattice orientations and the $\Z_4$ and $\Z_6$ rotation to
those depicted in figure~\ref{Fig:Z4-Z6lattice}. All the orbifold rotations listed in table~\ref{Tab:T6ZN+T6Z2Z2M-shifts}
can be obtained from these basic relations.

A $\Z_3$ symmetry with generator $\omega$
produces one orbifold invariant orbit consisting of three toroidal cycles at relative angles $\pm \frac{2\pi i}{3}$, 
and there exists  one orbit of $\OR\omega^n$ $(n=0,1,2)$ invariant O6-planes.
A $\Z_4$ symmetry provides orbifold invariant orbits of two toroidal cycles, and there exist two distinct 
orbits of O6-planes, $\OR\omega^{2k}$ and $\OR\omega^{2k+1}$ at angles $-(2k)\pi w_i$ and $-(2k+1)\pi w_i$ w.r.t. the 
$\OR$ invariant plane, which we denote by $\OR$ and $\OR\Z_4$
invariant O6-plane orbits. 
A $\Z_6$ symmetry again contains the orbits of $\Z_3$ invariant cycles, but has two distinct orbits 
 $\OR\omega^{2k}$ and $\OR\omega^{2k+1}$ of O6-planes at angles $-(2k)\pi w_i$ and $-(2k+1)\pi w_i$ w.r.t. the 
$\OR$ invariant plane, which we denote by one of their representatives as the $\OR$ and $\OR\Z_2$ invariant orbits.
For orbifolds with $\Z_2 \times \Z_2$ sub-symmetry, four different orbits of O6-planes $\OR$ and $\OR\Z_2^{(i)}$ 
with $i=1,2,3$ arise as will be detailed further in section~\ref{Sss:ThreeCycles+RRtcc}.

The one-cycle intersection number on the two-torus is antisymmetric in its subscripts and given by
\begin{equation}\label{Eq:Def-1cycle_Intersection}
I_{ab}^{(i)} \equiv n^i_a m^i_b - m^i_a n^i_b.
\end{equation}
In section~\ref{Ss:GeneralThresholds}, we use the fact that (at least in the defining region $0 \leqslant |\phi_{ab}^{(i)}| < 1$) 
the signs of intersection numbers and relative angles are identical,
\begin{equation*}
 \sgn(\phi_{ab}^{(i)}) = \sgn(I_{ab}^{(i)}),
\end{equation*}
and that for supersymmetric D6-brane configurations, the maximal angle comes with the opposite sign of the other two leading to
\begin{equation*}
\sum_{i=1}^3  \sgn(\phi_{ab}^{(i)}) = - \prod_{i=1}^3  \sgn(\phi_{ab}^{(i)}).
\end{equation*}
These relations permit to replace intersection numbers by their absolute values in the computation of 
gauge thresholds and beta function coefficients by means of the magnetic background field method
as briefly reviewed below in section~\ref{Ss:GeneralThresholds}.

The complex structure moduli dependent quantity
\begin{equation}\label{Eq:Def-Vab}
V_{ab}^{(i)} \equiv \left\{\begin{array}{cc}
\frac{R_1^{(i)}}{R_2^{(i)}} n^i_a n^i_b + \frac{R_2^{(i)}}{R_1^{(i)}} (m^i_a+b_i n^i_a) (m^i_b + b_i n^i_b)  & \Z_2({\bf a}, {\bf b})
\\
n^i_a n^i_b + m^i_a m^i_b   & \Z_4({\bf A},{\bf B})
\\
\frac{1}{\sqrt{3}}( 2 \, n^i_a n^i_b + n^i_a m^i_b + m^i_a n^i_b + 2 \, m^i_a m^i_b) & \Z_3({\bf A},{\bf B})
\end{array}\right\}
= \left\{\begin{array}{cc}
\frac{(L^{(i)}_a)^2}{{\rm Vol}(T^2_{(i)})} & \phi^{(i)}_{ab}=0
\\
I^{(i)}_{ab} \, \cot \left( \pi \phi^{(i)}_{ab} \right) & \phi^{(i)}_{ab} \neq 0
\end{array}\right.
\end{equation}
is symmetric in its subscripts and for $a=b$ computes the square of the length $L^{(i)}_a$  (in units of the two-torus volume ${\rm Vol}(T^2_{(i)})$)
of the one-cycle wrapped by the D$6_a$-brane along $T^2_{(i)}$.  Note in particular the following correspondences
between specific angles, intersection numbers and (length)${}^2$,
  \begin{equation*}
I^{(i)}_{ab} =0  \quad \Leftrightarrow \quad \pi \phi_{ab}^{(i)} = 0,
\qquad\qquad\qquad
 V_{ab}^{(i)} =0  \quad \Leftrightarrow \quad \pi \phi_{ab}^{(i)} = \pm \frac{\pi}{2},
 \end{equation*}
 which simplify the expressions for beta function coefficients and gauge thresholds.
 For later use, we define the dimensionless 
 real K\"ahler modulus
 \begin{equation*}
v_i \equiv \frac{{\rm Vol}(T^2_{(i)})}{\alpha'}  = \left\{\begin{array}{cc}
\frac{R_1^{(i)}R_2^{(i)}}{\alpha'} & \Z_2({\bf a}, {\bf b})
\\
\frac{r^2_{(i)}}{\alpha'} & \Z_4({\bf A}, {\bf B})
\\
\frac{\sqrt{3}}{2} \frac{r^2_{(i)}}{\alpha'} & \Z_3({\bf A}, {\bf B})
\end{array}\right. 
\end{equation*}
in slight abuse of the symbol $v_i$, which was used above also  for entries of the shift vector generating the orbifold rotation $\theta$.
Since the meaning should  be clear from the context throughout this article, we refrain from introducing a new symbol.

Last but not least, the intersection numbers and generalised volume forms involving some O6-plane are
in all computations weighted with the number $N_{\OR\theta^n\omega^m}^{(i)} = 2 \, (1-b_i)$ of parallel O6-planes on the two-torus $T^2_{(i)}$,
\begin{equation}\label{Eq:Def_I+V-tilde}
\tilde{I}_a^{\OR\theta^n\omega^m,(i)} \equiv 2 \, (1-b_i) \, I_{a,\OR\theta^n\omega^m}^{(i)},
\qquad
 \tilde{V}_a^{\OR\theta^n\omega^m,(i)} \equiv 2 \, (1-b_i) \, V_{a,\OR\theta^n\omega^m}^{(i)}.
\end{equation}
For use in later sections, it is useful to explicitly compute the intersection numbers for D6-branes perpendicular to some O6-plane.
In~\cite{Gmeiner:2009fb}, we already made use of the fact that 
\begin{equation}\label{Eq:Special-OR-Intersections}
\begin{aligned}
a \perp \OR\Z_2^{(l)} \text{ on } T^2_{(i)}:  \quad & | \tilde{I}_a^{\OR\Z_2^{(l)},(i)} | = 2  ,
\\
a \perp  \OR\Z_2^{(l)} \text{ on } T^2_{(i)} \times T^2_{(j)} : \quad &  \tilde{I}_a^{\OR\Z_2^{(l)},(i \cdot j)} = -4,
\end{aligned}
\end{equation}
where the minus sign in the second line arises due to supersymmetry.
In the same spirit, the (length)${}^2$ for one-cycles parallel to some O6-plane are given by 
\begin{equation*}
\begin{aligned}
a \pp \OR\Z_2^{(l)} \text{ on } T^2_{(i)}:  \quad & \tilde{V}_a^{\OR\Z_2^{(l)},(i)}  = \left\{\begin{array}{ccccc} 
\frac{2}{1-b_i} \frac{R_1^{(i)}}{R_2^{(i)}}   & \Z_2({\bf a},{\bf b}) & l = i\\
2 (1-b_i) \frac{R_2^{(i)}}{R_1^{(i)}}  & \Z_2({\bf a},{\bf b}) & l \neq i\\
2 & \Z_4({\bf A},{\bf B})\\
\frac{2}{\sqrt{3}} & \Z_6({\bf A}) & l = i; & \Z_6({\bf B}) & l \neq i\\
2 \sqrt{3} & \Z_6({\bf A}) & l \neq i; & \Z_6({\bf B}) & l = i 
\end{array}\right.
,
\end{aligned}
\end{equation*}
since $N_{\OR\Z_2^{(l)}}^{(i)}=2(1-b_i)$ on the $\Z_2$ invariant two-torus, $N_{\OR\Z_2^{(l)}}^{(i)}=1$ on the $\Z_3(\Z_6)$ invariant 
two-torus, and $N_{\OR\Z_2^{(l)}}^{(i)}=2$ if the O6-plane lies in the orbit formed by $(n^i_{\OR\Z_2^{(l)}},m^i_{\OR\Z_2^{(l)}})=(1,0),(0,1)$ 
on the {\bf A}-type $\Z_4$ invariant lattice, but $N_{\OR\Z_2^{(l)}}^{(i)}=1$ if the orbit contains $(n^i_{\OR\Z_2^{(l)}},m^i_{\OR\Z_2^{(l)}})=(1,1),(1,-1)$, 
and vice versa on the $\Z_4$ in variant {\bf B}-type lattice.

For the sake of brevity of the expressions pertaining to the lattice sums in the M\"obius strip contributions to the gauge thresholds
in section~\ref{Ss:GeneralThresholds}, we also introduce a weighted two-torus volume depending on the two-torus shape,
\begin{equation}\label{Eq:Def-v-tilde}
\tilde{v}_i \equiv \frac{v_i}{1-b_i}.
\end{equation}

The notation in this section fully agrees with the one in~\cite{Gmeiner:2009fb} and will be extended to bulk and exceptional three-cycles
in the following section.

\subsubsection{Three-cycles and RR tadpole cancellation}\label{Sss:ThreeCycles+RRtcc}

In this section, we briefly review the construction of fractional and rigid three-cycles, discuss their 
intersection numbers and implications for the normalisation of the beta function coefficients in terms of toroidal and exceptional intersection numbers
in the computation of the gauge thresholds in section~\ref{Ss:GeneralThresholds}.
We also comment on a rewritten version of the RR tadpole cancellation conditions, which 
serves as a cross-check and completion for the relative normalisation of the different
contributions to the gauge thresholds.
In the text, we focus on the technically most complicated case of  
$T^6/(\Z_2 \times \Z_{2M} \times \OR)$  with discrete torsion\footnote{For presentations of 
intersecting D6-branes on $T^6/(\Z_N \times \Z_M \times \OR)$ orbifolds without discrete torsion
see~\cite{Forste:2000hx}, 
for the first chiral models on $T^6/\Z_2 \times \Z_2$ and $T^6/\Z_2 \times \Z_4$ without torsion see~\cite{Cvetic:2001tj,Cvetic:2001nr} and~\cite{Honecker:2003vq,Honecker:2003vw}, 
respectively, and for early discussions of $T^6/\Z_2 \times \Z_2$ with discrete torsion on factorisable tori see~\cite{Blumenhagen:2005tn} and on non-factorisable 
tori~\cite{Forste:2007zb,Forste:2008ex},
which is completed and extended to all $T^6/\Z_2 \times \Z_{2M}$ orbifolds on factorisable tori with and without discrete torsion in~\cite{Forste:2010gw}.}
and refer to tables~\ref{tab:FractionalCycles},~\ref{Tab:NormIntersections} 
and~\ref{tab:Rewritten_RRtcc-for-thresholds} for a comparison with compactifications on
 $T^6/\OR$, $T^6/(\Z_2 \times \Z_{2M} \times \OR)$ without discrete torsion
  and $T^6/(\Z_N \times \OR)$, which have been studied to a greater extent
 in the literature before, see e.g.~\cite{Blumenhagen:1999ev,Honecker:2004kb,Gmeiner:2007zz}.

There exist two different basic building blocks to three-cycles on toroidal orbifolds:
The first one consists of the omnipresent bulk three-cycles, which are the superposition of 
all orbifold images of a given factorisable torus three-cycle
\begin{equation}\label{Eq:Def-bulk-3cycle}
 \Pi^{\rm bulk}_a =  4 \, \sum_{m=0}^{M-1} \Pi_{(\omega^m a)}^{\rm torus} 
\quad\text{ with }\quad 
\Pi_a^{\rm torus} \equiv \otimes_{i=1}^3 \left( n^i_a \,  \pi_{2i-1} + m^i_a \, \pi_{2i} \right),
\end{equation}
where $\omega$ is the generator of $\Z_{2M}$ and the factor 4 arises from the $\Z_2 \times \Z_2$
subgroup of $\Z_2 \times \Z_{2M}$.

In the presence of a $\Z_2^{(i)}$ sub-symmetry which leaves the two-torus $T^2_{(i)}$ invariant  and for discrete torsion  (and $2M \neq 4$)
parameterised by $\eta=-1$,
there exist exceptional three-cycles,
\begin{equation}\label{Eq:Def-ex-3cycle}
 \Pi_{a}^{\Z_2^{(i)}} =  2  \; (-1)^{\tau_a^{\Z_2^{(i)}}} \sum_{m=0}^{M-1}  \sum_{(\alpha\beta) \in T^2_{(j)} \times T^2_{(k)}}  c_{\alpha\beta}^{(i)} \,  
 \left(e_{\omega^m(\alpha\beta)}^{(i)} \otimes \pi_{(\omega^m a)}^{(i)} \right),
\end{equation}
where $\tau_a^{\Z_2^{(i)}} \in \{0,1\}$ parameterises the $\Z_2^{(i)}$ eigenvalue, $c_{\alpha\beta}^{(i)} =\pm 1$ depends on the combination of 
displacements and Wilson lines $(\vec{\sigma}_a,\vec{\tau}_a)$ along $T^2_{(j)} \times T^2_{(k)}$
with $(i,j,k)$ cyclic permutations of (1,2,3), and $(\alpha\beta)$ runs over four $\Z_2^{(i)}$ fixed points on $T^2_{(j)} \times T^2_{(k)}$.
For more details on the fixed points, exceptional two-cycles $e_{\omega^m(\alpha\beta)}^{(i)} $ and their sign prefactors $c_{\alpha\beta}^{(i)}$ the reader
is referred to  appendix~A of~\cite{Gmeiner:2009fb}. The global factor of 2 stems from the sum over 
images under the second independent $\Z_2^{(j)}$ symmetry inside $\Z_2 \times \Z_{2M}$ in the presence of discrete torsion (i.e. $\eta=-1$).

Fractional three-cycles on $T^6/\Z_2 \times \Z_{2M}$ with discrete torsion $(\eta=-1)$ are of the form
\begin{equation}\label{Eq:Def-Rigid-Cycle}
 \Pi_a = \frac{1}{4} \left( \Pi^{\rm bulk}_a  + \sum_{i=1}^3 \Pi^{\Z_2^{(i)}}_a  \right)
 .
\end{equation}
For $2M=2$, these three-cycles are completely rigid, i.e. have no adjoint matter and are stuck at the $\Z_2 \times \Z_2$ fixed points on each two-torus. 
For $2M=6,6'$, the three-cycles are also stuck at the $\Z_2 \times \Z_2$ fixed points, but adjoint matter can arise at intersections of a given torus cycle $a$ with its orbifold images 
$(\omega^m a)_{m=1,2}$. For a small number of combinations of wrapping numbers $(n^i_a,m^i_a)$ and discrete 
displacements and Wilson lines $(\vec{\sigma}_a,\vec{\tau}_a)$, no such adjoint matter arises, see appendix~B.2.1 of~\cite{Forste:2010gw} for a complete classification.
Since this detail is not relevant for the present discussion, we refer to D6-branes on three cycles of the form~(\ref{Eq:Def-Rigid-Cycle}) as 
`rigid'. 

The O6-planes on the same orbifold background are non-dynamical objects, which are also stuck at the $\Z_2 \times \Z_2$ fixed points, but only wrap 
a fraction of a bulk three-cycle,
\begin{equation*}
 \Pi_{O6}= \frac{1}{4} \, \Pi_{O6}^{\rm bulk} = \sum_{n=0}^1 \sum_{m=0}^{2M-1}  \eta_{\OR\theta^{n}\omega^{m}}  
N_{\OR\theta^{n}\omega^m} \, \Pi_{\OR\theta^{n}\omega^{m}}^{\rm torus}, 
\end{equation*}
where $\eta_{\OR\theta^{n}\omega^{m}} = \pm 1$ denotes an ordinary or exotic O6-plane, and the assignment is subject to the consistency condition
relating the choice of discrete torsion $\eta=\pm 1$ and the assignment of exotic O6-planes
(for a more detailed discussion see~\cite{Blumenhagen:2005tn,Forste:2010gw}),
\begin{equation}\label{Eq:Relation-etas}
\eta = \eta_{\OR} \prod_{i=1}^3  \eta_{\OR\Z_2^{(i)}}
\quad \text{ and } \quad
  \eta_{\OR\theta^{n}\omega^{m}}  =\left\{\begin{array}{cr} 
\eta_{\OR} & (n,m) = \text{(even,even)} \\
\eta_{\OR\Z_2^{(1)}} & \text{(even,odd)} \\ 
\eta_{\OR\Z_2^{(2)}} & \text{(odd,odd)} \\ 
\eta_{\OR\Z_2^{(3)}} & \text{(odd,even)}  
\end{array}\right.
.
\end{equation}
Note that $(n,m)$ here denote the exponents of the orbifold generators $\theta$ and $\omega$. 
This is distinguished from the one-cycle wrapping numbers $(n^i_a,m^i_a)$ throughout the article by keeping track of the sub- and superscripts,
and it will also be clear from the context.

For later convenience we also define the sign acting on the $\Z_2^{(i)}$ twisted cycles upon orientifolding,
\begin{equation}\label{Eq:Def-eta-Z2}
\eta_{\Z_2^{(i)}} \equiv \eta_{\OR} \cdot  \eta_{\OR\Z_2^{(i)}}.
\end{equation}

The three-cycles wrapped by D6-branes and O6-planes 
are tabulated and compared to the six-torus for all factorisable $T^6/\Z_N$ and $T^6/\Z_2 \times \Z_{2M}$ orbifolds without and with
discrete torsion in table~\ref{tab:FractionalCycles}.
\mathtabfix{
\begin{array}{|c||c|c||c|c||c|}\hline
\multicolumn{6}{|c|}{\text{\bf   Comparison of the bulk, exceptional and fractional 3-cycles on } T^6, T^6/\Z_N \; \text{\bf and } T^6/\Z_2 \times \Z_{2M}}
\\\hline\hline
T^6/ & \Pi_{\text{D}6_a} & \Pi_{O6} & \Pi_{\text{D}6_a}^{\rm bulk}=  & \Pi_{\text{D}6_a}^{\Z_2^{(i)}} =  & \Pi_{O6}^{\rm bulk}=
\\\hline\hline
\text{just } T^6 & \Pi^{\rm torus}_a & \Pi^{\rm torus}_{O6} & - & - & -
\\\hline
\Z_3 & \Pi^{\rm bulk}_a & \Pi^{\rm bulk}_{O6} & \sum_{n=0}^2 \Pi_{(\theta^n a)}^{\rm torus}
& - & \sum_{n=0}^{2} N_{\OR\theta^{n}} \,  \Pi_{\OR\theta^{n}}^{\rm torus}
\\\hline
\Z_{2N} & \frac{1}{2} \left( \Pi^{\rm bulk}_a  + \Pi^{\Z_2}_a  \right) &  \frac{1}{2} \,  \Pi^{\rm bulk}_{O6}
& 2 \sum_{n=0}^{N-1} \Pi_{(\theta^n a)}^{\rm torus}
& \begin{array}{c} (-1)^{\tau_a^{(i)}} \sum_{n=0}^{N-1} \sum_{(\alpha\beta)  \in T^2_{(j)} \times T^2_{(k)}}  \\ c_{\alpha\beta}^{(2)} \; \left(e_{\theta^n(\alpha\beta)}^{(2)} \otimes \pi_{(\theta^n a)}^{(2)} \right) \end{array}
& 2 \, \sum_{n=0}^{2N-1} N_{\OR\theta^{n}} \, \Pi_{\OR\theta^{n}}^{\rm torus}
\\\hline
\!\!\!\!\begin{array}{c} \Z_2 \times \Z_{2M} \\ \eta=1\end{array}\!\!\!\! 
&  \frac{1}{2} \,  \Pi^{\rm bulk}_a &  \frac{1}{4} \,  \Pi^{\rm bulk}_{O6}
& 4 \, \sum_{m=0}^{M-1} \Pi_{(\omega^m a)}^{\rm torus} & -
& \!\!\!\!\begin{array}{c}4  \sum_{n=0}^1 \sum_{m=0}^{2M-1}\\
 N_{\OR\theta^{n}\omega^m} \,\Pi_{\OR\theta^{n}\omega^{m}}^{\rm torus}\end{array}\!\!\!\!
\\\hline
\!\!\!\!\begin{array}{c} \Z_2 \times \Z_{2M} \\ \eta=-1 \\ \text{for } 2M \neq 4 \end{array}\!\!\!\!
&  \!\!\frac{1}{4} \left(\!\! \Pi^{\rm bulk}_a  + \sum_{i=1}^3 \Pi^{\Z_2^{(i)}}_a \!\! \right)\!\! & \frac{1}{4} \,  \Pi^{\rm bulk}_{O6}
& 4 \, \sum_{m=0}^{M-1} \Pi_{(\omega^m a)}^{\rm torus}
&  \begin{array}{c} 2 \; (-1)^{\tau_a^{(i)}}  \sum_{m=0}^{M-1} \sum_{(\alpha\beta) \in T^2_{(j)} \times T^2_{(k)}}  \\ c_{\alpha\beta}^{(i)} \; \left( e_{\omega^m(\alpha\beta)}^{(i)} \otimes \pi_{(\omega^m a)}^{(i)} \right)\end{array}
& \!\!\!\!\begin{array}{c} 
4   \sum_{n=0}^1 \sum_{m=0}^{2M-1} \\ \eta_{\OR\theta^{n}\omega^{m}}  
N_{\OR\theta^{n}\omega^m} \, \Pi_{\OR\theta^{n}\omega^{m}}^{\rm torus} \end{array}\!\!\!\!
\\\hline
\end{array}
}{FractionalCycles}{The fractional multiplicities for three-cycles wrapped by D6-branes, $\Pi_{\text{D}6_a}$, and O6-planes, $\Pi_{O6}$, 
for all relevant toroidal orbifold backgrounds are compared in the first columns. 
The explicit expressions of the bulk and exceptional contributions to each fractional three-cycle are given in the last three columns.
$\eta=1$ denotes orbifolds without discrete torsion, and $\eta=-1$ corresponds to orbifolds with discrete torsion.
}

In this article, we exclude the case $T^6/\Z_2 \times \Z_4$ with discrete torsion in order to avoid a more cumbersome notation. This is due to the fact that for $\Z_2 \times \Z_4$,
the discrete torsion factor does not affect the $\Z_2$ twisted sectors, and therefore there are no exceptional three-cycles at $\Z_2$ fixed points independently of the choice of 
discrete torsion.
The shape of the fractional three-cycles in table~\ref{tab:FractionalCycles} on which D6-branes on $\Z_2 \times \Z_4$ wrap {\it independently of the the choice of discrete torsion}
is the one listed for $\eta=1$, but the existence of an exotic O6-plane in the presence of discrete torsion
leads to the shape of the orientifold invariant three-cycle wrapped by the O6-planes given for $\eta=-1$. More details on this particular orbifold background can be 
found in~\cite{Forste:2010gw}.

The relative prefactors of fractional three-cycles of D6-branes and O6-planes carry over to the normalisation of the tree-level gauge couplings, 
the beta function coefficients and threshold corrections in terms of intersection numbers and (length)${}^2$. In this section, we focus on the intersection numbers and beta function coefficients. 
More details on the gauge threshold corrections are given in section~\ref{Ss:GeneralThresholds}, 
and the complete expressions for the K\"ahler metrics and gauge couplings  at one loop are
 presented in section~\ref{S:Kaehler_metrics+potential} for each of the bulk, fractional and rigid D6-branes introduced here.

The one-cycle intersection numbers in~(\ref{Eq:Def-1cycle_Intersection}) are generalised to intersection numbers for the bulk and exceptional three-cycles 
in~(\ref{Eq:Def-bulk-3cycle}),~(\ref{Eq:Def-ex-3cycle}), which read 
\begin{equation*}
\Pi_a^{\rm torus} \circ \Pi_b^{\rm torus} \equiv - I_{ab} =-\prod_{i=1}^3 I_{ab}^{(i)} ,
\qquad\qquad
e^{(i)}_{\alpha\beta} \circ e^{(j)}_{\gamma\delta} = -2 \; \delta^{ij} \delta_{\alpha\gamma}\delta_{\beta\delta}.
\end{equation*}
For $T^6/\Z_2 \times \Z_{2M}$ with discrete torsion, the combinatorial factor of $1/4M$ for intersection numbers of bulk three-cycles and $1/2M$ for the exceptional three-cycles
together with a simplification of the double sum over orbifold images using relations, e.g. $I_{(\omega^k a)(\omega^l b)}^{(i)} = I_{(\omega^{k-l} a)b}^{(i)} $, leads to the bulk 
and exceptional three-cycle intersection
numbers in terms of a single sum over orbifold images,
\begin{equation*}
\begin{aligned}
 \Pi_{a}^{\rm bulk} \circ  \Pi_{b}^{\rm bulk} &= \frac{1}{4M} \left(4 \, \sum_{m=0}^{M-1} \Pi_{(\omega^m a)}^{\rm torus} \right) \circ  \left(4 \, \sum_{m'=0}^{M-1} \Pi_{(\omega^{m'} b)}^{\rm torus} \right)
 = - 4 \, \sum_{m=0}^{M-1} I_{(\omega^m a)b}
 ,
\\
 \Pi_a^{\Z_2^{(i)}} \circ \Pi_b^{\Z_2^{(i)}} &\equiv - 4 \, \sum_{m=0}^{M-1} I_{(\omega^m a)b}^{\Z_2^{(i)}} = - 4 \, \sum_{m=0}^{M-1} I_{(\omega^m a)b}^{(i)} \; I_{(\omega^m a)b}^{\Z_2^{(i)},(j \cdot k)} 
 \\
& \text{with } \quad
 I_{(\omega^m a)b}^{\Z_2^{(i)},(j \cdot k)}  =  (-1)^{\tau_a^{\Z_2^{(i)}} + \tau_b^{\Z_2^{(i)}}} \!\!\! \sum_{(\alpha_a\beta_a), (\gamma_b\delta_b)  \in T^2_{(j)} \times T^2_{(k)}} 
 \left( c_{\alpha_a\beta_a}^{(i)} \, c_{\gamma_b\delta_b}^{(i)} \right) \, \delta_{(\omega^m \alpha_a)\gamma_b} \delta_{(\omega^m \beta_a)\delta_b}
 .
\end{aligned}
\end{equation*}
In contrast to earlier works~\cite{Gmeiner:2007zz,Gmeiner:2008xq,Gmeiner:2009fb,Forste:2010gw}, we perform the sum here on the first subscript. 
This is due to the fact that the contributions from intersections with O6-planes are most clearly arranged for the $T^6/\Z_2 \times \Z_2$ orbifold, and we can reduce our computations and notation to this particular background by treating $a \ldots (\omega^{M-1} a)$ as separate D6-branes. In other words, the $4M$ intersections of D6-brane $a$ with all $\OR\theta^n\omega^m$ invariant O6-planes
are replaced by $M$ sets of intersections of the branes $(\omega^m a)$ with the four orbits of O6-planes $\OR$ and $\OR\Z_2^{(i)}$ $(i=1,2,3)$.

For the weighted intersection numbers~(\ref{Eq:Def_I+V-tilde}) with O6-planes, the assignments~(\ref{Eq:Relation-etas}) of some exotic O6-plane need to be taken into account.
For some given D$6_a$-brane, all intersection numbers with different D$6_b$-branes and the O6-planes are given on the r.h.s. of table~\ref{Tab:NormIntersections} for each of the
toroidal orbifolds considered in this article.
\begin{table}[ht]
\renewcommand{\arraystretch}{1.3}
\begin{minipage}[b]{0.28\linewidth}\hspace{-5mm}
\hspace{-10mm}
\begin{equation*}
{\small 
\begin{array}{|c|c|}\hline
\multicolumn{2}{|c|}{\text{\bf Chiral spectrum}}
\\\hline\hline
\text{rep.} & \text{net chirality} \; \chi
\\\hline\hline
\!\!(\N_a,\ov{\N}_b)\!\! & \Pi_a \circ \Pi_b 
\\
\!\!(\N_a,\N_b)\!\!\!& \Pi_a \circ \Pi_{b'} 
\\
\!\!({\bf Anti}_a)\!\! &\!\! \frac{\Pi_a \circ \Pi_{a'} + \Pi_a \circ \Pi_{O6}}{2}
\\
\!\!({\bf Sym}_a)\!\! & \!\!  \frac{ \Pi_a \circ \Pi_{a'} - \Pi_a \circ \Pi_{O6}}{2}
\\\hline
\end{array}
}
\end{equation*}
\end{minipage}
\hspace{-1mm}
  \begin{minipage}[b]{0.73\linewidth}
     \begin{equation*}
{\small
\begin{array}{|c|c|c|}\hline
\multicolumn{3}{|c|}{\text{\bf  3-cycle intersection numbers on various orbifolds}}
\\\hline\hline
T^6/ & - \Pi_a \circ \Pi_b & - \Pi_a \circ \Pi_{O6}
\\\hline\hline
\text{just } T^6 & I_{ab} & \tilde{I}_{a}^{\OR}
\\\hline
\Z_3 & \sum_{n=0}^2 I_{(\theta^n a)b} & \sum_{n=0}^2 \tilde{I}_{(\theta^n a)}^{\OR}
\\\hline
\Z_{2N}  & \sum_{n=0}^{N-1} \frac{I_{(\theta^n a)b} +I_{(\theta^n a)b}^{\Z_2} }{2} 
&  \sum_{n=0}^{N-1} \frac{  \tilde{I}_{(\theta^n a)}^{\OR} +  \tilde{I}_{(\theta^n a)}^{\OR\Z_2} }{2}
\\\hline
\!\!\!\!\begin{array}{c} \Z_2 \times \Z_{2M} \\ \text{ with } \eta=1 \end{array}\!\!\!\!& \sum_{m=0}^{M-1}I_{(\omega^m a)b}
& \!\! \sum_{m=0}^{M-1} \frac{ \tilde{I}_{(\omega^m a)}^{\OR} + \sum_{i=1}^3 \tilde{I}_{(\omega^m a)}^{\OR\Z_2^{(i)}} }{2}\!\!
\\\hline
\!\!\!\!\!\!\begin{array}{c} \Z_2 \times \Z_{2M}\\ \text{ with } \eta=-1  \\ \text{for } 2M \neq 4 \end{array}\!\!\!\!\!
&\!\! \sum_{m=0}^{M-1}\frac{I_{(\omega^m a)b} + \sum_{i=1}^3 I_{(\omega^m a)b}^{\Z_2^{(i)}}}{4}\!\!
&  \!\!\sum_{m=0}^{M-1} \frac{\eta_{\OR}  \tilde{I}_{(\omega^m a)}^{\OR} + \sum_{i=1}^3 \eta_{\OR\Z_2^{(i)}} \tilde{I}_{(\omega^m a)}^{\OR\Z_2^{(i)}} }{4}\!\!
\\\hline
\end{array}
}
   \end{equation*}
\end{minipage}
\caption{Left: the multiplicities $\chi$ of chiral matter states at D6-brane intersections are given in terms of three-cycle intersection numbers.
Right: explicit expression for (minus) the three-cycle intersection numbers of the bulk and fractional and rigid D6-branes in table~\protect\ref{tab:FractionalCycles}.
The beta function coefficients are computed from the total amount of (chiral plus non-chiral)  matter at intersections of 
D6-branes. The total amount $\varphi^{ab} \supset |\chi^{ab}|$ of matter is given in terms of absolute values of individual contributions to the net-chiralities $\chi^{ab}$.}
\label{Tab:NormIntersections}
\end{table}

The total amount of (chiral plus non-chiral) matter at the intersections of two different stacks of D6-branes is given by the sum over the absolute values
of the contributions to the net-chirality from the various sectors $(\omega^m a)b$,
  \begin{equation}\label{Eq:expand-varphis}
\varphi^{ab} \equiv \sum_{m=0}^{M-1} \tilde{\varphi}^{(\omega^m a)b} 
 = \sum_{m=0}^{M-1} \Big| \frac{I_{(\omega^m a)b} + \sum_{i=1}^3 I_{(\omega^m a)b}^{\Z_2^{(i)}}}{4} \Big|
\quad
\text{ for }
\;
T^6/\Z_2 \times \Z_{2M}
\;
\text{ with discrete torsion},
   \end{equation}
and analogously for D6-branes at non-vanishing angles on all other orbifold backgrounds.
At this point it is important to notice that $\tilde{\varphi}^{(\omega^m a)b}$ counts the number of matter multiplets {\it localised at intersections $(\omega^m a)b$}.
The correct assignment of the point, at which matter exists, is essential for the correct computation of the holomorphic worldsheet instanton contributions to the 
Yukawa couplings and other $n$-point functions~\cite{Honecker:2011tbd}.
Most notably, there might not exist any matter state at the corner of some triangle formed by three D6-branes at non-vanishing angles
since \mbox{$I_{(\omega^m a)b} + \sum_{i=1}^3 I_{(\omega^m a)b}^{\Z_2^{(i)}} =0$}
(this happened e.g. for the $(\theta^n a)b$ sectors of the Standard Model examples on $T^6/\Z_6'$ in~\cite{Gmeiner:2007zz,Gmeiner:2008xq}, for which $\tilde{\varphi}^{(\theta^n a)b}=0$ 
for $n=0,1,2$ despite intersections of  the toroidal three-cycles, cf. table~\ref{tab:modelz6p+hidden-sectors_ax-Kaehler} below), and therefore the worldsheet instanton sum  for D6-branes on the six-torus~\cite{Cremades:2003qj,Cremades:2004wa} cannot be employed, but needs to be refined by 
taking into account the relative $\Z_2$ eigenvalues and discrete Wilson lines and displacements of the D6-branes under consideration.

The knowledge of the bifundamental and adjoint matter spectrum at non-trivial angles and the shape of the associated beta function coefficients for $SU(N_a)$ gauge groups, 
  \begin{equation}\label{Eq:Expand-beta_function}
\begin{aligned}
b_{SU(N_a)} =& \underbrace{N_a \left(-3+\varphi^{\Adj_a}\right)}+ \underbrace{\frac{N_a}{2} \, \left( \varphi^{\Sym_a} + \varphi^{\Anti_a} \right)}
 + \underbrace{\left( \varphi^{\Sym_a} -\varphi^{\Anti_a} \right)}
 +\underbrace{\sum_{b\neq a} \frac{N_b}{2} \left( \varphi^{ab} + \varphi^{ab'}\right) } 
 \\
 \equiv & \qquad\qquad
 b_{aa}^{\cal A} \qquad\quad + \quad\qquad b_{aa'}^{\cal A} \qquad\quad + \quad\qquad b_{aa'}^{\cal M} \qquad\quad +\quad \quad  \sum_{b\neq a} \left(b_{ab}^{\cal A} + b_{ab'}^{\cal A} \right) 
,
\end{aligned}
   \end{equation}
 is used in the computation of the gauge thresholds in order to determine the absolute normalisation of the annulus amplitudes, 
cf. section~\ref{Ss:GeneralThresholds} below and the detailed discussion in~\cite{Forste:2010gw}. In~(\ref{Eq:Expand-beta_function}),
 we imply the sum over orbifold images in the first index in analogy to~(\ref{Eq:expand-varphis}), e.g. 
 $b_{ab}^{\cal A}  \equiv  \sum_{m=0}^{M-1} \tilde{b}_{(\omega^m a)b}^{\cal A}$ with $\tilde{b}_{(\omega^m a)b}^{\cal A}$ the contribution to the beta function coefficient from 
 matter localised at the intersections of the orbifold image D6-brane $(\omega^m a)$ with D6-brane $b$.
 
The absolute normalisation of the M\"obius strip amplitudes for three non-trivial angles is obtained from the amount of antisymmetric and symmetric matter as read off by comparison with 
the net-chiralities in table~\ref{Tab:NormIntersections},
  \begin{equation}\label{Eq:Count-Anti-Sym}
\begin{aligned}
\varphi^{\Anti_a} + \varphi^{\Sym_a}  &=\sum_{m=0}^{M-1} \Bigl| \frac{I_{(\omega^m a)(\omega^m a)'} + \sum_{i=1}^3 I_{(\omega^m a)(\omega^m a)'}^{\Z_2^{(i)}}}{4} \Bigr|
\quad
\text{ for }
\;
T^6/\Z_2 \times \Z_{2M}
\;
\text{ with discrete torsion},
\\
\varphi^{\Sym_a} -\varphi^{\Anti_a} &=  \sum_{m=0}^{M-1}  \,  \hat{c}_{(\omega^m a)}^{\OR} \,  \eta_{\OR} \,  \Big| \frac{  \tilde{I}_{(\omega^m a)}^{\OR}
 + \sum_{i=1}^3 \eta_{\Z_2^{(i)}} \, \tilde{I}_{(\omega^m a)}^{\OR\Z_2^{(i)}} }{4}\Big|
,
\end{aligned}
   \end{equation}
where $\hat{c}_{(\omega^m a)}^{\OR} = - \sgn\left(\frac{I_{(\omega^m a)(\omega^m a)'} + \sum_{i=1}^3 I_{(\omega^m a)(\omega^m a)'}^{\Z_2^{(i)}}}{\tilde{I}_{(\omega^m a)}^{\OR} +\eta_{\Z_2^{(i)}} \, \sum_{i=1}^3 \tilde{I}_{(\omega^m a)}^{\OR\Z_2^{(i)}}  }  \right)$ depends on the relative sign of the intersection numbers from the annulus and M\"obius strip contributions for a given D6-brane image $(\omega^m a)$, 
and the signs $\eta_{\Z_2^{(i)}}$ are defined in equation~(\ref{Eq:Def-eta-Z2}).

For some vanishing angle, the absolute values of the entries in table~\ref{Tab:NormIntersections} do not provide the total amount of matter. The normalisation of the corresponding 
annulus and M\"obius strip amplitudes is instead inferred from a rewritten version of the RR tadpole cancellation conditions displayed in table~\ref{tab:Rewritten_RRtcc-for-thresholds},
where the overall normalisation is chosen such that the contributions from three non-trivial angles match with the result derived from the beta function coefficients.
\mathtabfix{
\begin{array}{|c||c|}\hline
\multicolumn{2}{|c|}{\text{\bf  Rewritten RR tadpole cancellation: gauge threshold contributions}}
\\\hline\hline
& 0= \Pi_{a} \star \left[ \sum_b N_b \left( \Pi_{b} + \Pi_{b'} \right) -4 \, \Pi_{O6}   \right] 
\\\hline\hline
T^6 & 
\begin{array}{c}  0= 
-\sum_b  N_b \sum_{i=1}^3 \left( V_{ab}^{(i)} \, I_{ab}^{(j\cdot k)} + V_{ab'}^{(i)} \, I_{ab'}^{(j\cdot k)} \right)
\\ +  4 \, \sum_{i=1}^3 \tilde{V}_a^{\OR,(i)} \, \tilde{I}_{a}^{\OR,(j\cdot k)}
\end{array}
\\\hline
 T^6/\Z_3 & 
\begin{array}{c}  0= \sum_{n=0}^2 \, \Bigl\{ 
-\sum_b  N_b \sum_{i=1}^3 \left( V_{(\theta^n a)b}^{(i)} \, I_{(\theta^n a)b}^{(j\cdot k)} + V_{(\theta^n a)b'}^{(i)} \, I_{(\theta^n a)b'}^{(j\cdot k)} \right)
\\ +  4 \, \sum_{i=1}^3 \tilde{V}_{(\theta^n a)}^{\OR,(i)} \, \tilde{I}_{(\theta^n a)}^{\OR,(j\cdot k)} \Bigr\}
\end{array}
\\\hline
 T^6/\Z_{2N} & 
\begin{array}{c}  0=  \sum_{n=0}^{N-1} \,  \Bigl\{
-\sum_b  \frac{N_b}{2} \sum_{i=1}^3 \left( V_{(\theta^n a)b}^{(i)} \, I_{(\theta^n a)b}^{(j\cdot k)} + V_{(\theta^n a)b'}^{(i)} \, I_{(\theta^n a)b'}^{(j\cdot k)} \right)
\\
\qquad +  2 \, \sum_{l=0,2}\sum_{i=1}^3 \tilde{V}_{(\theta^n a)}^{\OR\Z_2^{(l)},(i)} \, \tilde{I}_{(\theta^n a)}^{\OR\Z_2^{(l)},(j\cdot k)}
 \Bigr\}
 \\
0=  - \sum_{n=0}^{N-1} \, \sum_b  \frac{N_b}{2} \left( V_{(\theta^n a)b}^{(2)} \, I_{(\theta^n a)b}^{\Z_2,(1\cdot 3)} + V_{(\theta^n a)b'}^{(2)} \, I_{(\theta^n a)b'}^{\Z_2(1\cdot 3)} \right)
\end{array}
\\\hline
\begin{array}{c} T^6/\Z_2 \times \Z_{2M} \\ \eta=1\end{array} 
& 
\begin{array}{r}  0=  \sum_{m=0}^{M-1} \,  \Bigl\{ 
-\sum_b  N_b \sum_{i=1}^3 \left( V_{(\omega^m a)b}^{(i)} \, I_{(\omega^m a)b}^{(j\cdot k)} + V_{(\omega^m a)b'}^{(i)} \, I_{(\omega^m a)b'}^{(j\cdot k)} \right)
\\ +  2 \, \sum_{l=0}^3 \sum_{i=1}^3 \tilde{V}_{(\omega^m a)}^{\OR\Z_2^{(l)},(i)} \, \tilde{I}_{(\omega^m a)}^{\OR\Z_2^{(l)},(j\cdot k)}
 \Bigr\}
\end{array}
\\\hline
\begin{array}{c} T^6/\Z_2 \times \Z_{2M} \\ \eta=-1 \\ \text{for } 2M \neq 4\end{array} 
&  
\begin{array}{c}  0=  \sum_{m=0}^{M-1} \,  \Bigl\{ 
-\sum_b  \frac{N_b}{4} \sum_{i=1}^3 \left( V_{(\omega^m a)b}^{(i)} \, I_{(\omega^m a)b}^{(j\cdot k)} + V_{(\omega^m a)b'}^{(i)} \, I_{(\omega^m a)b'}^{(j\cdot k)} \right)
\\ + \sum_{l=0}^3  \sum_{i=1}^3 \eta_{\OR\Z_2^{(l)}} \,  \tilde{V}_{(\omega^m a)}^{\OR\Z_2^{(l)},(i)} \, \tilde{I}_{(\omega^m a)}^{\OR\Z_2^{(l)},(j\cdot k)} \Bigr\}
\\ 0=-  \sum_{m=0}^{M-1} \,\sum_b  \frac{N_b}{4} \sum_{i=1}^3 \left( V_{(\omega^m a)b}^{(i)} \, I_{(\omega^m a)b}^{\Z_2^{(i)},(j\cdot k)} + V_{(\omega^m a)b'}^{(i)} \, I_{(\omega^m a)b'}^{\Z_2^{(i)},(j\cdot k)} \right)
\end{array}
\\\hline
\end{array}
}{Rewritten_RRtcc-for-thresholds}{Rewritten form of the RR tadpole cancellation conditions by means of a symmetric contraction $\star$ of the three-cycles $\Pi_a$. For 
$T^6/\Z_{2N}$ and $T^6/\Z_2 \times \Z_{2M}$ with discrete torsion, the untwisted and twisted tadpoles are cancelled separately. 
These tadpoles are exactly those which cancel among the different gauge threshold amplitudes, cf. section~\protect\ref{Ss:GeneralThresholds}.
}
More details on the symmetric contraction $\star$ of the three-cycles $\Pi_a$ can be found in  appendix~A.4 of~\cite{Gmeiner:2009fb}.
Once the absolute normalisation of all amplitudes is determined by combining the known beta function coefficients with tadpole cancellation, 
the remaining beta function coefficients at some vanishing angle can be cross-checked 
with the method of Chan-Paton labels, for a detailed account on $T^6/\Z_{2N}$ backgrounds see appendix~A.2 in~\cite{Gmeiner:2009fb} and on $T^6/\Z_2 \times \Z_{2M}$
with discrete torsion see appendix~B.1 in~\cite{Forste:2010gw}.

The explicit expressions for the beta function coefficients allow for very compact expressions for the gauge thresholds due to massive strings, as we will see in the 
following section.

Since the perturbative formulas for the gauge couplings of $SO(2M_x)$ and $Sp(2M_x)$ gauge groups are very similar to the $SU(N_a)$ case, we briefly comment on 
the beta function coefficients for these (pseudo)real groups, while the more intricate discussion of anomaly-free $U(1)$s is relegated to 
section~\ref{Ss:Comments_on_U1}.
$SO(2M_x)$ and $Sp(2M_x)$ gauge groups  are generated by D$6_x$-branes wrapping three-cycles
of the form~(\ref{Eq:Def-Rigid-Cycle}) on the \mbox{$T^6/\Z_2 \times \Z_{2M}$} background with discrete torsion 
subject to the necessary condition that they are homologically their own orientifold image,
\begin{equation*}
\Pi_x \stackrel{!}{=} \Pi_x'
.
\end{equation*}
The sufficient condition requires that these three-cycles are parallel to  some O6-plane orbit
or perpendicular to it along some four-torus and parallel to it along the remaining two-torus.
For the six-torus and $T^6/\Z_N$ orbifolds, the distinction of these two cases has to be made, 
whereas for $T^6/\Z_2 \times \Z_{2M}$ each O6-plane is perpendicular to another O6-plane. 
For the latter, in~\cite{Forste:2010gw} we gave a complete classification of all $\OR$-invariant three-cycles in 
dependence of the choice of exotic O6-plane~(\ref{Eq:Def-eta-Z2}) and the shape 
of untilted or tilted two-tori backgrounds ($b_i=0$ and $\frac{1}{2}$, respectively). 
Since this classification is relevant for the examples in section~\ref{Sss:AntiSym} and
in section~\ref{Ss:Ex3_Angelantonj}, we repeat the result here in table~\ref{Tab:Conditions-on_b+t+s-SOSp-Z2Z2M}.
\begin{table}[h!]
\renewcommand{\arraystretch}{1.3}
  \begin{center}
\begin{equation*}
\begin{array}{|c|c|}\hline
\multicolumn{2}{|c|}{\text{\bf Existence of $\OR$ invariant 3-cycles on } T^6/\Z_2 \times \Z_{2M} \text{ \bf with discrete torsion}}
\\\hline\hline
\pp \text{ to O6-plane}  & (\eta_{\Z_2^{(1)}},\eta_{\Z_2^{(2)}},\eta_{\Z_2^{(3)}}) \stackrel{!}{=}
\\\hline\hline
\OR  & \left( -(-1)^{2(b_2\sigma^2\tau^2 + b_3\sigma^3 \tau^3)} , - (-1)^{2(b_1\sigma^1\tau^1 + b_3\sigma^3 \tau^3)} ,  -(-1)^{2(b_1\sigma^1\tau
_1 + b_2\sigma^2 \tau^2)} \right)
\\
\OR\Z_2^{(1)}  & \left( -(-1)^{2(b_2\sigma^2\tau^2 + b_3\sigma^3 \tau^3)} ,  (-1)^{2(b_1\sigma^1\tau^1 + b_3\sigma^3 \tau^3)} ,  (-1)^{2(b_1\sigma^1\tau^1 + b_2\sigma^2 \tau^2)} \right)
\\
\OR\Z_2^{(2)}  &  \left( (-1)^{2(b_2\sigma^2\tau^2 + b_3\sigma^3 \tau^3)} , - (-1)^{2(b_1\sigma^1\tau^1 + b_3\sigma^3 \tau^3)} ,  (-1)^{2(b_1\sigma^1\tau^1 + b_2\sigma^2 \tau^2)} \right)
\\
\OR\Z_2^{(3)}  &  \left( (-1)^{2(b_2\sigma^2\tau^2 + b_3\sigma^3 \tau^3)} ,  (-1)^{2(b_1\sigma^1\tau^1 + b_3\sigma^3 \tau^3)} ,  -(-1)^{2(b_1\sigma^1\tau^1 + b_2\sigma^2 \tau^2)} \right)
\\\hline
\end{array}
\end{equation*}
\end{center}
\caption{Conditions on the existence of $\OR$ invariant rigid  three-cycles on \mbox{$T^6/(\Z_2 \times \Z_{2M} \times \OR)$} with
discrete torsion for $2M \in \{2,6,6'\}$. Their existence depends on the choice $(\eta_{\OR\Z_2^{(i)}})_{i \in \{0\ldots 3\}}$
of some exotic O6-plane and the corresponding sign factors $(\eta_{\Z_2^{(i)}})$ defined in~(\protect\ref{Eq:Def-eta-Z2}), the shape of 
the two-tori $b_i \in \{0,\frac{1}{2}\}$ as well as the discrete displacements $\sigma^i$ and Wilson lines $\tau^i$ along $T^2_{(i)}$.
}
\label{Tab:Conditions-on_b+t+s-SOSp-Z2Z2M}
\end{table}
For a classification of $\OR$-invariant three-cycles on $T^6/\Z_6'$ see~\cite{Gmeiner:2007zz}, 
some comments and examples for $T^6/\Z_6$ can be found in~\cite{Honecker:2004kb,Gmeiner:2007we}, see also appendix A.3 of~\cite{Gmeiner:2009fb}.

The inverse of (the square of) the tree level gauge coupling given below in equation~(\ref{Eq:tree-gauge-coupling}) is reduced by the factor one-half, 
since the orientifold image brane $x'$ does not give a separate contribution.
The beta function coefficients for $SO(2M_x)$ and $Sp(2M_x)$ gauge groups read
\begin{equation}\label{Eq:Def-beta-SO+Sp}
\begin{aligned}
b_{SO/Sp(2M_x)} &= \underbrace{M_x \left( -3+\varphi^{\Sym_x} + \varphi^{\Anti_x} \right) } + \underbrace{ \left(  \varphi^{\Sym_x} - \varphi^{\Anti_x} -3 \, \xi_x \right)}
+ \underbrace{\sum_{b \neq x} \frac{N_b}{2}  \varphi^{xb}}
\\
&\equiv \qquad\qquad\qquad   b_{xx}^{\cal A}  \qquad\qquad +  \qquad\qquad  b_{xx}^{\cal M}  \qquad\qquad  +  \quad\quad  \sum_b b_{xb}^{\cal A}
\\
& \qquad\qquad\qquad\qquad\qquad\qquad\qquad  \text{ with } \qquad 
  \xi_x= \left\{\begin{array}{ccc} -1 & \text{ for } & SO(2M_x)  \\ 1 & \text{ for } & Sp(2M_x)
 \end{array}\right.
,
\end{aligned}
\end{equation}
and by comparison of the sum over bifundamental representations with the analogue in $b_{SU(N_a)}$ in equation~(\ref{Eq:Expand-beta_function}),
one sees that one can automatise the computation by multiplying the beta function coefficient of a hypothetical $SU(M_x)$ gauge factor wrapping the 
same three-cycle by one-half, i.e. loosely speaking ``$b_{SO/Sp(2M_x)} = \frac{1}{2} \, b_{SU(M_x)}$''. 
The same relative factor appears in the expansion of the gauge thresholds~\cite{Gmeiner:2009fb} summarised in the following section.

\subsection{$SU(N)$, $SO(2N)$ and $Sp(2N)$ gauge thresholds on fractional and rigid D6-branes}\label{Ss:GeneralThresholds}

In this section, we briefly review the magnetic background field method in order to compute the gauge thresholds,
and we comment on technical simplifications, which lead to compact expressions for each of the bulk, fractional and rigid D6-branes
on the different toroidal orbifold backgrounds under consideration. For concreteness, the discussion in this section 
focusses on $T^6/(\Z_2 \times \Z_{2M} \times \OR)$ with discrete torsion, but all necessary ingredients to compute the other 
backgrounds are contained and the final results stated for every single orbifold background and D6-brane configuration.

The gauge coupling of an $SU(N_a)$ (or $SO(2N_a)$ or $Sp(2N_a)$) gauge factor at energy scale $\mu$ is up to one-loop 
in string perturbation theory given by
\begin{equation}\label{Eq:Def-gauge-SUNa}
\frac{8 \pi^2}{g_a^2(\mu)} = \frac{8 \pi^2}{g_{a,{\rm string}}^2}
+\frac{b_a}{2} \ln\left(\frac{M_{\rm string}^2}{\mu^2}  \right) + \frac{\Delta_a}{2},
\end{equation}
where the tree-level value of (the square of) the gauge coupling is inversely proportional to the length of the three-cycle wrapped by 
the stack of D$6_a$-branes defined in equation~(\ref{Eq:Def-Vab}),
\begin{equation}\label{Eq:tree-gauge-coupling}
\frac{4 \pi}{g_{a,{\rm string}}^2} = \frac{1}{2\sqrt{2} k_a c_a} \frac{M_{\rm Planck}}{M_{\rm string}} \prod_{i=1}^3 \sqrt{V_{aa}^{(i)}}
\quad
\text{with}
\quad
c_a =
\left\{\!\!\begin{array}{cc}
1 & \text{bulk}\\
2 & \text{fract.}\\
4 & \text{rigid}
\end{array}\right.
\quad
\text{and}
\quad
k_a=
\left\{\!\!\begin{array}{cc}
1 &  SU(N_a) \\
2 & SO/Sp(2N_a)
\end{array}\right.
.
\end{equation}
The remaining two contributions, the beta function coefficient $b_a$ due to massless strings running in a loop and the gauge threshold $\Delta_a$ 
due to massive strings in the loop, are simultaneously obtained in a CFT computation, 
\begin{equation}\label{Eq:b-Delta}
b_a \, \ln\left(\frac{M_{\rm string}^2}{\mu^2}  \right) + \Delta_a = \sum_b \Bigl[ \mathcal{T}^A(D6_a,D6_b) + \mathcal{T}^A(D6_a,D6_{b'})   \Bigr] + \mathcal{T}^M (D6_a,O6),
\end{equation}
where $\mathcal{T}^A$ and $\mathcal{T}^M$ denote the gauge threshold amplitudes with annulus and M\"obius strip topology, respectively, and 
the sum runs over all D6-branes $b=a$ and $b\neq a$ and their orientifold images $b'$ in a given model.

For $T^6/\Z_2 \times \Z_{2M}$ with discrete torsion, the gauge threshold amplitudes consist of two types of sums, on the one hand 
the untwisted and $\Z_2^{(i)}$ twisted sector contributions in the annulus and cross-cap states of the $\OR$ and $\OR\Z_2^{(i)}$  invariant
O6-plane orbits,
and on the other hand a sum over orbifold images $(\omega^m a)$ under the $\Z_{2M}$ generator (for $2M \neq 2$)  for the D6-brane under consideration,
\begin{equation}
\begin{aligned}
\mathcal{T}^A(D6_a,D6_b) =&  \sum_{m=0}^{M-1}  \left(  T^{\unity}_{(\omega^m a)b} + \sum_{i=1}^3 T^{\Z_2^{(i)}}_{(\omega^m a)b} \right)
,\quad
\mathcal{T}^M(D6_a,O6) =&  \sum_{m=0}^{M-1} \left(  T^{\OR}_{(\omega^m a)} +  \sum_{i=1}^3  T^{\OR\Z_2^{(i)}}_{(\omega^m a)}  \right)
.
\end{aligned}
\end{equation}
These amplitudes are obtained from the tree channel vacuum annulus and M\"obius strip diagrams
 \begin{equation}\label{Eq:Vacuum-Amplitudes}
 \begin{aligned}
\mathcal{A} (D6_a,D6_b) \sim & \sum_{\rm sectors}\int_{0}^{\infty} dl \sum_{(\alpha,\beta)} (-1)^{2(\alpha+\beta)}
\frac{\vartheta\targ{\alpha}{\beta}(0,2il)}{\eta^3(2il)}
 A^{\rm sector}_{\rm compact} (\alpha,\beta;\{\phi^{(i)}\};2il)
 ,
\\
\mathcal{M} (D6_a,O6) \sim& \sum_{\rm sectors} \int_{0}^{\infty} dl \sum_{(\alpha,\beta)} (-1)^{2(\alpha+\beta)}
\frac{\vartheta\targ{\alpha}{\beta}
(0,2il-\frac{1}{2})}{\eta^3(2il-\frac{1}{2})} M^{\rm sector}_{\rm compact} (\alpha,\beta;\{\phi^{(i)}\};2il-\half)
,
\end{aligned}
\end{equation}
by gauging the non-compact oscillator contributions by a magnetic background field 
and expanding in a power series of the newly introduced magnetic field.  $\alpha,\beta \in \{ 0,1/2\} $ denote the different spin structures, and
the sum over sectors for the annulus amplitude contains the untwisted ($\unity$) and twisted ($\Z_2^{(i)}$ with $i \in \{1,2,3\}$) 
sectors as well as the sum over all orbifold images of the first D6-brane $(\omega^m a)_{m=0 \ldots M-1}$. For the M\"obius strip, instead of 
the twist sectors the sum is over the $\OR$ and $\OR\Z_2^{(i)}$ (with $i \in \{1,2,3\}$)
invariant O6-planes, and again a sum over orbifold images $(\omega^m a)_{m=0 \ldots M-1}$
is performed. In~\cite{Gmeiner:2009fb}, we had instead written the complete M\"obius strip contribution as a sum over all $\OR\theta^n\omega^m$ 
invariant O6-planes with $n \in \{0,1\}$ and $m \in \{0 \ldots 2M-1\}$, and we had allocated four different invariances $\OR\theta^p\omega^q$
with $p\in \{0,1\}$ and $q \in \{-k,-k+M\}$ to a string stretched between  D6-branes $a$ and the orientifold image $(\omega^k a')$.
These two ways of rewriting the sums give identical results, but the new convention in this article allows us to reduce the discussion to 
the $T^6/\Z_2 \times \Z_2$ background without and with discrete torsion, when all orbifold images $(\omega^m a)_{m=0 \ldots M-1}$ are treated as independent D6-branes.
By this trick, the gauge threshold contributions to the M\"obius strip for three non-vanishing angles can be explicitly rewritten in terms of 
annulus expressions for the ratios of Gamma functions plus constants and terms linear in the angles, cf. details in appendix~\ref{App:A}, and the gauge threshold contributions from antisymmetric and 
symmetric matter take the very simple forms in the last lines of table~\ref{tab:Six-torus-AntiSym-beta+thresholds},~\ref{tab:T6-Z2N-Anti+Sym-beta+thresholds},~\ref{tab:Z2Z2M-no_torsion-Anti+Sym-beta+thresholds} and~\ref{tab:Z2Z2M-torsion-AntiSym-beta+thresholds}
for bulk, fractional and rigid D6-branes on $T^6$, $T^6/\Z_{2N}$ and $T^6/\Z_2 \times \Z_{2M}$ without and with discrete torsion, respectively.

The passage from the vacuum to the gauge threshold amplitudes boils down to replacing the Jacobi theta functions of the 
non-compact fermionic contributions in~(\ref{Eq:Vacuum-Amplitudes}) by the second derivative w.r.t. the first argument (for details on the procedure see 
e.g.~\cite{Lust:2003ky,Akerblom:2007np,Gmeiner:2009fb} and references therein) of the same Jacobi theta functions,
\begin{equation}
\begin{aligned}
\mathcal{A} (D6_a,D6_b) &\longrightarrow \mathcal{T}^{A}(D6_a,D6_b),
\\
\frac{\vartheta\targ{\alpha}{\beta}(0,2il)}{\eta^3(2il)} &\longrightarrow 
\frac{\vartheta^{\prime\prime}\targ{\alpha}{\beta}(0,2il)}{\eta^3(2il)} ,
\end{aligned}
\end{equation}
while retaining the compact contributions $A^{\rm sector}_{\rm compact} (\alpha,\beta;\{\phi^{(i)}\};2il)$, and analogously for the 
amplitudes with M\"obius strip topology by replacing the argument $2il \rightarrow 2il - \frac{1}{2}$. 

For toroidal orbifold backgrounds,  the compact contributions $A^{\rm sector}_{\rm compact} (\alpha,\beta;\{\phi^{(i)}\};2il)$ and
$M^{\rm sector}_{\rm compact} (\alpha,\beta;\{\phi^{(i)}\};2il-\half)$ are known, see~\cite{Gmeiner:2009fb} for a complete list on $T^6/\Z_{2N}$ 
and~\cite{Lust:2003ky,Akerblom:2007np} for results on the six-torus without displacement and Wilson line moduli and~\cite{Blumenhagen:2007ip} for 
partial results on rigid intersecting D6-branes on $T^6/\Z_2 \times \Z_2$ with discrete torsion. 
They fall into three classes of supersymmetric
angles, $\sum_{i=1}^3\phi^{(i)}=0$, with three non-vanishing, one vanishing or three vanishing angles, and into three categories 
of sectors, the untwisted and $\Z_2^{(i)}$ twisted annulus and the M\"obius strips. After transformations of the integrands 
by means of resummations and Jacobi theta function identities, the gauge threshold amplitudes take the following form 
for $T^6/\Z_2 \times \Z_{2M}$ with discrete torsion (in the following $(i,j,k)$ are cyclic permutations of (1,2,3) whenever they appear
simultaneously in one term):\footnote{The annulus and M\"obius strip amplitudes for the six-torus and 
$T^6/\Z_{N}$ and $T^6/\Z_2 \times \Z_{2M}$ orbifolds without discrete torsions differ in the absolute normalisation, which can be read off
by using tables~\ref{tab:Rewritten_RRtcc-for-thresholds} and~\ref{Tab:NormIntersections}, which contain the contribution to the tadpole and $SU(N_a)$ 
{\it chiral} matter beta  function coefficient, respectively.
For the six-torus, $T^6/\Z_3$ and  $T^6/\Z_2 \times \Z_{2M}$ without discrete torsion, there are no twisted annulus contributions.
The gauge thresholds amplitudes for fractional D6-branes on $T^6/\Z_{2N}$ are explicitly tabulated in the notation of this article in the appendix of~\cite{Gmeiner:2009fb}.  
}
\begin{enumerate}
 \item
annulus topology, untwisted sector:
\begin{equation*}
\begin{aligned}
T^{\unity}_{ab}(\phi^{(1)}_{ab},\phi^{(2)}_{ab},\phi^{(3)}_{ab}) &= - \frac{N_b}{4} \;I_{ab}
\int_{0}^{\infty} dl \, l^{\varepsilon}  \; \sum_{i=1}^3 \frac{1}{\pi} \frac{\vartheta_1^{\, \prime}}{\vartheta_1}(\phi^{(i)}_{ab},2il)
,
\\
T^{\unity}_{ab}(0^{(i)},\phi^{(j)}_{ab},\phi^{(k)}_{ab})&= - \frac{N_b}{4} \;V_{ab}^{(i)} \; I_{ab}^{(j \cdot k)}
\int_{0}^{\infty} dl \, l^{\varepsilon}  \; {\mathcal L}^{(i)}_{ab}(v_i,V_{ab}^{(i)};l)
,
\\
T^{\unity}_{ab}(0^{(i)},0^{(j)},0^{(k)}) &=0
.
\end{aligned}
\end{equation*}
The first amplitude is ${\cal N}=1$ supersymmetric and depends on the complex structure moduli through the angles 
(only for the $\Z_2$ invariant lattice in figure~\ref{Fig:Z2-lattice}, cf. equation~(\ref{Eq:tan-angles})),
the second amplitude is ${\cal N}=2$ supersymmetric and depends on the K\"ahler modulus $v_i$ of the two-torus with vanishing relative angle, 
and the third one preserves ${\cal N}=4$ supersymmetry and hence vanishes.
\\
The dimensionally regularised integrals over the Jacobi theta functions $\vartheta_{\alpha}'/\vartheta_{\alpha}(\nu,2i l)$ and lattice contribution $\tilde{\mathcal L}^{(i)}(v_i,V^{(i)};l)$ 
with parameter $\varepsilon \!\!\to\!\!0$ are given explicitly below. 
 \item
annulus topology, $\Z_2^{(i)}$ twisted sector:
\begin{equation*}
\begin{aligned}
T^{\Z_2^{(i)}}_{ab}(\phi^{(1)}_{ab},\phi^{(2)}_{ab},\phi^{(3)}_{ab}) &= - \frac{N_b}{4} \;I_{ab}^{\Z_2^{(i)}}
\int_{0}^{\infty} dl \, l^{\varepsilon} \; \left(  \frac{1}{\pi} \frac{\vartheta_1^{\, \prime}}{\vartheta_1}(\phi^{(i)}_{ab},2il) + 
 \frac{1}{\pi} \frac{\vartheta_4^{\, \prime}}{\vartheta_4}(\phi^{(j)}_{ab},2il) + \frac{1}{\pi} \frac{\vartheta_4^{\, \prime}}{\vartheta_4}(\phi^{(k)}_{ab},2il)
\right)
,
\\
T^{\Z_2^{(i)}}_{ab}(0^{(i)},\phi^{(j)}_{ab},\phi^{(k)}_{ab})&=- \frac{N_b}{4} \;V_{ab}^{(i)} \; I_{ab}^{\Z_2^{(i)},(j \cdot k)}
\int_{0}^{\infty} dl \, l^{\varepsilon}  \; {\mathcal L}^{(i)}_{ab}(v_i,V_{ab}^{(i)};l)
,
\\
T^{\Z_2^{(i)}}_{ab}(\phi^{(i)}_{ab},0^{(j)},\phi^{(k)}_{ab})&=- \frac{N_b}{4} \;I_{ab}^{\Z_2^{(i)}}
\int_{0}^{\infty} dl \, l^{\varepsilon} \; \left(  \frac{1}{\pi} \frac{\vartheta_1'}{\vartheta_1}(\phi^{(i)}_{ab},2il) 
+ \frac{1}{\pi} \frac{\vartheta_4'}{\vartheta_4}(\phi^{(k)}_{ab},2il)
\right)
,
\\
T^{\Z_2^{(i)}}_{ab}(0^{(i)},0^{(j)},0^{(k)}) &=- \frac{N_b}{4} \;V_{ab}^{(i)} \; I_{ab}^{\Z_2^{(i)},(j \cdot k)}
\int_{0}^{\infty} dl \, l^{\varepsilon} \; {\mathcal L}^{(i)}_{ab}(v_i,V_{ab}^{(i)};l)
.
\end{aligned}
\end{equation*}
The first and third amplitude preserve ${\cal N}=1$ supersymmetry and depend on the complex structure moduli via the angles
(cf. comments above on the lattices).
The second and fourth amplitude preserve ${\cal N}=2$ supersymmetry and depend on the K\"ahler modulus $v_i$ of the two-torus where 
$\Z_2^{(i)}$ acts trivially.
\item
M\"obius strip topology (the $\OR \equiv \OR\Z_2^{(0)}$ invariant O6-plane is included in the notation by setting $l=0$): 
\begin{equation*}
\begin{aligned}\hspace{-6mm}
T^{\OR\Z_2^{(l)}}_{a}(\phi^{(1)}_{a,\OR\Z_2^{(l)}},\phi^{(2)}_{a,\OR\Z_2^{(l)}},\phi^{(3)}_{a,\OR\Z_2^{(l)}}) &=\tilde{I}_a^{\OR\Z_2^{(l)}}
\int_{0}^{\infty} dl \, l^{\varepsilon}  \; \sum_{i=1}^3 \frac{1}{\pi} \frac{\vartheta_1^{\, \prime}}{\vartheta_1}(\phi^{(i)}_{a,\OR\Z_2^{(l)}},2il-\frac{1}{2})
,
\\
T^{\OR\Z_2^{(l)}}_{a}(0^{(i)},\phi^{(j)}_{a,\OR\Z_2^{(l)}},\phi^{(k)}_{a,\OR\Z_2^{(l)}})&= \tilde{V}_a^{\OR\Z_2^{(l)},(i)} \; \tilde{I}_a^{\OR\Z_2^{(l)},(j \cdot k)} 
\int_{0}^{\infty} dl \, l^{\varepsilon}  \; {\mathcal L}^{(i)}_{a,\OR\Z_2^{(l)}}(\tilde{v}_i,2\tilde{V}_{a}^{\OR\Z_2^{(l)},(i)} ;l)
,
\\
T^{\OR\Z_2^{(l)}}_{a}(0^{(i)},0^{(j)},0^{(k)}) &=0
.
\end{aligned}
\end{equation*}
The first, second and third amplitude preserve ${\cal N}=1,2$ and 4 supersymmetry, respectively, and depend on the complex structure moduli via angles, the
weighted K\"ahler modulus $\tilde{v}_i$ or vanish. 
\end{enumerate}
The {\bf annulus} amplitudes for non-vanishing angles can be further evaluated using the relations (for $0<|\nu|<1$, see 
e.g.~\cite{Lust:2003ky,Akerblom:2007np,Blumenhagen:2007ip,Gmeiner:2009fb})
\begin{equation}\label{Eq:Theta-Expansions-Annulus}
\begin{aligned}
\frac{1}{\pi} \int_0^{\infty} dl \, l^{\varepsilon}  \frac{\vartheta_{\alpha}^{\, \prime}}{\vartheta_{\alpha}}(\nu,2il )
=&\delta_{\alpha \, 1}  \, \cot (\pi \nu) \, \int\limits_0^{\infty} dl
+ \left( \frac{1}{\varepsilon} +\gamma - \ln 2  \right) \, \left( \frac{{\rm sgn}(\nu)}{2} -\nu \right)
\\
&-\frac{1}{2} \ln \left(\frac{\Gamma(|\nu|)}{\Gamma(1-|\nu|)}\right)^{{\rm sgn}(\nu)}
+ \delta_{\alpha \, 4} \,  \left({\rm sgn} (\nu) - 2 \, \nu  \right)\, \ln (2)
\quad +{\mathcal O}(\varepsilon)
,
\end{aligned}
\end{equation}
where the first term on the r.h.s. provides a contribution to the tadpoles in table~\ref{tab:Rewritten_RRtcc-for-thresholds}
(remember that \mbox{$I_{ab} \sum_{i=1}^3 \cot(\pi \phi^{(i)}_{ab}) = \sum_{i=1}^3 V_{ab}^{(i)} I_{ab}^{(j \cdot k)}$}), 
the second one furnishes the contribution to the beta function coefficients (using $I_{ab} \sum_{i=1}^3 \left( \frac{{\rm sgn}(\phi_{ab}^{(i)})}{2} -\phi_{ab}^{(i)} \right)=-\frac{|I_{ab}|}{2}$, cf. the absolute values of the terms in table~\ref{Tab:NormIntersections}) 
when identifying
\begin{equation*}
\ln \left( \frac{M_{\rm string}}{\mu}  \right)^2  \equiv \frac{1}{\varepsilon} +\gamma - \ln 2,
\end{equation*}
and the finite terms in the second line of~(\ref{Eq:Theta-Expansions-Annulus}) constitute the contributions to the gauge thresholds due to massive strings.

For ${\cal N}=1$ supersymmetric sectors at three non-trivial angles, the gauge thresholds depend on the complex structure moduli of the two-tori via 
the relation~(\ref{Eq:tan-angles}) between the tangent of the angles and ratios of radii 
(more precisely, only a $\Z_2$ twist retains the complex structure modulus - a $\Z_3$ or $\Z_4$ symmetry extinguishes the modulus, cf. the sole dependence 
of the angles on torus wrapping numbers in~(\ref{Eq:tan-angles})). It should be noted here that the identification of scales might contain a proportionality constant,
which must be fixed in section~\ref{S:Kaehler_metrics+potential} when matching the string and field theoretical one-loop formulas for the gauge couplings.

For one vanishing angle, the integration over the lattice sum for arbitrary displacement ($0 \leqslant \sigma_{ab}^{(i)} \leqslant 1$) and Wilson line 
($0 \leqslant \tau_{ab}^{(i)} \leqslant 1$) moduli~\cite{Gmeiner:2009fb},  
\begin{equation}\label{Eq:KK-contributions-Annulus}
\begin{aligned}
V_{ab}^{(i)} \int_{0}^{\infty} dl \, l^{\varepsilon} \; {\mathcal L}^{(i)}_{ab}(v_i,V^{(i)}_{ab};l) =& V_{ab}^{(i)} \int_{0}^{\infty} dl
 +  \left( \frac{1}{\varepsilon} +\gamma - \ln 2  \right) \, \delta_{\sigma_{ab}^{(i)},0} \delta_{\tau_{ab}^{(i)},0} \\
& -  \delta_{\sigma_{ab}^{(i)},0} \delta_{\tau_{ab}^{(i)},0} \; \Lambda_{0,0}(v_i; V^{(i)}_{ab})
-\left( 1- \delta_{\sigma_{ab}^{(i)},0} \delta_{\tau_{ab}^{(i)},0} \right)\; \Lambda (\sigma^{(i)}_{ab},\tau^{(i)}_{ab},v_i) \\
&+{\mathcal O}(\varepsilon)
,
\end{aligned}
\end{equation}
is in the same way split into contributions to the tadpoles and beta function coefficients on the first line  
and gauge threshold contribution on the second line. This sector with one vanishing angle preserves ${\cal N}=2$ supersymmetry and depends on the 
K\"ahler modulus $v_i$ of the two-torus with vanishing angle between the D6-branes.
The functions $\Lambda(v_i)$ of this K\"ahler modulus are slightly differently defined from~\cite{Gmeiner:2009fb}:
\begin{equation}\label{Eq:Def-Lambdas}
\begin{aligned}
\Lambda_{0,0}(v;V) & \equiv  \ln \left( 2 \pi v V \,  \eta^4 (i v) \right)
,
\\
\Lambda(\sigma,\tau,v) &\equiv \ln\left|e^{-\pi \sigma^2 v/4}\frac{\vartheta_1 (\frac{\tau - i \sigma v}{2},i v)}{\eta (i v)}\right|^2
,
\end{aligned}
\end{equation}
in order to make the matching of the beta function coefficients as prefactors for vanishing relative displacements $\vec{\sigma}_{ab}$ or Wilson lines 
$\vec{\tau}_{ab}$ more obvious. The consistency of the two lattice contributions for $(\sigma,\tau) \to (0,0)$ can be checked using the product expansion
of the Jacobi-Theta-functions,
\begin{equation*}
\begin{aligned}
\frac{\vartheta_1(\nu,\tau)}{\eta(\tau)} &= 2 \, \sin(\pi \nu) \,  q^{\frac{1}{12}}
\prod_{n=1}^{\infty} \left(1-  2 \cos(2 \pi \nu)\,  q^{n} + q^{2n} \right)
\\
& \stackrel{\nu \to 0}{\longrightarrow}  \, 2 \pi \, \nu \;  \eta^2(\tau)+ {\cal O}(\nu^2)
,
\end{aligned}
\end{equation*}
together with identifying the divergent factor as the discrete change in the beta function coefficient,
\begin{equation*}
\ln \left| e^{-\pi \sigma^2 v/4} \,2\pi \frac{\tau - i \sigma v}{2} \right|^2 \stackrel{(\sigma,\tau) \to (0,0)}{\approx} - \ln \left(\frac{M_{\rm string}}{\mu}\right)^2 + \ln \left( 2 \pi v V \right)
.
\end{equation*}
In section~\ref{S:Kaehler_metrics+potential}, the lattice contributions will be related to the one-loop corrections to the holomorphic gauge kinetic functions
and to the K\"ahler metrics by using the decomposition
\begin{equation}\label{Eq:Lambdas-rewritten}
\begin{aligned}
\Lambda_{0,0}(v;V) & = \left[ 2   \ln  \eta (i v)  + c.c. \right] +  \ln \left( 2 \pi v V  \right)
,
\\
\Lambda(\sigma,\tau,v) &\equiv  \left[ \ln \left(e^{-\pi \sigma^2 v/4}\frac{\vartheta_1 (\frac{\tau - i \sigma v}{2},i v)}{\eta (i v)} \right)
+ c.c. \right]
.
\end{aligned}
\end{equation}

For rigid D6-branes on $T^6/\Z_2 \times \Z_{2M}$ orbifolds with discrete torsion, the relative displacements and Wilson lines 
$(\sigma_{ab}^{(i)},\tau_{ab}^{(i)})$ take only discrete values in $\{0,1\}$, but for bulk D6-branes on 
the six-torus or $T^6/\Z_2 \times \Z_{2M}$ without discrete torsion they are continuous open string moduli with values in [0,1] on each two-torus, 
and for fractional D6-branes on $T^6/\Z_{2N}$ there is one set of such open string moduli associated to the $\Z_2$-invariant two-torus.

Using the integrals~(\ref{Eq:Theta-Expansions-Annulus}) and~(\ref{Eq:KK-contributions-Annulus}) of Jacobi-theta functions and Kaluza-Klein and winding sums, 
all gauge threshold amplitudes with annulus topology can be evaluated explicitly.
The complete list of $SU(N_a)$ beta function coefficients from bifundamental and adjoint matter and the gauge threshold contributions 
from the same representations for all possible configurations of relative angles are given in table~\ref{tab:Z2Z2M-torsion-Bifundamentals-beta+thresholds} 
for the $T^6/\Z_2 \times \Z_{2M}$ orbifolds with discrete torsion. 
\mathtabfix{
\begin{array}{|c||c|c|}\hline
\multicolumn{3}{|c|}{b_{SU(N_a)} \text{ \bf  and gauge thresholds for bifundamental and adjoints: } T^6/\Z_2 \times \Z_{2M} \text{ \bf  with discrete torsion}}
\\\hline\hline
(\phi_{ab}^{(1)},\phi_{ab}^{(2)},\phi_{ab}^{(3)}) & 
\begin{array}{r} 
b_{SU(N_a)}^{\text{torsion}}= \sum_b b_{ab}^{\cal A} + \ldots \\ = \sum_b \frac{N_b}{2} \varphi^{ab} + \ldots 
\end{array} 
& \Delta_{SU(N_a)}^{\text{torsion}} = \sum_b N_b \, \tilde{\Delta}_{ab}^{\text{torsion}} + \ldots
\\\hline\hline
(0,0,0)  
& - \,  N_b \, \left(\prod_{n=1}^3 \delta_{\sigma^n_{ab},0} \delta_{\tau^n_{ab},0}\right) \sum_{i=1}^3  (-1)^{\tau^{\Z_2^{(i)}}_{ab}}  
&
\begin{array}{c}
- \sum_{i=1}^3 \left(- \frac{ I^{\Z_2^{(i)},(j \cdot k)}_{ab} N_b}{4} \,\delta_{\sigma^i_{ab},0} \delta_{\tau^i_{ab},0} \right) \, 
\Lambda_{0,0}(v_i;V_{ab}^{(i)})
\\
+ \sum_{i=1}^3 \frac{ I^{\Z_2^{(i)},(j \cdot k)}_{ab}  N_b}{4} \,\left(1-\delta_{\sigma^i_{ab},0} \delta_{\tau^i_{ab},0} \right) \,
\Lambda(\sigma_{ab}^i,\tau_{ab}^i,v_i)
\end{array}
\\\hline
(0,\phi,-\phi) 
&  \frac{N_b}{4} \,\delta_{\sigma^1_{ab},0} \, \delta_{\tau^1_{ab},0} \, \left( |I_{ab}^{(2 \cdot 3)} | - I_{ab}^{\Z_2^{(1)},(2 \cdot 3)} \right)
&
\begin{array}{c}
-  b_{ab}^{\cal A} \; \Lambda_{0,0}(v_1;V_{ab}^{(1)})
\\
+ \frac{\{ I^{(2 \cdot 3)}_{ab} +  I^{\Z_2^{(1)},(2 \cdot 3)}_{ab} \} N_b}{4} \,\left(1-\delta_{\sigma^1_{ab},0} \delta_{\tau^1_{ab},0} \right) \,
\Lambda(\sigma_{ab}^1,\tau_{ab}^1,v_1)
\\
+ \frac{N_b}{4} \, \ln(2) \left[ \left( I_{ab}^{\Z_2^{(2)}} -  I_{ab}^{\Z_2^{(3)}} \right) \left( \sgn(\phi) - 2 \, \phi \right)\right]
\end{array}
\\\hline
\begin{array}{c}
(\phi^{(1)},\phi^{(2)},\phi^{(3)})
\\
{\sum_{n=1}^3 \phi^{(n)}=0} 
\end{array}
& \frac{N_b}{8} \, \left( | I_{ab} | + \sgn(I_{ab}) \, \sum_{i=1}^3 I_{ab}^{\Z_2^{(i)}}  \right)
&
\begin{array}{c}
b_{ab}^{\cal A} \; \sgn(I_{ab}) \;
\sum_{i=1}^3 \ln \left(\frac{\Gamma(|\phi^{(i)}_{ab}|)}{\Gamma(1-|\phi^{(i)}_{ab}|)}\right)^{\sgn(\phi^{(i)}_{ab})}
\\
+ \frac{N_b}{4} \, \ln(2) \left[\sum_{i=1}^3 I_{ab}^{\Z_2^{(i)}} \left( \sgn(\phi^{(i)}_{ab}) - 2 \, \phi^{(i)}_{ab} + \sgn(I_{ab})\right)\right]
\end{array}
\\\hline
\end{array}
}{Z2Z2M-torsion-Bifundamentals-beta+thresholds}{Contributions to the $SU(N_a)$ beta function coefficients (middle column) and gauge thresholds (last column) 
from bifundamental and adjoint matter for all possible supersymmetric configurations, i.e. parallel D6-branes and at one vanishing or three non-vanishing angles, on the $T^6/\Z_2 \times \Z_{2M}$ orbifolds with discrete torsion and $2M \in \{2,6,6'\}$. The complete beta function coefficient and gauge threshold are obtained by summing over all $(\omega^m a)b$ sectors, cf. equations~(\protect\ref{Eq:Expand-beta_function}) and~(\protect\ref{Eq:Threh-SU(N)}). 
On the orbifolds with discrete torsion, there is no adjoint matter from the $aa$ sector, and for $a \neq b$ at vanishing angles there only exists one non-chiral pair of bifundamental representations for vanishing relative displacements and Wilson lines. 
For more details on the existence of adjoint matter see section~\protect\ref{Sss:adjoints} and table~\protect\ref{tab:Adjoint-Beta+Thresholds}. For later convenience of the notation, the beta function coefficient on parallel D6-branes can be decomposed
into its contributions from various $\Z_2^{(i)}$ sectors, $b^{{\cal A},(\vec{0})}_{ab} \equiv \sum_{i=1}^3b^{{\cal A},(i)}_{ab}$,
cf. e.g. table~\protect\ref{tab:Comparison-gaugekin-bifund}.
}

The complete  gauge thresholds $\Delta_{SU(N_a)}$ are obtained by summing over all sectors in the same way as the beta function coefficients~(\ref{Eq:Expand-beta_function}), i.e.
\begin{equation}\label{Eq:Threh-SU(N)}
\begin{aligned}
\Delta_{SU(N_a)} =&  \sum_b N_b 
\left( \tilde{\Delta}_{ab} + \tilde{\Delta}_{ab'}
\right) 
 +  \Delta_{a,\OR}
\qquad
  {\rm with}  \quad
 \tilde{\Delta}_{ab} =  \tilde{\Delta}_{ba}
\\
\text{and} \quad & 
\tilde{\Delta}_{ab} =  \sum _{m=0}^{M-1} \left( \tilde{\Delta}_{(\omega^m a)b}^{\unity} + \sum_{i=1}^3 \tilde{\Delta}_{(\omega^m a)b}^{\Z_2^{(i)}} \right)
,
\qquad
\Delta_{a,\OR} = \sum _{m=0}^{M-1}\left(\Delta_{(\omega^m a)}^{\OR} +  \sum_{i=1}^3 \Delta_{(\omega^m a)}^{\OR\Z_2^{(i)}}\right) 
.
\end{aligned}
\end{equation}
Analogously to the beta function coefficients for $SO(2M_x)$ or $Sp(2M_x)$ gauge groups in equation~(\ref{Eq:Def-beta-SO+Sp}), 
the gauge threshold contributions from orthogonal and symplectic gauge groups are given by
\begin{equation*}
\Delta_{SO/Sp(2M_x)} =  \sum_b N_b \; \tilde{\Delta}_{xb}  + \frac{1}{2} \, \Delta_{x,\OR},
\end{equation*}
which can be viewed as one-half of the formula for a hypothetical $SU(M_x)$ gauge factor wrapped on the same
three-cycle, i.e.  ``$\Delta_{SO/Sp(2M_x)} = \frac{1}{2} \Delta_{SU(M_x)}$'', see~\cite{Gmeiner:2009fb} for details.

The $SU(N_a)$ beta function and gauge threshold contributions from bifundamental and adjoint matter 
on $T^6/\Z_2 \times \Z_{2M}$ with discrete torsion in table~\ref{tab:Z2Z2M-torsion-Bifundamentals-beta+thresholds}
can be directly compared to those on the six-torus or $T^6/\Z_3$,  \mbox{$T^6/\Z_2 \times \Z_{2M}$} without discrete torsion
and $T^6/\Z_{2N}$ orbifolds in tables~\ref{tab:Six-torus-Bifund-beta+thresholds},~\ref{tab:Z2Z2M-no_torsion-Anti+Sym-beta+thresholds}
and~\ref{tab:T6-Z2N-Bifund-beta+thresholds} below. In section~\ref{S:Kaehler_metrics+potential}, we use these explicit formulas 
to derive the K\"ahler metrics, which are universal for all toroidal orbifolds and only depend on the number of non-vanishing angles, 
as well as the one-loop corrections to the holomorphic gauge kinetic function which also depends on the number of $\Z_2$ symmetries
and the choice of displacement and Wilson line moduli.

For the {\bf M\"obius strip} amplitudes, the expansion of the Jacobi theta functions depends on the range of the 
angle ($0 < |\nu| < \frac{1}{2}$ or $\frac{1}{2} < |\nu| < 1$ with vanishing result for $|\nu|=\frac{1}{2}$,
see e.g.~\cite{Lust:2003ky,Blumenhagen:2007ip,Gmeiner:2009fb} and details in appendix~\ref{App:A}),
\begin{equation}\label{Eq:Integ-MS-vartheta}
\begin{aligned}
\frac{1}{\pi} \,\int_0^{\infty} dl \, l^{\varepsilon} \; \frac{\vartheta_1^{\, \prime}}{\vartheta_1}(\nu,2il-\frac{1}{2} )
=& \cot (\pi \nu) \, \int_0^{\infty} dl
 +  \left( \frac{1}{\varepsilon} +\gamma - \ln 2  \right) \left( \frac{{\rm sgn}(\nu)\left[ 1 + 2 \, H(|2\nu|-1)
\right]}{4} -\nu \right)
\\
& -\frac{1}{4} \, \ln  \left( \frac{\Gamma(|2\nu|-H(|2\nu|-1))}{\Gamma(1 -|2\nu|+H( |2\nu|-1))}\right)^{\sgn(\nu)}
+  \left[ \nu -  \frac{ \sgn(\nu)}{2}   \right] \,  \ln (2)
\\
& +{\mathcal O}(\varepsilon)
,
\end{aligned}
\end{equation}
which is compactly written using the Heaviside step function
\begin{equation}\label{EqApp:Heavyside}
H (x) =\left\{\begin{array}{cc}
1 &  x > 0
\\
\frac{1}{2} & x=0
\\
0 & x < 0
\end{array}\right.
.
\end{equation}
The first line of~(\ref{Eq:Integ-MS-vartheta}) provides again the contribution to the tadpole (using \linebreak \mbox{$\tilde{I}_a^{\OR\Z_2^{(l)}} \sum_{i=1}^3
\cot(\phi_{a, \OR\Z_2^{(l)}}^{(i)}) = \sum_{i=1}^3 \tilde{V}_a^{ \OR\Z_2^{(l)},(i)} \tilde{I}_a^{\OR\Z_2^{(l)},(j \cdot k)}$})
and beta function coefficient, where for later convenience we define the sign factor
\begin{equation}\label{Eq:Def-sign_c}
c_a^{\OR\Z_2^{(l)}} \equiv   \left[2\, H \bigl( |2\phi^{(k)}_{a,\OR\Z_2^{(l)}}|-1 \bigr) - 1 \right] \in \{-1,0,1\}
\quad
\text{for}
\quad
l= 0 \ldots 3
,
\end{equation}
where again $\OR\Z_2^{(0)} \equiv \OR$, and the angle $\phi^{(k)}_{a,\OR\Z_2^{(l)}}$ that appears in the definition of the sign $c_a^{\OR\Z_2^{(l)}}$  is the one with maximal absolute value,
\begin{equation*}
0 < |\phi^{(i)}_{a,\OR\Z_2^{(l)}}|, |\phi^{(j)}_{a,\OR\Z_2^{(l)}}| < |\phi^{(k)}_{a,\OR\Z_2^{(l)}}| < 1
\quad
\text{and}
\quad
0 < |\phi^{(i)}_{a,\OR\Z_2^{(l)}}|, |\phi^{(j)}_{a,\OR\Z_2^{(l)}}| < \frac{1}{2}
,
\end{equation*}
where again $(i,j,k)$ is a cyclic permutation of (1,2,3).
The contribution to the beta function coefficient from all M\"obius amplitudes for a given orbifold invariant D$6_a$-brane orbit is thus
\begin{equation*}
b_{SU(N_a)} \supset \sum_{m=0}^{M-1}
\frac{ c_{(\omega^m a)}^{\OR}\, \eta_{\OR} \;  |\tilde{I}_{(\omega^m a)}^{\OR}|  + \sum_{i=1}^3  c_{(\omega^m a)}^{\OR\Z_2^{(i)}}\, \eta_{\OR\Z_2^{(i)}} \;  |\tilde{I}_{(\omega^m a)}^{\OR\Z_2^{(i)}}|}{4} 
, 
\end{equation*}
which is identical to the second line of~(\ref{Eq:Count-Anti-Sym}), as can be checked on a case-by-case basis, cf. appendix~\ref{App:A}
for details on the various signs $\sgn(\tilde{I}_a^{\OR\Z_2^{(l)}})$ and $ c_a^{\OR\Z_2^{(l)}}$ in dependence of the angles.

Finally, the second line on the r.h.s. of equation~(\ref{Eq:Integ-MS-vartheta}) constitutes the contribution to the gauge threshold.
In appendix~\ref{App:A}, we show in detail that the contribution to the logarithms of Gamma functions
from M\"obius strips involving any of the four $\OR\Z_2^{(l)}$ invariant O6-planes 
 can be brought to exactly the same form as the annulus contribution from the same D$6_a$-brane, 
\begin{equation}\label{Eq:Annulus-Moebius-Gamma-Rewritten}
\begin{aligned}
&-\frac{\eta_{\OR\Z_2^{(l)}} \, \tilde{I}_a^{\OR\Z_2^{(l)}}}{4} \sum_{i=1}^3 \ln  \left(\frac{\Gamma(|2\phi_{a,\OR\Z_2^{(l)}}^{(i)}|-H(|2\phi_{a,\OR\Z_2^{(l)}}^{(i)}|-1))}
{\Gamma(1 -|2\phi_{a,\OR\Z_2^{(l)}}^{(i)}|+H( |2\phi_{a,\OR\Z_2^{(l)}}^{(i)}|-1))}  \right)^{\sgn(\phi_{a,\OR\Z_2^{(l)}}^{(i)})} \!\!\!\!\!
\\
&\quad  - \eta_{\OR\Z_2^{(l)}} \, \tilde{I}_a^{\OR\Z_2^{(l)}} \sum_{i=1}^3 \left[  \frac{ \sgn(\phi_{a,\OR\Z_2^{(l)}}^{(i)})}{2} - \phi_{a,\OR\Z_2^{(l)}}^{(i)} \right] \,  \ln (2)
\\
=& \, 
\frac{c_a^{\OR\Z_2^{(l)}} \, \eta_{\OR\Z_2^{(l)}} \, |\tilde{I}_a^{\OR\Z_2^{(l)}}|}{4} \, 
\sgn(I_{aa'}) \; \sum_{i=1}^3 \ln \left(\frac{\Gamma(|\phi^{(i)}_{aa'}|)}{\Gamma(1-|\phi^{(i)}_{aa'}|)}\right)^{\sgn(\phi^{(i)}_{aa'})}
+  \frac{ \eta_{\OR\Z_2^{(l)}}  | \tilde{I}_a^{\OR\Z_2^{(l)}}| }{2} \, \ln (2)
,
\end{aligned}
\end{equation}
for every $l \in \{0 \ldots 3\}$ with  $c_a^{\OR\Z_2^{(l)}}$ defined in equation~(\ref{Eq:Def-sign_c}).
As a result, the annulus and M\"obius strip amplitudes from orientifold invariant D6-brane configurations at non-trivial angles can be summed up to give the 
simple expression in the last line of table~\ref{tab:Z2Z2M-torsion-AntiSym-beta+thresholds} for rigid D6-branes on the
$T^6/(\Z_2 \times \Z_{2M} \times \OR)$ orientifolds with discrete torsion in which the terms with Gamma functions have the beta function coefficients
of the corresponding  massless strings in the symmetric and antisymmetric representation as prefactors.

For one vanishing angle between the D$6_a$-brane and $\OR\Z_2^{(l)}$ invariant O6-plane, the Kaluza-Klein and winding sum of the 
annulus amplitude in~(\ref{Eq:KK-contributions-Annulus}) changes by $(v_i,V_{aa'}^{(i)}) \rightarrow (\tilde{v}_i, 2\tilde{V}_a^{\OR\Z_2^{(l)},(i)})$, where the weighted quantities $\tilde{v}_i$
and $\tilde{V}_a^{\OR\Z_2^{(l)},(i)}$ have been introduced in equation~(\ref{Eq:Def-v-tilde}) and~(\ref{Eq:Def_I+V-tilde}), respectively.
This leads to the expansion of the M\"obius strip contribution to the gauge threshold amplitude for one vanishing angle and arbitrary continuous
displacements and Wilson lines (for more details see~\cite{Gmeiner:2009fb}), 
\begin{equation*}
\begin{aligned}
2\tilde{V}_{a}^{\OR\Z_2^{(l)},(i)} \!\!\! \int_{0}^{\infty} \!\!\!\!\!\! dl \, l^{\varepsilon} \; {\mathcal L}^{(i)}_{a,\OR\Z_2^{(l)}}(\tilde{v}_i, 2\tilde{V}^{\OR\Z_2^{(l)},(i)}_{a};l) =& 
2\tilde{V}_{a}^{\OR\Z_2^{(l)},(i)} \, \int_{0}^{\infty} dl 
 +  \left( \frac{1}{\varepsilon} +\gamma - \ln 2  \right) \, \delta_{\sigma_{aa'}^{(i)},0} \delta_{\tau_{aa'}^{(i)},0} \\
& -  \delta_{\sigma_{aa'}^{(i)},0} \delta_{\tau_{aa'}^{(i)},0} \; \Lambda_{0,0}(\tilde{v}_i; 2\tilde{V}^{\OR\Z_2^{(l)},(i)}_{a})\\
&-\left( 1- \delta_{\sigma_{aa'}^{(i)},0} \delta_{\tau_{aa'}^{(i)},0} \right)\; \Lambda (\sigma^{(i)}_{aa'},\tau^{(i)}_{aa'},\tilde{v}_i)
\quad +{\mathcal O}(\varepsilon),
\end{aligned}
\end{equation*}
where again the tadpole and beta function coefficient are given in the first line and the gauge threshold due to massive strings in the second and
third line.
For the $T^6/(\Z_2 \times \Z_{2M} \times \OR)$ orientifold with discrete torsion, only  vanishing relative displacements and Wilson lines $(\sigma_{aa'}^{(i)},\tau_{aa'}^{(i)})=(0,0)$ among orientifold image D$6_a$ and D$6_{a'}$-branes occur. In this case, it is useful to expand 
\begin{equation*}
\Lambda_{0,0}(\tilde{v};2\, \tilde{V})=
\Lambda_{0,0}(v;V) + 2 \, \ln (2)
+ (4b) \, \ln \left(\frac{\eta(2iv)}{\vartheta_4(0,2iv)}   \right)
\end{equation*}
into an identical sum $\Lambda_{0,0}(v;V)$ as in the annulus amplitude plus a constant term $2 \ln(2)$ and $v$-dependent corrections in form of modular functions, which only appear for tilted tori
and have to our knowledge only been taken into account in the $T^6/\Z_{2N}$ context in~\cite{Gmeiner:2009fb} before.
The annulus contributions from untwisted and $\Z_2^{(i)}$ twisted sectors $aa'$ strings can now be combined with the M\"obius strip contributions from D$6_a$-branes parallel to some $\OR$ or $\OR\Z_2^{(i)}$ invariant O6-planes, see the first four lines in table~\ref{tab:Z2Z2M-torsion-AntiSym-beta+thresholds}.
\mathtabfix{
\begin{array}{|c||c|c|}\hline
\multicolumn{3}{|c|}{b_{SU(N_a)}  \text{ \bf and gauge thresholds  for (anti)symmetrics: } T^6/\Z_2 \times \Z_{2M} \text{ \bf  with discrete torsion}}
\\\hline\hline
(\phi_{aa'}^{(1)},\phi_{aa'}^{(2)},\phi_{aa'}^{(3)})
& \begin{array}{c}
b_{SU(N_a)}^{\text{torsion}} = b_{aa'}^{\cal A} + b_{aa'}^{\cal M} + \ldots \\
\frac{N_a}{2} \left( \varphi^{\Sym_a} + \varphi^{\Anti_a}\right)+ \left( \varphi^{\Sym_a} - \varphi^{\Anti_a}\right)  + \ldots 
\end{array}
&  \Delta_{SU(N_a)}^{\text{torsion}} = N_a \tilde{\Delta}_{aa'}^{\text{torsion}} + \Delta_{a,\OR}^{\text{torsion}} + \ldots
\\\hline\hline
\begin{array}{c} (0,0,0) \\ \pp \OR \end{array}  
&\begin{array}{c} 
-  \frac{N_a}{4}  \,  \sum_{i=1}^3  I_{aa'}^{\Z_2^{(i)},(j \cdot k)} \\ 
 - \frac{1}{2} \sum_{i=1}^3 \eta_{\OR\Z_2^{(i)}}  |\tilde{I}_a^{\OR\Z_2^{(i)},(j \cdot k)}|
\end{array} 
&
\begin{array}{c} 
-  \left( -  \frac{N_a}{4} \sum_{i=1}^3   I^{\Z_2^{(i)},(j \cdot k)}_{aa'}   \right)\, \Lambda_{0,0}(v_i,V_{aa'}^{(i)})
\\
+ \frac{1}{2}\sum_{i=1}^3  \eta_{\OR\Z_2^{(i)}}\,   |\tilde{I}_a^{\OR\Z_2^{(i)},(j \cdot k)}| \,
\Lambda_{0,0} (\tilde{v}_i, 2 \, \tilde{V}_{aa'}^{(i)})
 \end{array}  
\\\hline 
\begin{array}{c} (0,0,0) \\ \pp \OR \Z_2^{(i)} \end{array}   
 & \begin{array}{c} -  \frac{N_a}{4}  \,  \sum_{l=1}^3  I^{\Z_2^{(l),(m \cdot n)}}_{aa'} \\
  - \frac{1}{2} \left( \eta_{\OR}  \,   |\tilde{I}_a^{\OR,(j \cdot k)}|
+ \sum_{j \neq i} \eta_{\OR\Z_2^{(j)}}  \,   |\tilde{I}_a^{\OR\Z_2^{(j)} ,(i \cdot j)}|\right)
\end{array} \!\!\!
&
\begin{array}{c} 
-  \left( -  \frac{N_a}{4} \sum_{l=1}^3   I^{\Z_2^{(l)},(m \cdot n)}_{aa'}   \right)\, \Lambda_{0,0}(v_l,V_{aa'}^{(l)})\\
+  \frac{1}{2} \, \eta_{\OR} \,   |\tilde{I}_a^{\OR,(j \cdot k)}| \, \Lambda_{0,0} (\tilde{v}_i, 2 \, \tilde{V}_{aa'}^{(i)})\\
+  \frac{1}{2}  \sum_{j \neq i}  \eta_{\OR\Z_2^{(j)}}  \, |\tilde{I}_a^{\OR\Z_2^{(j)} ,(i \cdot j)}|\,\Lambda_{0,0} (\tilde{v}_k, 2 \, \tilde{V}_{aa'}^{(k)})
\end{array}
\\\hline 
\!\!\!
\begin{array}{c}(0^{(i)},\phi^{(j)},\phi^{(k)})\\ \pp \left( \OR + \OR\Z_2^{(i)} \right)\end{array}\!\!\!
&  \begin{array}{c} \frac{N_a}{4} \, \left( |I_{aa'}^{(j \cdot k)} |  - I_{aa'}^{\Z_2^{(i)},(j \cdot k)} \right)\\
-\frac{1}{2}  \,   \left(\eta_{\OR}  \, |\tilde{I}_a^{\OR,(j \cdot k)}| +  \eta_{\OR\Z_2^{(i)}} \, |\tilde{I}_a^{\OR\Z_2^{(i)} ,(j \cdot k)}|\right)  
\end{array}
&
\begin{array}{c}
- \left( b_{aa'}^{\cal A} +  b_{aa'}^{\cal M} \right) \;\Lambda_{0,0}(v_i;V_{aa'}^{(i)})
-( 4 \, b_i ) \;  b_{aa'}^{\cal M} \; \ln \left(\frac{\eta(2i v_i)}{\vartheta_4(0,2iv_i)}\right)\\
+ \Bigl[ \frac{N_b \left( I_{aa'}^{\Z_2^{(j)}} -  I_{aa'}^{\Z_2^{(k)}} \right) \left( \sgn(\phi^{(j)}_{aa'}) - 2 \, \phi^{(j)}_{aa'} \right)}{4} 
+ \frac{\eta_{\OR\Z_2^{(j)}} \, |\tilde{I}_a^{\OR\Z_2^{(j)}}|   + \eta_{\OR\Z_2^{(k)}} \,  |\tilde{I}_a^{\OR\Z_2^{(k)}}|}{2} \\
+ \eta_{\OR}  \, |\tilde{I}_a^{\OR,(j \cdot k)}| +  \eta_{\OR\Z_2^{(i)}} \, |\tilde{I}_a^{\OR\Z_2^{(i)} ,(j \cdot k)}|
\Bigr] \, \ln (2) 
\end{array}
\\\hline 
\!\!\!\!\!
\begin{array}{c}(0^{(i)},\phi^{(j)},\phi^{(k)})
\\ \pp \left( \OR\Z_2^{(j)} + \OR\Z_2^{(k)} \right)\end{array}\!\!\!\!\!
&  \begin{array}{c}  \frac{N_a}{4} \, \left( |I_{aa'}^{(j \cdot k)} |  - I_{aa'}^{\Z_2^{(i)},(j \cdot k)} \right)\\
 - \frac{1}{2} \,\left(   \eta_{\OR\Z_2^{(j)}}\, |\tilde{I}_a^{\OR\Z_2^{(j)} ,(j \cdot k)}| +\eta_{\OR\Z_2^{(k)}}\, |\tilde{I}_a^{\OR\Z_2^{(k)} ,(j \cdot k)}|    \right) 
\end{array}\!\!\!\!
&
\begin{array}{c}
- \left( b_{aa'}^{\cal A} +  b_{aa'}^{\cal M} \right) \; \Lambda_{0,0}(v_i;V_{aa'}^{(i)}) 
- (4 \, b_i) \;  b_{aa'}^{\cal M} \;\ln \left(\frac{\eta(2i v_i)}{\vartheta_4(0,2iv_i)}\right) \\
+ \Bigl[ \frac{N_a \left( I_{aa'}^{\Z_2^{(j)}} -  I_{aa'}^{\Z_2^{(k)}} \right) \left( \sgn(\phi^{(j)}_{aa'}) - 2 \, \phi^{(j)}_{aa'} \right)}{4} 
+  \frac{ \eta_{\OR} \, |\tilde{I}_a^{\OR}|   +  \eta_{\OR\Z_2^{(i)}} \,  |\tilde{I}_a^{\OR\Z_2^{(i)}}|}{2} \\
+ \eta_{\OR\Z_2^{(j)}}\, |\tilde{I}_a^{\OR\Z_2^{(j)} ,(j \cdot k)}| +\eta_{\OR\Z_2^{(k)}}\, |\tilde{I}_a^{\OR\Z_2^{(k)} ,(j \cdot k)}|
\Bigr] \, \ln (2)   
\end{array}
\\\hline
\!\!\!\!\!
\begin{array}{c}
(\phi^{(1)},\phi^{(2)},\phi^{(3)}) \\ {\sum_{n=1}^3 \phi^{(n)}=0} 
\end{array}
&\!\!\!\!\! \begin{array}{c}\frac{N_a}{8} \, \left( | I_{aa'} |
+ \sgn(I_{aa'} ) \sum_{i=1}^3 I_{aa'}^{\Z_2^{(i)}}  \right)
\\
+ \frac{1}{4} \left( c^{\OR}_a\,  \eta_{\OR}  \, |\tilde{I}_a^{\OR}|
+ \sum_{i=1}^3  c^{\OR\Z_2^{(i)}}_a \, \eta_{\OR\Z_2^{(i)}}\,  |\tilde{I}_a^{\OR\Z_2^{(i)}}| \right)
\end{array}\!\!\!\!\!\!
&
\begin{array}{c}
\bigl(b_{aa'}^{\cal A}  + b_{aa'}^{\cal M}
\bigr)  \;  \sgn(I_{aa'}) \; \sum_{i=1}^3 \ln \left(\frac{\Gamma(|\phi^{(i)}_{aa'}|)}{\Gamma(1-|\phi^{(i)}_{aa'}|)}\right)^{\sgn(\phi^{(i)}_{aa'})}\\
+  \Bigl[\frac{N_a \, \sum_{i=1}^3 I_{aa'}^{\Z_2^{(i)}} \left( \sgn(\phi^{(i)}_{aa'}) - 2 \, \phi^{(i)}_{aa'} + \sgn(I_{aa'})\right)}{4}
+\sum_{m=0}^3\frac{\eta_{\OR\Z_2^{(m)}} \, |\tilde{I}_{a}^{\OR\Z_2^{(m)} }|}{2} \Bigr]\, \ln (2)
\end{array}
\\\hline
\end{array}
}{Z2Z2M-torsion-AntiSym-beta+thresholds}{Contributions to $SU(N_a)$ beta function coefficients (middle column) and gauge thresholds (last column)
from antisymmetric and symmetric matter on orientifold image D$6_a$-branes in $T^6/(\Z_2 \times \Z_{2M} \times \OR)$ backgrounds with discrete torsion.
The special case of vanishing angles and identical three-cycles wrapped by orientifold image D$6_{a'}$-branes, $\Pi_a=\Pi_{a'}$, leads to
orthogonal or symplectic gauge groups and is discussed separately in section~\protect\ref{Sss:AntiSym}, see in particular  table~\protect\ref{tab:SO-Sp-groups}.                                                      
}

Tables~\ref{tab:Z2Z2M-torsion-Bifundamentals-beta+thresholds} and~\ref{tab:Z2Z2M-torsion-AntiSym-beta+thresholds} 
contain the complete gauge threshold result for all D6-brane configurations
(i.e. any intersecting angle, displacement, Wilson line and $\Z_2$ eigenvalue)
 on the $T^6/(\Z_2 \times \Z_{2M} \times \OR)$ background with discrete torsion.
Furthermore, since this background is the technically most challenging one, the complete results for D6-branes on the six-torus or $T^6/\Z_3$, the 
$T^6/\Z_2 \times \Z_{2M}$ orientifold without torsion and $T^6/\Z_{2N}$ can be deduced using the different numerical prefactors on the r.h.s. of table~\ref{Tab:NormIntersections} 
and in table~\ref{tab:Rewritten_RRtcc-for-thresholds}.
These results are presented in the following sections~\ref{Sss:BulkD6onT6} to~\ref{Sss:D6onT6Z2N}.
Together with the discussion of orthogonal and symplectic gauge factors on each orbifold background in section~\ref{Sss:AntiSym} and the Abelian gauge groups in section~\ref{Ss:Comments_on_U1}, this constitutes the most exhaustive possible treatment of all allowed gauge groups on D6-branes and possible
factorisable toroidal orbifolds of the type IIA string, which to our knowledge has not been dealt with before.

Building on the complete classification of gauge threshold amplitudes on all factorisable toroidal orbifold backgrounds in this section,
the decomposition into the holomorphic gauge kinetic function,
K\"ahler metrics for  open string matter fields and the K\"ahler potential for closed string moduli will be derived in full generality 
 in section~\ref{S:Kaehler_metrics+potential}.

\subsubsection{Bulk D6-branes on $T^6/\OR$ and $T^6/(\Z_3 \times \OR)$}\label{Sss:BulkD6onT6}

The gauge thresholds on the six-torus have been computed in~\cite{Lust:2003ky,Akerblom:2007np} for vanishing displacement and Wilson line moduli. 
We repeat here the results for three non-vanishing angles for completeness.
The formulas for one parallel direction with arbitrary continuous Wilson line or displacement in the annulus and M\"obius strip contribution have to our knowledge not 
been presented before, and also the explicit discussion of orthogonal and symplectic gauge factors in section~\ref{Sss:AntiSym} and Abelian groups in 
section~\ref{Ss:Comments_on_U1} is presented in this article for the first time.

On the six-torus, open strings on identical D6-branes preserve ${\cal N}=4$ supersymmetry and do not contribute to the gauge thresholds. 
D6-branes at one vanishing angle on $T^2_{(i)}$ preserve ${\cal N}=2$ supersymmetry and contribute to the Kaluza-Klein and winding sums in~(\ref{Eq:KK-contributions-Annulus}) which depend on the 
K\"ahler modulus $v_i$, and D6-branes at three angles preserve ${\cal N}=1$ supersymmetry and depend on the complex structure moduli through the angles in~(\ref{Eq:Theta-Expansions-Annulus}).
There exists only the untwisted annulus amplitude for D6-branes which are not their own orientifold image.
The result for all bifundamental and adjoint representations is displayed in table~\ref{tab:Six-torus-Bifund-beta+thresholds}.
\mathtabfix{
\begin{array}{|c||c|c|}\hline
\multicolumn{3}{|c|}{b_{SU(N_a)}  \text{ \bf and gauge thresholds  for bifundamental and adjoints: } T^6 \text{ and } T^6/\Z_3}
\\\hline\hline
(\phi_{ab}^{(1)},\phi_{ab}^{(2)},\phi_{ab}^{(3)}) & 
\begin{array}{r}
b_{SU(N_a)}^{\text{torus}} = \sum_b b_{ab}^{\cal A} + \ldots \\ = \sum_b \frac{N_b}{2} \varphi^{ab} + \ldots 
\end{array}
&  \Delta_{SU(N_a)}^{\text{torus}} = \sum_b N_b \,  \tilde{\Delta}_{ab}^{\text{torus}} + \ldots
\\\hline\hline
(0,0,0) &  - & - 
\\\hline
(0,\phi,-\phi) 
&   N_b \,\delta_{\sigma^1_{ab},0} \, \delta_{\tau^1_{ab},0} \, |I_{ab}^{(2 \cdot 3)} | 
&  
\begin{array}{c}
- b_{ab}^{\cal A} \;\Lambda_{0,0}(v_1;V_1)
\\
-  N_b \, | I_{ab}^{(2 \cdot 3)}| \, \left(1-\delta_{\sigma^1_{ab},0} \delta_{\tau^1_{ab},0} \right) \,
\Lambda(\sigma_{ab}^1,\tau_{ab}^1,v_1)
\end{array}
\\\hline
\begin{array}{c}
(\phi^{(1)},\phi^{(2)},\phi^{(3)})
\\
{\sum_{n=1}^3 \phi^{(n)}=0} 
\end{array}
& \frac{N_b}{2} \, | I_{ab} |
& b_{ab}^{\cal A} \; 
\sgn(I_{ab}) \; \sum_{i=1}^3 \ln \left(\frac{\Gamma(|\phi^{(i)}_{ab}|)}{\Gamma(1-|\phi^{(i)}_{ab}|)}\right)^{\sgn(\phi^{(i)}_{ab})}
\\\hline
\end{array}
}{Six-torus-Bifund-beta+thresholds}{Contributions to the $SU(N_a)$  beta function coefficients and gauge thresholds
from open strings in the bifundamental and adjoint representation on the six-torus and the $T^6/\Z_3$ orbifold with arbitrary
continuous displacement and Wilson line moduli $(\sigma_{ab}^i,\tau_{ab}^i)$.
Details on adjoint matter contributions are further discussed in section~\protect\ref{Sss:adjoints} and table~\protect\ref{tab:Adjoint-Beta+Thresholds}.
}
On the $T^6/\Z_3$ background, the D6-branes also wrap bulk three-cycles, but when using table~\ref{tab:Six-torus-Bifund-beta+thresholds},
the sum over orbifold images in the first index has to be performed, 
\begin{equation}\label{Eq:Expansion-T6-Z3}
\begin{aligned}
b_{SU(N_a)}^{T^6/\Z_3} =&  \sum_{k=0}^2 \left[ \sum_b  \left( \tilde{b}_{(\theta^k a)b}^{\cal A}  +  \tilde{b}_{(\theta^k a)b'}^{\cal A}\right) + \tilde{b}_{(\theta^k a)(\theta^k a)'}^{\cal M}
\right]
\\
& =  \sum_{k=0}^2 \left[ \sum_{b \neq a} \frac{N_b}{2} \left(\tilde{\varphi}^{(\theta^k a)b} + \tilde{\varphi}^{(\theta^k a)b'} \right)  \right]
 + N_a \sum_{k=1}^2 \tilde{\varphi}^{\Adj_{(\theta^k a)}} 
 \\
&\quad  + \sum_{k=0}^2 \left[ \frac{N_a}{2} \left( \tilde{\varphi}^{\Sym_{(\theta^k a)}} +\tilde{\varphi}^{\Anti_{(\theta^k a)}}  \right) 
 + \left( \tilde{\varphi}^{\Sym_{(\theta^k a)}} - \tilde{\varphi}^{\Anti_{(\theta^k a)}}  \right)  \right]
 ,
\\
 \Delta_{SU(N_a)}^{T^6/\Z_3} =& \sum_{k=0}^2 \left[ \sum_b N_b \,  \left( \tilde{\Delta}_{(\theta^k a)b}^{\text{torus}} +\tilde{\Delta}_{(\theta^k a)b'}^{\text{torus}} \right)
 + \Delta_{(\theta^k a),\OR}^{\text{torus}}  \right],
\end{aligned}
\end{equation}
where in the second line we used the fact that open $aa$ strings with endpoints on identical bulk D6-branes contribute $\tilde{\varphi}^{\Adj_a}=3$,
which cancels the contribution to the beta function coefficient from the vector multiplet.

There exists only one kind of $\OR$-invariant O6-plane on the six-torus, and D6-branes parallel to it preserve the full ${\cal N}=4$ supersymmetry,
or parallel along one two-torus the ${\cal N}=2$ supersymmetry stated above for bifundamental and adjoint matter. Contrariwise, if the D6-brane
is perpendicular to the $\OR$-invariant O6-plane along one or two tori, the orientifold symmetry breaks half of the supersymmetry.
The complete list of $SU(N_a)$ beta function coefficients and gauge threshold corrections due to matter in the symmetric and antisymmetric representations
on open strings stretched between orientifold image D6-branes $a$ and $a'$ on the six-torus is given in table~\ref{tab:Six-torus-AntiSym-beta+thresholds}.
\mathtabfix{
\begin{array}{|c||c|c|}\hline
\multicolumn{3}{|c|}{b_{SU(N_a)}  \text{ \bf and gauge thresholds  for symmetrics and antisymmetrics: } T^6 \text{ and } T^6/\Z_3}
\\\hline\hline
(\phi_{aa'}^{(1)},\phi_{aa'}^{(2)},\phi_{aa'}^{(3)})   & 
\begin{array}{c}
b_{SU(N_a)}^{\text{torus}}= b_{aa'}^{\cal A} + b_{aa'}^{\cal M} + \ldots \\
=\frac{N_a}{2} \left( \varphi^{\Sym_a} + \varphi^{\Anti_a}\right)+ \left( \varphi^{\Sym_a} - \varphi^{\Anti_a}\right)  + \ldots 
\end{array}
&  \Delta_{SU(N_a)}^{\text{torus}} = N_a \tilde{\Delta}_{aa'}^{\text{torus}} + \Delta_{a,O6}^{\text{torus}} + \ldots
\\\hline\hline
\begin{array}{c} (0,0,0) \\ \pp \OR \end{array}  
& -
& -
\\\hline 
\begin{array}{c} (0,0,0) \\ \perp \OR \\\text{ on } T_j \times T_k \end{array}  
& - \delta_{\sigma^i_{aa'},0} \delta_{\tau^i_{aa'},0}\, 2 \,  |\tilde{I}_a^{\OR,(j \cdot k)}|
& 
\begin{array}{c}
- b_{aa'}^{\cal M}  \; \Lambda_{0,0}(v_i,V_{aa'}^{(i)})
- (4b_i) \, b_{aa'}^{\cal M} \;  \ln \left(\frac{\eta(2iv_i)}{\vartheta_4(0,2iv_i)} \right)- 2 \,  b_{aa'}^{\cal M} \; \ln(2)\\
+ 2 \, |\tilde{I}_a^{\OR,(j \cdot k)}| \, \left( 1 - \delta_{\sigma_{aa'}^i,0}\delta_{\tau_{aa'}^i,0}\right) \Lambda(\sigma_{aa'}^i,\tau_{aa'}^i,\tilde{v}_i)
\end{array}
\\\hline 
\!\!\!
\begin{array}{c}(0^{(i)},\phi^{(j)},\phi^{(k)})
\\ \pp  \OR \\  \text{ on } T_i \end{array}\!\!\!
& \delta_{\sigma^i_{aa'}}\delta_{\tau^i_{aa'},0} \, \Bigl\{  N_a \,   |I_{aa'}^{(j \cdot k)} |  
-    2 \,  |\tilde{I}_a^{\OR,(j \cdot k)}|  \Bigr\}
 & 
\begin{array}{c}
- \left( b_{aa'}^{\cal A} + b_{aa'}^{\cal M} \right)\; \Lambda_{0,0}(v_i;V_i)
-  (4b_i) \, b_{aa'}^{\cal M} \; \ln \left(\frac{\eta(2iv_i)}{\vartheta_4(0,2iv_i)} \right)- 2 \,  b_{aa'}^{\cal M} \; \ln(2) \\
-  N_a \,  |I_{aa'}^{(j \cdot k)}| \, \left(1-\delta_{\sigma^i_{aa'},0} \delta_{\tau^i_{aa'},0} \right) \, \Lambda(\sigma_{aa'}^i,\tau_{aa'}^i,v_i)\\
+ 2 \, |\tilde{I}_a^{\OR,(j \cdot k)}| \, \left( 1 - \delta_{\sigma_{aa'}^i,0}\delta_{\tau_{aa'}^i,0}\right) \Lambda(\sigma_{aa'}^i,\tau_{aa'}^i,\tilde{v}_i)
\end{array}
 \\\hline
\!\!\!
\begin{array}{c}(0^{(i)},\phi^{(j)},\phi^{(k)})
\\ \perp  \OR  \\ \text{ on } T_i  \end{array}\!\!\!
&  \delta_{\sigma^i_{aa'},0}\delta_{\tau^i_{aa'},0}\,  N_a \, |I_{aa'}^{(j \cdot k)} |  
&
\begin{array}{c}
- b_{aa'}^{\cal A} \; \Lambda_{0,0}(v_i;V_i) \\
-  N_a \, | I_{aa'}^{(j \cdot k)}| \, \left(1-\delta_{\sigma^i_{aa'},0} \delta_{\tau^i_{aa'},0} \right) \, \Lambda(\sigma_{aa'}^i,\tau_{aa'}^i,v_i)\\
+ 2\,  |\tilde{I}_a^{\OR}| \, \ln(2)
\end{array}
\\\hline 
\!\!\!\!\!
\begin{array}{c}
(\phi^{(1)},\phi^{(2)},\phi^{(3)}) \\ {\sum_{n=1}^3 \phi^{(n)}=0} 
\end{array}
&   
\frac{N_a}{2} \, | I_{aa'} | + c^{\OR}_a \, |\tilde{I}_a^{\OR}| 
&
\begin{array}{c}
\left(b_{aa'}^{\cal A} + b_{aa'}^{\cal M} \right)\; \sgn(I_{aa'})
\sum_{i=1}^3 \ln \left(\frac{\Gamma(|\phi^{(i)}_{aa'}|)}{\Gamma(1-|\phi^{(i)}_{aa'}|)}\right)^{\sgn(\phi^{(i)}_{aa'})}\\
+ 2 \, |\tilde{I}_a^{\OR}| \, \ln(2)
\end{array}
\\\hline
\end{array}
}{Six-torus-AntiSym-beta+thresholds}{Beta function coefficients and gauge thresholds for $SU(N_a)$ gauge groups from open strings in the 
symmetric and antisymmetric representation on bulk D6-branes on the six-torus and $T^6/\Z_3$. The angles on the first line preserve ${\cal N}=4$,
on the second and third line ${\cal N}=2$ and on the last two lines ${\cal N}=1$ supersymmetry.
On the first two lines, for vanishing relative displacements and Wilson lines $(\sigma_{aa'}^i,\tau_{aa'}^i) = (0,0)$ along the two-tori (i.e. the D6-branes
are on top of or perpendicular to the O6-plane) the gauge group is enhanced to $SO(2N_a)$ and $Sp(2N_a)$, respectively. Details for the orthogonal and 
symplectic gauge factors are given in section~\protect\ref{Sss:AntiSym} and  table~\protect\ref{tab:SO-Sp-groups}.
}
For $T^6/\Z_3$, the sum over orbifold images $(\theta^k a)$ with $k=0,1,2$ needs to be performed for both the annulus and M\"obius strip contributions,
see equation~(\ref{Eq:Expansion-T6-Z3}).

\subsubsection{Fractional D6-branes on $T^6/(\Z_2 \times \Z_{2M} \times \OR)$ without discrete torsion}\label{Sss:D6onT6Z2Z2MwithoutTorsion}

D6-branes on the $T^6/\Z_2 \times \Z_{2M}$ orbifolds without discrete torsion only wrap the half-bulk three cycles displayed in table~\ref{tab:FractionalCycles}.
All annulus contributions to the gauge threshold amplitudes thus stem from the untwisted sector, and by comparison with 
the intersection numbers and rewritten RR tadpole cancellation conditions in tables~\ref{Tab:NormIntersections}
and~\ref{tab:Rewritten_RRtcc-for-thresholds}, the normalisation of the contributions to the $SU(N_a)$ beta function coefficients and gauge thresholds from open strings 
in the bifundamental or adjoint representation is shown to be identical to the six-torus (up to summation over orbifold images of the first D6-brane index in analogy to
equation~(\ref{Eq:expand-varphis})). 
The result is  listed in the upper part of table~\ref{tab:Z2Z2M-no_torsion-Anti+Sym-beta+thresholds}.
\mathtabfix{
\begin{array}{|c||c|c|}\hline
\multicolumn{3}{|c|}{b_{SU(N_a)}  \text{ \bf and gauge thresholds  for bifundamental and adjoints: } T^6/\Z_2 \times \Z_{2M} \text{ \bf  without discrete torsion}}
\\\hline\hline
(\phi_{ab}^{(1)},\phi_{ab}^{(2)},\phi_{ab}^{(3)}) & 
\begin{array}{c} b_{SU(N_a)}^{\text{no torsion}} = \sum_b b_{ab}^{\cal A} + \ldots \\ = \sum_b \frac{N_b}{2} \varphi^{ab} + \ldots 
\end{array}
&  \Delta_{SU(N_a)}^{\text{no torsion}}= \sum_b N_b \tilde{\Delta}_{ab} + \ldots
\\\hline\hline
(0,0,0) &  - & - 
\\\hline
(0,\phi,-\phi) 
& N_b \,\delta_{\sigma^1_{ab},0} \, \delta_{\tau^1_{ab},0} \, |I_{ab}^{(2 \cdot 3)} | 
&  
\begin{array}{c}
- b_{ab}^{\cal A}  \;  \Lambda_{0,0}(v_1;V_{ab}^{(1)})\\
- N_b \, | I_{ab}^{(2 \cdot 3)}| \, \left(1-\delta_{\sigma^1_{ab},0} \delta_{\tau^1_{ab},0} \right) \,
\Lambda(\sigma_{ab}^1,\tau_{ab}^1,v_1)
\end{array}
\\\hline
\begin{array}{c}
(\phi^{(1)},\phi^{(2)},\phi^{(3)})\\
{\sum_{n=1}^3 \phi^{(n)}=0} 
\end{array}
& \frac{N_b}{2} \, | I_{ab} |
& b_{ab}^{\cal A} \;  \sgn(I_{ab}) \sum_{i=1}^3 \ln \left(\frac{\Gamma(|\phi^{(i)}_{ab}|)}{\Gamma(1-|\phi^{(i)}_{ab}|)}\right)^{\sgn(\phi^{(i)}_{ab})}
\\\hline\hline\hline
\multicolumn{3}{|c|}{b_{SU(N_a)}  \text{ \bf and gauge thresholds  for (anti)symmetrics: } T^6/\Z_2 \times \Z_{2M} \text{ \bf  without discrete torsion}}
\\\hline\hline
(\phi_{aa'}^{(1)},\phi_{aa'}^{(2)},\phi_{aa'}^{(3)})  
& \begin{array}{c} 
b_{SU(N_a)}^{\text{no torsion}} = b_{aa'}^{\cal A} + b_{aa'}^{\cal M} + \ldots \\
\frac{N_a}{2} \left( \varphi^{\Sym_a} + \varphi^{\Anti_a}\right)+ \left( \varphi^{\Sym_a} - \varphi^{\Anti_a}\right)  + \ldots 
\end{array}
&  \Delta_{SU(N_a)}^{\text{no torsion}}= N_a \tilde{\Delta}_{aa'} + \Delta_{a,\OR} + \ldots
\\\hline\hline
\begin{array}{c} (0,0,0) \\ \pp \OR \end{array}  
& - \sum_{i=1}^3 \delta_{\sigma^i_{aa'},0} \delta_{\tau^i_{aa'},0}\,  | \tilde{I}_a^{\OR\Z_2^{(i)},(j \cdot k)} |
& \begin{array}{c} 
 \sum_{i=1}^3  \delta_{\sigma^i_{aa'},0} \delta_{\tau^i_{aa'},0}\,   |\tilde{I}_a^{\OR\Z_2^{(i)},(j \cdot k)}| \,
\Lambda_{0,0} (\tilde{v}_i, 2 \, \tilde{V}_{aa'}^{(i)})\\
+  \sum_{i=1}^3 \left(1- \delta_{\sigma^i_{aa'},0} \delta_{\tau^i_{aa'},0} \right) \,  | \tilde{I}_a^{\OR\Z_2^{(i)},(j \cdot k)}| \,
\Lambda(\sigma^i_{aa'},\tau^i_{aa'},\tilde{v}_i) 
\end{array}  
\\\hline 
\begin{array}{c} (0,0,0) \\ \pp \OR\Z_2^{(i)} \end{array}  
&\begin{array}{c} - \delta_{\sigma^i_{aa'},0} \delta^{\tau^i_{aa'},0}\,|\tilde{I}_a^{\OR,(j \cdot k)}| \\
- \sum_{j \neq i}   \delta_{\sigma^k_{aa'},0} \delta_{\tau^k_{aa'},0}\, |\tilde{I}_a^{\OR\Z_2^{(j)} ,(i \cdot j)}| \end{array}
&\begin{array}{c} 
 \delta_{\sigma^i_{aa'},0} \delta_{\tau^i_{aa'},0}\,   |\tilde{I}_a^{\OR,(j \cdot k)} | \, \Lambda_{0,0} (\tilde{v}_i, 2 \, \tilde{V}_{aa'}^{(i)})\\
+ \sum_{j \neq i} \delta_{\sigma^k_{aa'},0} \delta_{\tau^k_{aa'},0}\,| \tilde{I}_a^{\OR\Z_2^{(j)} ,(i \cdot j)}| \,\Lambda_{0,0} (\tilde{v}_k, 2 \, \tilde{V}_{aa'}^{(k)})
\\
+ \left( 1- \delta_{\sigma^i_{aa'},0} \delta_{\tau^i_{aa'},0} \right)\, |\tilde{I}_a^{\OR,(j \cdot k)}| \,\Lambda(\sigma^i_{aa'},\tau^i_{aa'},\tilde{v}_i)\\  
+ \sum_{j \neq i}  \left(1- \delta_{\sigma^k_{aa'},0} \delta_{\tau^k_{aa'},0} \right) \, |\tilde{I}_a^{\OR\Z_2^{(j)} ,(i \cdot j)}| 
\, \Lambda(\sigma^k_{aa'},\tau^k_{aa'},\tilde{v}_k) 
\end{array}
\\\hline 
\!\!\!
\begin{array}{c}(0_i,\phi_j,\phi_k)_{\phi_k = -\phi_j \neq \pm \frac{1}{2}} \\ \pp \left( \OR + \OR\Z_2^{(i)} \right)\\ \text{ on } T_i^2 \end{array}\!\!\!
& \begin{array}{c}\delta_{\sigma^i_{aa'}}\delta_{\tau^i_{aa'},0} \, \Bigl\{  N_a \,   |I_{aa'}^{(j \cdot k)} |  \\
-|\tilde{I}_a^{\OR,(j \cdot k)}| - |\tilde{I}_a^{\OR\Z_2^{(i)} ,(j \cdot k)}|  \Bigr\}
\end{array}
 & 
\begin{array}{c}
- \left( b_{aa'}^{\cal A} + b_{aa'}^{\cal M} \right) \; \Lambda_{0,0}(v_i;V_{aa'}^{(i)})
- (4 \, b_i) \;  b_{aa'}^{\cal M} \; \ln \left(\frac{\eta(2i v_i)}{(\vartheta_4(0,2iv_i)}\right) \\
+ \left(1-\delta_{\sigma^i_{aa'},0} \delta_{\tau^i_{aa'},0} \right) \times \\
\times \left( N_a \,  I_{aa'}^{(j \cdot k)} \,   \Lambda(\sigma_{aa'}^i,\tau_{aa'}^i,v_i)
+ \left[ | \tilde{I}_a^{\OR,(j \cdot k)}| + |\tilde{I}_a^{\OR\Z_2^{(i)} ,(j \cdot k)}|  \right] \, \Lambda(\sigma_{aa'}^i,\tau_{aa'}^i,\tilde{v}_i)\right)\\
+ \left(|\tilde{I}_a^{\OR\Z_2^{(j)}}|   +  |\tilde{I}_a^{\OR\Z_2^{(k)}}|
+ 2 \; \delta_{\sigma^i_{aa'}}\delta_{\tau^i_{aa'},0} \, \left\{ |\tilde{I}_a^{\OR,(j \cdot k)}| + |\tilde{I}_a^{\OR\Z_2^{(i)} ,(j \cdot k)}|  \right\}
 \right) \, \ln (2) 
\end{array}
 \\\hline
\!\!\!
\begin{array}{c}(0_i,\phi_j,\phi_k)_{\phi_k = -\phi_j \neq \pm \frac{1}{2}} \\ \pp \left( \OR\Z_2^{(j)} + \OR\Z_2^{(k)} \right)\\ \text{ on } T_i^2\end{array}\!\!\!
&\begin{array}{c} \delta_{\sigma^i_{aa'},0}\delta_{\tau^i_{aa'},0}\, \Bigl\{ N_a \, |I_{aa'}^{(j \cdot k)} |  \\
 - | \tilde{I}_a^{\OR\Z_2^{(j)} ,(j \cdot k)}|   - | \tilde{I}_a^{\OR\Z_2^{(k)} ,(j \cdot k)} |     \Bigr\}
\end{array}
&
\begin{array}{c}
- \left( b_{aa'}^{\cal A} +  b_{aa'}^{\cal M} \right) \; \Lambda_{0,0}(v_i;V_{aa'}^{(i)})
-( 4\, b_i) \; b_{aa'}^{\cal M} \; \ln \left(\frac{\eta(2i v_i)}{(\vartheta_4(0,2iv_i)}\right)\\
+  \left(1-\delta_{\sigma^i_{aa'},0} \delta_{\tau^i_{aa'},0} \right) \, \times \\
\times \left( N_a \,  I_{aa'}^{(j \cdot k)} \, \Lambda(\sigma_{aa'}^i,\tau_{aa'}^i,v_i)
+ \left[ |\tilde{I}_a^{\OR\Z_2^{(j)} ,(j \cdot k)}| + |\tilde{I}_a^{\OR\Z_2^{(k)} ,(j \cdot k)}|  \right] \, \Lambda(\sigma_{aa'}^i,\tau_{aa'}^i,\tilde{v}_i)\right)\\
+   \left(|\tilde{I}_a^{\OR}|   +  |\tilde{I}_a^{\OR\Z_2^{(i)}}|
+2 \; \delta_{\sigma^i_{aa'},0}\delta_{\tau^i_{aa'},0}\, \left\{ 
| \tilde{I}_a^{\OR\Z_2^{(j)} ,(j \cdot k)}| + | \tilde{I}_a^{\OR\Z_2^{(k)} ,(j \cdot k)} |  \right\} \right) \, \ln (2)   
\end{array}
\\\hline 
\!\!\!\!\!
\begin{array}{c}
(\phi^{(1)},\phi^{(2)},\phi^{(3)}) \\ {\sum_{n=1}^3 \phi^{(n)}=0} 
\end{array}
& \begin{array}{c}  \frac{N_a}{2} \, | I_{aa'} | \\ 
+\frac{c^{\OR}_a}{2}  |\tilde{I}_a^{\OR}| + \sum_{i=1}^3  \frac{c^{\OR\Z_2^{(i)}}_a}{2} | \tilde{I}_a^{\OR\Z_2^{(i)}}|
\end{array}
&
\begin{array}{c}
\bigl(b_{aa'}^{\cal A}   + b_{aa'}^{\cal M} \bigr) \;  \sgn(I_{aa'}) \; 
\sum_{i=1}^3 \ln \left(\frac{\Gamma(|\phi^{(i)}_{aa'}|)}{\Gamma(1-|\phi^{(i)}_{aa'}|)}\right)^{\sgn(\phi^{(i)}_{aa'})}\\
+\sum_{m=0}^3 |\tilde{I}_{a}^{\OR\Z_2^{(m)} }| \, \ln (2)
\end{array}
\\\hline
\end{array}
}{Z2Z2M-no_torsion-Anti+Sym-beta+thresholds}{$SU(N_a)$ beta function coefficients and gauge thresholds for half-bulk D6-branes on
$T^6/(\Z_2 \times \Z_{2M} \times \OR)$ without discrete torsion.
In the last line, the notation is shortened by setting $\OR\Z_2^{(0)} \equiv \OR$.
A stack of D6-branes on top of some O6-plane with vanishing relative displacements and Wilson lines 
of the orientifold image D6-branes everywhere, $(\vec{\sigma}_{aa'},\vec{\tau}_{aa'})=(0,0)$, leads to an $Sp(2N_a)$ gauge group
and is further discussed in section~\protect\ref{Sss:AntiSym} and table~\protect\ref{tab:SO-Sp-groups}. 
The contributions from adjoint matter are scrutinised in section~\protect\ref{Sss:adjoints} and table~\protect\ref{tab:Adjoint-Beta+Thresholds}.
}

In contrast to the six-torus, there exist four different orbits of $\OR\Z_2^{(l)}$ invariant O6-planes, where for the sake of a compact notation, we 
set $\OR \equiv \OR\Z_2^{(0)}$ and $l \in \{0 \ldots 3\}$. The M\"obius strip contributions to the $SU(N_a)$ beta function coefficients and gauge thresholds correspond, up to normalisation and up to the existence of {\it continuous} displacements and Wilson lines,
to those of the $T^6/\Z_2 \times \Z_{2M}$ orbifolds with discrete torsion  presented in detail above.
From the intersection numbers and rewritten RR tadpole cancellation conditions in tables~\ref{Tab:NormIntersections} and~\ref{tab:Rewritten_RRtcc-for-thresholds}, it is easy to derive the change of normalisation by a factor of two and the simplification in the signs for only ordinary O6-planes, $\eta_{\OR\Z_2^{(l)}} \equiv 1$ for all $l \in \{0 \ldots 3\}$ on orbifolds without discrete torsion. 
The complete expressions for $SU(N_a)$ beta function coefficients and gauge thresholds due to open strings in the symmetric and antisymmetric representation
are listed in the lower part of table~\ref{tab:Z2Z2M-no_torsion-Anti+Sym-beta+thresholds}.

\subsubsection{Fractional D6-branes on $T^6/(\Z_{2N}\times \OR)$}\label{Sss:D6onT6Z2N}

The beta function coefficients and gauge thresholds for fractional D6-branes on \mbox{$T^6/(\Z_{2N} \times \OR)$} backgrounds have been discussed in detail
in~\cite{Gmeiner:2009fb}. 
In order to be able to directly compare with the other factorisable orbifolds in this article and for the decomposition into holomorphic gauge kinetic function 
and K\"ahler metrics, we rewrite the results for bifundamental and adjoint matter in table~\ref{tab:T6-Z2N-Bifund-beta+thresholds} as sums of the untwisted and 
$\Z_2 \equiv \Z_2^{(2)}$ twisted annulus amplitudes and re-express the gauge thresholds by using the known form of the associated beta function coefficients.
\mathtabfix{
\begin{array}{|c||c|c|} \hline
\multicolumn{3}{|c|}{\rule[-3mm]{0mm}{8mm}
\text{\bf $b_{SU(N_a)}$ and gauge thresholds for bifundamentals and adjoints:} \quad T^6/\Z_{2N} \text{ with } \Z_2 \equiv \Z_2^{(2)}}
\\\hline\hline
(\phi_{ab}^{(1)},\phi_{ab}^{(2)},\phi_{ab}^{(3)})  
&\begin{array}{r} b^{\Z_2}_{SU(N_a)} = \sum_b b_{ab}^{\cal A} + \ldots \\ = \sum_b \frac{N_b}{2} \, \varphi^{ab} + \ldots
\end{array}
& \Delta^{\Z_{2N}}_{SU(N_a)} = N_b \, \tilde{\Delta}_{ab} + \ldots
\\\hline\hline
(0,0,0)
&  -\frac{ I^{\Z_2,(1 \cdot 3)}_{ab} N_b }{2} \, \delta_{\sigma^2_{ab},0} \delta_{\tau^2_{ab},0}
&
\begin{array}{c}
- b_{ab}^{\cal A} \; \Lambda_{0,0}(v_2;V_{ab}^{(2)})
\\
+ \frac{ I^{\Z_2,(1 \cdot 3)}_{ab}  N_b}{2} \,\left(1-\delta_{\sigma^2_{ab},0} \delta_{\tau^2_{ab},0} \right) \,
\Lambda(\sigma_{ab}^2,\tau_{ab}^2,v_2)
\end{array}
\\\hline\hline
(\phi,-\phi,0) 
&  \frac{N_b}{2} \,\delta_{\sigma^3_{ab},0} \, \delta_{\tau^3_{ab},0} \, | I_{ab}^{(1 \cdot 2)} |
&
\begin{array}{c}
- b_{ab}^{\cal A}  \;  
\Lambda_{0,0}(v_3;V_{ab}^{(3)})
\\
- \frac{N_b}{2} \,  |I_{ab}^{(1 \cdot 2)}| \, \left(1-\delta_{\sigma^3_{ab},0} \delta_{\tau^3_{ab},0} \right) \,
\Lambda(\sigma_{ab}^3,\tau_{ab}^3,v_3)
\\
- \frac{N_b I_{ab}^{\Z_2}}{2} \left( \sgn(\phi) - 2 \, \phi\right) \,  \ln (2)
\end{array}
\\\hline
(\phi,0,-\phi)
& \frac{N_b}{2} \,\delta_{\sigma^2_{ab},0} \, \delta_{\tau^2_{ab},0} \, \left(  | I_{ab}^{(1 \cdot 3)}| - I_{ab}^{\Z_2,(1\cdot 3)} \right)
&
\begin{array}{c}
- b_{ab}^{\cal A}  \; \Lambda_{0,0}(v_2;V_{ab}^{(2)})
\\
- \frac{N_b}{2} \,  \left\{ |I_{ab}^{(1 \cdot 3)}| - I_{ab}^{\Z_2,(1\cdot 3)} \right\} \, \left(1-\delta_{\sigma^2_{ab},0} \delta_{\tau^2_{ab},0} \right) \,
\Lambda(\sigma_{ab}^2,\tau_{ab}^2,v_2)
\end{array}
\\\hline
(0,\phi,-\phi) 
& \frac{N_b}{2} \,\delta_{\sigma^1_{ab},0} \, \delta_{\tau^1_{ab},0} \,|  I_{ab}^{(2 \cdot 3)} |
&
\begin{array}{c}
- b_{ab}^{\cal A}  \; \Lambda_{0,0}(v_1;V_{ab}^{(1)})
\\
- \frac{N_b}{2} \, |I_{ab}^{(2 \cdot 3)}| \, \left(1-\delta_{\sigma^1_{ab},0} \delta_{\tau^1_{ab},0} \right) \, \Lambda(\sigma_{ab}^1,\tau_{ab}^1,v_1)
\\
+ \frac{N_b I_{ab}^{\Z_2}}{2} \, \left( \sgn(\phi) - 2 \, \phi\right) \; \ln(2)
\end{array}
\\\hline\hline
\begin{array}{c}
(\phi^{(1)},\phi^{(2)},\phi^{(3)}) \\ \text{with } \sum_{k=1}^3 \phi^{(k)}=0 \end{array} 
& \frac{N_b}{4} \, \left( | I_{ab} |  +  \sgn(I_{ab}) \, I_{ab}^{\Z_2} \right) 
& 
\begin{array}{c}
b_{ab}^{\cal A}  \, \sgn(I_{ab}) \, 
\sum_{i=1}^3 \ln \left(\frac{\Gamma(|\phi^{(i)}_{ab}|)}{\Gamma(1-|\phi^{(i)}_{ab}|)}\right)^{\sgn(\phi^{(i)}_{ab})}
\\
+ \frac{N_b}{2} \,I_{ab}^{\Z_2} \,  \left(\sgn(I_{ab}) + \sgn(\phi^{(2)}_{ab}) -2 \phi^{(2)}_{ab}  \right) \; \ln (2)
\end{array}
\\\hline
    \end{array}
}{T6-Z2N-Bifund-beta+thresholds}{
$SU(N_a)$ beta function coefficients and gauge thresholds from bifundamental and adjoint matter on fractional D6-branes on the 
$T^6/\Z_{2N}$ orbifold. The $\Z_2 \equiv \Z_2^{(2)}$ subgroup is chosen to leave the second two-torus invariant. The entries on the first 
and third line preserve ${\cal N}=2$ supersymmetry, all other entries have only ${\cal N}=1$.
Details on adjoint matter are discussed in section~\protect\ref{Sss:adjoints} and table~\protect\ref{tab:Adjoint-Beta+Thresholds}.
}

The $SU(N_a)$ beta function coefficients and gauge thresholds for antisymmetric and symmetric matter on $T^6/(\Z_{2N} \times \OR)$ are given in table~\ref{tab:T6-Z2N-Anti+Sym-beta+thresholds} in a greatly simplified form compared to~\cite{Gmeiner:2009fb}, i.e. we have factorised out the beta function coefficients as prefactors of gauge thresholds and used the 
relation~(\ref{Eq:Annulus-Moebius-Gamma-Rewritten}) for the logarithms of Gamma functions in the M\"obius strip contributions. 
\mathtabfix{
\begin{array}{|c||c|c|} \hline
\multicolumn{3}{|c|}{\rule[-3mm]{0mm}{8mm}
\text{\bf  $b_{SU(N_a)}$ and  gauge thresholds for symmetrics and antisymmetrics:} \quad T^6/\Z_{2N} \; \text{\bf with } \Z_2 \equiv \Z_2^{(2)}}
\\\hline\hline
 (\phi_{aa'}^{(1)},\phi_{aa'}^{(2)},\phi_{aa'}^{(3)})
&\!\!\!\!\! \begin{array}{c} 
b_{SU(N_a)}^{\Z_{2N}} = b_{aa'}^{\cal A} + b_{aa'}^{\cal M} + \ldots= \\ 
\frac{N_a}{2} \left(\varphi^{\Sym_a} + \varphi^{\Anti_a} \right) +\left(\varphi^{\Sym_a} - \varphi^{\Anti_a} \right)  + \ldots
\end{array}\!\!\!\!\!& 
\Delta_{SU(N_a)}^{\Z_{2N}} = N_a \, \tilde{\Delta}_{aa'} + \Delta_{a,O6} + \ldots
\\\hline\hline
\begin{array}{c} (0,0,0) \\ \pp \OR \end{array} 
 &
\begin{array}{c}  - \delta_{\sigma^2_{aa'},0} \delta_{\tau^2_{aa'},0} \; \times \\
 \bigl( \frac{N_a I^{\Z_2,(1 \cdot 3)}_{aa'}}{2} + |\tilde{I}_a^{\OR\Z_2,(1 \cdot 3)}| \bigr)
\end{array}
&\!\!\!\!
\begin{array}{c}
- \left( b_{aa'}^{\cal A} + b_{aa'}^{\cal M} \right)  \; \Lambda_{0,0}(v_2;V_{aa'}^{(2)})
- (4 \,b_2) \;  b_{aa'}^{\cal M} \; \ln \left(\frac{\eta(2i v_2)}{\vartheta_4(0,2i v_2)}\right)
- 2 \; b_{aa'}^{\cal M} \; \ln (2)
\\
+ \left(1-\delta_{\sigma^2_{aa'},0} \delta_{\tau^2_{aa'},0} \right) 
\Bigl[\frac{ I^{\Z_2,(1 \cdot 3)}_{aa'}  N_a}{2} \,\Lambda(\sigma_{aa'}^2,\tau_{aa'}^2,v_2)
+|\tilde{I}_a^{\OR\Z_2,(1 \cdot 3)}| \, \Lambda(\sigma_{aa'}^2,\tau_{aa'}^2,\tilde{v}_2) \Bigr]
\\
\end{array}\!\!\!\!
\\\hline
\begin{array}{c} (0,0,0) \\ \pp \OR\Z_2^{(2)} \end{array} 
 &
\begin{array}{c} - \delta_{\sigma^2_{aa'},0} \delta_{\tau^2_{aa'},0} \;  \times 
\\ \bigl( \frac{ N_a \, I^{\Z_2,(1 \cdot 3)}_{aa'}}{2} + |\tilde{I}_a^{\OR,(1 \cdot 3)}| \bigr)
\end{array}
&\!\!\!\!
\begin{array}{c}
- \left( b_{aa'}^{\cal A} +  b_{aa'}^{\cal M} \right) \; \Lambda_{0,0}(v_2;V_{aa'}^{(2)})
- (4 \, b_2) \; b_{aa'}^{\cal M} \; \ln \left(\frac{\eta(2i v_2)}{\vartheta_4(0,2i v_2)}\right)
- 2 \;  b_{aa'}^{\cal M} \; \ln(2)
\\
+\left(1-\delta_{\sigma^2_{aa'},0} \delta_{\tau^2_{aa'},0} \right) \, \Bigl[ \frac{ I^{\Z_2,(1 \cdot 3)}_{aa'}  N_a}{2} \,
\Lambda(\sigma_{aa'}^2,\tau_{aa'}^2,v_2)
+|\tilde{I}_a^{\OR,(1 \cdot 3)}| \, \Lambda(\sigma_{aa'}^2,\tau_{aa'}^2,\tilde{v}_2) \Bigr]
\end{array}\!\!\!\!
\\\hline
\begin{array}{c} (0,0,0) \\ \perp \OR  \text{ on } T_1 \times T_2 \end{array} 
&
\begin{array}{c}  - \Bigl( \frac{N_a \, I^{\Z_2,(1 \cdot 3)}_{aa'} \, \delta_{\sigma^2_{aa'},0} \delta_{\tau^2_{aa'},0}}{2}\\ 
+ |\tilde{I}_a^{\OR,(1 \cdot 2)}| + |\tilde{I}_a^{\OR\Z_2,(2 \cdot 3)}| \Bigr)
\end{array}
&
\begin{array}{c}
- b_{aa'}^{\cal A} \; \Lambda_{0,0}(v_2;V_{aa'}^{(2)}) +|\tilde{I}_a^{\OR,(1 \cdot 2)}|\, \Lambda_{0,0}(\tilde{v}_3;2 \,\tilde{V}_{aa'}^{(3)})
+ |\tilde{I}_a^{\OR\Z_2,(2 \cdot 3)}| \,  \Lambda_{0,0}(\tilde{v}_1;2 \, \tilde{V}_{aa'}^{(1)})
\\
+ \frac{ I^{\Z_2,(1 \cdot 3)}_{aa'}  N_a}{2} \,\left(1-\delta_{\sigma^2_{aa'},0} \delta_{\tau^2_{aa'},0} \right) \,\Lambda(\sigma_{aa'}^2,\tau_{aa'}^2,v_2)
\end{array}
\\\hline
\begin{array}{c} (0,0,0) \\ \perp \OR  \text{ on } T_2 \times T_3 \end{array} 
 &
\begin{array}{c}  -\frac{ N_a \, I^{\Z_2,(1 \cdot 3)}_{aa'} \, \delta_{\sigma^2_{aa'},0} \delta_{\tau^2_{aa'},0}}{2}\\ 
+ \tilde{I}_a^{\OR,(2 \cdot 3)} + \tilde{I}_a^{\OR\Z_2,(1 \cdot 2)} 
\end{array}
&
\begin{array}{c}
- b_{aa'}^{\cal A} \; \Lambda_{0,0}(v_2;V_{aa'}^{(2)}) + |\tilde{I}_a^{\OR,(2 \cdot 3)}| \,  \Lambda_{0,0}(\tilde{v}_1;2 \, \tilde{V}_{aa'}^{(1)})
+ |\tilde{I}_a^{\OR\Z_2,(1 \cdot 2)}| \,  \Lambda_{0,0}(\tilde{v}_3;2 \, \tilde{V}_{aa'}^{(3)})
\\
+ \frac{ I^{\Z_2,(1 \cdot 3)}_{aa'}  N_a}{2} \,\left(1-\delta_{\sigma^2_{aa'},0} \delta_{\tau^2_{aa'},0} \right) \,\Lambda(\sigma_{aa'}^2,\tau_{aa'}^2,v_2)
\end{array}
\\\hline\hline
\begin{array}{c} (\phi,-\phi,0)
\\ \pp \OR  \text{ on } T_3 \end{array}  
&  
 \frac{N_a}{2} \, |I_{aa'}^{(1 \cdot 2)}| - | \tilde{I}_a^{\OR,(1 \cdot 2)}| 
&
\begin{array}{c}
- \left( b_{aa'}^{\cal A}  + b_{aa'}^{\cal M} \right) \; \Lambda_{0,0}(v_3;V_{aa'}^{(3)})
- (4 \, b_3) \;  b_{aa'}^{\cal M} \; \ln \left(\frac{\eta(2i v_3)}{\vartheta_4(0,2i v_3)}\right)
\\
+ \bigl( - \frac{N_a I_{aa'}^{\Z_2}}{2} \left( \sgn(\phi) - 2 \, \phi\right) + |\tilde{I}_a^{\OR\Z_2}| + 2 \,| \tilde{I}_a^{\OR,(1 \cdot 2)}|  
\bigr) \, \ln (2)\\
\end{array}
\\\hline
\begin{array}{c} (\phi,-\phi,0)
 \\ \perp \OR  \text{ on } T_3  \end{array} 
&  \frac{N_a}{2}  \, | I_{aa'}^{(1 \cdot 2)}|  - | \tilde{I}_a^{\OR\Z_2,(1 \cdot 2)}|
&
\begin{array}{c}
- \left( b_{aa'}^{\cal A} + b_{aa'}^{\cal M} \right)  \; \Lambda_{0,0}(v_3;V_{aa'}^{(3)}) 
- (4 \, b_3) \; b_{aa'}^{\cal M} \;\ln \left(\frac{\eta(2i v_3)}{\vartheta_4(0,2i v_3)}\right) \\
+ \bigl( - \frac{N_a I_{aa'}^{\Z_2}}{2} \left( \sgn(\phi) - 2 \, \phi\right) + |  \tilde{I}_a^{\OR} | + 2 \, | \tilde{I}_a^{\OR\Z_2,(1 \cdot 2)}|
\bigr) \,  \ln (2) 
\end{array}
\\\hline\hline
\begin{array}{c} (\phi,0,-\phi)_{\phi \neq \pm \frac{1}{2}} \\ \pp \left(\OR + \OR\Z_2^{(2)}\right) \\ \text{ on } T_2 \end{array} 
&
 \begin{array}{c}
\delta_{\sigma^2_{aa'},0} \, \delta_{\tau^2_{aa'},0} 
 \bigl[ \frac{N_a \,\left( | I_{aa'}^{(1 \cdot 3)} | - I_{aa'}^{\Z_2,(1\cdot 3)} \right)}{2} \\
-| \tilde{I}_a^{\OR,(1 \cdot 3)}| -| \tilde{I}_a^{\OR\Z_2,(1 \cdot 3)} | \bigr]
\end{array} 
&
\begin{array}{c}
-\left(  b_{aa'}^{\cal A}  +  b_{aa'}^{\cal M} \right) \; \Lambda_{0,0}(v_2;V_{aa'}^{(2)})
- (4 \, b_2) \; b_{aa'}^{\cal M}  \; \ln\left(\frac{\eta(2i v_2)}{\vartheta_4(0,2iv_2)} \right)  
\\
+ \left(1-\delta_{\sigma^2_{aa'},0} \delta_{\tau^2_{aa'},0} \right) \times \Bigl[
\frac{N_a \, ( I_{aa'}^{(1 \cdot 2)} + I_{aa'}^{\Z_2,(1\cdot 3)} )}{2} \, \, \Lambda(\sigma_{aa'}^2,\tau_{aa'}^2,v_2)
\\
- \left( \tilde{I}_a^{\OR,(1 \cdot 3)} + \tilde{I}_a^{\OR\Z_2,(1 \cdot 3)} \right) \,\Lambda(\sigma_{aa'}^2,\tau_{aa'}^2, \tilde{v}_2) \Bigr]
\\
- 2 \,  b_{aa'}^{\cal M} \, \ln(2)
\end{array}
\\\hline
\begin{array}{c} (\phi,0,-\phi)_{\phi \neq \pm \frac{1}{2}} \\ \perp \left(\OR + \OR\Z_2^{(2)}\right) \\ \text{ on } T_2 \end{array} 
& 
 \frac{N_a \,\left( | I_{aa'}^{(1 \cdot 3)}| - I_{aa'}^{\Z_2,(1\cdot 3)} \right)}{2}\delta_{\sigma^2_{aa'},0} \, \delta_{\tau^2_{aa'},0}
&
\begin{array}{c}
- b_{aa'}^{\cal A}  \; \Lambda_{0,0}(v_2;V_{aa'}^{(2)})
\\
+ \frac{N_a \,  \left( I_{aa'}^{(1 \cdot 2)} + I_{aa'}^{\Z_2,(1\cdot 3)} \right)}{2} \, \left(1-\delta_{\sigma^2_{aa'},0} \delta_{\tau^2_{aa'},0} \right) \, \Lambda(\sigma_{aa'}^2,\tau_{aa'}^2,v_2)\\
+\bigl(|\tilde{I}_a^{\OR}| +  |\tilde{I}_a^{\OR\Z_2}|  \bigr) \, \ln(2)
\end{array}
\\\hline\hline
\begin{array}{c} (0,\phi,-\phi) \\ \pp \OR  \text{ on } T_1 \end{array} 
& \frac{N_a}{2} \, |I_{aa'}^{(2 \cdot 3)}| -  |\tilde{I}_a^{\OR,(2 \cdot 3)}|
&
\begin{array}{c}
- \left( b_{aa'}^{\cal A} + b_{aa'}^{\cal M} \right) \; \Lambda_{0,0}(v_1;V_{aa'}^{(1)})
- (4 \, b_1) \;  b_{aa'}^{\cal M} \; \ln  \left(\frac{\eta(2iv_1)}{\vartheta_4(0,2iv_1)}\right)
\\
+ \bigl( \frac{N_a \, I_{aa'}^{\Z_2}}{2} \left( \sgn(\phi) - 2 \, \phi\right) + |\tilde{I}_a^{\OR\Z_2}|
+ 2 \, |\tilde{I}_a^{\OR,(2 \cdot 3)}|
\bigr) \,  \ln (2)
\end{array}
\\\hline
\begin{array}{c} (0,\phi,-\phi) \\ \perp \OR  \text{ on } T_1  \end{array} 
& \frac{N_a}{2}\, |I_{aa'}^{(2 \cdot 3)}| -  |\tilde{I}_a^{\OR\Z_2,(2 \cdot 3)}|
&
\begin{array}{c}
- \left( b_{aa'}^{\cal A} + b_{aa'}^{\cal M} \right)   \; \Lambda_{0,0}(v_1;V_{aa'}^{(1)})
- (4 \, b_1) \; b_{aa'}^{\cal M} \;\ln  \left(\frac{\eta(2iv_1)}{\vartheta_4(0,2iv_1)}\right) 
\\
+ \bigl( \frac{N_a \, I_{aa'}^{\Z_2}}{2} \left( \sgn(\phi) - 2 \, \phi\right) + |  \tilde{I}_a^{\OR} |
 +2 \, |\tilde{I}_a^{\OR\Z_2,(2 \cdot 3)}|
\bigr) \,  \ln (2)
\end{array}
\\\hline\hline
\!\!\!\!\!
\begin{array}{c}
(\phi^{(1)},\phi^{(2)}, \phi^{(3)})
\\
0<|\phi^{(i)}|,|\phi^{(j)}| \leq |\phi^{(k)}|<1
\\
\sgn(\phi^{(i)}) = \sgn(\phi^{(j)})\\
 \neq \sgn(\phi^{(k)})
\end{array}
& \begin{array}{c}\frac{N_a \left( |I_{aa'}|  + \sgn(I_{aa'}) I_{aa'}^{\Z_2} \right)}{4} \\
+ \frac{c_a^{\OR} \,|\tilde{I}_a^{\OR}| + \, c_a^{\OR\Z_2} \,|\tilde{I}_a^{\OR\Z_2}|}{2}
\end{array}
& 
\begin{array}{c}
\bigl( b_{aa'}^{\cal A}   + b_{aa'}^{\cal M} \bigr) 
\; \sgn(I_{aa'}) \; \sum_{i=1}^3 \ln \left(\frac{\Gamma(|\phi^{(i)}_{aa'}|)}{\Gamma(1-|\phi^{(i)}_{aa'}|)}\right)^{\sgn(\phi^{(i)}_{aa'})}\\
+ \bigl(\frac{N_a \,I_{aa'}^{\Z_2}}{2} \left(\sgn(I_{aa'}) +\sgn(\phi^{(2)}_{aa'}) -2 \phi^{(2)}_{aa'}  \right)
+ |\tilde{I}_a^{\OR}| + |\tilde{I}_a^{\OR\Z_2}| \bigr) 
\, \ln(2)
\end{array}
\\\hline
    \end{array}
}{T6-Z2N-Anti+Sym-beta+thresholds}{$SU(N_a)$ beta function coefficients and gauge thresholds from symmetric and antisymmetric matter on
$T^6/(\Z_{2N} \times \OR)$. If the D$6_a$-brane is parallel or perpendicular to one of the O6-planes with vanishing relative displacement and 
Wilson line, $(\sigma_{aa'}^2,\tau_{aa'}^2)=(0,0)$, along the $\Z_2$ invariant two-torus $T^2_{(2)}$,  the gauge group is enhanced to $Sp(2N_a)$,
cf. section~\protect\ref{Sss:AntiSym} and table~\protect\ref{tab:SO-Sp-groups}.
}

\subsubsection{Example I: the adjoints of $SU(N_a)$}\label{Sss:adjoints}

In order to further evaluate the amount of adjoint matter on each factorisable toroidal orbifold background, we can use the fact, that the toroidal intersection numbers
of orbifold image D6-branes for every single case can be derived using the relations~(\ref{Eq:Torus-wrapping-adjoint}),
\begin{equation}\label{Eq:Inters-for-Adjoints}
\begin{aligned}
T^6/\Z_3: & \quad I_{(\theta a)a} = - I_{(\theta^2 a)a}= - \prod_{i=1}^3 \left[(n_a^i)^2 + n_a^i m_a^i + (m_a^i)^2   \right],
 \\ 
T^6/\Z_4: & \quad I_{(\theta a)a} = -0^{(2)} \cdot  \prod_{i=1,3} \left[(n_a^i)^2 +  (m_a^i)^2  \right]  ,
 \\ 
T^6/\Z_6: & \quad I_{(\theta a)a} = - I_{(\theta^2 a)a}= \prod_{i=1}^3 \left[(n_a^i)^2 + n_a^i m_a^i + (m_a^i)^2   \right],
 \\ 
 T^6/\Z_6': & \quad I_{(\theta a)a} = I_{(\theta^2 a)a}=  - 0^{(3)} \cdot \prod_{i=1,2} \left[(n_a^i)^2 + n_a^i m_a^i + (m_a^i)^2   \right],
 \\ 
 T^6/\Z_2 \times \Z_4: & \quad I_{(\omega a)a} = - 0^{(1)} \cdot \prod_{i=2,3} \left[ (n_a^i)^2 +  (m_a^i)^2  \right],
 \\ 
 T^6/\Z_2 \times \Z_6: & \quad I_{(\omega a)a} =I_{(\omega^2 a)a} = - 0^{(1)} \cdot \prod_{i=2,3} \left[(n_a^i)^2 + n_a^i m_a^i + (m_a^i)^2  \right],
 \\ 
T^6/\Z_2 \times \Z_6' :& \quad I_{(\omega a)a} = - I_{(\omega^2 a)a} = \prod_{i=1}^3 \left[(n_a^i)^2 + n_a^i m_a^i + (m_a^i)^2   \right],
\end{aligned}
\end{equation}
and the explicit expressions for the contributions to the gauge thresholds are obtained from
the relative angles, which are given by $2\pi \vec{v}$ and $2\pi \vec{w}$ for $T^6/\Z_{2N}$ and $T^6/\Z_2 \times \Z_{2M}$ orbifolds,
respectively, with the shift vectors listed in table~\ref{Tab:T6ZN+T6Z2Z2M-shifts}.
The results for all factorisable toroidal orbifold backgrounds are compared in table~\ref{tab:Adjoint-Beta+Thresholds}.
\mathtabfix{\vspace{-10mm}
\begin{array}{|c|c||c|c|}\hline
\multicolumn{4}{|c|}{\text{\bf Beta function coefficients $b_{SU(N_a)}$ and gauge thresholds $\Delta_{SU(N_a)}$ for the adjoints}}
\\\hline\hline
k & \begin{array}{c} ( \vec{\phi}_{(\theta^k a)a}) \\ \text{or } \theta \leftrightarrow \omega  \end{array} &  \begin{array}{c} b_{SU(N_a)} = \sum _k b_{(\theta^k a)a}^{\cal A} + \ldots  \\
=   N_a (-3 + \sum_k \varphi^{\Adj_{a,k}} ) + \ldots  \end{array}
& \Delta_{SU(N_a)} = N_a \sum_k \tilde{\Delta}_{(\theta^k a)a} + \ldots 
\\\hline\hline
\multicolumn{4}{|c|}{T^6 \text{ and } T^6/\Z_3}
\\\hline\hline
0 & (0,0,0) & - & - 
\\\hline
1 + 2 & \mp (\frac{1}{3},-\frac{2}{3},\frac{1}{3}) & N_a \, |I_{(\theta a)a}| &  - 6 \, b_{(\theta a)a} \ln (2)
\\\hline\hline\hline
\multicolumn{4}{|c|}{T^6/\Z_{2N}}
\\\hline\hline
0 & (0,0,0) & -2 \, N_a & 2 \, N_a \, \Lambda_{0,0}(v_2;V_{aa}^{(2)})
\\\hline
\multicolumn{4}{|c|}{T^6/\Z_4}
\\\hline
1 & (\frac{1}{2},0,-\frac{1}{2}) & \frac{N_a \, \left( | I_{(\theta a)a}^{(1 \cdot 3)} | -  I_{(\theta a)a}^{\Z_2,(1 \cdot 3)}   \right)}{2} & -b_{(\theta a)a}^{\cal A} \, \Lambda_{0,0}(v_2; V_{aa}^{(2)})
\\\hline
\multicolumn{4}{|c|}{T^6/\Z_6}
\\\hline
1 + 2 & \pm (\frac{1}{3},-\frac{2}{3},\frac{1}{3}) & 
 \frac{N_a \, \left( |I_{(\theta a)a}| + \sgn(I_{(\theta a)a}) \cdot I_{(\theta a)a}^{\Z_2} \right)}{2} &  - \left( 6 \, b_{(\theta a)a}^{\cal A} + \frac{2 \, N_a \, I_{(\theta a)a}^{\Z_2}}{3} \right)\, \ln(2)
\\\hline
\multicolumn{4}{|c|}{T^6/\Z_6'} 
\\\hline
1 + 2 & \pm (\frac{1}{3},-\frac{1}{3},0) & 
N_a  \,  | I_{(\theta a)a}^{(1 \cdot 2)} | 
&  -2 \, b_{(\theta a)a}^{\cal A} \; \Lambda_{0,0}(v_3; V_{aa}^{(3)}) 
-N_a \, \frac{I_{(\theta a)a}^{\Z_2} }{3} \, \ln(2)
\\\hline\hline\hline
\multicolumn{4}{|c|}{T^6/\Z_2 \times \Z_{2M} \text{ without discrete torsion}}
\\\hline\hline
0 & (0,0,0) & - & - 
\\\hline
\multicolumn{4}{|c|}{T^6/\Z_2 \times \Z_4 \text{ without and with discrete torsion}}
\\\hline
1 & (0,\frac{1}{2},-\frac{1}{2}) & N_a \, |I_{(\omega a)a}^{(2 \cdot 3)} | 
& - b_{(\omega a)a}^{\cal A} \; \Lambda_{0,0} (v_1;V_{aa}^{(1)})
\\\hline
\multicolumn{4}{|c|}{T^6/\Z_2 \times \Z_6 \text{ without discrete torsion}}
\\\hline
1 + 2 & \pm(0,\frac{1}{3},-\frac{1}{3}) & 2 N_a \,  |I_{(\omega a)a}^{(2 \cdot 3)} | 
&  - 2 \, b_{(\omega a)a}^{\cal A} \; \Lambda_{0,0} (v_1;V_{aa}^{(1)})
\\\hline
\multicolumn{4}{|c|}{T^6/\Z_2 \times \Z_6' \text{ without discrete torsion}}
\\\hline
1 + 2 & \pm(-\frac{2}{3},\frac{1}{3},\frac{1}{3}) &  N_a \, | I_{(\omega a)a} |
& - 6 \, b_{(\omega a)a}^{\cal A} \, \ln (2)
\\\hline\hline\hline
\multicolumn{4}{|c|}{T^6/\Z_2 \times \Z_{2M} \text{ with discrete torsion}}
\\\hline\hline
0 & (0,0,0) & -3 \, N_a &   N_a \sum_{i=1}^3 \Lambda_{0,0}(v_i;V_{aa}^{(i)})
\\\hline
\multicolumn{4}{|c|}{T^6/\Z_2 \times \Z_6 \text{ with discrete torsion}}
\\\hline
1 + 2 & \pm(0,\frac{1}{3},-\frac{1}{3}) & 
 \frac{N_a  \left(  | I _{(\omega a)a}^{(2 \cdot 3)} | -I _{(\omega a)a}^{\Z_2,(2 \cdot 3)}   \right)}{2}
& 
- 2 \, b_{(\omega a)a}^{\cal A} \; \Lambda_{0,0}(v_1;V_{aa}^{(1)}) + \frac{N_a \, I_{(\omega a)a}^{\Z_2}}{6} \, \ln (2)
\\\hline
\multicolumn{4}{|c|}{T^6/\Z_2 \times \Z_6' \text{ with discrete torsion}}
\\\hline
1 + 2 & \pm(-\frac{2}{3},\frac{1}{3},\frac{1}{3}) & 
\frac{N_a  \left(   | I _{(\omega a)a} | + \sgn(I _{(\omega a)a} ) \sum_{i=1}^3 I _{(\omega a)a}^{\Z_2^{(i)}}   \right)}{4}
& \begin{array}{c}
\left( -6 \, b_{(\omega a)a}^{\cal A}  + \frac{N_a}{3} \, \sum_{i=1}^3 I_{(\omega a)a}^{\Z_2^{(i)}}  \right)\; \ln(2)
\end{array}
\\\hline
\end{array}
}{Adjoint-Beta+Thresholds}{$SU(N_a)$ beta function coefficients and gauge thresholds due to massless and massive open strings
transforming in the adjoint representation for all factorisable toroidal orbifold backgrounds. The toroidal intersection numbers 
are given in equation~(\protect\ref{Eq:Inters-for-Adjoints}). The corresponding K\"ahler metrics for adjoints on identical D6-branes 
are given in equation~(\protect\ref{Eq:Def-Kaehler_adjoints}), while those at intersections of orbifold images $(\omega^{k} a)a$ 
have the same form as the K\"ahler metrics for bifundamental matter in table~\protect\ref{tab:Comparison-Kaehler-bifund}. 
Similarly, the $v_i$ dependent
one-loop contributions to the holomorphic gauge kinetic function are given in equation~(\protect\ref{Eq:Def-1loop_hol}) and 
table~\protect\ref{tab:Comparison-gaugekin-bifund}. The angle dependent loop corrections to the gauge kinetic function are again identical to the contributions 
from bifundamentals in~(\protect\ref{Eq:3angle-f-angle}).
}

One can directly read off from table~\ref{tab:Adjoint-Beta+Thresholds} that the $aa$-sector on the six-torus, $T^6/\Z_3$ and $T^6/\Z_2 \times \Z_{2M}$ 
without discrete torsion preserves ${\cal N}=4$ supersymmetry (i.e. there exist three matter multiplets in the adjoint representation) and therefore
does not contribute to the beta function coefficient and gauge threshold. The $aa$-sector on $T^6/\Z_{2N}$ preserves only ${\cal N}=2$ and on
$T^6/\Z_2 \times \Z_{2M}$ with discrete torsion ${\cal N}=1$ supersymmetry, which corresponds to one and no adjoint matter multiplet, respectively,
as well as a non-vanishing gauge threshold which depends on the K\"ahler moduli $v_i$ via the Kaluza-Klein and winding sums
in the first line of~(\ref{Eq:Def-Lambdas}) for both ${\cal N}=1,2$. 
This completes the classification of open strings transforming in the adjoint representation on the six-torus and the $T^6/\Z_2 \times \Z_2$ orbifolds 
without and with discrete torsion.

The $T^6/\Z_4$ and $T^6/\Z_2 \times \Z_4$ orbifolds have one sector $(\theta a)a$ or $(\omega a)a$, respectively, of intersections of orbifold image D6-branes
which can provide additional adjoint matter. While on $T^6/\Z_4$ the (non)existence of matter depends on the $\Z_2$ invariance of the intersection point 
of the D$6_{(\theta a)}$-  and D$6_a$-brane
and the combination of discrete displacements and Wilson lines $(\sigma_a^i,\tau_a^i)_{i \in \{1,3\}}$ along the directions where $\Z_2$ acts non-trivially,
on $T^6/\Z_2 \times \Z_4$ there exists always one adjoint matter multiplet per intersection of orbifold image D$6_{(\omega a)}$- and D$6_a$-branes.
In both cases, the gauge threshold contribution is due to massive strings at the same intersection points and consists of a Kaluza-Klein and winding sum
along the two-torus where the orbifold images D6-branes are parallel to each other.

On the $T^6/\Z_N$ orbifolds with $N\in \{3,6,6'\}$ as well as the $T^6/\Z_2 \times \Z_{2M}$ orbifolds with $2M \in \{6,6'\}$, 
the $(\theta a)a$ and $(\theta^2 a)a$ [or $(\omega a)a$ and $(\omega^2 a)a$] sectors are paired up to provide the two helicity states and scalar degrees
of freedom of one massless chiral multiplet with a given $\Z_2$ transformation behaviour per intersection.
The contributions from adjoints at intersections of orbifold images in table~\ref{tab:Adjoint-Beta+Thresholds} can be classified along two different lines: 
on $T^6/\Z_6'$ and $T^6/\Z_2 \times \Z_6$ there is one vanishing angle leading to a K\"ahler modulus dependence of the gauge threshold along this
two torus, whereas for $T^6/\Z_3$, $T^6/\Z_6$ and $T^6/\Z_2 \times \Z_6$ the orbifold images intersect non-trivially along all three tori and the
Gamma functions can be evaluated explicitly at the intersection angles, cf. table~\ref{Tab:Gamma-values}.
\begin{table}[ht]
\begin{center}
\begin{equation*}
\begin{array}{|c||c|c|c|c|c|c|c|}\hline
\phi & 1/6 &1/3 & 1/2 & 2/3 & 5/6 
\\\hline\hline
\frac{ \Gamma(\phi)}{\Gamma(1-\phi)} &  5 & 2 & 1 & \frac{1}{2} & \frac{1}{5}
\\\hline
\end{array}
\end{equation*}
\end{center}
\caption{Ratios of Gamma functions for special values of intersection angles commonly appearing in D6-brane
models on toroidal orbifolds.}
\label{Tab:Gamma-values}
\end{table}
On the other hand, the $T^6/\Z_3$ and $T^6/\Z_2 \times \Z_{2M}$ orbifolds without discrete torsion only have contributions from the untwisted annulus amplitude
to the gauge thresholds, and the beta function coefficient appears as a global prefactor, whereas the $T^6/\Z_{2N}$ and $T^6/\Z_2 \times \Z_{2M}$ orbifolds with
discrete torsion ($2N,2M \in \{6,6'\}$) have $\Z_2$ twisted annulus amplitudes contributing to the gauge threshold. Since in the latter case, the beta function coefficient
cannot be factored out of the  gauge thresholds, we conclude that (some of) the massive open string modes transform differently from the massless modes under 
the $\Z_2$ transformations.

\subsubsection{Example 2: (anti)symmetric matter of $SO(2M_x)$ and $Sp(2M_x)$ gauge groups}\label{Sss:AntiSym}

The orientifold invariant D6-branes on the six-torus, $T^6/\Z_3$ and $T^6/\Z_2 \times \Z_{2M}$ orbifolds without discrete torsion wrap bulk three-cycles 
which are either parallel or perpendicular (along some four-torus) to some O6-plane.
For $T^6/\Z_{2N}$ and $T^6/\Z_2 \times \Z_{2M}$ orbifolds with discrete torsion, in addition the exceptional contributions to the fractional or rigid three-cycles
have to be mapped to themselves in order to obtain orientifold invariant three-cycles. 
A classification of these three-cycles for the $T^6/\Z_6'$ background is given in~\cite{Gmeiner:2007zz},~\cite{Honecker:2004kb,Gmeiner:2007we} contain some 
examples and comments for $T^6/\Z_6$ and some explicit examples on these two orbifolds are discussed in appendix A.3 of~\cite{Gmeiner:2009fb};
finally for $T^6/\Z_2 \times \Z_{2M}$ with discrete torsion the complete classification in table~\ref{Tab:Conditions-on_b+t+s-SOSp-Z2Z2M}
is reproduced from~\cite{Forste:2010gw}; for $T^6/\Z_4$ there exists to our knowledge no discussion of $\OR$ invariant fractional D6-branes.

The beta function coefficients and gauge thresholds from the $xx$ sector of orientifold invariant D$6_x$-branes on all factorisable toroidal orbifolds 
background are listed in table~\ref{tab:SO-Sp-groups} together with the amount of supersymmetry preserved by the D$6_x$-brane and the type of gauge group
and matter content from the $xx$-sector.
\mathtabfix{
\begin{array}{|c||c|c||c|c|}\hline
\multicolumn{5}{|c|}{\text{\bf Beta function coefficients and gauge thresholds for $SO(2M_x)$ and $Sp(2M_x)$}}
\\\hline\hline
(\phi_{xx'}^{(1)},\phi_{xx'}^{(2)},\phi_{xx'}^{(3)}) & \begin{array}{c} b_{SO/Sp(2M_x)} = \\ b_{xx'}^{\cal A} + b_{xx'}^{\cal M} + \ldots \end{array}
& \Delta_{SO(2M_x)} = M_x \tilde{\Delta}_{xx'} + \frac{1}{2} \Delta_{x,\OR} + \ldots
&\begin{sideways} \!\!\!\!\!\!\!\! \text{SUSY} \end{sideways} & \!\!\!\!\begin{array}{c} \text{(exotic O-plane)} \\ \text{gauge group} \\ + \text{ matter} \end{array}\!\!\!\!
\\\hline\hline
\multicolumn{5}{|c|}{T^6 \text{ and } T^6/\Z_3 }
\\\hline
\begin{array}{c}   (0,0,0)  \\ \pp \OR \end{array}  &  - &  - 
&\begin{sideways}  \!\!\!\!\!\!\!\! ${\cal N}=4$  \end{sideways} & \begin{array}{c} SO(2M_x)  \\ + 3 \, \Adj_x  \end{array}
\\\hline
\begin{array}{c}   (0,0,0)  \\ \perp \OR \text{ on } \\ T^2_{(j)} \times T^2_{(k)}\end{array}  & -4   & 4 \, \Lambda_{0,0}(v_i, V_{xx'}^{(i)})
+ 8 \, \ln (2) + 16 \, b_i \, \ln \left( \frac{\eta(2iv_i)}{\vartheta_4(0,2iv_i)} \right)
& \begin{sideways} \!\!\!\!\!\!\!\! ${\cal N}=2$ \end{sideways} &  \begin{array}{c} Sp(2M_x)  \\ +  \Sym_x  \\ + 2 \, \Anti_x \end{array}
\\\hline\hline
\multicolumn{5}{|c|}{T^6/\Z_{2N} }
\\\hline
\!\!\!\!\begin{array}{c}  (0,0,0) \\ \pp \OR \text{ or } \OR\Z_2^{(2)} \end{array}\!\!\!\!  & - 2M_x -2 
& (2M_x+2) \, \Lambda_{0,0}(v_2, V_{xx'}^{(2)}) + 4 \, \ln(2) + 8 \, b_2 \, \ln \left( \frac{\eta(2iv_2)}{\vartheta_4(0,2iv_2)} \right)
& \begin{sideways} \!\!\!\!\!\!\!\! ${\cal N}=2$ \end{sideways}&  \begin{array}{c} Sp(2M_x)  \\ +  \Sym_x   \end{array} 
\\\hline
\!\!\!\!\begin{array}{c}  (0,0,0) \\ \perp  \OR \text{ and } \OR\Z_2^{(2)} \\ \text{ on } T^2_{(2)} \end{array}\!\!\!\!  & - 2M_x - 4 & 
\begin{array}{c} 2 M_x \, \Lambda_{0,0}(v_2, V_{xx'}^{(2)}) + 2 \, \sum_{i=1,3} \Lambda_{0,0}(v_i, V_{xx'}^{(i)}) 
\\ + 8 \,  \ln (2) + 8 \sum_{i=1,3} b_i \;\ln \left( \frac{\eta(2iv_i)}{\vartheta_4(0,2iv_i)} \right) \end{array}  
& \begin{sideways} \!\!\!\!\!\!\!\! ${\cal N}=1$ \end{sideways}&  \begin{array}{c} Sp(2M_x)  \\ +  \Anti_x   \end{array}  
\\\hline\hline
\multicolumn{5}{|c|}{T^6/\Z_{2}\times \Z_{2M} \text{ without discrete torsion}}
\\\hline
(0,0,0) & -6 &  2 \sum_{i=1}^3 \Lambda_{0,0}(v_i, V_{xx'}^{(i)}) + 12 \, \ln (2) + 8 \, \sum_{i=1}^3 b_i \;  \ln \left( \frac{\eta(2iv_i)}{\vartheta_4(0,2iv_i)} \right)
 & \begin{sideways} \!\!\!\!\!\!\!\! ${\cal N}=1$ \end{sideways}&   \begin{array}{c} Sp(2M_x)  \\ + 3\, \Anti_x   \end{array}  
\\\hline\hline
\multicolumn{5}{|c|}{T^6/\Z_{2}\times \Z_{2M} \text{ with discrete torsion}}
\\\hline
\begin{array}{c} (0,0,0) \\  \pp \OR\Z_2^{(k)} \\ k=0 \ldots 3  \end{array} 
& -3 M_x -3 &  \!\!\!\!\!\!\begin{array}{c}
\sum_{i=1}^3 (M_x + 1 ) \; \Lambda_{0,0}(v_i, V_{xx'}^{(i)})\\ + 6 \, \ln(2)
+ 4 \sum_{i=1}^3  b_i \;  \ln \left( \frac{\eta(2iv_i)}{\vartheta_4(0,2iv_i)} \right) \end{array}\!\!\!\!\!\!
& \begin{sideways} \!\!\!\!\!\!\!\! ${\cal N}=1$ \end{sideways}&  \begin{array}{c} (\eta_{\OR\Z_2^{(k)}}=-1) \\ Sp(2M_x)  \end{array}
\\\hline
\end{array}
}{SO-Sp-groups}{Beta function coefficients and gauge thresholds for D$6_x$-branes which are their own orientifold image D$6_{x'}$. 
The amount of supersymmetry preserved for each D6-brane is given in the fourth column, and in the last column the type of (pseudo)real
gauge group and matter content from the $xx$-sector is given.
The special intersection numbers with O6-planes~(\protect\ref{Eq:Special-OR-Intersections}) have been inserted.
Additional matter in the symmetric or antisymmetric representation generically exists at the intersection of orbifold image D6-branes, but
a systematic case-by-case study for any orbifold and lattice orientation goes beyond the scope of this article. An example for each of the
two distinct classes of $Sp(2N_x)$ groups on the {\bf ABa} lattice on $T^6/\Z_6'$ is presented in section~\ref{S:Z6p-Example}.
}

While the gauge group and matter content  from the $xx$-sector is independent of the background lattice orientations  - but the bulk wrapping numbers and 
discrete Wilson lines and displacements of  an invariant three-cycle depend on the lattice - 
 the matter content at intersections of orbifold image D$6_{(\theta^k x)}$-branes depends
on the $\OR$-invariance of the intersection points and therefore the background lattice.
Some examples on $T^6/\Z_6$ and $T^6/\Z_6'$  are given in appendix A.3 of~\cite{Gmeiner:2009fb}.
A complete analysis is time-consuming and at this point not very illuminating, but we have given the fully generic prescription for evaluating  examples
in the previous sections, and for $T^6/\Z_6'$ the examples of $Sp(2)_c$ and $Sp(6)_{h_3}$ gauge groups perpendicular and parallel along $T^2_{(2)}$
to the O6-planes, respectively, are evaluated in section~\ref{S:Z6p-Example}.

\subsection{Comments on anomaly-free $U(1)$ gauge groups}\label{Ss:Comments_on_U1}

For $U(1)_a \subset U(N_a)$ gauge couplings, the normalisation of the tree level gauge coupling~(\ref{Eq:tree-gauge-coupling}) changes
compared to the non-Abelian $SU(N_a) \subset U(N_a)$ factor (see e.g.~\cite{Ghilencea:2002da,Gmeiner:2008xq}),
\begin{equation}\label{Eq:Def-gauge-U1a}
\frac{1}{g^2_{U(1)_a,{\rm tree}}}= \frac{2N_a}{g_{a,{\rm tree}}^2}
,
\end{equation}
and for massless linear combinations of Abelian gauge factors, $U(1)_X = \sum_i x_i U(1)_i$, the tree level gauge coupling 
is a sum of contributions from different D$6_i$-branes,
\begin{equation}\label{Eq:Def-gauge-U1X}
\frac{1}{g^2_X} = \sum_i x_i^2 \frac{1}{g^2_{U(1)_i}}
.
\end{equation}
At one-loop, the beta function coefficient contains the same factor $2N_a$ compared to the $SU(N_a)$ case, and adjoint matter of $U(N_a)$
is uncharged under $U(1)_a$, whereas symmetric and antisymmetric matter has twice the charge of states transforming in the fundamental representation.
The result for a single (unphysical) $U(1)_a$ gauge factor,
\begin{equation}\label{Eq:beta-U1a}
\begin{aligned}
b_{U(1)_a}=& \, 2 \, N_a \, \left( \underbrace{N_a \left(\varphi^{\Sym_a}+\varphi^{\Anti_a} \right)}
+ \underbrace{\left(\varphi^{\Sym_a}-\varphi^{\Anti_a} \right)} + \underbrace{\sum_{b\neq a}  \frac{N_b}{2} \left( \varphi^{ab} + \varphi^{ab'}\right)}  
\right) 
\\
\equiv &  \, 2 \, N_a \, \left( \quad\qquad 2 \, b_{aa'}^{\cal A} \qquad\qquad + \quad\qquad  b_{aa'}^{\cal M}  \quad\qquad + \quad  \sum_{b \neq a} 
\left(b_{ab}^{\cal A} + b_{ab'}^{\cal A} \right) \quad  \right)
,
\end{aligned}
\end{equation}
can be directly compared to the $SU(N_a)$ beta function coefficient in~(\ref{Eq:Expand-beta_function}).

For massless $U(1)_X$ factors, the beta function coefficient consists of a sum over the beta function coefficients per D$6_i$-brane
plus corrections from bifundamental matter on two such D6-branes~\cite{Gmeiner:2008xq},
\begin{equation}\label{Eq:beta-U(1)X}
b_{U(1)_X} =
\sum_i  x_i^2 \; b_{U(1)_i} + 2 \, \sum_{i < j}\,  N_i \,  N_j \,  x_i \,  x_j \,  \left(-\varphi^{ij} + \varphi^{ij'}\right)
.
\end{equation}
Also the gauge threshold corrections to a single $U(1)_a$ factor
due to massive strings running in the loop can be expressed in terms of the building blocks 
$\tilde{\Delta}$ for the $SU(N_a)$ case~\cite{Forste:2010gw},
\begin{equation}\label{Eq:Threh-U(1)a}
\begin{aligned}
\Delta_{U(1)_a} &=2 \, N_a \left(2 \, N_a  \tilde{\Delta}_{aa'}  +   \Delta_{a,\OR}
+  \sum_{b\neq a} N_b  \left( \tilde{\Delta}_{ab} + \tilde{\Delta}_{ab'}
\right)  \right)
\\
&= 2 \, N_a \, \left( \Delta_{SU(N_a)} + N_a \, \left(\tilde{\Delta}_{aa'}   -  \tilde{\Delta}_{aa} \right) \right)
,
\end{aligned}
\end{equation}
and finally the gauge thresholds for massless linear combinations $U(1)_X$ are given by
\begin{equation}\label{Eq:Threh-U(1)X}
\Delta_{U(1)_X} = \sum_i x_i^2 \, \Delta_{U(1)_i} + 4 \sum_{i<j} \,  N_i \, N_j \, x_i \, x_j \, \left(-\tilde{\Delta}_{ij} + \tilde{\Delta}_{ij'} \right) 
.
\end{equation}
This completes the necessary input data for deriving the holomorphic gauge kinetic function of Abelian gauge factors,
which are included in the general discussion in the following section.

\section{K\"ahler metrics, K\"ahler potential and holomorphic gauge kinetic function at one loop}\label{S:Kaehler_metrics+potential}

In the previous section, the corrections to the gauge couplings were computed via string one-loop amplitudes
leading to formula~(\ref{Eq:Def-gauge-SUNa}). In field theory, the gauge couplings are given by~\cite{Shifman:1986zi,Shifman:1991dz}
\begin{equation}\label{Eq:Gauge-FieldTheory}
\begin{aligned}
\frac{8 \pi^2}{g_a^2(\mu)} =& 8 \pi^2 \; \Re({\rm f}_a) +\frac{b_a}{2} \ln\left(\frac{M_{\rm Planck}^2}{\mu^2}  \right)
+ \frac{b_a + 2 \, C_2(\Adj_a)}{2} \; {\cal K} \\ & + C_2(\Adj_a) \; \ln [g_a^{-2}(\mu^2)] - \sum_a C_2({\bf R}_a) \; \ln \det K_{{\bf R}_a}(\mu^2)
,
\end{aligned}
\end{equation}
where ${\rm f}_a$ is the holomorphic gauge kinetic function, $K_{{\bf R}_a}$ the K\"ahler metric of the representation ${\bf R}_a$
under the gauge group and $C_2({\bf R}_a)$ the quadratic Casimir with \linebreak \mbox{${\bf R}_a \in \{(\N_a,\ov{\N}_b), (\N_a,\N_b), \Anti_a,\Sym_a,\Adj_a\}$}
 (or some complex conjugate) of $U(N_a) \times U(N_b)$ for D-brane models, and $SO(2M_x)$ or $Sp(2M_x)$ in case of orientifold invariant D-branes. 
${\cal K}$ denotes the K\"ahler potential and $b_a$ the beta function coefficient of the gauge group $G_a \in \{SU(N_a),SO(2M_a),Sp(2M_a),U(1)_a\}$.

\subsection{Tree level gauge kinetic function}\label{Ss:Tree-GaugeKin}

The holomorphic gauge kinetic function and K\"ahler metrics are obtained by matching the expressions~(\ref{Eq:Def-gauge-SUNa}) 
and~(\ref{Eq:Gauge-FieldTheory}) stepwise, namely first at tree level, 
\begin{equation}\label{Eq:tree-Mass_ratios}
\frac{1}{g^2_{a,{\rm tree}}} = \frac{(2\pi)^{3/2}}{c_a \, k_a} \left(S \prod_{l=1}^3 U_l \right)^{1/4} \, 
\prod_{i=1}^3 \sqrt{V_{aa}^{(i)}}
\stackrel{!}{=} \Re({\rm f}_a^{\text{tree}})
,
\end{equation}
where $h_{21}^{\rm bulk}=3$ has been used and $S$ and $U_l$ are the four dimensional dilaton and bulk complex structure moduli defined in 
equation~(\ref{Eq:Def-S+Ui=3}) below. 
Modifications for $h_{21}^{\rm bulk}=1,0$ are discussed in section~\ref{Ss:Modifications-less-complex} below.
The factor $\prod_{i=1}^3 \sqrt{V_{aa}^{(i)}}$, which is the ratio of the  three-cycle volume to the square root of the total compact volume
and depends on the ratios of two-torus radii
(cf. the definition in equation~(\ref{Eq:Def-Vab})), reduces to a holomorphic expression of the complex structure moduli upon supersymmetry as follows.
A generic three-cycle can be decomposed into orientifold even and odd components,
\begin{equation}\label{Eq:def-XY}
\Pi_a = \sum_{i=0}^{h_{21}} \left( \tilde{X}^i_a \; \Pi_i^{\rm even} + \tilde{Y}_a^i \; \Pi_i^{\rm odd} \right)
,
\qquad
\Pi_{a'} = \sum_{i=0}^{h_{21}} \left( \tilde{X}^i_a \; \Pi_i^{\rm even} - \tilde{Y}_a^i \; \Pi_i^{\rm odd} \right)
,
\end{equation}
and in terms of the corresponding wrapping numbers $(\tilde{X}^i_a , \tilde{Y}_a^i)$,
the bulk supersymmetry conditions can be cast into the form
\begin{equation}\label{Eq:Bulk-SUSY-General}
\sum_{i=0}^{h_{21}^{\rm bulk}} \tilde{Y}_a^i \, f_i(r_k)=0,
\qquad
\sum_{i=0}^{h_{21}^{\rm bulk}} \tilde{X}_a^i\, g_i(r_k)>0,
\end{equation}
where $f_i(r_k)$ and $g_i(r_k)$ are functions of the ratio of radii $r_k$ of the three two-tori $T^2_{(k)}$
that can be rewritten in terms of linear dependences on the dilaton and bulk complex structure moduli, cf. equation~(\ref{Eq:Def-S+Ui=3}) below.
For a suitable choice of the global prefactor in the functions $f_i(r_k)$ and $g_i(r_k)$, one can show on a case-by-case 
basis that the (length)${}^2$ of the bulk three-cycles can be written as
\begin{equation}\label{Eq:Cycles-Decomposition_even+odd}
\prod_{i=1}^3 V_{aa}^{(i)} = \left[\sum_{i=0}^{h_{21}^{\rm bulk}} \tilde{X}_a^i\, g_i(r_k) \right]^2 
+ \left[\sum_{i=0}^{h_{21}^{\rm bulk}} \tilde{Y}_a^i \, f_i(r_k) \right]^2
,
\end{equation}
where the second term drops out due to supersymmetry and the first term is the square of a holomorphic function
which is linear in the dilaton and bulk complex structure moduli
and depends on the choice of the orbifold and orientifold invariant background lattice.
At this point, we discuss the six-torus and its $T^6/\Z_2$ and  $T^6/\Z_2 \times \Z_2$ orbifolds in detail, while the orbifolds  
$T^6/\Z_4$ and $T^6/\Z_2 \times \Z_4$ as well as $T^6/\Z_6'$ and
$T^6/\Z_2 \times \Z_6$ with one bulk complex structure modulus each and $T^6/\Z_6$ and $T^6/\Z_2 \times \Z_6'$ without bulk complex structures
are relegated to appendix~\ref{App:V-Z4} to~\ref{App:V-Z6}.

For arbitrary untilted {\bf a}-type and tilted {\bf b}-type lattices ($b_i \in \{0,\frac{1}{2}\}$), the $\OR$-even and odd bulk three cycles on $T^6/\Z_2 \times \Z_2$ 
are given by (with $(i,j,k)$ cyclic permutations of (1,2,3), see e.g.~\cite{Blumenhagen:2004xx,Gmeiner:2005vz,Forste:2010gw})
\begin{equation}\label{Eq:Def-Z2Z2-bulk_even+odd-cycles}
\begin{aligned}
\Pi^{\rm even}_{0}&= \left(\prod_{i=1}^3 \frac{1}{1-b_i} \right)\left(\Pi_{135} - \sum_{i=1}^3 b_i \; \Pi_{2i;2j-1;2k-1}
+ \sum_{k=1}^3 b_i b_j \Pi_{2i;2j;2k-1} -b_1b_2b_3 \, \Pi_{246}\right)
,
\\
\Pi^{\rm even}_{i}&=\frac{1}{1-b_i} \Bigl( \Pi_{2i-1;2j;2k} - b_i \; \Pi_{246}   \Bigr)
,
\\
\Pi^{\rm odd}_0 &=\Pi_{246}
,
\\
\Pi^{\rm odd}_i &=\frac{1}{(1-b_j)(1-b_k)} \Bigl( \Pi_{2i;2j-1;2k-1} - b_j \; \Pi_{2i;2j;2k-1} - b_k \; \Pi_{2i;2j-1;2k} + b_j b_k \; \Pi_{246}\Bigr)
,
\end{aligned}
\end{equation}
with non-vanishing intersection numbers 
\begin{equation*}
\Pi^{\rm even}_{K} \circ \Pi^{\rm odd}_L =- 4 \, \left(\prod_{i=1}^3 \frac{1}{1-b_i} \right) \delta_{KL}
\qquad
\text{for}
\qquad
K,L \in \{0 \ldots 3\}
.
\end{equation*}
The coefficients of orientifold even and odd bulk three-cycles are as usual given by 
(see e.g.~\cite{Blumenhagen:2004xx,Gmeiner:2005vz})
\begin{equation}
\begin{aligned}
 \tilde{X}_a^0 &\equiv \prod_{i=1}^3 n^i_a
,
\qquad\qquad\qquad
\tilde{X}_a^i \equiv n^i_a \left( m^j_a+ b_jn^j_a\right) \left( m^k_a + b_kn^k_a\right)
,
\\
\tilde{Y}_a^0 &\equiv  \prod_{i=1}^3 \left( m^i_a + b_i n^i_a \right)
,
\qquad
\tilde{Y}_a^i \equiv \left( m^i_a  +  b_i n^i_a\right)  n^j_a n^k_a
,
\end{aligned}
\end{equation}
and the bulk supersymmetry conditions read 
\begin{equation}\label{Eq:SUSY_Z2Z2_bulk}
\tilde{Y}_a^0 - \sum_{i=1}^3 \frac{1}{r_jr_k} \tilde{Y}_a^i =0
,
\qquad
\tilde{X}_a^0 -  \sum_{i=1}^3 (r_jr_k) \tilde{X}_a^i  >0
,
\end{equation}
which leads to the expression for the (length)${}^2$ of a bulk three-cycle
\begin{equation*}
\begin{aligned}
\prod_{i=1}^3 V_{aa}^{(i)} &= \prod_{i=1}^3 \frac{1}{r_i} \left( (n^i_a)^2 + r_i^2 (m^i_a + b_i n^i_a)^2 \right)
\\
&=\frac{1}{(r_1r_2r_3)}\left[ \left(\tilde{X}_a^0 - \sum_{i=1}^3 (r_jr_k)\tilde{X}_a^i \right)^2
+ (r_1r_2r_3)^2\left(\tilde{Y}_a^0 - \sum_{i=1}^3 \frac{1}{r_jr_k} \tilde{Y}_a^i \right)^2 \right]
\\
&\stackrel{\text{SUSY}}{=}\left(  \frac{1}{\sqrt{r_1r_2r_3}} \, \tilde{X}_a^0 \; - \; \sum_{i=1}^3 \sqrt{\frac{r_jr_k}{r_i}} \,  \tilde{X}_a^i \right)^2
,
\end{aligned}
\end{equation*}
where on the second line, the identities $\tilde{X}_a^0 \tilde{X}_a^i = \tilde{Y}_a^j  \tilde{Y}_a^k$
and $\tilde{Y}_a^0  \tilde{Y}_a^i =\tilde{X}_a^j \tilde{X}_a^k$ have been used.
In terms of the notation in equations~(\ref{Eq:Bulk-SUSY-General}) and~(\ref{Eq:Cycles-Decomposition_even+odd}),
the functions are $\vec{f}(r_1,r_2,r_3)=\left(\sqrt{r_1r_2r_3},-\sqrt{\frac{r_i}{r_jr_k} }\right)$ and 
$\vec{g}(r_1,r_2,r_3)=\left(\frac{1}{\sqrt{r_1r_2r_3}},-\sqrt{\frac{r_jr_k}{r_i}} \right)$ for $T^6/\Z_2 \times \Z_2$ with
$(i,j,k)$ cyclic permutations of (1,2,3).

Defining the four dimensional field theoretical dilaton and complex structure moduli fields at tree level via the ratios of radii
and the stringy dilaton $e^{\phi_4}=\frac{e^{\phi_{10}}}{\sqrt{v_1v_2v_3}}$,
\begin{equation}\label{Eq:Def-S+Ui=3}
S \sim \frac{e^{-\phi_4}}{\sqrt{r_1r_2r_3}}
,
\qquad
U_i \sim e^{-\phi_4}  \, \sqrt{\frac{r_jr_k}{r_i}}
,
\end{equation}
the gauge couplings for supersymmetric D6-branes thus take the form
\begin{equation}\label{Eq:tree-level-fhol_Z2Z2}
\Re({\rm f}_{SU(N_a)}^{\text{tree}}) \stackrel{!}{=} \frac{1}{g^2_{a,\text{tree}}}
 \sim \frac{1}{k_ac_a} \left(S  \tilde{X}_a^0- \sum_{i=1}^3 U_i \tilde{X}_a^i\right)
\qquad
\text{on}
\quad
T^6/(\Z_2 \times \Z_2 \times \OR)
,
\end{equation}
with the constants $c_a$ and $k_a$ related to the type of D6-brane and gauge group, respectively, as 
defined in equation~(\ref{Eq:tree-gauge-coupling}).

Other toroidal orbifolds have a reduced number of bulk complex structures due to some underlying
$\Z_3$ or $\Z_4$ symmetry leading to similar expressions as~(\ref{Eq:tree-level-fhol_Z2Z2}) for the 
tree level holomorphic gauge kinetic function with the sum running over $h_{21}^{\rm bulk}=1,0$. 
The corresponding modifications are discussed in section~\ref{Ss:Modifications-less-complex} below.

\subsection{One-loop results for $T^6$ and $T^6/\Z_2$ and $T^6/\Z_2 \times \Z_2$ with $\eta=\pm 1$}\label{Ss:loop-KaehlerAdj}

The one-loop corrections to both formulas~(\ref{Eq:Def-gauge-SUNa}) 
and~(\ref{Eq:Gauge-FieldTheory}) can be decomposed according to the open string sectors with
identical ($a=b$) or different endpoints $(a \neq b)$, cf. the decomposition of gauge thresholds in~(\ref{Eq:Threh-SU(N)}). The case $a\neq b$ also 
includes orbifold ($b=(\theta^k a)$ or $(\omega^k a)$) and orientifold ($b=(\theta^k a')$ or $(\omega^k a')$) image D6-branes. 
The following discussion focuses on $SU(N_a)$ gauge groups on the $T^6/\Z_2 \times \Z_2$
orientifolds without and with discrete torsion as well as the six-torus and $T^4/\Z_2 \times T^2$. This covers all 
cases of bulk, fractional and rigid D6-branes.
Additional $\Z_3$ or $\Z_4$ symmetries lead to a reduced number of bulk complex structures and thus a modified universal global prefactor $f(S,U_l)$ of the 
open string K\"ahler metrics as detailed in section~\ref{Ss:Modifications-less-complex} below.
Comments on other types of gauge groups $SO(2M)$, $Sp(2M)$ and $U(1)_X$ are given in section~\ref{Ss:SO+Sp_Kaehler+GaugeKinetic} 
and~\ref{Ss:U1_Kaehler+GaugeKinetic}, respectively.
\begin{itemize}
\item
Strings with endpoints on identical branes $aa$ provide the vector multiplet in the adjoint representation
of $SU(N_a)$ and three, one or no chiral multiplet in the adjoint representation, i.e. $\varphi^{\Adj_a}=3,1,0$ for $T^6/\Z_2 \times \Z_2$ 
without torsion, $T^4_{(1 \cdot 3)}/\Z_2 \times T^2_{(2)}$ and $T^6/\Z_2 \times \Z_2$ with discrete torsion, respectively.

The result from the one-loop string computation needs to be matched by all those field theory contributions describing the dynamics of the same 
fields,
\begin{equation*}
\begin{aligned}
 \frac{b_{aa}^{\cal A}}{2}  \;  \ln \left(\frac{M_{\rm string}}{\mu}\right)^2 + \frac{N_a \tilde{\Delta}_{aa}}{2}
\stackrel{!}{=} & \,  \frac{b_{aa}^{\cal A}}{2}   \; \ln \left(\frac{M_{\rm Planck}}{\mu}\right)^2 
+ \frac{b_{aa}^{\cal A} + 2 C_2(\Adj_a)}{2}   \; {\cal K} +\\
&  + C_2(\Adj_a) \, \ln \, [g^{-2}_{a}(\mu^2)]  -  \sum_{i=1}^{\varphi^{\Adj_a}} C_2(\Adj_a)  \, \ln K_{\Adj_a}^{(i)} \\
& + 8 \pi^2 \, \Re(\delta_a \, {\rm f}_{SU(N_a)}^{\rm 1-loop})
,
\end{aligned}
\end{equation*}
where ${\rm f}_{SU(N_a)}^{\text{1-loop}} = {\rm f}_{SU(N_a)}^{\text{tree}} + \sum_b \delta_b \, {\rm f}_{SU(N_a)}^{\text{1-loop}}$ constitutes the perturbatively exact
result for the holomorphic gauge kinetic function and the terms $C_2(\Adj_a) \left({\cal K} +  \ln \, [g^{-2}_{a}(\mu^2)]\right)$ on the r.h.s.
only occur for the case of identical D6-branes. The beta function coefficients and gauge thresholds for the $aa$ sector are given in table~\ref{tab:Adjoint-Beta+Thresholds}.
 The K\"ahler metrics for the adjoint matter multiplets on identical D6-branes are derived in an
iterative procedure since we explicitly insert the tree level gauge coupling in the logarithm on the r.h.s. and only take into account the K\"ahler potential for
the bulk closed string moduli and the dilaton,
\begin{equation}\label{Eq:Matching_aa_sector}
\begin{aligned}
0
\stackrel{!}{=}& \,
\frac{-3 + \varphi^{\Adj_a}}{2}  \; \left[  \ln \left(\frac{M_{\rm Planck}}{M_{\rm string}}\right)^2 +  {\cal K}_{\text{bulk}} \right]
+  {\cal K}_{\text{bulk}} + \ln \, [g^{-2}_{a,{\rm tree}}]  \\ &-  \sum_{i=1}^{\varphi^{\Adj_a}}  \ln K_{\Adj_a}^{(i)} 
+ \frac{8 \pi^2}{N_a} \, \Re(\delta_a \, {\rm f}_{SU(N_a)}^{\rm 1-loop}) - \frac{\tilde{\Delta}_{aa}}{2}
,
\end{aligned}
\end{equation}
where the value of the quadratic Casimir of the adjoint representation, $C_2(\Adj_a)= 2N_a \, C_2(\N_a) = N_a$,
 has been inserted.

Using the form of tree level gauge coupling in equation~(\ref{Eq:tree-Mass_ratios}) and the definition of the dilaton and bulk complex structure moduli
 in~(\ref{Eq:Def-S+Ui=3}) together with the standard ansatz for the K\"ahler potential for the closed string bulk fields,
\begin{equation}\label{Eq:Def-K_bulk}
{\cal K}_{\text{bulk}} = - \ln S - \sum_{i=1}^3 \ln U_i - \sum_{i=1}^3 \ln v_i
,
\end{equation}
 the first line in~(\ref{Eq:Matching_aa_sector}) can be rewritten as (with $k_a=1$ for $SU(N_a)$)
\begin{eqnarray}
& \frac{-3 + \varphi^{\Adj_a}}{2}  \; \left[  \ln \left(\frac{M_{\rm Planck}}{M_{\rm string}}\right)^2 +  {\cal K}_{\text{bulk}} \right] +  {\cal K}_{\text{bulk}} + \ln \, [g^{-2}_{a,{\rm tree}}]  \nonumber\\
&=\frac{-3+\varphi^{\Adj_a}}{2}  \; \left[  \ln \left( S\prod_{i=1}^3 U_i \right)^{1/2} - \ln \!\!\left( S \prod_{i=1}^3 U_i \, v_i \right)\right] 
- \ln \left(S \prod_{i=1}^3 U_i \, v_i \right) \nonumber\\
& \quad + \ln \left[ \left(S \prod_{i=1}^3 U_i \right)^{1/4}  \frac{ (2\pi)^{3/2} \prod_{i=1}^3\sqrt{V_{aa}^{(i)}}}{c_a} \right]
\nonumber\\
&= - \frac{ \varphi^{\Adj_a}}{4} \ln \!\left(\!S\prod_{i=1}^3 U_i\!\right) + \frac{1- \varphi^{\Adj_a}}{2}   \ln \!\left( \prod_{i=1}^3 v_i\!\right) + \ln \left(\frac{ (2\pi)^{3/2} \prod_{i=1}^3 \sqrt{V_{aa}^{(i)}}}{c_a}\right) 
.
\label{Eq:1st-line-aa}
\end{eqnarray}
In order to rewrite the second line in~(\ref{Eq:Matching_aa_sector}), the gauge threshold contributions can be read off from  table~\ref{tab:Adjoint-Beta+Thresholds},
\begin{equation*}
\frac{\tilde{\Delta}_{aa}}{2} = \frac{2}{c_a} \, \left\{\begin{array}{cl}
0 & T^6 \text{ and } T^6/\Z_2 \times \Z_{2M} \text{ with } \eta=1
\\
 \Lambda_{0,0}(v_2;V_{aa}^{(2)}) & T^6/\Z_{2N} 
 \\
\sum_{i=1}^3  \Lambda_{0,0}(v_i;V_{aa}^{(i)})  & T^6/\Z_2 \times \Z_{2M} \text{ with } \eta=-1
\end{array}\right.
,
\end{equation*}
with the lattice sums defined in~(\ref{Eq:Def-Lambdas}) and $c_a=1,2,4$ for bulk, fractional and rigid D6-branes, respectively.
With the ansatz for the K\"ahler metrics analogous to the six-torus~\cite{Akerblom:2007np} (with $(ijk)$ cyclic permutations of (123) and $i$ the index 
associated to the continuous displacement and Wilson line modulus on $T^2_{(i)}$),
\begin{equation}\label{Eq:Def-Kaehler_adjoints}
K_{\Adj_a}^{(i)} = \frac{\sqrt{2 \pi}}{c_a} \, \frac{f(S,U_l)}{v_i} \sqrt{\frac{V_{aa}^{(j)} V_{aa}^{(k)}}{V_{aa}^{(i)}}}
\qquad 
\text{with}
\qquad
f(S,U_l) =\left( S \prod_{l=1}^3 U_l \right)^{-1/4}
,
\end{equation}
and the one-loop contribution to the holomorphic gauge kinetic function,
\begin{equation}\label{Eq:Def-1loop_hol}
\delta_a \, {\rm f}_{SU(N_a)}^{\text{1-loop}}= \frac{N_a}{\pi^2 c_a} \times \left\{\begin{array}{cl}
0 & T^6 \text{ and } T^6/\Z_2 \times \Z_{2M} \text{ with } \eta=1
\\
\ln \eta(iv_2) & T^6/\Z_{2N} 
 \\
\sum_{i=1}^3  \ln \eta(iv_i)   & T^6/\Z_2 \times \Z_{2M} \text{ with } \eta=-1
\end{array}\right.
,
\end{equation}
the second line of~(\ref{Eq:Matching_aa_sector}) can be recast for the various torus and orbifold backgrounds as follows.
\begin{itemize}
\item
On $T^6$ and $T^6/\Z_2 \times \Z_2$ without discrete torsion, the second line reads
\begin{equation*}
\begin{aligned}
&-  \sum_{i=1}^{\varphi^{\Adj_a}}  \ln K_{\Adj_a}^{(i)}  + \frac{8 \pi^2}{N_a} \, \Re(\delta_a \, {\rm f}_{SU(N_a)}^{\rm 1-loop}) - \frac{\tilde{\Delta}_{aa}}{2}
\\
&= - \ln \left[ \frac{(2\pi)^{3/2} \sqrt{ V_{aa}^{(1)} V_{aa}^{(2)} V_{aa}^{(3)} }}{c_a^3 v_1v_2v_3 (S \prod_{l=1}^3 U_l)^{3/4}} \right] + \emptyset + \emptyset
\\
&= - \ln \frac{(2\pi)^{3/2} \sqrt{\prod_{i=1}^3 V_{aa}^{(i)}}}{c_a^3} + \ln \left(\prod_{i=1}^3 v_i \right) + \frac{3}{4} \ln \left( S \prod_{l=1}^3 U_l \right)
,
\end{aligned}
\end{equation*}
and exactly cancels the first line~(\ref{Eq:1st-line-aa}) of~(\ref{Eq:Matching_aa_sector}) since $c_a=1$ for toroidal three-cycles, $k_a=1$ for $SU(N_a)$ gauge groups and because there exist
three adjoint multiplets, \mbox{$\varphi^{\Adj_a}=3$}, related to the continuous displacement and Wilson line moduli per two-torus. The ansatz for the bulk K\"ahler 
potential~(\ref{Eq:Def-K_bulk}), adjoint K\"ahler metrics on identical D6-branes~(\ref{Eq:Def-Kaehler_adjoints}) and one-loop contribution to the holomorphic gauge kinetic function~(\ref{Eq:Def-1loop_hol}) are thus mutually consistent.
\item
On $T^4_{(1\cdot 3)}/\Z_2 \times T^2_{(2)}$, there exists one adjoint multiplet, $\varphi^{\Adj_a}=1$,
related to the Wilson line and displacement modulus on the two-torus $T^2_{(2)}$ without $\Z_2$ twist,
and the gauge threshold receives lattice sum contributions from this two-torus,  
\begin{equation*}
\begin{aligned}
&-  \sum_{i=1}^{\varphi^{\Adj_a}}  \ln K_{\Adj_a}^{(i)}  + \frac{8 \pi^2}{N_a} \, \Re(\delta_a \, {\rm f}_{SU(N_a)}^{\rm 1-loop}) - \frac{\tilde{\Delta}_{aa}}{2}
\\
&= - \ln \left[ \frac{\sqrt{2\pi}}{c_a \, v_2 \, (S \prod_{l=1}^3 U_l)^{1/4}} \sqrt{\frac{V_{aa}^{(1)} V_{aa}^{(3)}}{V_{aa}^{(2)}}} 
\right]
+ \frac{8}{c_a} \ln \eta(iv_2) 
- \frac{2}{c_a}  \, \left[ 4 \, \ln \eta (iv_2) + \ln \left(2\pi v_2 V_{aa}^{(2)}\right)  \right]
\\
&= \frac{1}{4} \ln \left( S \prod_{l=1}^3 U_l \right) - \ln \frac{(2\pi)^{3/2}  \prod_{i=1}^3 \sqrt{V_{aa}^{(i)}} }{c_a}
.
\end{aligned}
\end{equation*}
Due to $c_a=2$ for fractional D6-branes on orbifolds with one $\Z_2$ (sub)symmetry and $k_a=1$ 
for $SU(N_a)$, the ansatz for the bulk K\"ahler potential~(\ref{Eq:Def-K_bulk}) and K\"ahler metrics~(\ref{Eq:Def-Kaehler_adjoints}) and holomorphic gauge kinetic 
function~(\ref{Eq:Def-1loop_hol}) are again mutually consistent.
\item
On $T^6/\Z_2 \times \Z_2$ with discrete torsion, the $aa$ sector does not contain any chiral multiplet,  $\varphi_{\Adj_a}$=0, due to 
the discrete character of the displacements and Wilson lines, and  with $c_a=4$ for rigid D6-branes on orbifolds with $\Z_2 \times \Z_2$ (sub)symmetry
and discrete torsion,
\begin{equation*}
\begin{aligned}
&-  \sum_{i=1}^{\varphi^{\Adj_a}}  \ln K_{\Adj_a}^{(i)}  + \frac{8 \pi^2}{N_a} \, \Re(\delta_a \, {\rm f}_{SU(N_a)}^{\rm 1-loop}) - \frac{\tilde{\Delta}_{aa}}{2}
\\
&= \emptyset + \frac{8}{c_a} \sum_{i=1}^3  \ln \eta(iv_i) 
-\frac{2}{c_a}  \sum_{i=1}^3 \left[4 \, \ln \eta (iv_i) + \ln \left(2\pi v_i V_{aa}^{(i)}\right)  \right]
\\
&= - \frac{1}{2} \ln \left(\prod_{i=1}^3 v_i \right) - \ln \left( (2\pi)^{3/2} \prod_{i=1}^3 \sqrt{V_{aa}^{(i)}} \right) 
,
\end{aligned}
\end{equation*}
the second line of~(\ref{Eq:Matching_aa_sector}) cancels the first line~(\ref{Eq:1st-line-aa}) as expected.
\end{itemize}
This completes the discussion of the $aa$ sector for the six-torus and its orbifolds $T^4/\Z_2 \times T^2$ and $T^6/\Z_2 \times \Z_2$
without and with discrete torsion.
Changes due to the existence of a reduced number of bulk complex structure moduli on other toroidal orbifolds amount to 
modifications in the prefactor $f(S,U_l)$ in~(\ref{Eq:Def-Kaehler_adjoints}) and changes in the prefactors of the dilaton and 
bulk complex structure contributions to the K\"ahler potential~(\ref{Eq:Def-K_bulk})
as discussed in section~\ref{Ss:Modifications-less-complex}.
\item
For strings with endpoints on two different D6-branes $a$ and $b$ and $b \neq (\omega^k a), (\omega^l a')$ for any $k,l$, the generic form of matching 
the string and field theory calculation is simpler since the term 
$C_2(\Adj_a) \left({\cal K} + \ln [g_a^{-2}(\mu)^2]\right)$ has already been taken care of in the $aa$ sector,
\begin{equation*}
\begin{aligned}
 \frac{b_{ab}^{\cal A}}{2}  \;  \ln \left(\frac{M_{\rm string}}{\mu}\right)^2 + \frac{N_a \tilde{\Delta}_{ab}}{2}
\stackrel{!}{=} & \,  \frac{b_{ab}^{\cal A}}{2} \; \ln \left(\frac{M_{\rm Planck}}{\mu}\right)^2 
+ \frac{b_{ab}^{\cal A}}{2}  \;  {\cal K}  +\\
&  -  \sum_{i=1}^{\varphi^{(\N_a,\ov{\N}_b)}} N_b \, C_2(\N_a)  \, \ln K_{(\N_a,\ov{\N}_b)}^{(i)} 
 + 8 \pi^2 \, \Re(\delta_b \, {\rm f}_{SU(N_a)}^{\rm 1-loop})
.
\end{aligned}
\end{equation*}
Using the value $C_2(\N_a)=\frac{1}{2}$ for the quadratic Casimir of the fundamental representation, this matching condition can be rewritten as 
\begin{equation}\label{Eq:Matching_ab_sector}
\begin{aligned}
0
\stackrel{!}{=}& \, \frac{b_{ab}^{\cal A}}{2} \left( \ln \left(\frac{M_{\rm Planck}}{M_{\rm string}}\right)^2 + {\cal K}_{\text{bulk}} \right) 
-\sum_{i=1}^{\varphi^{(\N_a,\ov{\N}_b)}} \!\!\frac{N_b}{2} \, \ln K_{(\N_a,\ov{\N}_b)}^{(i)}  +  8 \pi^2 \, \Re(\delta_b \, {\rm f}_{SU(N_a)}^{\rm 1-loop})
-\frac{\Delta_{ab}}{2}
\\
=& \, - \frac{b_{ab}^{\cal A}}{2} \ln \left[\left(S \prod_{i=1}^3 U_i \right)^{1/2} \left(\prod_{i=1}^3 v_i \right)\right] 
-\!\!\sum_{i=1}^{\varphi^{(\N_a,\ov{\N}_b)}} \!\!\frac{N_b}{2} \ln K_{(\N_a,\ov{\N}_b)}^{(i)}  + 8 \pi^2 \, \Re(\delta_b \, {\rm f}_{SU(N_a)}^{\rm 1-loop})
-\frac{\Delta_{ab}}{2} 
,  
\end{aligned}
\end{equation}
where on the second line the relation of the mass scales in terms of the  dilaton and bulk 
complex structure moduli defined in~(\ref{Eq:Def-S+Ui=3})  as well as the bulk K\"ahler potential~(\ref{Eq:Def-K_bulk}) 
have been used. For the six-torus considered in~\cite{Akerblom:2007np}, all multiplets in a given representations 
are on equal footing, and the sum over logarithms of K\"ahler metrics on 
the second line  of~(\ref{Eq:Matching_ab_sector}) boils down to $- b_{ab}^{\cal A} \, \ln K_{(\N_a,\ov{\N}_b)}$.
Since on the six-torus, also the gauge thresholds from open strings in the bifundamental representation
in table~\ref{tab:Six-torus-Bifund-beta+thresholds} have the beta function coefficient
$b_{ab}^{\cal A}$ as a global prefactor, the K\"ahler metrics can be read off in a straight forward manner. 
For orbifolds containing some $\Z_2^{(i)}$ symmetry, the gauge thresholds contain additional terms proportional to $I_{ab}^{\Z_2^{(i)}}$
times the angle $\phi^{(i)}_{ab}$ plus constants consisting of sign factors
as displayed in tables~\ref{tab:Z2Z2M-torsion-Bifundamentals-beta+thresholds} and~\ref{tab:T6-Z2N-Bifund-beta+thresholds}
for bifundamental and adjoint matter, and terms proportional to the intersection numbers with the O6-planes
for symmetric or antisymmetric matter on any torus or orbifold background as displayed in 
tables~\ref{tab:Z2Z2M-torsion-AntiSym-beta+thresholds},~\ref{tab:Six-torus-AntiSym-beta+thresholds},~\ref{tab:Z2Z2M-no_torsion-Anti+Sym-beta+thresholds} and~\ref{tab:T6-Z2N-Anti+Sym-beta+thresholds}. We will first determine the K\"ahler metrics from the terms
proportional to the beta function coefficients and argue further below that all the additional terms enter the one-loop
corrections to the gauge kinetic functions.
This leads to K\"ahler metrics which only depend on the (non-)vanishing of the angles, but are universal for all
orbifold backgrounds and all  (bifundamental, adjoint, symmetric, antisymmetric) matter representations. 
The non-trivial information on the background and matter representations is encoded in the beta function coefficients
and modifies the one-loop contributions to the holomorphic gauge kinetic function.

The gauge threshold corrections from $ab$ sectors computed in section~\ref{S:gauge-thres-revisit} 
and the corresponding open string K\"ahler metrics and corrections to the holomorphic gauge kinetic functions
fall into two distinct classes:
\begin{enumerate}
\item 
The D6-branes $a$ and $b$ under consideration are parallel along at least one two-torus.
This includes the case with completely parallel D6-branes with different choices of $\Z_2$ eigenvalues,
displacements or Wilson lines on $T^6/\Z_{2N}$ and \mbox{$T^6/\Z_2 \times \Z_2$} with discrete torsion.
Following the argumentation of~\cite{Akerblom:2007np}, the functional dependence 
of the K\"ahler metrics on the bulk moduli and the (length)${}^2$ $V_{ab}^{(i)}$ on the two-torus $T^2_{(i)}$ 
where the branes are parallel is given by (with $(ijk)$ a cyclic permutation of (123)) 
\begin{equation}\label{Eq:Kaehler-general-1angle}
K_{(\N_a.\ov{\N}_b)}^{(i)}  = f(S,U_l) \;   \sqrt{\frac{ 2 \pi V_{ab}^{(i)}}{v_jv_k}}
,
\end{equation}
where  $f(S,U_l)$ is defined in~(\ref{Eq:Def-Kaehler_adjoints}).
The generic form of the functional dependence can be seen from the fact that beta function coefficient $b_{ab}^{\cal A}$
for one vanishing angle indeed appears in front of the lattice sums in tables~\ref{tab:Z2Z2M-torsion-Bifundamentals-beta+thresholds},~\ref{tab:Z2Z2M-no_torsion-Anti+Sym-beta+thresholds},~\ref{tab:Six-torus-Bifund-beta+thresholds} and ~\ref{tab:T6-Z2N-Bifund-beta+thresholds} for
$T^6/\Z_2 \times \Z_{2M}$ with and without discrete torsion, the six-torus and $T^6/\Z_{2N}$, respectively.
On $T^6/\Z_{2N}$, the case of parallel D6-branes is of exactly the same type, whereas for 
 $T^6/\Z_2 \times \Z_{2M}$ with discrete torsion the beta function coefficient needs to be decomposed into three contributions,
 reflecting the fact that the three different kinds of possible non-chiral bifundamental matter pairs correspond to the two-torus
 label $i$ or in other words to opposite $\Z_2^{(j)}$ and $\Z_2^{(k)}$ and identical $\Z_2^{(i)}$ eigenvalues. 
 Our present result is in contrast to~\cite{Blumenhagen:2007ip}, where the expression~(\ref{Eq:Def-Kaehler_adjoints}) 
 was proposed also for bifundamental matter. However, in the derivation of~(\ref{Eq:Def-Kaehler_adjoints})  the term 
 $C_2(\Adj_a) \left( {\cal K} + \ln [g_a^{-2}(\mu^2)] \right)$ was assigned to the $aa$ sector, and it cannot be again used for the $ab$ contributions.
 The K\"ahler metrics for bifundamental matter on parallel or intersecting D6-branes in various toroidal orbifold backgrounds 
 are summarised in table~\ref{tab:Comparison-Kaehler-bifund}.
\mathtabfix{
\begin{array}{|c||c|c|c|}\hline
\multicolumn{4}{|c|}{\text{\bf K\"ahler metrics for bifundamental matter $(\N_a,\ov{\N}_b)$ on various orbifolds}}
\\\hline\hline
(\phi_{ab}^{(1)},\phi_{ab}^{(2)},\phi_{ab}^{(3)}) & \begin{array}{c} T^6 \quad \text{ and} \\ T^6/\Z_2 \times \Z_{2M} \\ \text{without torsion}
\end{array} &T^6/\Z_{2N}   &\begin{array}{c} T^6/\Z_2 \times \Z_{2M} \\ \text{with discrete torsion} \end{array}
\\\hline\hline
(0,0,0) & - & f(S,U_l) \;   \sqrt{\frac{ 2 \pi V_{ab}^{(2)}}{v_1v_3}} & \begin{array}{c}
 f(S,U_l) \;   \sqrt{\frac{ 2 \pi V_{ab}^{(i)}}{v_jv_k}} \\ \text{\footnotesize $(ijk) \simeq (1,2,3)$ \text{ cyclic}}
\end{array}
\\\hline
\begin{array}{c} (0^{(i)},\phi^{(j)},\phi^{(k)}) \\{\phi^{(j)}=-\phi^{(k)}} \end{array} & 
\multicolumn{3}{|c|}{ f(S,U_l) \;   \sqrt{\frac{ 2 \pi V_{ab}^{(i)}}{v_jv_k}} }
\\\hline 
\begin{array}{c} (\phi^{(1)},\phi^{(2)},\phi^{(3)}) \\ {\sum_{i=1}^3 \phi^{(i)}=0} \end{array}
& \multicolumn{3}{|c|}{
f(S,U_l)  \,  \sqrt{\prod_{i=1}^3 \frac{1}{ v_i} \left(\frac{\Gamma(|\phi^{(i)}_{ab}|)}{\Gamma(1-|\phi^{(i)}_{ab}|)}\right)^{-\frac{\sgn(\phi^{(i)}_{ab})}{\sgn(I_{ab})} }}
}
\\\hline
\end{array}
}{Comparison-Kaehler-bifund}{Comparison of the K\"ahler metrics for bifundamental matter. The functional dependence of the K\"ahler
metrics is solely determined by  the number of vanishing angles and the assignment of multiplets to the two-torus $T^2_{(i)}$ where the D6-branes
are parallel, independently of the orbifold background. For two or three intersection angles, the K\"ahler metrics of adjoint matter at orbifold intersections have the same
shape, whereas the K\"ahler metrics for adjoint matter on identical D6-branes differ and are given by~(\protect\ref{Eq:Def-Kaehler_adjoints}).
}

The one-loop contributions to the holomorphic gauge kinetic function from $ab$ sectors,
which depend on the two-torus volumes $v_i$,  can be straightforwardly read off from the lattice sums in 
tables~\ref{tab:Z2Z2M-torsion-Bifundamentals-beta+thresholds},~\ref{tab:Z2Z2M-no_torsion-Anti+Sym-beta+thresholds},~\ref{tab:Six-torus-Bifund-beta+thresholds}
and~\ref{tab:T6-Z2N-Bifund-beta+thresholds} for the various cases of open strings with endpoints on different D6-branes $a$ and $b$, 
and the result is summarised in table~\ref{tab:Comparison-gaugekin-bifund}.
\mathsidetabfix{
\begin{array}{|c||c|c|c|}\hline
\multicolumn{4}{|c|}{\text{\bf 1-loop contribution to the gauge kinetic function } \delta_b \,{\rm f}_{SU(N_a)}^{\text{1-loop}}(v_i)
\text{ \bf from bifundamental sectors on various orbifolds}}
\\\hline\hline
(\phi_{ab}^{(1)},\phi_{ab}^{(2)},\phi_{ab}^{(3)}) & \begin{array}{c} T^6 \quad \text{ and} \\ T^6/\Z_2 \times \Z_{2M} \\ \text{without torsion}
\end{array} & T^6/\Z_{2N} &\begin{array}{c} T^6/\Z_2 \times \Z_{2M} \\ \text{with discrete torsion} \end{array}
\\\hline\hline
(0,0,0) & - &  \begin{array}{c}- \frac{b_{ab}^{\cal A}}{4\pi^2} \ln \eta(iv_2) \\ - \frac{\tilde{b}_{ab}^{\cal A}}{8\pi^2}
(1-\delta_{\sigma^2_{ab},0}\delta_{\tau^2_{ab},0}) \times \\ \times \ln \left(e^{-\pi (\sigma^2_{ab})^2 v_2/4}\frac{\vartheta_1 (\frac{\tau^2_{ab} - i \sigma^2_{ab} v_2}{2},i v_2)}{\eta (i v_2)} \right) \end{array}
&
 \begin{array}{c}- \sum_{i=1}^3 \frac{b_{ab}^{{\cal A},(i)}}{4\pi^2} \ln \eta(iv_i) \\ -\sum_{i=1}^3  \frac{\tilde{b}_{ab}^{{\cal A},(i)}}{8\pi^2}
(1-\delta_{\sigma^i_{ab},0}\delta_{\tau^i_{ab},0}) \times \\ \times \ln \left(e^{-\pi (\sigma^i_{ab})^2 v_i/4}\frac{\vartheta_1 (\frac{\tau^i_{ab} - i \sigma^i_{ab} v_i}{2},i v_i)}{\eta (i v_i)} \right) \end{array}
\\\hline
(0^{(i)},\phi^{(j)},\phi^{(k)})_{\phi^{(j)}=-\phi^{(k)}} &\multicolumn{3}{|c|}{
 - \frac{b_{ab}^{\cal A}}{4\pi^2} \ln \eta(iv_i) 
 - \frac{\tilde{b}_{ab}^{\cal A}}{8\pi^2}
(1-\delta_{\sigma^i_{ab},0}\delta_{\tau^i_{ab},0}) \ln \left(e^{-\pi (\sigma^i_{ab})^2 v_i/4}\frac{\vartheta_1 (\frac{\tau^i_{ab} - i \sigma^i_{ab} v_i}{2},i v_i)}{\eta (i v_i)} \right) 
}
\\\hline 
(\phi^{(1)},\phi^{(2)},\phi^{(3)}) & \multicolumn{3}{|c|}{-} 
\\\hline
\end{array}
}{Comparison-gaugekin-bifund}{Comparison of the two-torus volume $v_i$ dependent
one-loop contributions $\delta_b {\rm f}_{SU(N_a)}^{\text{1-loop}}(v_i)$ to the holomorphic gauge kinetic function from bifundamental
and adjoint sectors. The corresponding beta-function coefficients are given in the second column of table~\protect\ref{tab:Six-torus-Bifund-beta+thresholds}  for the six-torus, 
table~\protect\ref{tab:Z2Z2M-no_torsion-Anti+Sym-beta+thresholds} for $T^6/\Z_2 \times \Z_{2M}$ 
without discrete torsion, table~\protect\ref{tab:T6-Z2N-Bifund-beta+thresholds} for $T^6/\Z_{2N}$ and table~\protect\ref{tab:Z2Z2M-torsion-Bifundamentals-beta+thresholds} for $T^6/\Z_2 \times \Z_{2M}$ with discrete torsion. 
The contributions from adjoint matter at intersection of orbifold images are of the same form, whereas those
on identical D6-branes differ and are given in~(\protect\ref{Eq:Def-1loop_hol}).
}
In the presence of some $\Z_2$ symmetry, the gauge threshold for D6-branes parallel along $T^2_{(i)}$ and at angles
along $T^2_{(j)} \times T^2_{(k)}$ contains additional terms
\begin{equation}\label{Eq:2angle-Delta-terms}
\Delta_{SU(N_a)} \supset - \frac{N_b}{c_a} \sum_{l=j,k} I_{ab}^{\Z_2^{(l)}} \left(2 \, \phi^{(l)}_{ab} - \sgn(\phi^{(l)}_{ab}) \right)
\; \ln 2
, 
\end{equation}
cf. tables~\ref{tab:Z2Z2M-torsion-Bifundamentals-beta+thresholds} and~\ref{tab:T6-Z2N-Bifund-beta+thresholds} 
for $T^6/\Z_2 \times \Z_{2M}$ with discrete torsion and $T^6/\Z_{2N}$, respectively, where in the latter case $I_{ab}^{\Z_2^{(l)}} \equiv 0$ for $l \neq 2$. 
It is tempting to assign the terms~(\ref{Eq:2angle-Delta-terms}) to some dependence of the K\"ahler metrics 
on the $\Z_2$ transformation properties of the matter localisations. However, it is possible to have~(\ref{Eq:2angle-Delta-terms})
non-vanishing, while there is no massless matter state in the corresponding $ab$ sector as discussed below for examples with
three non-vanishing angles in the Standard Model on $T^6/\Z_6'$.
The term~(\ref{Eq:2angle-Delta-terms}) is thus interpreted as an angle-dependent one-loop contribution to the 
holomorphic gauge kinetic function $\delta_b \, {\rm f}^{\text{1-loop}}_{SU(N_a)}(\phi^{(i)}_{ab})$,
\begin{equation}\label{Eq:2angle-f-angle}
\Re \left(\delta_b \, {\rm f}^{\text{1-loop}}_{SU(N_a)}(\phi^{(i)}_{ab})\right) = -
\frac{N_b}{16 \pi^2 \, c_a} \sum_{l=j,k} I_{ab}^{\Z_2^{(l)}} \left(2 \, \phi^{(l)}_{ab} - \sgn(\phi^{(l)}_{ab}) \right)
\; \ln 2
,
\end{equation}
which implicitly depends on the complex structure moduli through \mbox{$\arctan(\phi^{(l)}_{ab}) \sim r_l$}.
\item
If the D6-branes are at angles on all three tori, the gauge threshold computation provides the functional dependence, cf. e.g.~\cite{Akerblom:2007np},
\begin{equation}\label{Eq:Kaehler-general-3angles}
K_{(\N_a,\ov{\N}_b)} = f(S,U_l)  \;    \sqrt{\prod_{i=1}^3 \frac{1}{ v_i} \left(\frac{\Gamma(|\phi^{(i)}_{ab}|)}{\Gamma(1-|\phi^{(i)}_{ab}|)}\right)^{-\frac{\sgn(\phi^{(i)}_{ab})}{\sgn(I_{ab})} }}
, 
\end{equation}
for all toroidal orbifold backgrounds, where all angles are chosen in the range \mbox{$0<|\phi_{ab}^{(i)}| <1$} and obey the bulk supersymmetry 
condition $\sum_{i=1}^3 \phi^{(i)}_{ab}=0$, and $f(S,U_l)$ has been defined in~(\ref{Eq:Def-Kaehler_adjoints}). 
The sign factor $\sgn(I_{ab})$ in the exponential is essential for obtaining the same K\"ahler potential from the computations of the $ab$ and its inverse $ba$ sector
and has to our knowledge not properly been taken into account before.
The comparison of all bifundamental K\"ahler metrics is given in table~\ref{tab:Comparison-Kaehler-bifund}.

The dependence on the two-torus volumes $v_i$ is fully contained in the K\"ahler metrics~(\ref{Eq:Kaehler-general-3angles}),
and there is no $v_i$ dependent contribution to the holomorphic gauge kinetic functions.
In the presence of $\Z_2$ symmetries, however, the gauge threshold contains additional terms proportional to 
the intersection angles $\phi^{(i)}_{ab}$,
\begin{equation}\label{Eq:3angle-Delta-terms}
\Delta_{SU(N_a)} \supset -\frac{N_b}{c_a} \sum_{i=1}^3 I_{ab}^{\Z_2^{(i)}} \left( 2 \, \phi^{(i)}_{ab} - \sgn(\phi^{(i)}_{ab}) - \sgn(I_{ab})
\right) \, \ln 2
,
\end{equation} 
cf. tables~\ref{tab:Z2Z2M-torsion-Bifundamentals-beta+thresholds} and~\ref{tab:T6-Z2N-Bifund-beta+thresholds}.
These terms have the identical shape as~(\ref{Eq:2angle-Delta-terms}) for one vanishing angle along $T^2_{(i)}$ 
when taking into account that this leads  to \mbox{$\sgn(I_{ab})=0$} and $I_{ab}^{\Z_2^{(i)}}=0$.
The case of the six-torus and the  $T^6/\Z_2 \times \Z_{2M}$ orbifolds without torsion can be formally included by setting 
$I_{ab}^{\Z_2^{(i)}} \equiv 0$ for all $i$. Taking a look at the Standard Model example on $T^6/\Z_6'$ in 
table~\ref{tab:modelz6p+hidden-sectors_ax-Kaehler} reveals that there are no massless states in the $xy \in \{
a(\theta^k b)_{k=0,1,2}, a(\theta c),c(\theta^2 d)\}$ sectors, while the angles $\phi^{(2)}_{xy}$ 
in table~\ref{tab:modelz6p+hidden-sectors_ax-Kaehler} and intersection numbers 
$I_{xy}^{\Z_2^{(2)}}$ listed explicitly in~\cite{Gmeiner:2009fb} are non-vanishing. The term~(\ref{Eq:3angle-Delta-terms})
can thus not be assigned to a change in the normalisation of K\"ahler metrics due to different matter localisations, 
but in complete analogy to~(\ref{Eq:2angle-f-angle}) for one vanishing angle, it is instead interpreted as an
angle dependent one-loop correction to the holomorphic gauge kinetic function,
\begin{equation}\label{Eq:3angle-f-angle}
\Re \left(\delta_b \, {\rm f}^{\text{1-loop}}_{SU(N_a)}(\phi^{(i)}_{ab})\right) = -
\frac{N_b}{16 \pi^2 \, c_a} \sum_{i=1}^3 I_{ab}^{\Z_2^{(i)}} \left(2 \, \phi^{(i)}_{ab} - \sgn(\phi^{(i)}_{ab}) - \sgn(I_{ab}) \right)
\; \ln 2
,
\end{equation}
which can be formally taken to hold for all supersymmetric configurations with vanishing or non-vanishing angles
upon setting  $\Z_2$ invariant intersection numbers or sign factors to zero as described above.
\end{enumerate}
\item
At the intersection of orbifold image D6-branes $a$ and $(\omega^k a)$, the matter states transform in the adjoint representation.
The matching in equation~(\ref{Eq:Matching_ab_sector}) needs to be modified by using the corresponding quadratic Casimir, $C_2(\Adj_a)=N_a$,
\begin{equation}\label{Eq:Matching_aTHETAa_sector}
\begin{aligned}
0
\stackrel{!}{=}& 
\, - \frac{b_{(\omega^k a)a}^{\cal A} + b_{(\omega^{-k} a)a}^{\cal A}}{2} \ln \left[\left(S \prod_{i=1}^3 U_i \right)^{1/2}  \left(\prod_{i=1}^3 v_i \right)\right] -\sum_{i=1}^{\varphi^{\Adj_a}} \!\! N_a \ln K_{(\Adj_a)}^{(i)}  \\
& + 8 \pi^2 \, \Re(\left[\delta_{(\theta^k a)} +\delta_{(\theta^{-k} a)}\right] \, {\rm f}_{SU(N_a)}^{\rm 1-loop}) -\frac{\Delta_{(\theta^k a)a} + \Delta_{(\theta^{-k} a)a}}{2}
,  
\end{aligned}
\end{equation}
where we have used that for orbifolds other than $T^6/\Z_4$ or $T^6/\Z_2 \times \Z_4$, the two degrees of freedom 
of a complex boson or Weyl fermion in the adjoint representation stem from the combination of one sector
$(\theta^k a)a$ plus its inverse $(\theta^{-k} a)a$.
As used in table~\ref{tab:Adjoint-Beta+Thresholds}, the gauge threshold and beta function contributions from inverse sectors are identical,
$\Delta_{(\theta^k a)a} = \Delta_{(\theta^{-k} a)a}$ and $b_{(\omega^k a)a}^{\cal A} = b_{(\omega^{-k} a)a}^{\cal A}$, which means that 
the factor of two in the quadratic Casimir of the adjoint  in~(\ref{Eq:Matching_aTHETAa_sector})  compared to the bifundamental in~(\ref{Eq:Matching_ab_sector})
 is absorbed by the combination of inverse
sectors, and therefore the K\"ahler metrics and $v_i$ dependent loop corrections to the holomorphic gauge kinetic function are 
identical to those of bifundamentals at some non-vanishing intersection angles in tables~\ref{tab:Comparison-Kaehler-bifund} 
and~\ref{tab:Comparison-gaugekin-bifund}. Also the angle dependent loop corrections to the holomorphic gauge kinetic 
function match~(\ref{Eq:3angle-f-angle}).
\item
Last but not least, for orientifold image D6-branes $a$ and $a'$ the symmetric and antisymmetric representations $\Sym_a$ and $\Anti_a$
of $SU(N_a)$ have to be taken into account,
\begin{equation}\label{Eq:Matching_aaprime_sector}
\begin{aligned}
0
\stackrel{!}{=}& \frac{b_{aa'}^{\cal A} + b_{aa'}^{\cal M}}{2}  
\left[ \ln\left(\frac{M_{\rm Planck}}{M_{\rm string}}\right)^2 + {\cal K}_{\text{bulk}} \right] 
 - \sum_{i=1}^{\varphi^{\Anti_a}} \frac{N_a-2}{2} \ln K_{\Anti_a}^{(i)} - \sum_{j=1}^{\varphi^{\Sym_a}} \frac{N_a+2}{2} \ln K_{\Sym_a}^{(j)} 
\\
&  + 8 \pi^2 \, \Re(\delta_{a'} \, {\rm f}_{U(1)_a}^{\rm 1-loop})-\frac{\Delta_{aa'} +\Delta_{a,\OR}}{2}
\\
=& -\frac{b_{aa'}^{\cal A} + b_{aa'}^{\cal M}}{2} \ln \left[\left(S \prod_{i=1}^3 U_i \right)^{1/2} \left(\prod_{i=1}^3 v_i \right)\right] 
 - \frac{N_a}{2} \left(\sum_{i=1}^{\varphi^{\Anti_a}}  \ln K_{\Anti_a}^{(i)}+ \sum_{j=1}^{\varphi^{\Sym_a}} \ln K_{\Sym_a}^{(j)}\right)
\\
&-\left( \sum_{j=1}^{\varphi^{\Sym_a}} \ln K_{\Sym_a}^{(j)}-\sum_{i=1}^{\varphi^{\Anti_a}}  \ln K_{\Anti_a}^{(i)}\right)
  + 8 \pi^2 \, \Re(\delta_{a'} \, {\rm f}_{SU(N_a)}^{\rm 1-loop})-\frac{\Delta_{aa'} +\Delta_{a,\OR}}{2}
,  
\end{aligned}
\end{equation}
where on the last two lines the K\"ahler metrics have been grouped into contributions from the annulus diagram
with global prefactor $\frac{N_a}{2}$ and those from the M\"obius strip without this factor. 
The gauge threshold contributions in tables~\ref{tab:Z2Z2M-torsion-AntiSym-beta+thresholds},~\ref{tab:Six-torus-AntiSym-beta+thresholds},~\ref{tab:Z2Z2M-no_torsion-Anti+Sym-beta+thresholds} and~\ref{tab:T6-Z2N-Anti+Sym-beta+thresholds}
can be brought to a global form depending on the number of non-vanishing angles,
where our notation is adopted to the $T^6/(\Z_2 \times \Z_{2M} \times \OR)$ orientifold with discrete torsion, 
and modifications for other torus and orbifold backgrounds boil down to setting some constants to zero. 
While for rigid D6-branes, the relative displacements and Wilson lines vanish identically, i.e. 
$ \delta_{\sigma^i_{aa'},0}\delta_{\tau^i_{aa'},0} \equiv 1$ (cf. table~\ref{tab:Z2Z2M-torsion-AntiSym-beta+thresholds}), we keep the notation such that fractional and bulk D6-branes
are taken into account in the following as well.
\begin{itemize}
\item
D6-brane parallel to some $\OR\Z_2^{(m)}$ plane with $m \in \{0\ldots 3\}$ on $T^6/\Z_2 \times \Z_{2M}$ with discrete torsion
and $a \neq a'$ due to the exceptional three-cycles contribute
\begin{equation}\label{Eq:Threh_antisym_parallel}
\begin{aligned}
\frac{\Delta_{aa'} + \Delta_{a,\OR}}{2} =& - \frac{1}{2} \,\sum_{i=1}^3 \left( b_{aa'}^{{\cal A},(i)} \, \Lambda_{0,0}(v_i; V_{aa'}^{(i)}) 
+b_{aa'}^{{\cal M},(i)} \, \Lambda_{0,0}(\tilde{v}_i; 2\tilde{V}_{aa'}^{(i)}) \right)\\
&  - \frac{1}{2} \,\sum_{i=1}^3\left(1-\delta_{\sigma^i_{aa'},0}\delta_{\tau^i_{aa'},0} \right) \, 
\left( \tilde{b}_{aa'}^{{\cal A},(i)} \, \Lambda(\sigma^i_{aa'},\tau^i_{aa'},v_i) + \tilde{b}_{aa'}^{{\cal M},(i)}  \,
\Lambda(\sigma^i_{aa'},\tau^i_{aa'}, \tilde{v}_i)
\right)
\end{aligned}
\end{equation}
to the gauge threshold, cf. table~\ref{tab:Z2Z2M-torsion-AntiSym-beta+thresholds}. This is modified for other backgrounds as follows:
\begin{itemize}
\item 
$T^6$: The annulus amplitude preserves ${\cal N}=4$ supersymmetry with $b_{aa'}^{{\cal A},(i)} \equiv 0$ for $i=1,2,3$. 
For the M\"obius strip two cases need to be distinguished, cf. table~\ref{tab:Six-torus-AntiSym-beta+thresholds}.
\begin{itemize}
\item
If the D6-brane $a$ is parallel to the $\OR$ plane, also the M\"obius strip contribution preserves ${\cal N}=4$ 
and consequently $b_{aa'}^{{\cal M},(i)} \equiv 0$ for all $i$. For $(\vec{\sigma}_{aa'},\vec{\tau}_{aa'}) = (\vec{0},\vec{0})$,
the gauge group is enhanced to $SO(2N_a)$, cf. tables~\ref{tab:SO-Sp-groups} and~\ref{tab:SO-Sp-GaugeKin+KaehlerMetric}, otherwise the symmetric and 
antisymmetric matter states receive a mass, and the matching condition~(\ref{Eq:Matching_aaprime_sector}) is trivially fulfilled.
\item
If the D6-brane $a$ is parallel to some $\OR\Z_2^{(m)}$ plane with $m \in \{1,2,3\}$, the M\"obius strip amplitude 
only preserves ${\cal N}=2$ supersymmetry, i.e. $b_{aa'}^{{\cal M},(m)} \neq 0$ and $b_{aa'}^{{\cal M},(n)} \equiv 0 \equiv
b_{aa'}^{{\cal M},(p)}$, where $(m,n,p)$ is a cyclic permutation of (1,2,3). The gauge group is of $U(N_a)$ type for 
$(\sigma^m_{aa'},\tau^m_{aa'}) \neq (0,0)$ and $Sp(2M_a)$ otherwise, cf. tables~\ref{tab:SO-Sp-groups} and~\ref{tab:SO-Sp-GaugeKin+KaehlerMetric}.
\end{itemize}
\item
$T^6/\Z_{2N}$: The annulus amplitude is ${\cal N}=2$ supersymmetric with $b_{aa'}^{{\cal A},(1)}\equiv 0 \equiv b_{aa'}^{{\cal A},(3)}$. 
For $(\sigma_{aa'}^2,\tau_{aa'}^2) =(0,0)$ the gauge group is of $Sp(2N_a)$ type, cf. tables~\ref{tab:SO-Sp-groups} and~\ref{tab:SO-Sp-GaugeKin+KaehlerMetric},
and $U(N_a)$ otherwise.
The M\"obius strip amplitude belongs to one of the two cases, cf. table~\ref{tab:T6-Z2N-Anti+Sym-beta+thresholds}.  
\begin{itemize}
\item
For $m \in \{0,2\}$ the M\"obius strip amplitude preserves the same ${\cal N}=2$ supersymmetry, i.e. 
$b_{aa'}^{{\cal M},(1)} \equiv 0 \equiv b_{aa'}^{{\cal M},(3)}$.
\item
For $m \in \{1,3\}$ the M\"obius strip amplitude is only ${\cal N}=1$ supersymmetric with 
$b_{aa'}^{{\cal M},(2)} \equiv 0$ but $b_{aa'}^{{\cal M},(1)} ,b_{aa'}^{{\cal M},(3)} \neq 0$ and $b_{aa'}^{{\cal A},(2)} \neq0$, and the Wilson lines and 
displacements on $T^2_{(1)} \times T^2_{(3)}$ only take discrete values, i.e. $\sigma^1_{aa'}=\tau_{aa'}^1=\sigma^3_{aa'}=\tau_{aa'}^3=0$.
\end{itemize}
\item
$T^6/\Z_2 \times \Z_{2M}$ without discrete torsion: The annulus amplitude is (up to normalisation) identical to $T^6$ with 
$b_{aa'}^{{\cal A},(i)} \equiv 0$ for $i=1,2,3$, cf.~\ref{tab:Z2Z2M-no_torsion-Anti+Sym-beta+thresholds}.
\end{itemize}
On the $T^6/\Z_2 \times \Z_{2M}$ orbifold with discrete torsion, all displacements and Wilson lines are discrete, i.e.
$(\vec{\sigma}_{aa'},\vec{\tau}_{aa'}) = (\vec{0},\vec{0})$, and therefore the second line in~(\ref{Eq:Threh_antisym_parallel})
vanishes. 

The two-torus volume $v_i$ dependent one-loop contributions to the holomorphic gauge kinetic functions are now easily read off as displayed in the first two rows 
of table~\ref{tab:Comparison-gaugekin-antisym}. 
\mathsidetabfix{
\begin{array}{|c||c|c|c|c|}\hline
\multicolumn{5}{|c|}{\text{\bf Comparison of one-loop contribution to the gauge kinetic function } \delta_{a'} \, {\rm f}_{SU(N_a)}^{\text{1-loop}}(v_i)
\text{ \bf from (anti)symmetric sectors on various orbifolds}}
\\\hline\hline
(\phi_{aa'}^{(1)},\phi_{aa'}^{(2)},\phi_{aa'}^{(3)}) &  T^6  & \begin{array}{c} T^6/\Z_2 \times \Z_{2M} \\ \text{without torsion}
\end{array} & T^6/\Z_{2N} &\begin{array}{c} T^6/\Z_2 \times \Z_{2M} \\ \text{with discrete torsion} \end{array}
\\\hline\hline
\begin{array}{c} (0,0,0) \\ \pp \OR \end{array} & - 
& \begin{array}{c}- \sum_{i=1}^3 \frac{b_{aa'}^{{\cal M},(i)}}{4\pi^2} \ln \eta(i \tilde{v}_i)
\\-\sum_{i=1}^3 \Biggl[\frac{\tilde{b}_{aa'}^{{\cal M},(i)}}{8\pi^2} (1-\delta_{\sigma^i_{aa'},0}\delta_{\tau^i_{aa'},0}) \times \\
\times \ln \left(e^{-\pi (\sigma^i_{aa'})^2 \tilde{v}_i/4}\frac{\vartheta_1 (\frac{\tau^i_{aa'} - i \sigma^i_{aa'} \tilde{v}_i}{2},i \tilde{v}_i)}{\eta (i \tilde{v}_i)} \right) \Biggr] 
\end{array} 
& \begin{array}{c}- \frac{b_{aa'}^{\cal A}}{4\pi^2} \ln \eta(iv_2) -\frac{b_{aa'}^{\cal M}}{4\pi^2} \ln \eta(i \tilde{v}_2) 
\\ - (1-\delta_{\sigma^2_{aa'},0}\delta_{\tau^2_{aa'},0}) \times \\ \times \Biggl[
\frac{\tilde{b}_{aa'}^{\cal A}}{8\pi^2}
 \ln \left(e^{-\pi (\sigma^2_{aa'})^2 v_2/4}\frac{\vartheta_1 (\frac{\tau^2_{aa'} - i \sigma^2_{aa'} v_2}{2},i v_2)}{\eta (i v_2)} \right) \\
+ \frac{\tilde{b}_{aa'}^{\cal M}}{8\pi^2}
\ln \left(e^{-\pi (\sigma^2_{aa'})^2 \tilde{v}_2/4}\frac{\vartheta_1 (\frac{\tau^2_{aa'} - i \sigma^2_{aa'} \tilde{v}_2}{2},i \tilde{v}_2)}{\eta (i \tilde{v}_2)} \right) \Biggr]
\end{array} 
& \begin{array}{c}- \sum_{i=1}^3 \frac{b_{aa'}^{{\cal A},(i)}}{4\pi^2} \ln \eta(iv_i) \\
- \sum_{i=1}^3 \frac{b_{aa'}^{{\cal M},(i)}}{4\pi^2} \ln \eta(i \tilde{v}_i)
\end{array} 
\\\hline
\begin{array}{c} (0,0,0) \\ \pp \OR \Z_2^{(i)} \end{array} & 
\begin{array}{c}- \frac{b_{aa'}^{\cal M}}{4\pi^2} \ln \eta(i \tilde{v}_i) 
\\- \frac{\tilde{b}_{aa'}^{\cal M}}{8\pi^2} (1-\delta_{\sigma^i_{aa'},0}\delta_{\tau^i_{aa'},0}) \times \\
\times \ln \left(e^{-\pi (\sigma^i_{aa'})^2 \tilde{v}_i/4}\frac{\vartheta_1 (\frac{\tau^i_{aa'} - i \sigma^i_{aa'} \tilde{v}_i}{2},i \tilde{v}_i)}{\eta (i \tilde{v}_i)} \right)
\end{array} 
& \text{same as } \pp \OR 
& \begin{array}{c} \boxed{i=2: }  \quad \text{ same as } \pp \OR 
\\\hline
\boxed{i=1,3: } \qquad -\frac{b_{aa'}^{\cal A}}{4\pi^2} \ln \eta(iv_2)\\
 -\frac{b_{aa'}^{{\cal M},(1)}}{4\pi^2} \ln \eta(i \tilde{v}_1) -\frac{b_{aa'}^{{\cal M},(3)}}{4\pi^2} \ln \eta(i \tilde{v}_3) 
\end{array} 
& \text{same as } \pp \OR 
\\\hline
\!\!\!\!\!
\begin{array}{c}(0^{(i)},\phi^{(j)},\phi^{(k)})\\ \pp \left( \OR + \OR\Z_2^{(i)} \right)\end{array}\!\!\! &
\multicolumn{2}{|c|}{
\begin{array}{c}- \frac{b_{aa'}^{\cal A}}{4\pi^2} \ln \eta(iv_i)- \frac{b_{aa'}^{\cal M}}{4\pi^2} \ln \eta(i \tilde{v}_i)
 \\ - \frac{\tilde{b}_{aa'}^{\cal A}}{8\pi^2} (1-\delta_{\sigma^i_{aa'},0}\delta_{\tau^i_{aa'},0}) 
\ln \left(e^{-\pi (\sigma^i_{aa'})^2 v_i/4}\frac{\vartheta_1 (\frac{\tau^i_{aa'} - i \sigma^i_{aa'} v_i}{2},i v_i)}{\eta (i v_i)} \right) 
\\- \frac{\tilde{b}_{aa'}^{\cal M}}{8\pi^2} (1-\delta_{\sigma^i_{aa'},0}\delta_{\tau^i_{aa'},0}) 
\ln \left(e^{-\pi (\sigma^i_{aa'})^2 \tilde{v}_i/4}\frac{\vartheta_1 (\frac{\tau^i_{aa'} - i \sigma^i_{aa'} \tilde{v}_i}{2},i \tilde{v}_i)}{\eta (i \tilde{v}_i)} \right) 
\end{array} 
}
& \begin{array}{c} \boxed{i=2: } \quad \text{ same as } (0,0,0) \pp \OR 
\\\hline
\boxed{i=1,3: }
-  \frac{b_{aa'}^{{\cal A}}}{4\pi^2} \ln \eta(iv_i)  -  \frac{b_{aa'}^{{\cal M}}}{4\pi^2} \ln \eta(i \tilde{v}_i)
\end{array} 
&  \begin{array}{c}-  \frac{b_{aa'}^{{\cal A}}}{4\pi^2} \ln \eta(iv_i) \\
-  \frac{b_{aa'}^{{\cal M}}}{4\pi^2} \ln \eta(i \tilde{v}_i)
\end{array} 
\\\hline
\!\!\!\!\!
\begin{array}{c}(0^{(i)},\phi^{(j)},\phi^{(k)}) \\ \pp \left( \OR\Z_2^{(j)} + \OR\Z_2^{(k)} \right)\end{array}\!\!\!\!\! & 
\begin{array}{c}- \frac{b_{aa'}^{\cal A}}{4\pi^2} \ln \eta(iv_i) \\ - \frac{\tilde{b}_{aa'}^{\cal A}}{8\pi^2}
(1-\delta_{\sigma^i_{aa'},0}\delta_{\tau^i_{aa'},0}) \times \\ 
\times \ln \left(e^{-\pi (\sigma^i_{aa'})^2 v_i/4}\frac{\vartheta_1 (\frac{\tau^i_{aa'} - i \sigma^i_{aa'} v_i}{2},i v_i)}{\eta (i v_i)} \right) \end{array} 
& \text{same as } \pp \left(\OR + \OR\Z_2^{(i)} \right) 
&  \begin{array}{c} \boxed{i=2: } \qquad - \frac{b_{aa'}^{\cal A}}{4\pi^2} \ln \eta(iv_2)
\\\hline
\boxed{i=1,3: } \quad \text{same as } \pp \left(\OR + \OR\Z_2^{(i)} \right) 
\end{array} 
& \begin{array}{c} \text{same as } \\ \pp \left(\OR + \OR\Z_2^{(i)} \right) \end{array}
\\\hline
(\phi^{(1)},\phi^{(2)},\phi^{(3)}) & \multicolumn{4}{|c|}{-}  
\\\hline
\end{array}
}{Comparison-gaugekin-antisym}{Comparison of the two-torus volume $v_i$ dependent
one-loop contributions $\delta_{a'} {\rm f}_{SU(N_a)}^{\text{1-loop}}(v_i)$ to the holomorphic gauge kinetic function from 
(anti)symmetric sectors. The corresponding beta-function coefficients are given in the second column of table~\ref{tab:Six-torus-AntiSym-beta+thresholds}  for the six-torus, 
table~\protect\ref{tab:Z2Z2M-no_torsion-Anti+Sym-beta+thresholds} for $T^6/\Z_2 \times \Z_{2M}$ 
without torsion, table~\protect\ref{tab:T6-Z2N-Anti+Sym-beta+thresholds} for $T^6/\Z_{2N}$ and table~\protect\ref{tab:Z2Z2M-torsion-AntiSym-beta+thresholds} for $T^6/\Z_2 \times \Z_{2M}$ with discrete torsion.
For the six-torus, the three non-trivial cases are identical since $b_{aa'}^{\cal A}=0$ for $(0,0,0) \pp \OR\Z_2^{(i)}$ and $b_{aa'}^{\cal M}=0$ for 
$(0^{(i)},\phi^{(j)},\phi^{(k)}) \pp \left( \OR\Z_2^{(j)} + \OR\Z_2^{(k)} \right)$, cf. table~\protect\ref{tab:Six-torus-AntiSym-beta+thresholds}. 
Similarly, for $T^6/\Z_{2N}$ the case $(0^{(i)},\phi^{(j)},\phi^{(k)}) \pp \left( \OR\Z_2^{(j)} + \OR\Z_2^{(k)} \right)$ is identical to $(0^{(i)},\phi^{(j)},\phi^{(k)}) \pp \left( \OR + \OR\Z_2^{(i)} \right)$ 
since $b_{aa'}^{\cal M} \equiv 0$ for the former, cf.  table~\protect\ref{tab:T6-Z2N-Anti+Sym-beta+thresholds}.
}
The K\"ahler metrics for symmetric or antisymmetric matter take the form
\begin{equation}\label{Eq:K_AntiSym_parallel}
K_{\Anti_a}^{(i)} =  f(S,U_l) \;   \sqrt{\frac{ 2 \pi V_{aa'}^{(i)}}{v_jv_k}}
\qquad
\text{ for vanishing angles}
,
\end{equation}
which is identical to the one for bifundamental matter on parallel D6-branes~(\ref{Eq:Kaehler-general-1angle}).
Here we have used the fact that the beta function coefficients in tables~\ref{tab:Z2Z2M-torsion-AntiSym-beta+thresholds},~\ref{tab:Six-torus-AntiSym-beta+thresholds},~\ref{tab:Z2Z2M-no_torsion-Anti+Sym-beta+thresholds} and~\ref{tab:T6-Z2N-Anti+Sym-beta+thresholds}
can be evaluated on a case-by case basis for vanishing angles leading to 
\begin{equation*}
(\varphi^{\Anti_a},\varphi^{\Sym_a}) = \left\{\begin{array}{cl} (2,0) & T^6/\Z_2 \times \Z_{2M} \text{ with } \eta=-1
\\ (4,0) & T^6/\Z_{2N} 
\end{array}\right.
.
\end{equation*}
In all other cases, the symmetric or antisymmetric matter is either massive due to some non-vanishing displacement or Wilson line,
or the gauge group is enhanced to $SO(2N_a)$ or $Sp(2N_a)$ as discussed in section~\ref{Sss:AntiSym}.

The gauge threshold contains a constant contribution which we assign to the holomorphic gauge kinetic function,
\begin{equation}\label{Eq:ConstContr}
\Re(\delta_{a'} {\rm f}^{\text{1-loop}}_{SU(N_a)}(c)) =\left\{\begin{array}{c}
\frac{1}{2\pi^2}  \ln \frac{2 \, \left(v_1 v_3V^{(1)}_{aa'} V^{(3)}_{aa'}\right)^{1/4}}{\sqrt{v_2V^{(2)}_{aa'}}} \qquad T^6/\Z_{2N} \text{ and } a \perp \OR \text{ on }T^2_{(2)}
\\
  - \frac{b^{\cal M}_{aa'}}{8 \pi^2} \, \ln 2
= \left\{\begin{array}{cc} \frac{1}{4\pi^2} \ln 2
& T^6/\Z_2 \times \Z_{2M} \text{ with } \eta=-1 \\
\frac{1}{2 \pi^2} \ln 2& T^6/\Z_{2N} \text{ and } a \pp \OR \text{ on }T^2_{(2)}
\\ 0 & \text{otherwise}
\end{array}\right.
\end{array}\right.  
,
\end{equation}
where in a slight abuse of notation we have included a logarithmic dependence on the two-torus volumes for $T^6/\Z_{2N}$ and $a \perp \OR$
along $T^2_{(2)}$. All other dependences on these variables are contained in the usual Dedekind eta and Jacobi theta functions. 
\item
A D6-brane parallel  to some $\OR\Z_2^{(m)}$ and $\OR\Z_2^{(n)}$ plane with $m,n \in \{0\ldots 3\}$ along one two-torus $T^2_{(i)}$
and at non-trivial angles on the remaining four-torus (with $(m,n,p,q)$ some cyclic permutation of (0,1,2,3))
contributes the following to the gauge thresholds,
\begin{equation}\hspace{-15mm}
\begin{aligned}
\frac{\Delta_{aa'} + \Delta_{a,\OR}}{2} =&- \frac{1}{2} \, \left( b_{aa'}^{{\cal A}} \, \Lambda_{0,0}(v_i; V_{aa'}^{(i)})  +b_{aa'}^{{\cal M}}\; \Lambda_{0,0}(\tilde{v}_i; 2\tilde{V}_{aa'}^{(i)}) \right)
\\
& - \frac{1}{2} \, \left(1-\delta_{\sigma^i_{aa'},0}\delta_{\tau^i_{aa'},0} \right) \, \left( \tilde{b}_{aa'}^{{\cal A}} \, \Lambda(\sigma^i_{aa'},\tau^i_{aa'},v_i)
+\tilde{b}_{aa'}^{{\cal M}}\; \Lambda(\sigma^i_{aa'},\tau^i_{aa'},\tilde{v}_i)\right)
\\
& + \left[- \frac{N_a}{2} \, \sum_{l \neq i} \frac{I_{aa'}^{\Z_2^{(l)}}}{c_a} \left(2 \, \phi^{(l)}_{aa'} - \sgn(\phi^{(l)}_{aa'})  \right) 
+ \frac{\eta_{\OR\Z_2^{(p)}} |\tilde{I}_a^{\OR\Z_2^{(p)}}| + \eta_{\OR\Z_2^{(q)}} |\tilde{I}_a^{\OR\Z_2^{(q)}}|}{c_a} \right]\ln 2
.
\end{aligned}
\end{equation}
The second line is again absent for rigid D6-branes on $T^6/\Z_2 \times \Z_{2M}$ with discrete torsion, cf. table~\ref{tab:Z2Z2M-torsion-AntiSym-beta+thresholds}, while 
other orbifold backgrounds have the following simplifications. 
\begin{itemize}
\item 
$T^6$: there is no contribution from $\Z_2$ fixed points, i.e. $I_{aa'}^{\Z_2^{(l)}} \equiv 0$ for \mbox{$l \in \{1,2,3\}$},
and only one regular O6-plane exists, i.e. $\eta_{\OR} \equiv 1$ and $\eta_{\OR\Z_2^{(l)}} \equiv 0$ for all $l \in \{1,2,3\}$.
Two cases need to be distinguished, cf. table~\ref{tab:Six-torus-AntiSym-beta+thresholds}.
\begin{itemize}
\item
For the D6-branes parallel to the $\OR$-plane along $T^2_{(i)}$, the two contributions $b_{aa'}^{\cal A}$ and $b_{aa'}^{\cal M}$ to
the beta function coefficients are non-vanishing, but $\eta_{\OR\Z_2^{(p)}} |\tilde{I}_a^{\OR\Z_2^{(p)}}| + \eta_{\OR\Z_2^{(q)}} |\tilde{I}_a^{\OR\Z_2^{(q)}}=0$.
\item
For the D6-branes perpendicular to the $\OR$-plane along $T^2_{(i)}$, the M\"obius strip contribution to the
beta function vanishes, $b_{aa'}^{\cal M}=0$, but the intersections with the O6-planes contribute a constant to the 
gauge threshold, $\eta_{\OR\Z_2^{(p)}} |\tilde{I}_a^{\OR\Z_2^{(p)}}| + \eta_{\OR\Z_2^{(q)}} |\tilde{I}_a^{\OR\Z_2^{(q)}}| =|\tilde{I}_a^{\OR}| + |\tilde{I}_a^{\OR\Z_2^{(2)}}|$. 
\end{itemize}
\item
$T^6/\Z_{2N}$: only $\Z_2^{(2)}$ forms a subgroup of $\Z_{2N}$, i.e. $I_{aa'}^{\Z_2^{(1)}} \equiv 0 \equiv I_{aa'}^{\Z_2^{(3)}}$,
and only two regular O6-planes exist, i.e. $\eta_{\OR} \equiv 1 \equiv \eta_{\OR\Z_2^{(2)}}$ and  $\eta_{\OR\Z_2^{(1)}} \equiv 0
\equiv \eta_{\OR\Z_2^{(3)}}$. Three cases have to be distinguished, cf. table~\ref{tab:T6-Z2N-Anti+Sym-beta+thresholds}.
\begin{itemize}
\item
If the D6-branes are parallel to the $\OR$-plane along $T^2_{(2)}$, the constant term from the O6-plane intersections vanishes,
\mbox{$\eta_{\OR\Z_2^{(p)}} |\tilde{I}_a^{\OR\Z_2^{(p)}}| + \eta_{\OR\Z_2^{(q)}} |\tilde{I}_a^{\OR\Z_2^{(q)}}|=0$}.
\item
For the D6-branes perpendicular to the $\OR$-plane along $T^2_{(2)}$, the M\"obius strip does not contribute to the
beta function, $b_{aa'}^{\cal M}=0$, but the O6-plane intersections contribute to the gauge threshold,
\mbox{$\eta_{\OR\Z_2^{(p)}} |\tilde{I}_a^{\OR\Z_2^{(p)}}| + \eta_{\OR\Z_2^{(q)}} |\tilde{I}_a^{\OR\Z_2^{(q)}}|=
|\tilde{I}_a^{\OR}| + |\tilde{I}_a^{\OR\Z_2^{(2)}}|$}.
\item
If the stack of D6-branes is parallel or perpendicular to the $\OR$-plane along $T^2_{(i)}$ with $i \in \{1,3\}$, the
displacement and Wilson line only take discrete values, i.e. $\delta_{\sigma_{aa'}^i,0}\delta_{\tau_{aa'}^i,0}=1$,
and only one of the O6-plane intersections contributes, 
$\eta_{\OR\Z_2^{(p)}} |\tilde{I}_a^{\OR\Z_2^{(p)}}| + \eta_{\OR\Z_2^{(q)}} |\tilde{I}_a^{\OR\Z_2^{(q)}}|=
|\tilde{I}_a^{\OR\Z_2^{(2)}}|$ or $|\tilde{I}_a^{\OR}|$ for parallel or perpendicular to $\OR$, respectively.
\end{itemize}
\item
$T^6/\Z_2 \times \Z_{2M}$ without torsion: as for $T^6$ there is no fixed point contribution, i.e. $I_{aa'}^{\Z_2^{(l)}} \equiv 0$
for all $l$, and all O6-planes are regular, i.e. $\eta_{\OR} \equiv 1 \equiv \eta_{\OR\Z_2^{(l)}}$ for all $l$,
cf. table~~\ref{tab:Z2Z2M-no_torsion-Anti+Sym-beta+thresholds}.
\end{itemize}
The $v_i$ dependent contributions to the holomorphic gauge kinetic function  from this sector are displayed in the third and fourth row in table~\ref{tab:Comparison-gaugekin-antisym} for all orbifolds under consideration.
The functional dependence of the K\"ahler metrics in this sector is identical to~(\ref{Eq:K_AntiSym_parallel}),
\begin{equation}\label{Eq:K_AntiSym_1angle}
K_{\Anti_a/\Sym_a}^{(i)} =  f(S,U_l) \;   \sqrt{\frac{ 2 \pi V_{aa'}^{(i)}}{v_jv_k}}
\qquad
\text{ for one vanishing angle}
,
\end{equation}
and additionally there is an angle dependent contribution  identical to~(\ref{Eq:2angle-f-angle}) plus a constant contribution
 to the holomorphic gauge kinetic function,
\begin{equation}\label{Eq:aap2angle-f-angle}
\begin{aligned}
\Re \left(\delta_{a'} \, {\rm f}^{\text{1-loop}}_{SU(N_a)}(\phi^{(i)}_{ab})\right) =& \frac{1}{8 \pi^2 \, c_a} \Biggl(-
\frac{N_{a}}{2} \sum_{l=j,k} I_{aa'}^{\Z_2^{(l)}} \left(2 \, \phi^{(l)}_{aa'} - \sgn(\phi^{(l)}_{aa'}) \right)
\\ 
& \hspace{13mm} + \eta_{\OR\Z_2^{(p)}} |\tilde{I}_a^{\OR\Z_2^{(p)}}| + \eta_{\OR\Z_2^{(q)}} |\tilde{I}_a^{\OR\Z_2^{(q)}}|
\Biggr)
.
\end{aligned}
\end{equation}
This formula holds again for all torus and orbifold backgrounds when setting some of the intersection numbers to zero as detailed above. Examples with non-vanishing~(\ref{Eq:aap2angle-f-angle}) and no massless matter are given by the $dd'$ and $d(\theta^2 d')$
sectors of the Standard Model on $T^6/\Z_6'$ in table~\ref{tab:modelz6p+hidden-sectors_ax-Kaehler}, supporting again the 
correct assignment of the terms to the holomorphic gauge kinetic function.
\item
If the stack of D6-branes is at three non-trivial angles to the O6-planes, the gauge threshold computation gives 
\begin{equation}\hspace{-15mm}\label{Eq:aap-Delta-3angles}
\begin{aligned}
\frac{\Delta_{aa'} + \Delta_{a,\OR}}{2} =&\, - \frac{ b_{aa'}^{\cal A} +b_{aa'}^{\cal M} }{2}  \, \sum_{i=1}^3 \, \ln \left(\frac{\Gamma(|\phi^{(i)}
_{aa'}|)}{\Gamma(1-|\phi^{(i)}_{aa'}|)}\right)^{-\frac{\sgn(\phi^{(i)}_{aa'})}{\sgn(I_{aa'})}}
\\
& + \left[- \frac{N_a}{2} \sum_{i=1}^3 \frac{I_{aa'}^{\Z_2^{(i)}}}{c_a} \left( 2 \, \phi^{(i)}_{aa'} - \sgn(\phi^{(i)}_{aa'}) - \sgn(I_{aa'})\right) 
+\sum_{m=0}^3 \frac{\eta_{\OR\Z_2^{(m)}} |\tilde{I}_z^{\OR\Z_2^{(m)}}|}{c_a}
\right]
\ln 2
.
\end{aligned}
\end{equation}
This formula holds for all toroidal orbifold backgrounds discussed in this article, provided the
vanishing of some $\Z_2$ fixed or orientifold invariant intersection points is taken into account as detailed
for the case of one vanishing angle.

The sectors with three non-trivial angles do not contribute to the $v_i$-dependent part of the holomorphic gauge kinetic function.
The K\"ahler metrics for symmetric and antisymmetric matter at $aa'$ intersections are given by
\begin{equation}
K_{\Anti_a/\Sym_a} =  f(S,U_l) \; \sqrt{\prod_{i=1}^3 \frac{1}{v_i} \,
\left(\frac{\Gamma(|\phi^{(i)}_{aa'}|)}{\Gamma(1-|\phi^{(i)}_{aa'}|)} \right)^{-\frac{\sgn(\phi^{(i)}_{aa'})}{\sgn(I_{aa'})}}}
,
\end{equation}
and the additional angle-dependent and constant terms in the gauge threshold~(\ref{Eq:aap-Delta-3angles})
are assigned to the loop-correction to the holomorphic gauge kinetic function,
\begin{equation}\hspace{-15mm}\label{Eq:aap-f-3angles}
\begin{aligned}
\Re \left(\delta_{a'} \, {\rm f}^{\text{1-loop}}_{SU(N_a)}(\phi^{(i)}_{aa'})\right) =&
\frac{1}{8\pi^2 \, c_a}
 \Biggr(- \frac{N_a}{2} \sum_{i=1}^3 I_{aa'}^{\Z_2^{(i)}} \left( 2 \, \phi^{(i)}_{aa'} - \sgn(\phi^{(i)}_{aa'}) - \sgn(I_{aa'})\right) 
\\
& \hspace{15mm} +\sum_{m=0}^3 \eta_{\OR\Z_2^{(m)}} |\tilde{I}_a^{\OR\Z_2^{(m)}}| \Biggl) \, \ln 2
.
\end{aligned}
\end{equation}
The case with one vanishing angle~(\ref{Eq:aap2angle-f-angle}) can again be included by noticing that some of the intersection numbers
vanish.
\end{itemize}
\end{itemize}

For later comparison with the
case of a single $U(1)_a \subset U(N_a)$ factor in section~\ref{Sss:Single_U1_GaugeKinetic}, we can formally decompose the 
one-loop correction to the holomorphic gauge kinetic function from the sector $aa'$ into its annulus and M\"obius strip contributions,
\begin{equation}
\delta_{a'} \, {\rm f}^{\text{1-loop}}_{SU(N_a)} 
\equiv \delta_{a'} \, {\rm f}^{\text{1-loop},{\cal A}}_{SU(N_a)} + \delta_{a'} \, {\rm f}^{\text{1-loop}, {\cal M}}_{SU(N_a)}
.
\end{equation}
The complete list of such $v_i$-dependent contributions for any torus or orbifold background considered in this article
is listed in table~\ref{tab:Comparison-gaugekin-antisym}. In the angle plus constant term~(\ref{Eq:aap-f-3angles}), the
annulus and M\"obius strip contributions correspond to the first and second line, respectively.

\subsubsection{Complexification and one-loop redefinition of the closed string moduli}\label{Ss:loop-redef-S+Ui}

{\bf Complexification of moduli by axions}\\
The ansatz for the tree level gauge kinetic function~(\ref{Eq:tree-gauge-coupling}) and gauge threshold amplitudes~(\ref{Eq:b-Delta}) uses the geometric moduli only.
However, the ${\cal N}=1$ field theory depends on the complexifications of the dilaton $S$ and complex structures $U_l$ and K\"ahler moduli $v_i$,
\begin{equation}
S^c = S + i \, \xi_0,
\qquad
U_l^c=U_l + i \,  \xi_l,
\qquad
T^c_i=v_i + i \, b_i ,
\end{equation}
with the RR axions $\xi_{L,L=0 \ldots h_{21}} = \int_{L^{\rm th}-{\cal R} \text{ even 3-cycle}} C_3$ and the NSNS axions \linebreak
\mbox{$b_{i,i=1\ldots h_{11}}= \int_{i^{\rm th}-{\cal R} \text{ odd 2-cycle}} B_2$}, see~\cite{Grimm:2004ua} for the derivation of IIA orientifolds
on smooth Calabi-Yau spaces and Appendix A of~\cite{Forste:2010gw} for the evaluation of the closed string spectrum on all type IIA  toroidal orbifolds considered here.

The rewritten form (\ref{Eq:Lambdas-rewritten}) of the lattice sums and the transformation of  Dedekind eta and Jacobi theta functions
under complex conjugation,
\begin{equation*}
\ov{\eta(\tau) }= \eta(-\ov{\tau}) 
,
\qquad\qquad
\overline{\vartheta_i (\nu,\tau)} =\vartheta_i (\ov{\nu},-\ov{\tau})
\qquad
i=1\ldots 4,
\end{equation*}
justifies the ansatz, e.g.~\cite{Akerblom:2007uc,Blumenhagen:2007ip}, to replace the moduli by their complexifications in the tree-level and
all one-loop corrections to the holomorphic gauge kinetic functions in equations~(\ref{Eq:tree-level-fhol_Z2Z2}),~(\ref{Eq:tree-f-Z6p}) and~(\ref{Eq:tree-f-Z6})
 and tables~\ref{tab:Comparison-gaugekin-bifund} and~\ref{tab:Comparison-gaugekin-antisym}, if at the same time
the displacement and Wilson line moduli are paired into complex scalars,
\begin{equation}
\Sigma^i_a=\sigma^i_a + i \, \frac{\tau^i_a}{v_i},
\end{equation}
leading to the complexification $-i \frac{T^c_i \Sigma^i_{ab}}{2}$ of the argument $\frac{\tau^i_{ab} - i \sigma^i_{ab}}{2}$ of the Jacobi theta function $\vartheta_1(\frac{\tau^i_{ab} - i \sigma^i_{ab}}{2},iv_i)$ in the one-loop contributions to the holomorphic gauge kinetic functions from D6-branes with some non-vanishing relative displacement or Wilson line.

{\bf One-loop field redefinitions}\\
In  type IIA compactifications, the four dimensional dilaton and complex structure moduli participate in the generalised Green-Schwarz mechanism for anomalous $U(1)$ gauge factors.
This leads to one-loop redefinitions of the field theory expressions,
\begin{equation}
S \to S + \delta_{GS} \, S,
\qquad
U_l \to U_l + \delta_{GS} \, U_l,
\end{equation}
under gauge transformations of the anomalous (massive) Abelian gauge factors.
This field redefinition cannot be seen in the matching of gauge thresholds computed by CFT methods
with the standard ${\cal N}=1$ supergravity expressions above.
There exist different proposals for the field redefinitions in the literature, which we discuss below.
\begin{itemize}
\item
In~\cite{Akerblom:2007uc,Blumenhagen:2007ip}, it was proposed that the one-loop redefinitions are given by
\begin{equation}\label{Eq:wrong-field-redefinitions}
\delta_{GS} \, S= \mp \frac{1}{16 \pi^2 c_a} \sum_b N_b \tilde{Y}_b^0 \sum_{i=1}^3 \phi^{(i)}_b \, \ln v_i
,
\quad
\delta_{GS} \, U_l=\pm \frac{1}{16 \pi^2 c_a} \sum_b N_b \tilde{Y}_b^l \sum_{i=1}^3 \phi^{(i)}_b \, \ln v_i
,
\end{equation}
with the orientifold odd three-cycle wrapping numbers $ \tilde{Y}_b^l$ defined in equation~(\ref{Eq:def-XY}),
providing a mixing with the bulk K\"ahler moduli.
It was furthermore argued that the angle dependent contributions~(\ref{Eq:3angle-f-angle}) to the holomorphic gauge kinetic function
might  be the result of a one-loop redefinition of exceptional complex structure moduli.
We were unable to reproduce the proposed transformations for the toroidal orbifolds considered in this article for the 
following reasons.
\begin{enumerate}
\item
The derivation of these transformations relies on chiral matter only existing at three intersection angles
and consequently the same type of K\"ahler metrics~(\ref{Eq:Kaehler-general-3angles}) with different angles as arguments for all chiral states. 
However, in the presence of $\Z_2$ symmetries, additional chiral matter states can arise at one vanishing
intersection angle with a different shape of the K\"ahler metrics~(\ref{Eq:Kaehler-general-1angle}). A 
prominent example of this kind are two generations of right-handed quarks from the $ac$ sector in the $T^6/\Z_6'$
example in table~\ref{tab:modelz6p+hidden-sectors_ax-Kaehler} versus the third generation from the $a(\theta^2 c)$ sector below.
\item
For orbifolds with $\Z_3$ or $\Z_4$ subsymmetry, the bulk three-cycle wrapping numbers are sums of toroidal wrapping
numbers over all images $(\omega^k a)$ of $a$. While the net-intersection number for exceptional three-cycles is given by
$\frac{1}{c_a} \sum_{m=0}^{M-1} I_{(\omega^m a)b}^{\Z_2^{(i)}}$, cf. table~\ref{Tab:NormIntersections}, the angle dependent 
one-loop contribution to the gauge thresholds (\ref{Eq:3angle-Delta-terms})
 requires to simultaneously transform the angles \mbox{$\phi^{(i)}_{(\omega^m a)b}=\phi^{(i)}_{(\omega^m a)b} -2\pi m w_i$ mod 1}. 
 This is in contradiction to factorising out the bulk wrapping numbers in~(\ref{Eq:wrong-field-redefinitions}), especially since $\phi^{(2)}_{x(\omega^k y)}=0$
 occurs for some orbifold images $(\omega^k y)$, e.g. for $x(\omega^k y)\in \{b(\theta^2 d), cd, bd'\}$ of the $T^6/\Z_6'$ example in section~\ref{S:Z6p-Example}.
 \item
 The combination of $ \tilde{Y}_b^l$ times the angle $\phi^{(i)}_b$ in~(\ref{Eq:wrong-field-redefinitions}) leads to identical expressions for 
 orientifold image D6-branes $b$ and $b'$ with opposite $U(1)$ charges. However, the anomaly of some Abelian gauge factor is in direct correspondence to the 
 chirality of the massless states and thus also the signs of the corresponding $U(1)$ charges. 
 \end{enumerate}
In conclusion, the proposed field redefinitions~(\ref{Eq:wrong-field-redefinitions}) cannot apply to toroidal orbifolds with $\Z_2$ subsectors or non-trivial orbifold image cycles,
and are therefore also likely to not  occur for the six-torus, for which already a contradiction concerning chirality arises.
\item
In~\cite{Angelantonj:2009yj}, the one-loop field redefinitions are expressed as 
\begin{equation}\label{Eq:other-field-redefinitions}
\delta_{GS} \, S=  \lambda_S \sum_b N_b \tilde{Y}_b^0 \, \Lambda^{(b)}
,
\qquad
\delta_{GS} \, U_l=\lambda_l \sum_b N_b \tilde{Y}_b^l \, \Lambda^{(b)}
,
\end{equation}
in terms of the gauge transformation parameters $\Lambda^{(b)}$ of anomalous massive $U(1)_b$ gauge factors,
where $\lambda_S, \lambda_l$ are normalisation constants which depend on the chiral spectrum. 
They are encoded in the anomaly matrix with components~\cite{Angelantonj:2009yj}, 
\begin{equation}\label{Eq:Def-AnomalyMatrix-Entries}
C_{ab} = \frac{1}{4 \pi^2} \, {\rm tr} (Q_a^2 Q_b)
,
\qquad
C_{aa} = \frac{1}{12 \pi^2} \, {\rm tr} (Q_a^3)
,
\end{equation}
which is computed from the chiral spectrum in table~\ref{Tab:NormIntersections} on the left hand side.
The expressions~(\ref{Eq:other-field-redefinitions}) are consistent with summations over orbifold image D6-branes as well as with the distinction of 
orientifold image D6-branes with opposite $U(1)$ charge assignments.\\
We give the anomaly matrices for all examples on $T^6/\Z_2 \times \Z_2$ with discrete torsion and on $T^6/\Z_6'$
in the corresponding sections, but leave the evaluation of the constants $\lambda_S,\lambda_l$ to the interested reader
since gauge transformations of anomalous $U(1)$s are not relevant for the perturbative treatment of the effective action performed in this article.
\end{itemize}

\subsubsection{Modifications for $T^6/\Z_{2N}$ and $T^6/\Z_2 \times \Z_{2M}$ with $\eta=\pm 1$ and $h_{21}^{\rm bulk}=0,1$}\label{Ss:Modifications-less-complex}

In section~\ref{Ss:Tree-GaugeKin}, the holomorphic gauge kinetic function at tree level was derived for $h_{21}^{\rm bulk}=3$,
which is valid for the six-torus and its orbifolds $T^6/\Z_2$ and $T^6/\Z_2 \times \Z_2$ with and without discrete torsion.
All other orbifolds in table~\ref{Tab:T6ZN+T6Z2Z2M-shifts} have a reduced number of bulk complex structures due to some
$\Z_3$ or $\Z_4$ subsymmetry.

As detailed in appendix~\ref{App:V-for-Z6+Z6p}, the $T^6/\Z_6'$ and $T^6/\Z_2 \times \Z_6$ orbifolds project out two of 
the three bulk complex structure moduli yielding the modified definitions of the field theoretical dilaton and complex structure, 
\begin{equation}\label{Eq:Def-S+Ui=1}
S \sim \frac{e^{-\phi_4}}{\sqrt{r}}
,
\qquad
U \sim e^{-\phi_4}  \, \sqrt{r}
,
\end{equation}
where $r$ is the ratio of radii on the one two-torus where only a $\Z_2$ symmetry acts.
This leads to the relation 
\begin{equation}\label{Eq:tree-f-Z6p}
\Re({\rm f}_{SU(N_a)}^{\text{tree}}) \stackrel{!}{=} \frac{1}{g^2_{a,\text{tree}}}
 \sim \frac{1}{k_ac_a} \left(S  \tilde{X}_a^0- U \tilde{X}_a^1 \right)
\qquad
\text{on}
\quad
T^6/(\Z_6' \times \OR) \text{ and } T^6/(\Z_2 \times \Z_6 \times \OR)
,
\end{equation}
with the bulk wrapping numbers $\tilde{X}_a^0,\tilde{X}_a^1$ for all six inequivalent lattices given in table~\ref{Tab:T6Z6p+T6Z2Z6-OR_even+odd}
of appendix~\ref{App:V-Z6p}.
The corresponding results for $T^6/\Z_4$ and $T^6/\Z_2 \times \Z_4$ with also one bulk complex structure modulus
are given in appendix~\ref{App:V-Z4}, table~\ref{Tab:T6Z4+T6Z2Z4-OR_even+odd}.

The $T^6/\Z_6$ and $T^6/\Z_2 \times \Z_6'$ orbifolds have no bulk complex structure modulus, $S \sim  e^{-\phi_4}$,  and thus 
\begin{equation}\label{Eq:tree-f-Z6}
\Re({\rm f}_{SU(N_a)}^{\text{tree}}) \stackrel{!}{=} \frac{1}{g^2_{a,\text{tree}}}
 \sim \frac{1}{k_ac_a} \, S  \tilde{X}_a^0
\qquad
\text{on}
\quad
T^6/(\Z_6 \times \OR) \text{ and } T^6/(\Z_2 \times \Z_6' \times \OR)
,
\end{equation}
with $\tilde{X}_a^0$ listed in table~\ref{Tab:T6Z6+T6Z2Z6p-OR_even+odd} in appendix~\ref{App:V-Z6}
for the different background lattices.

At one-loop, the ratio of the mass scales $M_{\rm Planck}/M_{\rm string} \sim e^{-\phi_4}$ is recovered by 
replacing the product of bulk moduli in~(\ref{Eq:Def-Kaehler_adjoints}) as follows,
\begin{equation}\label{Eq:Modified_fSU}
f(S,U_l) = \left( S \prod_{l=1}^{h_{21}^{\rm bulk}} U_l \right)^{-\alpha/4}
\qquad
\text{with}
\quad 
\alpha=\left\{\begin{array}{cr} 1 & h_{21}^{\rm bulk}=3 \\ 2 & 1 \\ 4 & 0 \end{array}\right.
.
\end{equation}
The matching of the one-loop field and string theory results for the gauge coupling
with the terms involving the K\"ahler potential requires the simultaneous replacement
\begin{equation}\label{Eq:Def-K_bulk-modified}
{\cal K}_{\text{bulk}} = - \alpha \ln S - \alpha \sum_{l=1}^{h_{21}^{\rm bulk}} \ln U_l - \sum_{i=1}^3 \ln v_i
.
\end{equation}
Both modifications~(\ref{Eq:Modified_fSU}) and~(\ref{Eq:Def-K_bulk-modified}) can also be understood 
from the fact that for $r_2,r_3$ fixed constants, the definition~(\ref{Eq:Def-S+Ui=3}) of the bulk 
complex structures and dilaton on the factorisable six-torus leads to $S \propto U_1$ and $U_2 \propto U_3$.
If also $r_1$ is a fixed constant, all four moduli are related, $S \propto U_1 \propto U_2 \propto U_3$.
Inserting these relations in~(\ref{Eq:Def-Kaehler_adjoints}) and~(\ref{Eq:Def-K_bulk}) leads to the exponent $\alpha$ in~(\ref{Eq:Modified_fSU}) 
and~(\ref{Eq:Def-K_bulk-modified}), respectively.

\subsection{K\"ahler metrics and holomorphic gauge kinetic functions for $SO(2M)$ and $Sp(2M)$}\label{Ss:SO+Sp_Kaehler+GaugeKinetic}

For D6-branes which are their own orientifold images, $x=x'$, the quadratic Casimir of the adjoint representation in 
equation~(\ref{Eq:Gauge-FieldTheory}) is given by $C_2(\Adj_x)= M_x + \xi_x$, where $\xi_x=\pm 1$ for $Sp(2M_x)$ and $SO(2M_x)$, respectively.
With the beta function coefficients given by~(\ref{Eq:Def-beta-SO+Sp}) and the multiplicities of symmetric and antisymmetric representations 
listed in table~\ref{tab:SO-Sp-groups}, the matching of the string and field theory computation reads
\begin{equation}\label{Eq:Matching_xx_sector_SO+Sp}
\begin{aligned}
0
\stackrel{!}{=}& \, - \frac{M_x(-3 + \varphi^{\Sym_x}+\varphi^{\Anti_x}) + (\varphi^{\Sym_x}-\varphi^{\Anti_x} - 3 \, \xi_x)}{2} 
\; \ln\left[ \left(S \prod_{i=1}^3 U_i \right)^{1/2} \prod_{j=1}^3 v_j \right] \\
& + (M_x+\xi_x) \, \ln \left[\left(S \prod_{i=1}^3 U_i \right)^{-3/4} \frac{(2\pi)^{3/2} \prod_{i=1}^3 
\sqrt{V_{xx}^{(i)}}}{ \left(\prod_{k=1}^3 v_k \right) \, c_xk_x}\right]
- \varphi^{\Anti_x} (M_x-1)  \ln K_{\Anti_x} \\ &-  \varphi^{\Sym_x} (M_x+1)  \ln K_{\Sym_x}
+ 8 \pi^2 \, \Re(\delta_x \, {\rm f}_{SO/Sp(2M_x)}^{\rm 1-loop}) - \frac{M_x\tilde{\Delta}_{xx}+ \Delta_{x,\OR}/2}{2}
,
\end{aligned}
\end{equation}
where the relations for the mass scales, bulk K\"ahler potential and tree level gauge couplings have been inserted.
The matching condition can be evaluated on a case-by case basis, for which the result is summarised in 
table~\ref{tab:SO-Sp-GaugeKin+KaehlerMetric}. The derivation requires a distinction of the following cases.
\mathtabfix{
\begin{array}{|c||c|c||c|}\hline
\multicolumn{4}{|c|}{\text{\bf 1-loop gauge kinetic functions and K\"ahler metrics for $SO(2M_x)$ and $Sp(2M_x)$}}
\\\hline\hline
(\phi_{xx'}^{(1)},\phi_{xx'}^{(2)},\phi_{xx'}^{(3)}) & \begin{sideways} \!\!\!\!\ \text{SUSY} \end{sideways} 
& \delta_x \, {\rm f}^{\text{1-loop}}_{SO/Sp(2M_x)} & K_{\Anti_x}^{(i)} , K_{\Sym_x}^{(i)}  
\\\hline\hline
\multicolumn{4}{|c|}{T^6 \text{ and } T^6/\Z_3 }
\\\hline
\begin{array}{c}   (0,0,0)  \\ \pp \OR \end{array} 
&\begin{sideways}  \!\!\!\!\!\!\!\! ${\cal N}=4$  \end{sideways} 
& - & K_{\Anti_x}^{(i),i=1,2,3} =f(S,U_l) \;    \frac{(2\pi)^{1/2}}{2^{1/3} \, v_i}
\sqrt{ \frac{V_{xx}^{(j)}V_{xx}^{(k)} }{V_{xx}^{(i)}}}
\\\hline
\begin{array}{c}   (0,0,0)  \\ \perp \OR \text{ on } \\ T^2_{(j)} \times T^2_{(k)}\end{array}  
& \begin{sideways} \!\!\!\!\!\!\!\! ${\cal N}=2$ \end{sideways} 
& \frac{1}{\pi^2}\ln \left( 2^{2/3} \, \eta(i\tilde{v}_i) \right)
& \begin{array}{c}
 K_{\Sym_x}^{(i)} =f(S,U_l) \;   \frac{\sqrt{2\pi}}{2^{1/3} \, v_i} \sqrt{\frac{V_{x}^{(j)} V_{x}^{(k)}}{V_{x}^{(i)}} } 
 \\
K_{\Anti_x}^{(j)} =f(S,U_l) \;   \frac{\sqrt{2\pi}}{2^{1/3} \, v_j} \sqrt{\frac{V_{x}^{(i)} V_{x}^{(k)}}{V_{x}^{(j)}} }
\\
 K_{\Anti_x}^{(k)} =f(S,U_l) \;   \frac{\sqrt{2\pi}}{2^{1/3} \, v_k} \sqrt{\frac{V_{x}^{(i)} V_{x}^{(j)}}{V_{x}^{(k)}} }
\end{array}
\\\hline\hline
\multicolumn{4}{|c|}{T^6/\Z_{2N} }
\\\hline
\!\!\!\!\begin{array}{c}  (0,0,0) \\ \pp \OR \text{ or } \OR\Z_2^{(2)} \end{array}\!\!\!\!  
& \begin{sideways} \!\!\!\!\!\!\!\! ${\cal N}=2$ \end{sideways} 
& \begin{array}{c} \frac{1}{2\pi^2} \Bigl( M_x \ln  \left( 2^{1/6}\, \eta(i v_2)\right)\\ +  
\ln \left(2^{2/3}\,\eta(i\tilde{v}_2)\right)\Bigr) \end{array} 
& K_{\Sym_x}^{(2)} = f(S,U_l) \;       \frac{\sqrt{2\pi}}{2^{4/3} \, v_2} \sqrt{\frac{V_{x}^{(1)} V_{x}^{(3)}}{V_{x}^{(2)}} } 
\\\hline
\!\!\!\!\begin{array}{c}  (0,0,0) \\ \perp  \OR \text{ and } \OR\Z_2^{(2)} \\ \text{ on } T^2_{(2)} \end{array}\!\!\!\! 
& \begin{sideways} \!\!\!\!\!\!\!\! ${\cal N}=1$ \end{sideways}
&  \begin{array}{c}\frac{1}{2\pi^2} \Bigl( M_x \ln \left( 2^{1/6}\,  \eta(iv_2) \right) \\
+ \sum_{j=1,3} \ln \left( 2^{11/12} \, \eta(i \tilde{v}_j) \right)\Bigr) \end{array}
& K_{\Anti_x}^{(2)} =f(S,U_l) \;   \frac{\sqrt{2\pi}}{2^{4/3} \, v_2} \sqrt{\frac{V_{xx}^{(1)}V_{xx}^{(3)}}{V_{xx}^{(2)}}}
\\\hline\hline
\multicolumn{4}{|c|}{T^6/\Z_{2}\times \Z_{2M} \text{ without discrete torsion}}
\\\hline
(0,0,0)
 & \begin{sideways} \!\!\!\!\!\!\!\! ${\cal N}=1$ \end{sideways}
& \frac{1}{2\pi^2} \Bigl( M_x \ln 2^{-1/2} +  \sum_{i=1}^3 \ln \left(2 \, \eta(i\tilde{v}_i)\right)\Bigr)
& K_{\Anti_x}^{(i),i=1,2,3} =f(S,U_l) \;    \frac{\sqrt{2\pi}}{2^{4/3}  \, v_i} \sqrt{\frac{V_{xx}^{(j)}V_{xx}^{(k)}}{V_{xx}^{(i)}}}
\\\hline\hline
\multicolumn{4}{|c|}{T^6/\Z_{2}\times \Z_{2M} \text{ with discrete torsion}}
\\\hline
\begin{array}{c} (0,0,0) \\  \pp \OR\Z_2^{(k)} \\ k=0 \ldots 3  \end{array} 
& \begin{sideways} \!\!\!\!\!\!\!\! ${\cal N}=1$ \end{sideways}
& \begin{array}{c} \frac{1}{4 \pi^2} \,
\Bigl( M_x \sum_{i=1}^3 \ln \left(\sqrt{2}\eta(i v_i)\right) \\ +  \sum_{i=1}^3 \ln \left(2 \eta(i \tilde{v}_i)\right) \Bigr)
\end{array}
& - 
\\\hline
\end{array}
}{SO-Sp-GaugeKin+KaehlerMetric}{One-loop contributions to the holomorphic gauge kinetic functions and K\"ahler metrics from the 
$xx$ sector of orientifold invariant D6-branes with $SO(2M_x)$ ($T^6$ and $x\pp \OR$) or $Sp(2M_x)$ (otherwise) gauge factors.
In all cases with non-vanishing $\delta_x \, {\rm f}^{\text{1-loop}}_{Sp(2M_x)}$ and $K_{\Anti_x/\Sym_x}^{(j)}$ for some $j$,
there is an ambiguity of assigning constants, which has been fixed here by requiring the (up to the normalisation factor $1/c_x$)
identical form~(\protect\ref{Eq:Kaehlermetric_SO}) for all K\"ahler metrics.
}
\begin{itemize}
\item
For $T^6$ and $T^6/\Z_3$, two cases need to be distinguished.
\begin{itemize}
\item
For $x \pp \OR$, one has $(\xi_x, \varphi^{\Sym_x},\varphi^{\Anti_x})= (-1,0,3)$ with
vanishing gauge threshold correction, $\frac{M_x\tilde{\Delta}_{xx}+ \Delta_{x,\OR}/2}{2} =0$, due to the underlying ${\cal N}=4$ supersymmetry.
The string to field theory matching condition reduces to (with $c_x=1$ for bulk D6-branes and $k_x=2$ for $SO(2M_x)$)
\begin{equation}
\begin{aligned}
0 \stackrel{!}{=}& \,
 (M_x-1) \, \ln \left[\left(S \prod_{i=1}^3 U_i \right)^{-3/4} \frac{(2\pi)^{3/2} \prod_{i=1}^3 \sqrt{V_{xx}^{(i)}}}
{2 \, \left(\prod_{k=1}^3 v_k \right)}\right]
- 3 (M_x-1)  \ln K_{\Anti_x} 
,
\end{aligned}
\end{equation}
from which the K\"ahler metrics of the antisymmetrics are inferred, 
\begin{equation}\label{Eq:Kaehlermetric_SO}
K_{\Anti_x}^{(i)} = \left(S \prod_{l=1}^3 U_l \right)^{-1/4} \frac{(2\pi)^{1/2}}{2^{1/3} \, c_x \, v_i}
\sqrt{ \frac{V_{xx}^{(j)}V_{xx}^{(k)} }{V_{xx}^{(i)}}}
\quad \text{ with } \quad
(ijk)=(123) \text{ cyclic}
.
\end{equation}
Due to ${\cal N}=4$, there is no one-loop contribution to the holomorphic gauge kinetic function.
\item
For $x \perp \OR$ along some $T^4_{(j \cdot k)}$, the parameters are  $(\xi_x, \varphi^{\Sym_x},\varphi^{\Anti_x})=(1,1,2)$
and the gauge threshold $\frac{M_x\tilde{\Delta}_{xx}+ \Delta_{x,\OR}/2}{2} =2 \, \Lambda_{0,0}(\tilde{v}_i;2 \tilde{V}_{xx}^{(i)})$
stems from the O6-plane breaking ${\cal N}=4$ to ${\cal N}=2$ supersymmetry. The terms proportional to $M_x$ in~(\ref{Eq:Matching_xx_sector_SO+Sp})
are consistent with the normalisation~(\ref{Eq:Kaehlermetric_SO}) of the K\"ahler metrics for both antisymmetric and symmetric
representations, and the remaining constants are assigned to $\delta_x \, {\rm f}^{\text{1-loop}}_{Sp(2M_x)}$, cf. 
table~\ref{tab:SO-Sp-GaugeKin+KaehlerMetric}.
\end{itemize}
\item
For $T^6/\Z_{2N}$, again two distinct cases exist.
\begin{itemize}
\item
For $x \pp \OR$ or $\OR\Z_2^{(2)}$, one has the constants $(\xi_x, \varphi^{\Sym_x},\varphi^{\Anti_x})=(1,1,0) $ and the 
gauge threshold contribution
$\frac{M_x\tilde{\Delta}_{xx}+ \Delta_{x,\OR}/2}{2} = M_x \Lambda_{0,0}(v_2;V_{xx}^{(2)}) + \Lambda_{0,0}(\tilde{v}_2;2\tilde{V}_{xx}^{(2)})$
from an ${\cal N}=2$ supersymmetric sector. The contribution to the holomorphic gauge kinetic function and the K\"ahler metric are displayed in 
table~\ref{tab:SO-Sp-GaugeKin+KaehlerMetric}.
\item
For $x \perp \OR$ along  $T^2_{(2)}$, the parameters are  $(\xi_x, \varphi^{\Sym_x},\varphi^{\Anti_x})=(1,0,1) $ and the
gauge threshold 
$\frac{M_x\tilde{\Delta}_{xx}+ \Delta_{x,\OR}/2}{2} = M_x \Lambda_{0,0}(v_2;V_{xx}^{(2)}) + \sum_{i=1,3}\Lambda_{0,0}(\tilde{v}_i;2 \tilde{V}_{xx}^{(i)}) $ belong to an ${\cal N}=1$ supersymmetric sector, cf. table~\ref{tab:SO-Sp-GaugeKin+KaehlerMetric}
for the K\"ahler metric and contribution to the holomorphic gauge kinetic function.
\end{itemize}
\item
For $T^6/\Z_2 \times \Z_{2M}$ without torsion, the parameters $(\xi_x, \varphi^{\Sym_x},\varphi^{\Anti_x})=(1,0,3) $
and  $\frac{M_x\tilde{\Delta}_{xx}+ \Delta_{x,\OR}/2}{2} = \sum_{i=1}^3 \Lambda_{0,0}(\tilde{v}_i;2 \tilde{V}_{xx}^{(i)})$
belong to an ${\cal N}=1$ supersymmetric sector,  cf. the penultimate line in
table~\ref{tab:SO-Sp-GaugeKin+KaehlerMetric} for the field theory result.
\item
For $T^6/\Z_2 \times \Z_{2M}$ with discrete torsion the parameters are \mbox{$(\xi_x, \varphi^{\Sym_x},\varphi^{\Anti_x})=(1,0,0)$}
and the gauge threshold reads
$\frac{M_x\tilde{\Delta}_{xx}+ \Delta_{x,\OR}/2}{2} = \frac{M_x}{2} \sum_{i=1}^3 \Lambda_{0,0}(v_i;V_{xx}^{(i)})+
\frac{1}{2} \sum_{i=1}^3 \Lambda_{0,0}(\tilde{v}_i;2 \tilde{V}_{xx}^{(i)})$. The sector is ${\cal N}=1$ supersymmetric without 
massless matter. 
The assignment of all constants to the holomorphic gauge kinetic function $\delta_x \, {\rm f}^{\text{1-loop}}_{Sp(2M_x)}$
on the last line in table~\ref{tab:SO-Sp-GaugeKin+KaehlerMetric} is unique - in contrast to those cases which
contain both, some K\"ahler metric and contribution to $\delta {\rm f}^{\text{1-loop}}$.
\end{itemize}
In summary, the K\"ahler metrics for antisymmetric and symmetric matter from the $xx$ sector of orientifold
invariant D6-branes have the universal shape~(\ref{Eq:Kaehlermetric_SO}), and the one-loop contribution 
to the holomorphic gauge kinetic function has the same global prefactor $1/c_x$ as at tree level. The powers of 2 in the 
annulus contribution stem from the factor $\ln(1/c_xk_x)$ in the logarithm of the tree level gauge coupling, while the powers of 
2 in the M\"obius strip contribution to $\delta_{x} \, {\rm f}^{\text{1-loop}}_{Sp(2M_x)}$ can be traced back to a combination 
of $\ln(1/c_xk_x)$ with $\ln 4$ per lattice sum $\Lambda_{0,0}(\tilde{v}_i; 2\tilde{V}_{xx}^{(i)})$.

The present discussion covers only the $xx$ sector of orientifold invariant D6-branes. The $x(\omega^k x)$ sectors require 
a discussion of the transformation properties of the intersection points under the orbifold and orientifold projection. Since the latter
depend on the choice of the background lattice, a complete case-by-case study goes beyond the scope of this article. 
The branes $c$ and $h_3$ in the Standard Model example on $T^6/\Z_6'$ with gauge groups $Sp(2)_c$ and $Sp(6)_{h_3}$
are of the type $c \pp \OR$ and $h_3 \perp \OR$ on $T^2_{(2)}$
discussed in this section, and in section~\ref{S:Z6p-Example} details on the $x(\omega^k x)$ sectors are given for these two examples.

\subsection{Gauge kinetic functions for anomaly-free $U(1)$s}\label{Ss:U1_Kaehler+GaugeKinetic}

Abelian gauge symmetries are ubiquitous in D-brane models, they appear as anomaly-free linear combinations in the physical 
hyper charge and $B-L$ symmetry and provide kinetic mixing in the open string sector~\cite{Abel:2003ue,Abel:2008ai} as well as with the closed string RR photons,
thereby providing candidates for a `dark photon', see e.g.~\cite{Camara:2011jg} and references therein. 
Anomalous $U(1)$ symmetries remain as global symmetries in perturbation theory which can be broken non-perturbatively by instanton 
effects, see e.g.~\cite{Ibanez:2006da,Blumenhagen:2006xt,Blumenhagen:2009qh} for D2-instantons in D6-brane models. 
The kinetic mixing happens at the one-loop level, and we derive here the perturbatively exact holomorphic gauge kinetic functions for a single 
$U(1)_a$ on a stack of D$6_a$-branes and for an anomaly-free linear combination $U(1)_X$ from various stacks of D6-branes
in section~\ref{Sss:Single_U1_GaugeKinetic} and~\ref{Sss:LinearCombined_U1_GaugeKinetic}, respectively.

\subsubsection{Holomorphic gauge kinetic function for a single $U(1)$ factor}\label{Sss:Single_U1_GaugeKinetic}

The generic formula~(\ref{Eq:Gauge-FieldTheory}) for the field theoretical gauge coupling at one loop is modified 
for a single $U(1)_a \subset U(N_a)$ gauge factor by replacing the quadratic Casimir by the product of the dimension of the 
representation times the (charge)${}^2$,
\begin{equation}
C_2({\bf R}_a) \to Q_a^2 \, {\rm dim}({\bf R}_a)
\quad
\text{with}
\quad 
Q_a=\left\{\begin{array}{c} 0 \\ 1 \\ 2  
\end{array}\right.
\quad
\text{and}
\quad
{\rm dim}({\bf R}_a) = \left\{\begin{array}{cr} N_a^2 & {\bf R}_a= (\Adj_a)\\ N_aN_b & (\N_a,\ov{\N}_b) \\ \frac{N_a(N_a\pm 1)}{2} &  (\Sym_a/\Anti_a) 
\end{array}\right.
,
\end{equation}
and the corresponding beta function coefficient is given in equation~(\ref{Eq:beta-U1a}) in section~\ref{Ss:Comments_on_U1}.
The holomorphic gauge kinetic function at tree level is encoded in equation~(\ref{Eq:Def-gauge-U1a}),
\begin{equation}\label{Eq:single_U1_tree}
{\rm f}^{\text{tree}}_{U(1)_a} = 2 N_a \; {\rm f}^{\text{tree}}_{SU(N_a)} 
,
\end{equation}
and the one-loop contributions to the holomorphic gauge kinetic function are inclosed in the gauge threshold~(\ref{Eq:Threh-U(1)a})
for a single $U(1)_a$ gauge factor. The matching of the result of the one-loop string computation with the field theory formula
is decomposed into open string sectors analogously to the $SU(N_a)$ case.
\begin{itemize}
\item
$aa$ strings with endpoints on identical D6-branes neither contribute to the beta function coefficient~(\ref{Eq:beta-U1a}) nor
the gauge threshold~(\ref{Eq:Threh-U(1)a}), and since $Q_a=0$ for matter in the adjoint representation $\Adj_a$ of $U(N_a)$ 
this matches the expected vanishing contribution to the field theory result. This argument applies also to strings ending on orbifold 
image D6-branes $a$ and $(\omega^k a)$.
\item
Strings with endpoints on different D6-branes $a$ and $b$ with $b \neq (\omega^k a'),(\omega^k a)$ contribute $2N_a$ times the result of the string calculation of the
 $SU(N_a)$ case to the beta function coefficient~(\ref{Eq:beta-U1a}) and gauge threshold~(\ref{Eq:Threh-U(1)a}) of $U(1)_a$.
Since on the field theory side, the quadratic Casimir is replaced by the charge of a fundamental 
representation, $N_b \, C_2(\N_a)= \frac{N_b}{2} \to N_aN_b$, the matching~(\ref{Eq:Matching_ab_sector}) in the $ab$ sector of $SU(N_a)$ is 
reproduced by $U(1)_a$, and the K\"ahler metrics for bifundamental representations in table~\ref{tab:Comparison-Kaehler-bifund} are recovered.
The one-loop contributions to the holomorphic gauge kinetic function of $U(1)_a$ are given by,
\begin{equation}
\delta_b \, {\rm f}^{\text{1-loop}}_{U(1)_a} = 2 N_a \; \delta_b \, {\rm f}^{\text{1-loop}}_{SU(N_a)}
\qquad\quad \text{ for } \quad b \neq (\omega^k a'),(\omega^k a)
,
\end{equation}
with the one-loop corrections to the holomorphic gauge kinetic function of $SU(N_a)$ presented 
in table~\ref{tab:Comparison-gaugekin-bifund} for any orbifold 
considered in this article.
\item
For strings with endpoints on orientifold image D6-branes $a$ and $a'$, the contributions from the annulus and M\"obius strip diagrams have to
be taken into account separately. Due to the $U(1)_a$ charge $Q_a=2$ of symmetric and antisymmetric representations of $U(N_a)$, 
the annulus contributes $4N_a$ times the $SU(N_a)$ annulus result to the beta function coefficient and gauge threshold, while the 
M\"obius strip contributes $2N_a$ times the $SU(N_a)$ M\"obius strip result, cf. equations~(\ref{Eq:beta-U1a}) and~(\ref{Eq:Threh-U(1)a}).
The one-loop contribution to the holomorphic gauge kinetic function from the $aa'$ sector is thus given by
\begin{equation}
\begin{aligned}
\delta_{a'} \, {\rm f}^{\text{1-loop}}_{U(1)_a} 
&\equiv \delta_{a'} \, {\rm f}^{\text{1-loop}, {\cal A}}_{U(1)_a} + \delta_{a'} \, {\rm f}^{\text{1-loop}, {\cal M}}_{U(1)_a}
\\
&= 4N_a \, \delta_{a'} \, {\rm f}^{\text{1-loop}, {\cal A}}_{SU(N_a)} + 2N_a \delta_{a'} \, {\rm f}^{\text{1-loop}, {\cal M}}_{SU(N_a)}
,
\end{aligned}
\end{equation}
where the individual annulus and M\"obius strip contributions to the $SU(N_a)$ case can be read off from
 table~\ref{tab:Comparison-gaugekin-antisym}. The result can be checked explicitly by comparing the 
matching for the string and field theoretic computations for the $U(1)_a$ case,
\begin{equation}\label{Eq:Matching_aaprime_sector_U1a}
\begin{aligned}
0
\stackrel{!}{=}& 
2 N_a \Biggl\{\frac{2 b_{aa'}^{\cal A} + b_{aa'}^{\cal M}}{2} \left[ \ln\left(\frac{M_{\rm Planck}}{M_{\rm string}}\right)^2 + {\cal K}_{\text{bulk}} \right] 
\\
& \qquad - N_a \left( \sum_{i=1}^{\varphi^{\Anti_a}} \ln K_{\Anti_a}^{(i)}+ \sum_{i=1}^{\varphi^{\Sym_a}}  \ln K_{\Sym_a}^{(i)} \right)
- \left(\sum_{i=1}^{\varphi^{\Sym_a}}  \ln K_{\Sym_a}^{(i)} - \sum_{i=1}^{\varphi^{\Anti_a}} \ln K_{\Anti_a}^{(i)} \right)
\\
& \qquad + 8 \pi^2 \, \Re(\delta_{a'} \, {\rm f}_{U(1)_a}^{\rm 1-loop})-\frac{2 N_a \left( 2 \Delta_{aa'} +\Delta_{a,\OR} \right) }{2}
\Biggr\}
,  
\end{aligned}
\end{equation}
with the $SU(N_a)$ case in~(\ref{Eq:Matching_aaprime_sector}).

\end{itemize}

The total one-loop correction to the holomorphic gauge kinetic function of a single (not necessarily anomaly-free) $U(1)_a$ factor 
can be summarised as follows,
\begin{equation}\label{Eq:Delta_U1i_total}
\begin{aligned}
\delta_{\text{total}} \, {\rm f}^{\text{1-loop}}_{U(1)_a} \equiv& \; \delta_{a'} \, {\rm f}^{\text{1-loop}}_{U(1)_a}
+ \sum_{b\neq a,a'} \delta_b \,  {\rm f}^{\text{1-loop}}_{U(1)_a}
\\
=& \, 2 N_a \left( 2\, \delta_{a'} \, {\rm f}^{\text{1-loop}, {\cal A}}_{SU(N_a)} +  \delta_{a'} \, {\rm f}^{\text{1-loop}, {\cal M}}_{SU(N_a)}
+ \sum_{b\neq a,a'} \delta_b \,  {\rm f}^{\text{1-loop}}_{SU(N_a)} \right)
.
\end{aligned}
\end{equation}
This form will serve as a basic building block for the anomaly-free linear combination of several individual Abelian gauge factors
in the following section.

\subsubsection{Gauge kinetic function for anomaly-free linear combinations of $U(1)$s}\label{Sss:LinearCombined_U1_GaugeKinetic}

Anomaly-free massless Abelian gauge factors $U(1)_X$ such as the hyper charge or $B-L$ symmetry are typically linear combinations of
several $U(1)_i \subset U(N_i)$ factors, \mbox{$U(1)_X = \sum_i x_i U(1)_i$}, with charge assignments
\begin{equation}
Q_X = \sum_i x_i \, Q_{U(1)_i}
.
\end{equation}
The tree level gauge coupling~(\ref{Eq:Def-gauge-U1X}) in section~\ref{Ss:Comments_on_U1} leads to the 
holomorphic gauge kinetic function at tree-level,
\begin{equation}\label{Eq:linear_comb_U1_tree}
{\rm f}^{\text{tree}}_{U(1)_X} = \sum_i x_i^2 \;  {\rm f}^{\text{tree}}_{U(1)_i} 
,
\end{equation}
with ${\rm f}^{\text{tree}}_{U(1)_i}$ given in~(\ref{Eq:single_U1_tree}),
and the beta function coefficients~(\ref{Eq:beta-U(1)X}) and gauge threshold corrections~(\ref{Eq:Threh-U(1)X}) contain the 
analogous summation over contributions from each single $U(1)_i$ factor weighted by $x_i^2$ plus mixed terms proportional to $x_ix_j$
with $i<j$ that arise at one loop. The matching of string and field theory expressions is simplified by rewriting the sum over K\"ahler 
metrics analogously,
\begin{equation}
\begin{aligned}
\sum_{{\bf R}_a} Q_{X,{\bf R}_a}^2 {\rm dim}({\bf R}_a) \, \ln K_{{\bf R}_a} =&
\sum_{i < j} N_i N_j \left[ \left(x_i - x_j\right)^2 \ln K_{(\N_i,\ov{\N}_j)} +\left(x_i + x_j\right)^2 \ln K_{(\N_i,\N_j)} \right]
\\ & + \sum_i  2 N_i x_i^2 \left[ (N_i-1) \ln K_{\Anti_i} +  (N_i+ 1) \ln K_{\Sym_i}\right]
\\
=& \sum_i x_i^2 N_i \Biggl( \sum_{j \neq i}  N_j  \left[  \ln K_{(\N_i,\ov{\N}_j)}  + \ln K_{(\N_i,\N_j)}  \right] 
\\ &\qquad  + 2 N_i \, (\ln K_{\Anti_i} + \ln K_{\Sym_i}) +  2 \, (\ln K_{\Sym_i} - \ln K_{\Anti_i}) \Biggr)
\\
 & + 2 \sum_{i<j} N_iN_j x_ix_j \left[ - \ln K_{(\N_i,\ov{\N}_j)}  + \ln K_{(\N_i,\N_j)}  \right]
, 
\end{aligned}
\end{equation}
where the sum on the third and fourth line exactly matches the single $U(1)_i$ contributions from the beta function coefficient
and gauge threshold. 
The complete one-loop correction to the holomorphic gauge kinetic function of the linear combination $U(1)_X$ takes the form
\begin{equation}\label{Eq:linear_comb_U1_loop}
\delta_{\text{total}} \, {\rm f}^{\text{1-loop}}_{U(1)_X} = \sum_i x_i^2  \,\delta_{\text{total}} {\rm f}^{\text{1-loop}}_{U(1)_i} 
+ 4 \sum_{i<j} x_i x_j N_i \left(- \delta_{j} \, {\rm f}^{\text{1-loop}}_{SU(N_i)} + \delta_{j'} \, {\rm f}^{\text{1-loop}}_{SU(N_i)}\right)
,
\end{equation}
where in the sum over $i<j$, we have used the fact that $\tilde{\Delta}_{ij}=\tilde{\Delta}_{ji}$, cf. e.g.~\cite{Lust:2003ky,Akerblom:2007np,Blumenhagen:2007ip,Gmeiner:2009fb}, and therefore
$N_iN_j\tilde{\Delta}_{ij}=N_i \Delta_{ij}=N_j \Delta_{ji}$, which carries over to the holomorphic part as 
$N_i \, \delta_j \, {\rm f}^{\text{1-loop}}_{U(1)_i} = N_j \, \delta_i \, {\rm f}^{\text{1-loop}}_{U(1)_j}$.  
The total one-loop correction to the holomorphic gauge kinetic function has been defined in~(\ref{Eq:Delta_U1i_total})
for a single $U(1)_i$, and the individual
contributions are given in table~\ref{tab:Comparison-gaugekin-bifund} and~\ref{tab:Comparison-gaugekin-antisym} for all orbifold backgrounds considered in this paper.\\
This completes the discussion of massless and massive Abelian gauge factors. An example of the massless $B-L$ symmetry 
in the Standard Model on $T^6/\Z_6'$ is given in section~\ref{S:Z6p-Example}.

\section{Example on $T^6/\Z_2 \times \Z_2$: Magnetised D9-branes vs. D6-branes at angles}\label{S:Compare-Example}

Up to now, we have shown that our results for the gauge thresholds, K\"ahler metrics and holomorphic gauge kinetic functions of bulk, fractional and rigid D6-branes fit - up to subtleties in the one-loop field redefinitions of the dilaton and complex structure moduli and the extra terms from
$\Z_2$ fixed and orientifold invariant points in the holomorphic gauge kinetic function at one loop -
 with those existing in the literature for the six-torus and $T^6/\Z_2 \times \Z_2$ without torsion
with vanishing displacement and Wilson line moduli~\cite{Lust:2003ky,Akerblom:2007uc} 
 and the partial results on $T^6/\Z_2 \times \Z_2$ with discrete torsion~\cite{Blumenhagen:2007ip}.
In this section, we further test the consistency of our results for rigid D6-branes in Type IIA string theory on $T^6/(\Z_2 \times \Z_2 \times \OR)$ 
with discrete torsion by matching with the T-dual D9- and D5-brane models in Type IIB string theory on $T^6/(\Z_2 \times \Z_2 \times \Omega)$ 
without torsion that were introduced in~\cite{Angelantonj:2009yj}. 
We find that, using rigid D6-branes at $\Z_2 \times \Z_2$ singularities, the gauge groups of the T-dual rigid D5-branes are of $U(N)$ type, 
which can be related to the $Sp(2N)$ groups in~\cite{Angelantonj:2009yj} by a recombination of rigid orientifold image D6-branes to fractional D6-branes
stuck at one type of $\Z_2$ singularity only.

The correspondence of our notation with the one by Angelantonj {\it et al.}~\cite{Angelantonj:2009yj} is displayed in table~\ref{tab:Comparison-Magnetised}:
%
\mathtabfix{
\begin{array}{|ccc|}\hline
\multicolumn{3}{|c|}{\text{\bf Comparison of magnetised D9-branes and intersecting D6-branes on } T^6/\Z_2 \times \Z_2}
\\\hline\hline
\text{Angelantonj et al.} & \quad \stackrel{\text{T-duality}}{\Leftrightarrow} \quad &  \text{this article}
\\\hline\hline
\text{no $B$-field} & & \text{{\bf aaa}-torus}
\\
g, f, h & \text{Orbifold generators} & \Z_2^{(1)} (\omega), \Z_2^{(2)}(\theta\omega), \Z_2^{(3)}(\theta)
\\
O9 & \text{O-planes} & \OR (\eta_{\OR}=1) 
\\
O5_1 (\epsilon_1=1) & & \OR\Z_2^{(1)} (\eta_{\OR\Z_2^{(1)}}=1) 
\\
O5_2 (\epsilon_2=1) & & \OR\Z_2^{(2)} (\eta_{\OR\Z_2^{(2)}}=1) 
\\
O5_3 (\epsilon_3=-1) & \text{exotic O-plane} & \OR\Z_2^{(3)} (\eta_{\OR\Z_2^{(3)}}=-1)
\\
 H^{(i)}_a=\frac{m^i_a}{n^i_a \, R_1^{(i)} R_2^{(i)} } &\text{Magnetic Flux} \Leftrightarrow \text{Angle}
& \tan(\pi \phi^{(i)}_a) = \frac{m_i^a}{n_i^a} \frac{R_2^{(i)}}{R_1^{(i)}} 
\\
S & \text{Dilaton} & S
\\
T_i \text{ (K\"ahler)} & \text{Bulk moduli} & U_i \text{ (Complex structures)} 
\\
U_i \text{ (Complex structures)}  & & T_i \text{ (K\"ahler)} 
\\
M_i^{l} \text{ (K\"ahler)}  & \text{Except. moduli} &  W_i^l\text{ (Complex structures)} 
\\
\left.\begin{array}{c}  (X_1^{a_i},X_2^{a_i},X_3^{a_i}) \in \\ \{(1,1,1),(1,-1,-1)\} \end{array} \right\} 
&\begin{array}{c}  \text{D-branes} \\ \Z_2 \text{ eigenvalues} \end{array} 
& \left\{\begin{array}{c} \left((-1)^{\tau_{a_i}^{\Z_2^{(1)}}},(-1)^{\tau_{a_i}^{\Z_2^{(2)}}},(-1)^{\tau_{a_i}^{\Z_2^{(3)}}} \right) \\ \in \{(+,+,+),(+,-,-)\}  \end{array} \right.
\\
\epsilon_l^{(a)} \in \{0,1\} & 
\begin{array}{c} 
\text{fixed point choices}
 \end{array}
& \left\{\begin{array}{c} \sigma^i_a \in \{0,1\}  \text{ (displacements)}\\ \tau^i_a \equiv 0 \text{ (no Wilson lines)} \end{array}\right.
\\\hline
\end{array}
}{Comparison-Magnetised}{
Comparison of the notation for magnetised D9- and D5-branes on $T^6/(\Z_2 \times \Z_2\times \Omega)$ without discrete torsion in~\protect\cite{Angelantonj:2009yj}
and the T-dual D6-branes at angles on the untilted {\bf aaa}-torus of the $T^6/(\Z_2 \times \Z_2\times \OR)$ 
background with discrete torsion.
}
the T-dual of the exotic $O5_3$-plane is the $\OR\Z_2^{(3)}$ invariant exotic O6-plane, and the discrete choices of fixed point contributions correspond to displacements $\sigma^i$ away from the origin of the two-torus $T^2_{(i)}$. Discrete Wilson lines are not taken into
account in these examples.

For the bulk three-cycles on $T^6/\Z_2 \times \Z_2$, the notation is given in section~\ref{Ss:Tree-GaugeKin} in 
formulas~(\ref{Eq:Def-Z2Z2-bulk_even+odd-cycles}) to~(\ref{Eq:SUSY_Z2Z2_bulk}) with
$b_1=b_2=b_3=0$ for the {\bf aaa}-torus.
For vanishing discrete Wilson lines, the exceptional three-cycles can be written as
\begin{equation*}
\Pi^{\Z_2^{(k)}} = (-1)^{\tau_a^{\Z_2^{(k)}}}
 \sum_{\alpha\beta \in F_k^a} \left( n^k_a \; \varepsilon_{\alpha\beta}^{(k)} + m^k_a \; \tilde{\varepsilon}_{\alpha\beta}^{(k)}  \right)
,
\end{equation*}
where $F_k^a$ denotes the sets of $\Z_2^{(k)}$ fixed points on $T^2_{(i)} \times T^2_{(j)}$ through which the toroidal three-cycles pass.
The inclusion of discrete Wilson lines introduces additional relative signs among the different fixed point contributions, see~\cite{Forste:2010gw} for details,
but this possibility will be neglected in the following.
For the given choice of an exotic O6-plane $\OR\Z_2^{(3)}$, the overall three-cycle of the O6-planes is given by
(remember the number $N_{O6}=8$ of parallel O6-planes on the {\bf aaa}-torus),
\begin{equation*}
\Pi_{O6} = 2 \; \left( \Pi_{135}^{\rm bulk} - \Pi_{146}^{\rm bulk} - \Pi_{236}^{\rm bulk} + \Pi_{245}^{\rm bulk} \right)
,
\end{equation*}
and the exceptional three-cycles transform under the orientifold symmetry on the {\bf aaa}-torus as follows
\begin{equation}\label{Eq:OR-images-Ex-cycles-Examples}
(\varepsilon_{\beta\gamma}^{(1)},\tilde{\varepsilon}_{\beta\gamma}^{(1)}) \stackrel{\OR}{\longrightarrow}
(-\varepsilon_{\beta\gamma}^{(1)},\tilde{\varepsilon}_{\beta\gamma}^{(1)})
,
\qquad
(\varepsilon_{\alpha\gamma}^{(2)},\tilde{\varepsilon}_{\alpha\gamma}^{(2)}) \stackrel{\OR}{\longrightarrow}
(-\varepsilon_{\alpha\gamma}^{(2)},\tilde{\varepsilon}_{\alpha\gamma}^{(2)})
,
\qquad
(\varepsilon_{\alpha\beta}^{(3)},\tilde{\varepsilon}_{\alpha\beta}^{(3)}) \stackrel{\OR}{\longrightarrow}
(\varepsilon_{\alpha\beta}^{(3)},-\tilde{\varepsilon}_{\alpha\beta}^{(3)})
,
\end{equation}
where each $\Z_2^{(i)}$ fixed point $\alpha\beta \in T^2_{(j)} \times T^2_{(k)}$ remains inert on the {\bf aaa}-torus.
In other words, the orientifold projection on the bulk and exceptional three-cycles on the {\bf aaa}-torus
is fixed completely by
\begin{equation}\label{Eq:Orientifold-Z2-Eigenvalues-Examples}
\begin{aligned}
(n^i_a,m^i_a) \stackrel{\OR}{\longrightarrow} (n^i_a,-m^i_a),
\qquad\qquad
(\tau_a^{\Z_2^{(1)}},\tau_a^{\Z_2^{(2)}},\tau_a^{\Z_2^{(3)}})  \stackrel{\OR}{\longrightarrow}
(\tau_a^{\Z_2^{(1)}}+1,\tau_a^{\Z_2^{(2)}}+1,\tau_a^{\Z_2^{(3)}}). 
\end{aligned}
\end{equation}

The supersymmetry conditions on the bulk three-cycles are given in equation~(\ref{Eq:SUSY_Z2Z2_bulk}).
Supersymmetry of the rigid three-cycle implies that only exceptional three-cycles in a certain set  $\{F_k^a\}_{k=1,2,3}$ of points 
traversed by the bulk three-cycle contribute, for more details see appendix~\ref{App:Magnetised}.

The bulk RR tadpole cancellation condition  for an exotic $\OR\Z_2^{(3)}$-plane and the 
{\bf aaa}-torus can be brought to the form,
\begin{equation*}
\sum_x N_x \, \vec{X}_x = 16 \, \left( 1,-1,-1,1\right)^T
,
\end{equation*}
while the exceptional contributions to the RR tadpoles have to cancel solely among the rigid D6-branes,
\begin{equation*}
\sum_x N_x m^i_x \sum_{\alpha\beta \in F_i^x} \tilde{\varepsilon}_{\alpha\beta}^{(i)}
=0
\quad\text{ for } \quad i=1,2
,
\qquad\qquad\quad
\sum_x N_x n^3_x \sum_{\alpha\beta \in F_3^x}\varepsilon_{\alpha\beta}^{(3)}
=0
.
\end{equation*}
Net-chiralities are computed via the intersection numbers of rigid three-cycles,
\begin{equation*}\hspace{-6mm}
\begin{aligned}
\Pi^{\rm rigid}_a \circ \Pi^{\rm rigid}_b &= - \frac{1}{4} \left( \vec{X}_a \cdot \vec{Y}_b - \vec{Y}_a \cdot \vec{X}_b
+ \sum_{i=1}^3 (-1)^{\tau_{ab}^{\Z_2^{(i)}}} I_{ab}^{(i)} 
 \sum_{\tiny \begin{array}{c} \alpha_a\beta_a \in F_a^i \\ \alpha_b\beta_b \in F_b^i \end{array} }
\!\!\!\!\! \delta_{\alpha_a\alpha_b} \, \delta_{\beta_a\beta_b} \; \right)
 \equiv - \frac{1}{4} \left(I_{ab}+ \sum_{i=1}^3I_{ab}^{\Z_2^{(i)}}
\right)
,
\\
\Pi^{\rm rigid}_a \circ \Pi_{O6} &= 2 \; \vec{Y}_a \cdot \vec{X}_{O6}
\hspace{90mm}
 \equiv - \frac{1}{4} \sum_{i=0}^3 \eta_{\OR\Z_2^{(i)}}  \tilde{I}_{a}^{\OR\Z_2^{(i)}}
,
\end{aligned}
\end{equation*}
where the first line holds for vanishing discrete Wilson lines. 
For orientifold image D6-branes $a$ and $b=a'$, the sum over fixed points gives 
$\sum_{\tiny \begin{array}{c} \alpha_a\beta_a \in F_a^i \\ \alpha_b\beta_b \in F_{a'}^i \end{array}} \delta_{\alpha_a\alpha_b} \, \delta_{\beta_a\beta_b}=4$, 
and the differences in $\Z_2^{(i)}$ eigenvalues due to the orientifold projection
are given in~(\ref{Eq:Orientifold-Z2-Eigenvalues-Examples}).

The non-chiral part of the spectrum is computed from the beta function coefficients in 
tables~\ref{tab:Z2Z2M-torsion-Bifundamentals-beta+thresholds} and~\ref{tab:Z2Z2M-torsion-AntiSym-beta+thresholds}.
For all examples, the individual (toroidal and $\Z_2^{(i)}$ fixed point) intersection numbers are given in appendix~\ref{App:Magnetised}.

\subsection{Examples 1 and 2 by Angelantonj {\it et al.} revisited}\label{Ss:Ex1+2_Angelantonj}

\subsubsection{Example 1}\label{Sss:Ex1}

The T-dual to the first magnetised D9-brane example in~\cite{Angelantonj:2009yj} contains four rigid D6-branes wrapping the same bulk three-cycle
but with all four possible different combinations of $\Z_2^{(i)}$ eigenvalues. The D6-brane configuration is displayed in 
table~\ref{tab:T-dual-Angelant-1+2}.
\mathtabfix{
\begin{array}{|c|c|c|c|c|c|c|c|c|}\hline
\multicolumn{8}{|c|}{\text{\bf T-dual D6-brane configurations of examples 1 \& 2 on } T^6/\Z_2 \times \Z_2}
\\\hline\hline
\rotatebox{90}{\!\!\!\!\!\!\!\text{D$6_x$-brane}} & (\vec{\phi}_x) 
& \!\!\!\!\begin{array}{c} \text{Torus wrappings} \\
(n_x^1,m_x^1; n_x^2,m_x^2; n_x^3,m_x^3) \end{array}\!\!\!\! & (\vec{X}_x);(\vec{Y}_x) & (\vec{\sigma}_x) & (-1)^{\tau^{\Z_2}_x} & (\vec{\tau}_x)
& (\vec{V}_{xx})
\\\hline\hline
\begin{array}{c} a_1 \\ a_2 \\ a_3 \\ a_4 \end{array}
&  \begin{array}{c} (\phi^{(1)},\phi^{(2)},\phi^{(3)}) \\ {\footnotesize \sum_{i=1}^3 \phi^{(i)}=0}\end{array}
& (1,1;1,1;1,-1)& \left(\begin{array}{c} 1 \\ -1 \\ -1 \\1 \end{array}\right) ; \left(\begin{array}{c} -1 \\ 1 \\ 1 \\-1 \end{array} \right) & (\vec{0}) 
& \begin{array}{c} (+++)  \\(+--) \\ (-+-) \\ (--+)\end{array} & (\vec{0})
& \!\!\!\left(\!\!\begin{array}{c} \frac{1}{r_1} + r_1 \\ \frac{1}{r_2} + r_2 \\ \frac{1}{r_3} + r_3 \end{array}\!\!\right)\!\!\!
\\\hline\hline
\begin{array}{c} a_1' \\ a_2' \\ a_3' \\ a_4' \end{array}
& \!\!\!(-\phi^{(1)},-\phi^{(2)}, -\phi^{(3)})\!\!\!
& (1,-1;1,-1;1,1) &\left(\begin{array}{c} 1 \\ -1 \\ -1 \\1 \end{array}\right) ; \left(\begin{array}{c} 1 \\ -1 \\ -1 \\1 \end{array}\right) 
&  (\vec{0}) & \begin{array}{c} (--+) \\(-+-) \\ (+--)  \\ (+++) \end{array} & (\vec{0})
& \!\!\!\left(\!\!\begin{array}{c} \frac{1}{r_1} + r_1 \\ \frac{1}{r_2} + r_2 \\ \frac{1}{r_3} + r_3 \end{array}\!\!\right)\!\!\!
\\\hline
\end{array}
}{T-dual-Angelant-1+2}{
The four stacks of D6-branes $a_i$ on $T^6/\Z_2 \times \Z_2$ with discrete torsion
 which give the T-dual to the first magnetised D9-brane model in~\protect\cite{Angelantonj:2009yj}.
The bulk RR tadpoles cancel for $N_{a_1} = \ldots N_{a_4}=4$ resulting in the gauge group $\prod_{i=1}^4 U(4)_{a_i}$.
The three-cycles are supersymmetric if the complex structure on the last two-torus is related to the other two by $r_3=\frac{r_1+r_2}{1-r_1r_2}$.}
The explicit rigid three-cycles are given in equation~(\ref{Eq:-Pi3-Ex1+2}) of appendix~\ref{App:Mag-Ex1+2}.
The bulk RR tadpole cancellation condition requires $N_{a_i}=4$ for $i =1\ldots 4$,
and net-chiralities are given by the intersection numbers of the rigid D6-branes,
\begin{equation*}
\begin{aligned}
\Pi_{a_i} \circ \Pi_{a_j} &=0
,
\\
\Pi_{a_i} \circ \Pi_{a_j}' &= -2 \, \left( 1+ (-1)^{\tau^{\Z_2^{(1)}}_{a_ia_j}}\;  \delta_{(\sigma^2\sigma^3)_{a_ia_j},0} 
+ (-1)^{\tau^{\Z_2^{(2})}_{a_ia_j}} \;  \delta_{(\sigma^1\sigma^3)_{a_ia_j},0}
+(-1)^{\tau^{\Z_2^{(3})}_{a_ia_j}} \; \delta_{(\sigma^1\sigma^2)_{a_ia_j},0}
\right)
\\
&=\left\{\begin{array}{cc} -8 & a_i=a_j \\  0 & a_i \neq a_j
\end{array}\right.
\qquad \text{ with } \qquad
(\vec{\sigma}_{a_ia_j})=(\vec{0})
,
\\
\Pi_{a_i} \circ \Pi_{O6} &= -8
,
\end{aligned}
\end{equation*}
where on the second line we abbreviated $\delta_{(\sigma^m\sigma^n)_{a_ia_j},0} \equiv \delta_{\sigma^m_{a_ia_j},0} \, \delta_{\sigma^n_{a_ia_j},0}$.
The individual toroidal and $\Z_2^{(i)}$ invariant intersection numbers 
are displayed in table~\ref{tab:IntersectionNumbers_Ex1+2} of appendix~\ref{App:Mag-Ex1+2}
and can be checked for consistency with the net-chiralities.
As a result, the gauge group $\prod_{i=1}^4 U(4)_{a_i}$ arises with the complete (chiral + non-chiral) massless matter spectrum
consisting of non-chiral pairs of bifundamentals in the $a_ia_{j, j>i}$ sectors 
plus eight antisymmetric representations per D6-brane stack as listed in table~\ref{tab:Spectrum-Tdual_Ex1},
\begin{table}[h]
\begin{equation*}
\begin{array}{|c|c|}\hline
\multicolumn{2}{|c|}{\text{\bf Massless matter spectrum of Ex. 1 on $T^6/\Z_2 \times \Z_2$ with gauge group } \prod_{i=1}^4 U(4)_{a_i}}
\\\hline\hline
a_ia_j & \left[  (\4,\ov{\4},\1,\1) + (\1,\1,\4,\ov{\4}) + (\4,\1,\ov{\4},\1) + (\1,\4,\1,\ov{\4})
 + (\4,\1,\1,\ov{\4}) + (\1,\4,\ov{\4},\1)+ c.c.\right]
\\
a_ia_j' & +   8 \times \left[ (\ov{\6}_{\ov{\Anti}},\1,\1,\1) + (\1,\ov{\6}_{\ov{\Anti}},\1,\1) + (\1,\1,\ov{\6}_{\ov{\Anti}},\1) + (\1,\1,\1,\ov{\6}_{\ov{\Anti}})\right]
\\\hline
\end{array}
\end{equation*}
\caption{The matter spectrum of the first example with magnetised D9-branes on $T^6/\Z_2 \times \Z_2$ without torsion
in~\protect\cite{Angelantonj:2009yj} is (up to the renaming of D-branes, cf.~(\protect\ref{Eq:Relabel-Branes-Ex1})) correctly reproduced
by the T-dual configuration of intersecting rigid D6-branes  on $T^6/\Z_2 \times \Z_2$ with discrete torsion
given in table~\protect\ref{tab:T-dual-Angelant-1+2}. The left column shows the sector from which the representations arise.}
\label{tab:Spectrum-Tdual_Ex1}
\end{table}
which matches with the original spectrum of~\cite{Angelantonj:2009yj} 
upon renaming of the D6-branes and their orientifold images,
\begin{equation}\label{Eq:Relabel-Branes-Ex1}
(a_1,a_2,a_3,a_4)_{\text{here}} \simeq (a,\bar{a}, b',\bar{b}')_{\text{Angelantonj {\it et al.}}}
.
\end{equation}
The anomaly matrix matches as expected the result reported in~\cite{Angelantonj:2009yj},
\begin{equation*}
\left( \, C_{a_ia_j} \, \right) = \frac{32}{\pi^2} \left(\begin{array}{cccc} -1 & 0 & 0 & 0 \\  0 & -1 & 0 & 0 \\ 0 & 0 & -1 & 0 \\ 0 & 0 & 0 & -1
\end{array}\right)
,
\end{equation*}
up to re-labelling of the $\OR$ images $(b',\bar{b}')$ as $(a_3,a_4)$.

The K\"ahler metrics and $v_i$ dependent one-loop contributions to the holomorphic gauge kinetic function 
$ \delta_{y} \,{\rm f}^{\text{1-loop}}_{SU(4)_{a_1}}$ with $y \in \{a_j,a_j'\}$
of the first gauge factor are given in table~\ref{tab:Z2Z2ex1_Kaehler}.
\mathtabfix{
\begin{array}{|c||c|c|c|c|}\hline
\muc{5}{|c|}{\text{\bf Example 1 on $T^6/\Z_2 \times \Z_2$: K\"ahler metrics and gauge kinetic functions involving brane } a_1}
\\\hline\hline
y & (\vec{\phi}_{a_1y}) & b_{a_1y}^{\cal A} & K_{(\4_{a_1},\ov{\N}_y)} & \delta_y \,  {\rm f}_{SU(4)_{a_1} }^{\text{1-loop}}(v_i)
\\\hline\hline
a_1 & (0,0,0) & -12 & -  &   \sum_{i=1}^3 \frac{1}{\pi^2} \ln \eta(iv_i)
\\\hline
a_{j,j>1} & (0,0,0) & 4 &   \begin{array}{c}   f(S,U_l) \; \sqrt{ \frac{2\pi \left(\frac{1}{r_1} + r_1\right)}{v_2v_3}} \\
f(S,U_l) \; \sqrt{ \frac{2\pi \left(\frac{1}{r_2} + r_2\right)}{v_1v_3}} \\ 
 f(S,U_l) \; \sqrt{ \frac{2\pi \left(\frac{1}{r_3} + r_3\right)}{v_1v_2}}   \end{array}
& 
\begin{array}{cr} - \frac{1}{\pi^2} \ln \eta(iv_{1})  & j=2 \\   - \frac{1}{\pi^2} \ln \eta(iv_{2})  & 3 \\ - \frac{1}{\pi^2} \ln \eta(iv_{3}) & 4   \end{array}
\\\hline
a_1' & (-2\phi^{(1)},-2\phi^{(2)},-2\phi^{(3)}) 
& \begin{array}{c} b_{a_1a_1'}^{\cal A} + b_{a_1a_1'}^{\cal M}  \\ =16-8 \end{array} 
& \begin{array}{c} K_{\Anti_{a_1}}=   \frac{f(S,U_l)}{\sqrt{v_1v_2v_3}} \times \\ \sqrt{
\frac{\Gamma(2\phi^{(1)})\Gamma(2\phi^{(2)})\Gamma(1-2\phi^{(1)}-2\phi^{(2)})}{\Gamma(1-2\phi^{(1)})\Gamma(1-2\phi^{(2)})\Gamma(2\phi^{(1)}+2\phi^{(2)})} 
} \end{array} & -
\\\hline
a_{j,j>1}' & (-2\phi^{(1)},-2\phi^{(2)},-2\phi^{(3)}) & 0 & - & - 
\\\hline
\end{array}
}{Z2Z2ex1_Kaehler}{Relative angles, beta function coefficients, K\"ahler metrics and $v_i$ dependent one-loop contributions to the 
holomorphic gauge kinetic function involving D6-brane $a_1$ of example 1 on $T^6/\Z_2 \times \Z_2$ with discrete torsion, 
which is T-dual to the first magnetised D9-brane example in~\cite{Angelantonj:2009yj}.
The bifundamental representations arise on parallel D6-branes whereas the antisymmetric matter states arise at 
three non-vanishing intersections.
}
%
The complete $v_i$-dependent one-loop contribution to $SU(4)_{a_1}$,
\begin{equation*}
\delta_{\rm total} {\rm f}^{\text{1-loop}}_{SU(4)_{a_1}} (v_i) = \sum_{j=1}^4 \delta_{a_j}  {\rm f}^{\text{1-loop}}_{SU(4)_{a_1}}(v_i) =0
,
\end{equation*}
vanishes due to cancellations between $j=1$ and $j=2,3,4$. 
In accord with~\cite{Angelantonj:2009yj}, the perturbative holomorphic gauge kinetic function is given by its tree-level
value,
\begin{equation*}
{\rm f}^{\text{1-loop}}_{SU(4)_{a_1}} ={\rm f}^{\text{tree}}_{SU(4)_{a_1}} + {\rm const} \sim S + U_1 + U_2 - U_3 + {\rm const}
.
\end{equation*}
The additional angle dependent terms on the first line in~(\ref{Eq:3angle-f-angle}) and~(\ref{Eq:aap-f-3angles}) drop 
out upon summation, $\sum_{j=1}^4 I_{a_1a_j'}^{\Z_2^{(k)}}=0$ for any $k\in \{1,2,3\}$, as can be read off from the individual $\Z_2$ invariant 
intersection numbers in table~\ref{tab:IntersectionNumbers_Ex1+2} upon using the relative $\Z_2$ eigenvalues in table~\ref{tab:Ex_1-3_Relative-Z2}
of  appendix~\ref{App:Mag-Ex1+2}. The constant term consisting of the intersection numbers with O6-planes listed in  table~\ref{tab:IntersectionNumbers_Ex1+2} gives
\begin{equation*}
{\rm const}= \frac{1}{2\pi^2} \, \ln 2
.
\end{equation*}
This completes the discussion of the perturbatively exact holomorphic gauge kinetic function for $SU(4)_{a_1}$.
The constant factor was to our knowledge not included in~\cite{Angelantonj:2009yj}.\\
The K\"ahler metrics in table~\ref{tab:Z2Z2ex1_Kaehler} for the complete massless spectrum
have not been computed in~\cite{Angelantonj:2009yj} and are, to our knowledge, listed here for the first time.

\subsubsection{Comments on example 2}

The second example in~\cite{Angelantonj:2009yj} relies on the same rigid D6-branes as the previous example, 
where the gauge group is partially broken by vevs of some bifundamental representations as follows:
each gauge factor is decomposed as $U(4)_{a_i} \to U(2)_{a_i} \times \widetilde{U(2)}_{a_i}$ 
with the splitting of the corresponding representations.
\begin{equation*}
\begin{aligned}
U(4)_{a_i} & \to U(2)_{a_i} \times \widetilde{U(2)}_{a_i}
\\\hline
{\bf 16}_{\Adj} & \to (\4_{\Adj},\1) + (\1,\4_{\Adj}) + [ (\2,\ov{\2}) + c.c.]
\\
\4 & \to (\2,\1)+ (\1,\2)
\\
\6_{\Anti} & \to (\1_{\Anti},\1) + (\1,\1_{\Anti}) + (\2,\2)
\\
{\bf 10}_{\Sym} & \to (\3_{\Sym},\1) + (\1,\3_{\Sym}) + (\2,\2)
\end{aligned}
\end{equation*}
Under this decomposition, the representations in table~\ref{tab:Spectrum-Tdual_Ex1} split as, 
\begin{equation*}
\begin{aligned}
\prod_{i=1}^4 U(4)_{a_i} & \to \left(\prod_{i=1}^4 U(2)_{a_i} \right) \times  \left(\prod_{i=1}^4  \widetilde{U(2)}_{a_i}\right)
\\\hline
({\bf 16}_{\Adj},\1,\1,\1) & \to (\4_{\Adj},\1,\1,\1;\1,\1,\1,\1) + (\1,\1,\1,\1;\4_{\Adj},\1,\1,\1) + [(2,1,1,1;\ov{2},1,1,1)+ c.c.]  
\\
(\4,\ov{\4},\1,\1) & \to (\2,\ov{\2},\1,\1;\1,\1,\1,\1) + (2,1,1,1;1,\ov{2},1,1) + (1,\ov{2},1,1;2,1,1,1) + (1,1,1,1;2,\ov{2},1,1)
\\
(\ov{\6}_{\ov{\Anti}},\1,\1,\1) & \to (\1_{\ov{\Anti}},\1,\1,\1;\1,\1,\1,\1) + (\1,\1,\1,\1;\1_{\ov{\Anti}},\1,\1,\1) + (\ov{\2},\1,\1,\1;\ov{\2},\1,\1,\1)
\end{aligned}
\end{equation*}
and the last four gauge factors are broken to the diagonal one, $\left(\prod_{i=1}^4  \widetilde{U(2)}_{a_i}\right) \to \widetilde{U(2)}_{\rm diag}$,
by giving suitable vevs to bifundamental matter states. The states which have to survive the projection in order to reproduce the 
 spectrum of the second example in~\cite{Angelantonj:2009yj} are highlighted in bold letters in table~\ref{tab:Spectrum-Tdual_Ex2}.
\begin{table}[h]
\begin{equation*}
\begin{array}{|c|c|c|}\hline
\multicolumn{3}{|c|}{\text{\bf Gauge breaking from example 1 to 2 on } T^6/\Z_2 \times \Z_2}
\\\hline
\multicolumn{2}{|c|}{ \prod_{i=1}^4 U(4)_{a_i} \rightarrow } &  \left(\prod_{i=1}^4 U(2)_{a_i}\right) \times \widetilde{U(2)}_{\rm diag} 
\\\hline\hline
a_ia_j & \left[  (\4,\ov{\4},\1,\1)+(\1,\1,\4,\ov{\4}) +  c.c.\right]
&\left[  (\2,\ov{\2},\1,\1;\1)+(\1,\1,\2,\ov{\2};\1) +  c.c.\right]
\\
& + \left[(\4,\1,\ov{\4},\1) + (\1,\4,\1,\ov{\4}) + c.c.\right]
&+ \left[(2,1,\ov{2},1;1) + (1,2,1,\ov{2};1) + c.c.\right]
\\ 
& + \left[(\4,\1,\1,\ov{\4}) + (\1,\4,\ov{\4},\1) + c.c.\right]
& + \left[(2,1,1,\ov{2};1) + (1,2,\ov{2},1;1) + c.c.\right]
\\
a_ia_j' & +   8 \times \left[ (\ov{\6}_{\ov{\Anti}},\1,\1,\1) + (\1,\ov{\6}_{\ov{\Anti}},\1,\1) \right]
&  +8 \times \left[ (\ov{\1}_{\ov{\Anti}},\1,\1,\1;\1) + (\1,\ov{\1}_{\ov{\Anti}},\1,\1;\1) \right]
\\
&& + 8 \times \left[(\ov{\2},\1,\1,\1;\ov{\2}) + (\1,\ov{\2},\1,\1;\ov{\2})\right] 
\\
&& + 16 \times (\1,\1,\1,\1;\ov{\1}_{\ov{\Anti}})
\\
& + 8 \times \left[ (\1,\1,\ov{\6}_{\ov{\Anti}},\1) + (\1,\1,\1,\ov{\6}_{\ov{\Anti}})\right]
&+ 8 \times \left[ (\1,\1,\ov{\1}_{\ov{\Anti}},\1;\1) + (\1,\1,\1,\ov{\1}_{\ov{\Anti}};\1)\right]
\\
&& + 8 \times \left[(\1,\1,\ov{\2},\1;\ov{\2}) + (\1,\1,\1,\ov{\2};\ov{\2})\right] 
\\
&&+ 16 \times (\1,\1,\1,\1;\ov{\1}_{\ov{\Anti}})
\\\hline
\end{array}
\end{equation*}
\caption{The matter representations in the second example of~\protect\cite{Angelantonj:2009yj} on $T^6/\Z_2 \times \Z_2$
originate from giving vevs to some bifundamental representations in example 1.
On the r.h.s., the spectrum in bold letters corresponds to the one listed in~\protect\cite{Angelantonj:2009yj}.
}
\label{tab:Spectrum-Tdual_Ex2}
\end{table}
The vevs are chosen such that the diagonal Abelian gauge factor $ \widetilde{U(2)}_{\rm diag}$ effectively wraps a bulk three-cycle.

\subsection{Example 3 by Angelantonj {\it et al.} revisited}\label{Ss:Ex3_Angelantonj}

The third example in~\cite{Angelantonj:2009yj} has three different kinds of D6-branes 
$a_i$, $b_j$ and $c_k$ with \mbox{$i\in \{1\ldots 4\}$} and \mbox{$j,k \in \{1,2\}$}. The explicit 
bulk and fixed point configurations are given in table~\ref{tab:T-dual-Angelant-3}, and the bulk RR tadpoles cancel for 
$N_{a_i}=2$, $N_{b_j}=N_{c_k}=4$.\footnote{In table 10 of~\cite{Angelantonj:2009yj}v3, only 
bifundamental matter in $a_ia_j$ sectors with $(i,j)=(1,2)$ and $(3,4)$ is listed, for which a matching of the spectrum requires a discrete displacement
$(\vec{\sigma}_{a_j,j=3,4})=(1,0,0)$ along $T^2_{(1)}$, and correspondingly not only the beta function coefficients
 $b^{\cal A}_{a_1a_j, j\in \{3,4\}}=0$ in the analogon to table~\ref{tab:Z2Z2ex3_Kaehler} vanish, but also the one-loop gauge threshold contributions 
$\delta_{a_j,j \in \{3,4\}} {\rm f}_{SU(2)_{a_1}}^{\text{1-loop}}(v_j)$
leading in total to a non-vanishing one-loop contribution $\sum_{j=1}^4 \delta_{a_j}{\rm f}^{\text{1-loop}}_{SU(2)_{a_1}}(v_i)
= \frac{1}{2\pi} \sum_{i=2}^3 \ln \eta(i v_i)$. The displacement $(\vec{\sigma}_{a_i,i \in \{3,4\}})=(1,0,0)$
 modifies also the $a_ib_j$ and $a_ic_j$ sectors for $i \in \{3,4\}$ in table~\ref{tab:Ex3-Full-Spectrum} and the corresponding entries in the
anomaly-matrix~(\ref{Eq:Anomaly-Matrix-Ex3}) as follows: 
the $a_ib_j$ sectors contribute $(\1,\1,\ov{\2},\1;\4,\1;\1,\1) + (\1,\1,\ov{\2},\1;\1,\4;\1,\1) 
+ (\1,\1,\1,\ov{\2};\1,\4;\1,\1) + (\1,\1,\1,\ov{\2};\4,\1;\1,\1)$ and the $a_ic_j$ sectors
$(\1,\1,\ov{\2},\1;\1,\1;\4,\1) + (\1,\1,\ov{\2},\1;\1,\1; \1,\4) + (\1,\1,\1,\ov{\2};\1,\1;\1,\4)
+(\1,\1,\1,\ov{\2};\1,\1; \4,\1)$ to the massless spectrum, and the associated anomaly matrix entries read 
{\tiny 
$\left(\begin{array}{cc} C_{a_3b_1} & C_{a_3b_2} \\ C_{a_4b_1} &  C_{a_4b_2}\end{array}\right)
= -\left(\begin{array}{cc} C_{b_1a_3} & C_{b_1a_4} \\ C_{b_2a_3} &  C_{b_2a_4}\end{array}\right)
=\left(\begin{array}{cc} C_{a_3c_1} & C_{a_3c_2} \\ C_{a_4c_1} &  C_{a_4c_2}\end{array}\right)
= -\left(\begin{array}{cc} C_{c_1a_3} & C_{c_1a_4} \\ C_{c_2a_3} &  C_{c_2a_4}\end{array}\right)
=\left(\begin{array}{cc} \frac{1}{2} & \frac{1}{2} \\ \frac{1}{2} & \frac{1}{2}  \end{array}\right)$}.
However, as communicated to me by some authors of~\cite{Angelantonj:2009yj}, the spectrum in table 10 of~\cite{Angelantonj:2009yj}v3 belongs to 
a D-brane configuration without any discrete displacement and Wilson line parameters turned on, and the missing states in the $a_ia_j$ sector 
with $i \in \{1,2\}$ and $j \in \{3,4\}$ can be found in table~2~of~\cite{Camara:2010zm}.}
\mathtabfix{
\begin{array}{|c|c|c|c|c|c|c|c|c|}\hline
\multicolumn{8}{|c|}{\text{\bf T-dual D6-brane configurations of example 3 on } T^6/\Z_2 \times \Z_2}
\\\hline\hline
\rotatebox{90}{\!\!\!\!\!\!\!\text{D$6_x$-brane}} & (\vec{\phi}_x) 
& \!\!\!\!\begin{array}{c} \text{Torus wrappings} \\ (n_x^1,m_x^1; n_x^2,m_x^2; n_x^3,m_x^3) \end{array}\!\!\!\! 
& (\vec{X}_x);(\vec{Y}_x) & (\vec{\sigma}_x) & (-1)^{\tau^{\Z_2}_x} & (\vec{\tau}_x)& (\vec{V}_{xx})
\\\hline\hline
\begin{array}{c} a_1 \\ a_2 \\ a_3 \\ a_4 \end{array}
&  \begin{array}{c} (\phi^{(1)},\phi^{(2)},\phi^{(3)}) \\ {\footnotesize \sum_{i=1}^3 \phi^{(i)}=0}\end{array}
& (1,2;1,1;1,-1) 
& \left(\begin{array}{c} 1 \\ -1 \\ -2 \\ 2 \end{array}\right) ;  \left(\begin{array}{c} -2 \\ 2 \\ 1 \\ -1 \end{array}\right)
& (\vec{0}) 
& \begin{array}{c} (+++) \\ (+--) \\ (-+-) \\(--+) \end{array} & (\vec{0})
& \!\!\!\left(\!\!\begin{array}{c} \frac{1}{r_1} + 4r_1 \\ \frac{1}{r_2} + r_2 \\ \frac{1}{r_3} + r_3 \end{array}\!\!\right)\!\!\!
\\\hline
\begin{array}{c} a_1' \\ a_2' \\ a_3' \\ a_4' \end{array}
&   (-\phi^{(1)},-\phi^{(2)},-\phi^{(3)}) & (1,-2;1,-1;1,1) 
& \left(\begin{array}{c} 1 \\ -1 \\ -2 \\ 2 \end{array}\right) ;  \left(\begin{array}{c} 2 \\ -2 \\ -1 \\ 1 \end{array}\right)
& (\vec{0}) 
 & \begin{array}{c} (--+) \\ (-+-) \\ (+--) \\(+++) \end{array} & (\vec{0})
& \!\!\!\left(\!\!\begin{array}{c} \frac{1}{r_1} + 4r_1 \\ \frac{1}{r_2} + r_2 \\ \frac{1}{r_3} + r_3 \end{array}\!\!\right)\!\!\!
\\\hline\hline
\begin{array}{c} b_1 \\ b_2  \end{array} & (0,0,0) & (1,0;1,0;1,0) 
&\left(\begin{array}{c} 1 \\ 0 \\ 0 \\ 0  \end{array}\right) ;  \left(\begin{array}{c} 0 \\ 0 \\ 0 \\ 0  \end{array}\right)
& (\vec{0}) &  \begin{array}{c}  (+++) \\ (+--)\end{array} & (\vec{0})
& \!\!\!\left(\!\!\begin{array}{c} \frac{1}{r_1} \\ \frac{1}{r_2} \\ \frac{1}{r_3}  \end{array}\!\!\right)\!\!\!
\\\hline
\begin{array}{c} b_2' (\equiv b_3) \\ b_1'(\equiv b_4)  \end{array} & (0,0,0) & (1,0;1,0;1,0) 
& \left(\begin{array}{c} 1 \\ 0 \\ 0 \\ 0  \end{array}\right) ;  \left(\begin{array}{c} 0 \\ 0 \\ 0 \\ 0  \end{array}\right)
& (\vec{0}) &  \begin{array}{c} (-+-) \\ (--+) \end{array} & (\vec{0})
& \!\!\!\left(\!\!\begin{array}{c} \frac{1}{r_1} \\ \frac{1}{r_2} \\ \frac{1}{r_3}  \end{array}\!\!\right)\!\!\!
\\\hline\hline
\begin{array}{c} c_1 \\ c_2  \end{array} & (0,\frac{1}{2},-\frac{1}{2}) & (1,0;0,1;0,-1) 
& \left(\begin{array}{c} 0 \\ -1 \\ 0 \\ 0  \end{array}\right) ;  \left(\begin{array}{c} 0 \\ 0 \\ 0 \\ 0  \end{array}\right)
& (\vec{0}) &  \begin{array}{c}  (+++) \\(+--) \end{array} & (\vec{0})
& \!\!\!\left(\!\!\begin{array}{c} \frac{1}{r_1} \\  r_2 \\  r_3 \end{array}\!\!\right)\!\!\!
\\\hline
\begin{array}{c} c_1'(\equiv c_3) \\ c_2' (\equiv c_4) \end{array} & (0,\frac{1}{2},-\frac{1}{2}) & (1,0;0,1;0,-1)
& \left(\begin{array}{c} 0 \\ -1 \\ 0 \\ 0  \end{array}\right) ;  \left(\begin{array}{c} 0 \\ 0 \\ 0 \\ 0  \end{array}\right)
& (\vec{0}) &  \begin{array}{c}  (-+-) \\(--+) \end{array} & (\vec{0})
& \!\!\!\left(\!\!\begin{array}{c} \frac{1}{r_1} \\  r_2 \\  r_3 \end{array}\!\!\right)\!\!\!
\\\hline
\end{array}
}{T-dual-Angelant-3}{
The eight stacks of D6-branes $a_i$, $b_j$, $c_k$ and their orientifold images which are T-dual to the 
third magnetised D9/D5-brane model of~\protect\cite{Angelantonj:2009yj}. A close inspection of the rigid three-cycles reveals that 
none of the D6-brane can be chosen to be orientifold invariant. RR tadpole cancellation leads to the gauge group 
$\left(\prod_{i=1}^4 U(2)_{a_i} \right) \times \left(\prod_{j=1}^2 U(4)_{b_j} \right) \times  \left(\prod_{k=1}^2 U(4)_{c_k} \right)$,
and the matter spectrum matches the one in~\protect\cite{Angelantonj:2009yj} with corrections in~\protect\cite{Camara:2010zm} taken into account.
For the D$6_{c_j}$-branes, the orientifold image D-branes listed here have a different shape 
$(\vec{\tau}_{c_j'}^{ \, \Z_2}) = (\vec{\tau}_{c_j}^{ \, \Z_2})  +(1,0,1)$
because we performed a simultaneous sign-flip of the toroidal wrapping numbers on $T^2_{(2)} \times T^2_{(3)}$ for the sake of a more compact notation. 
The $\left(\prod_{j=1}^2 Sp(4)_{b_j} \right) \times  \left(\prod_{k=1}^2 Sp(4)_{c_k} \right)$ gauge group in~\protect\cite{Angelantonj:2009yj}
is obtained after recombination of orientifold image D6-branes $b_j+b_j'$ and $c_k+c_k'$ as detailed in the text.
Supersymmetry requires the relation $r_3=\frac{2 \,r_1+r_2}{1- 2 \, r_1r_2}$ for the shapes 
of the three two-tori.}
In contrast to the four unitary times four symplectic gauge factor listed in~\cite{Angelantonj:2009yj}, it turns out that the full gauge group
on rigid D6-branes consists of eight unitary gauge factors,
\begin{equation*}
\left( \prod_{i=1}^4 U(2)_{a_i} \right) \times \left( \prod_{i=1}^2  U(4)_{b_i} \right) \times \left( \prod_{i=1}^2  U(4)_{c_i} \right)
,
\end{equation*}
as can be explicitly verified by close inspection of the exceptional parts of the three-cycles $\Pi_{b_j}$, $\Pi_{c_k}$ and their orientifold
images given in~(\ref{Eq:3-cycles-Ex3}) and~(\ref{Eq:OR-3-cycles-Ex3}) in appendix~\ref{App:Mag-Ex3}. 
Table~\ref{Tab:Conditions-on_b+t+s-SOSp-Z2Z2M} also shows that for the present choice of exotic O6-plane with
$(\eta_{\Z_2^{(1)}},\eta_{\Z_2^{(2)}},\eta_{\Z_2^{(3)}}) =(1,1,-1)$ on the {\bf aaa}-torus, i.e. $b_1=b_2=b_3=0$, all
orientifold invariant rigid three-cycles will have their bulk parts parallel to the $\OR\Z_2^{(3)}$ invariant O6-plane,
whereas the D$6_{b_j}$ are parallel to the $\OR$-invariant plane and the D$6_{c_k}$-branes are parallel to the $\OR\Z_2^{(1)}$-invariant plane.
The symplectic gauge factors~\cite{Angelantonj:2009yj} arise by recombination processes of orientifold image D6-branes as detailed further below.   

The net-chiralities are obtained from the following intersection numbers,
\begin{equation*}
\begin{aligned}
\Pi_{a_i} \circ \Pi_{a_j} &=\Pi_{b_i} \circ \Pi_{b_j} =\Pi_{c_i} \circ \Pi_{c_j}=
\Pi_{c_i} \circ \Pi_{O6} = \Pi_{c_j} \circ \Pi_{O6}=0
,
\\
\Pi_{a_i} \circ \Pi_{b_j} &= \frac{1}{2} \, \left( -1 + (-1)^{\tau_{a_ib_j}^{\Z_2^{(1)}}}
 + \left[(-1)^{\tau_{a_ib_j}^{\Z_2^{(2)}}}-(-1)^{\tau_{a_ib_j}^{\Z_2^{(3)}}} \right] \, \delta_{\sigma^1_{a_ib_j},0} \right)
\\
&= \left\{\begin{array}{lr}
0 & (-1)^{\tau_{a_ib_j}^{\Z_2^{(k)}}} = (+,+,+),(+,-,-)\\
-1 +\delta_{\sigma^1_{a_ib_j},0} & (-,+,-)\\
-1 -\delta_{\sigma^1_{a_ib_j},0} &  (-,-,+)
\end{array}\right.
,
\end{aligned}
\end{equation*}
\begin{equation*}
\begin{aligned}
\Pi_{a_i} \circ \Pi_{c_j} &=\frac{1}{2} \, \left( -1 + (-1)^{\tau_{a_ic_j}^{\Z_2^{(1)}}}
 + \left[ -(-1)^{\tau_{a_ic_j}^{\Z_2^{(2)}}} + (-1)^{\tau_{a_ic_j}^{\Z_2^{(3)}}} \right] \, \delta_{\sigma^1_{a_ic_j},0} \right)
\\&=\left\{\begin{array}{lr}
0 & (-1)^{\tau_{a_ic_j}^{\Z_2^{(k)}}} = (+,+,+), (+,-,-)\\
-1 -\delta_{\sigma^1_{a_ic_j},0} & (-,+,-)\\
-1 +\delta_{\sigma^1_{a_ic_j},0} &  (-,-,+)
\end{array}\right.
,
\\
\Pi_{b_i} \circ \Pi_{c_j} &= \frac{\delta_{\sigma^1_{b_ic_j},0}}{2} \; \left( - (-1)^{\tau_{b_ic_j}^{\Z_2^{(2)}}} +  (-1)^{\tau_{b_ic_j}^{\Z_2^{(3)}}}
\right)
= \delta_{\sigma^1_{b_ic_j},0} \times \left\{\begin{array}{lr}
-1 &  (-1)^{\tau_{b_ic_j}^{\Z_2^{(k)}}} =   (-+-)
\\ 0 &(+++), (+--) 
\\ 1 &(--+)
\end{array}\right.
,
\end{aligned}
\end{equation*}
\begin{equation*}
\begin{aligned}
\Pi_{a_i} \circ \Pi_{a_j}' &= -2 \; \left( 2 +2 \,(-1)^{\tau_{a_ia_j}^{\Z_2^{(1)}}} \,\delta_{(\sigma^2 +\sigma^3)_{a_ia_j},0}
+ (-1)^{\tau_{a_ia_j}^{\Z_2^{(2)}}}\, \delta_{(\sigma^1+\sigma^3)_{a_ia_j},0} + (-1)^{\tau_{a_ia_j}^{\Z_2^{(3)}}}\, \delta_{(\sigma^1 +\sigma^2)_{a_ia_j},0}
\right)
\\
&=\left\{\begin{array}{lr}
-12& (-1)^{\tau_{a_i a_j}^{\Z_2^{(i)}}} = (+++) \text{ and } \sigma^1_{a_ia_j}=\sigma^2_{a_ia_j}=\sigma^3_{a_ia_j}=0\\
-4 & (+--) \text{ and } \sigma^1_{a_ia_j}=\sigma^2_{a_ia_j}=\sigma^3_{a_ia_j}=0\\
0 & (-+-),(--+) \text{ and }  \sigma^2_{a_ia_j}=\sigma^3_{a_ia_j}=0\\
\end{array}\right.
,
\\
\Pi_{a_i} \circ \Pi_{O6} &= -12
,
\end{aligned}
\end{equation*}
and the full matter spectrum is computed using the individual torus and $\Z_2^{(i)}$ invariant
intersection numbers in table~\ref{tab:IntersectionNumbers_Ex3} of appendix~\ref{App:Mag-Ex3}.
It consists of different kinds of bifundamental and antisymmetric matter representations as
listed in table~\ref{tab:Ex3-Full-Spectrum}.
\mathtabfix{
\begin{array}{|l|c|}\hline
\multicolumn{2}{|c|}{\text{\bf Matter spectrum of Ex. 3 on $T^6/\Z_2 \times \Z_2$ with gauge group } 
\left( \prod_{i=1}^4 U(2)_{a_i} \right) \times \left( \prod_{j=1}^2  U(4)_{b_j} \right) \times \left( \prod_{k=1}^2  U(4)_{c_k} \right)
}
\\\hline\hline
a_ia_j & \left[ (\2,\ov{\2},\1,\1;\1,\1;\1,\1) + (\1,\1,\2,\ov{\2};\1,\1;\1,\1) +  (\2,\1,\ov{\2},\1;\1,\1;\1,\1) +  (\1,\2,\1,\ov{\2};\1,\1;\1,\1)
+ (\2,\1,\1,\ov{\2};\1,\1;\1,\1) + (\1,\2,\ov{\2},\1;\1,\1;\1,\1)
+ c.c. \right]
\\
a_ia_j' &  + 4 \times \left[(\ov{\2},\ov{\2},\1,\1;\1,\1;\1,\1) + (\1,\1,\ov{\2},\ov{\2};\1,\1;\1,\1) \right] 
\\
& + 12 \times \left[ (\1_{\ov{\Anti}},\1,\1,\1;\1,\1;\1,\1) + (\1,\1_{\ov{\Anti}},\1,\1;\1,\1;\1,\1) + (\1,\1,\1_{\ov{\Anti}},\1;\1,\1;\1,\1) + (\1,\1,\1,\1_{\ov{\Anti}};\1,\1;\1,\1) \right]
\\\hline
a_ib_j & + 2 \times (\1,\1,\ov{\2},\1;\1,\4;\1,\1) + 2 \times  (\1,\1,\1,\ov{\2};\4,\1;\1,\1)
\\
a_ib_j' & + 2 \times \left[ (\ov{\2},\1,\1,\1;\ov{\4},\1;\1,\1) + (\1,\ov{\2},\1,\1;\1,\ov{\4};\1,\1)\right]
\\\hline
a_ic_j & + 2 \times (\1,\1,\ov{\2},\1;\1,\1;\4,\1)  + 2 \times (\1,\1,\1,\ov{\2};\1,\1;\1,\4)
\\
a_ic_j' & + 2 \times \left[ (\ov{\2},\1,\1,\1;\1,\1;\ov{\4},\1) + (\1,\ov{\2},\1,\1;\1,\1;\1,\ov{\4})
\right]
\\\hline
b_ib_j & + \left[ (\1,\1,\1,\1;\4,\ov{\4};\1,\1) +c.c. \right]
\\
b_ib_j'& + \left[(\1,\1,\1,\1;\4,\4;\1,\1) + (\1,\1,\1,\1;\6_{\Anti},\1;\1,\1) + (\1,\1,\1,\1;\1,\6_{\Anti};\1,\1) + c.c. \right] 
\\\hline
b_ic_j & + \emptyset 
\\
b_ic_j' & + (\1,\1,\1,\1;\4,\1;\1,\4) + (\1,\1,\1,\1;\1,\4;\4,\1)
+ (\1,\1,\1,\1;\ov{\4},\1;\ov{\4},\1) + (\1,\1,\1,\1;\1,\ov{\4};\1,\ov{\4})
\\\hline
c_ic_j & + \left[ (\1,\1,\1,\1;\1,\1;\4,\ov{\4}) +c.c. \right] 
\\
c_ic_j' & +\left[ (\1,\1,\1,\1;\1,\1;\4,\4) +(\1,\1,\1,\1;\1,\1;\6_{\Anti},\1) + (\1,\1,\1,\1;\1,\1;\1,\6_{\Anti}) + c.c. \right]
\\\hline
\end{array}
}{Ex3-Full-Spectrum}{The  complete massless matter spectrum on intersecting rigid D6-branes on $T^6/\Z_2 \times \Z_2$ with discrete torsion
which is T-dual to the magnetised D9/D5-brane example 3 in~\protect\cite{Angelantonj:2009yj} including the corrections in~\cite{Camara:2010zm}.
The spectra match up to reordering and complex conjugation of all $a_3$ and $a_4$ states and upon the recombination of the D$6_{b_j}$ and D$6_{c_j}$-branes
with their orientifold images to the reduced gauge group $\left( \prod_{j=1}^2  Sp(4)_{b_j} \right) \times \left( \prod_{k=1}^2  Sp(4)_{c_k} \right)$ 
 as discussed in the text.  
}

Since the rigid $b_i$ and $c_j$-branes do support the symplectic gauge factors listed in~\cite{Angelantonj:2009yj} before recombination
to fractional non-rigid orientifold invariant D6-branes, it is not surprising that the matter spectrum
presented here does not fully agree with the literature. The $a_i a_j$, $a_i a_j'$,  $a_i b_j$, $a_i b_j'$, $a_i c_j$ and $a_i c_j'$ sectors match with~\cite{Angelantonj:2009yj} 
(up to the conjugate representation on all $a_3$ and $a_4$ branes just like in the first example and up to the complex instead of real representations on $b_j$ and $c_j$). 
The multiplicities in all $b_ib_j$, $b_ib_j'$, $c_ic_j$ and $c_ic_j'$ sectors are by a factor of two bigger than those given in~\cite{Angelantonj:2009yj}, which is consistent with 
the following breaking pattern of the gauge groups upon the recombination of two rigid to two fractional to a single bulk D6-brane along path (a), where the spectrum in~\cite{Angelantonj:2009yj}
corresponds to 
\begin{equation*}
\begin{aligned}
\boxed{\begin{array}{c}  U(4)_{c_1} \times U(4)_{c_2} \\
c_ic_j: \quad  \left[ (\1,\1,\1,\1;\1,\1;\4,\ov{\4}) +c.c. \right]\\
c_ic_j': \quad \Bigl[ (\1,\1,\1,\1;\1,\1;\4,\4) +\qquad \\ \qquad+(\1,\1,\1,\1;\1,\1;\6_{\Anti},\1) + \\\qquad+ (\1,\1,\1,\1;\1,\1;\1,\6_{\Anti}) + c.c. \Bigr]
\end{array}} & \quad \stackrel{\text{(1b)}}{\longrightarrow} \quad 
\boxed{\begin{array}{c}  U(4)_{\tilde{C}}\\ \tilde{C}\tilde{C}: \quad  \left[(\1,\1,\1,\1;\1,\1;\6_{\Anti}) + c.c. \right] \end{array}}
\\
\text{\footnotesize (1a)} \downarrow \qquad\quad & \hspace{25mm} \downarrow \text{\footnotesize (2b)}
\\
\boxed{\begin{array}{c} Sp(4)_{\tilde{c}_1} \times Sp(4)_{\tilde{c}_2}\\
\tilde{c}_i \tilde{c_i}: \quad  (\1,\1,\1,\1;\1,\1;\6_{\Anti},\1)\\ \qquad\qquad + (\1,\1,\1,\1;\1,\1;\1,\6_{\Anti}) \\
\tilde{c}_1 \tilde{c_2}: \quad  2 \times (\1,\1,\1,\1;\1,\1;\4,\4)
\end{array}} 
& \quad \stackrel{\text{(2a)}}{\longrightarrow} \quad 
 \boxed{\begin{array}{c} Sp(4)_{C} \\  CC: \quad (\1,\1,\1,\1;\1,\1;\6_{\Anti})\end{array}}
\end{aligned}
\end{equation*}
performing step (1a) on the spectrum in table~\ref{tab:Ex3-Full-Spectrum} by recombining $\Pi_{\tilde{c}_i}=\Pi_{c_i}+\Pi_{c_i'}$. Performing instead step (1b)
with $\Pi_{\tilde{C}} =  \Pi_{c_1} + \Pi_{c_2}$ or $\Pi_{\tilde{C}} =  \Pi_{c_1} + \Pi_{c_2'}$ leads to fractional D6-branes along 
$T^2_{(1)} \times \left(T^4_{(2 \cdot 3)}/\Z_2^{(1)}\right)$
which are T-dual to the well-known fractional D5-branes in~\cite{Bianchi:1990yu,Gimon:1996rq}. Finally step (2a) or (2b) leads to a pure bulk D6-brane wrapping 
$\Pi_C = \sum_{i=1}^2 \left(\Pi_{c_i} + \Pi_{c_i'}\right)$, which is the T-dual to a D5-brane at an arbitrary position in the bulk.
The discussion for the recombination of the $b_i$-branes follows the same lines with the T-duality directions chosen along the $\Re(z_{2i-1})$ axes instead of the $\Im(z_{2i})$ axes
above.

The spectrum in table~\ref{tab:Ex3-Full-Spectrum} satisfies all consistency checks involving net-chiralities and the cancellation 
of the eight different $U(N_x)^3$ anomalies. Details on the match to~\cite{Angelantonj:2009yj} are 
given in appendix~\ref{App:Mag-Ex3} based on a close inspection of  the rigid three-cycles and RR tadpole cancellation conditions
in the presence of an exotic $\OR\Z_2^{(3)}$-plane.

The anomaly matrix with entries defined in~(\ref{Eq:Def-AnomalyMatrix-Entries}) reads
{\small 
\begin{equation}\label{Eq:Anomaly-Matrix-Ex3}
\left(\begin{array}{c|c|c}  
 C_{a_ia_j} & C_{a_ib_j}  & C_{a_ic_j} \\\hline
 C_{b_ia_j} & C_{b_ib_j} & C_{b_ic_j}\\\hline
 C_{c_ia_j} & C_{c_ib_j}  & C_{c_ic_j} 
\end{array}\right)
= \frac{4}{\pi^2} \left(\begin{array}{cccc|cc|cc} 
-3 & -1 & 0 & 0 & -1 & 0 & -1 & 0 \\ 
-1 & -3 & 0 & 0 & 0 & -1 & 0 & -1 \\ 
0 & 0 & -3 & -1 & 0 & 1 & 1 & 0 \\
0 & 0 & -1 & -3 & 1 & 0 & 0 & 1 \\\hline
-1 & 0 & 0 & -1 & 0 & 0 & -1 & 1 \\
0 & -1 &  -1 & 0 &  0 & 0 & 1 & -1\\\hline
-1 & 0 &  -1 & 0 &  -1 & 1 & 0 & 0\\
0 & -1 &  0 & -1 &  1 & -1 & 0 & 0
\end{array}\right)
,
\end{equation}
}
and the $a_ia_j$ sectors in the upper left corner agree with the complete anomaly matrix in~\cite{Angelantonj:2009yj},
whereas the other entries appear here since we consider the gauge group $\left(\prod_{i=1}^2 U(4)_{b_i} \right) \times \left(\prod_{j=1}^2 U(4)_{c_j} \right)$ before 
the breaking to $\left(\prod_{i=1}^2 Sp(4)_{b_i} \right) \times \left(\prod_{j=1}^2 Sp(4)_{c_j} \right)$.

The K\"ahler metrics and the $v_i$-dependent one-loop contributions to the holomorphic
gauge kinetic function $\delta_y {\rm f}^{\text{1-loop}}_{SU(2)_{a_1}}(v_i)$
involving brane $a_1$ are listed in table~\ref{tab:Z2Z2ex3_Kaehler}.
\mathtabfix{
\begin{array}{|c||c|c|c|c|}\hline
\muc{5}{|c|}{\text{\bf Example 3 on $T^6/\Z_2 \times \Z_2$: K\"ahler metrics and gauge kinetic functions involving brane } a_1}
\\\hline\hline
y & (\vec{\phi}_{a_1y}) & b_{a_1y}^{\cal A} & K_{(\4_{a_1},\ov{\N}_y)} & \delta_y {\rm f}_{SU(2)_{a_1}}^{\text{1-loop}}(v_i)
\\\hline\hline
a_1 & (0,0,0) & -6 & - &   \sum_{i=1}^3 \frac{1}{2\pi^2} \ln \eta(iv_i)
\\\hline
a_{j,j=2,3,4} & (0,0,0) & 2 &   f(S,U_l) \sqrt{\frac{ 2\pi \left(\frac{1}{r_{j-1}} + 4 \, r_{j-1}\right)}{v_{j \, \text{mod}\, 3} \;\, v_{j+1 \,\text{mod}\, 3}}}
&  - \frac{1}{2\pi^2} \ln \eta(iv_{j-1})
\\\hline
a_1' & (-2\phi^{(1)},-2\phi^{(2)},-2\phi^{(3)}) &\!\!\!\begin{array}{c} b_{ay}^{\cal A} + b_{ay}^{\cal M} \\ =12 -12 \end{array}\!\!\! 
& \!\!\!\begin{array}{c} K_{\Anti_a} =  \frac{f(S,U_l)}{\sqrt{v_1v_2v_3}} \times \\
\; \sqrt{\frac{\Gamma(2\phi^{(1)}) \, \Gamma(2\phi^{(2)}) \,  \Gamma(1-2|\phi^{(3)}|) }{\Gamma(1-2\phi^{(1)}) \, \Gamma(1-2\phi^{(2)}) \, \Gamma(2|\phi^{(3)}|)} }  \end{array}\!\!\! & - 
\\\hline
a_2' & (-2\phi^{(1)},-2\phi^{(2)},-2\phi^{(3)}) & 4 & \frac{f(S,U_l)}{\sqrt{v_1v_2v_3}} 
\; \sqrt{\frac{\Gamma(2\phi^{(1)}) \, \Gamma(2\phi^{(2)}) \,  \Gamma(1-2|\phi^{(3)}|) }{\Gamma(1-2\phi^{(1)}) \, \Gamma(1-2\phi^{(2)}) \, \Gamma(2|\phi^{(3)}|)} }
& - 
\\\hline
a_{j,j=3,4}' & (-2\phi^{(1)},-2\phi^{(2)},-2\phi^{(3)}) & 0 & - & -
\\\hline
b_{k,k=1,2,2'} & (-\phi^{(1)},-\phi^{(2)},-\phi^{(3)}) & 0 & - & -
\\\hline
b_1 ' & (-\phi^{(1)},-\phi^{(2)},-\phi^{(3)}) & 4 &   \frac{f(S,U_l)}{\sqrt{v_1v_2v_3}} 
\; \sqrt{\frac{\Gamma(\phi^{(1)}) \, \Gamma(\phi^{(2)}) \,  \Gamma(1-|\phi^{(3)}|) }{\Gamma(1-\phi^{(1)}) \, \Gamma(1-\phi^{(2)}) \, \Gamma(|\phi^{(3)}|)} }
& -
\\\hline
c_{l,l=1,2,2'} &  (-\phi^{(1)},\frac{1}{2}-\phi^{(2)},-\frac{1}{2}-\phi^{(3)}) & 0 & - & -
\\\hline
c_1' &  (-\phi^{(1)},\frac{1}{2}-\phi^{(2)},-\frac{1}{2}-\phi^{(3)}) & 4 &  \frac{f(S,U_l)}{\sqrt{v_1v_2v_3}} 
\; \sqrt{\frac{\Gamma(\phi^{(1)}) \, \Gamma(\frac{1}{2}-\phi^{(2)}) \,  \Gamma(\frac{1}{2}+|\phi^{(3)}|) }{\Gamma(1-\phi^{(1)}) \, \Gamma(\frac{1}{2}+\phi^{(2)}) \, \Gamma(\frac{1}{2}-|\phi^{(3)}|)} }
& -
\\\hline
\end{array}
}{Z2Z2ex3_Kaehler}{Relative angles, beta function coefficients, K\"ahler metrics and two-torus volume
$v_i$ dependent one-loop contributions to the holomorphic
gauge kinetic function involving D6-brane $a_1$ of example 3 on $T^6/\Z_2 \times \Z_2$ with discrete torsion.}
The tree-level value of the holomorphic gauge kinetic functions for the $a_i$ branes, 
\begin{equation*}
 {\rm f}^{\text{tree}}_{SU(2)_{a_i}} \sim S +  U_1 + 2 \, U_2- 2 \, U_3
,
\end{equation*}
agrees with~\cite{Angelantonj:2009yj}.

The holomorphic gauge kinetic function for $a_1$ in this example does not receive a $v_i$-moduli dependent one-loop correction
due to the cancellation from all $a_i$ sectors,
\begin{equation*}
\delta_{\rm total} {\rm f}^{\text{1-loop}}_{SU(2)_{a_1}}= \sum_{j=1}^4 \delta_{a_j}  {\rm f}^{\text{1-loop}}_{SU(2)_{a_1}}(v_i) + {\rm const}
= {\rm const}
,
\end{equation*}
in agreement with~\cite{Angelantonj:2009yj}. 
The K\"ahler metrics for matter with $SU(2)_{a_1}$ charge in table~\ref{tab:Z2Z2ex3_Kaehler} are to our knowledge given here for the first time.

The angle dependent contributions in~(\ref{Eq:3angle-f-angle}) and~(\ref{Eq:aap-f-3angles}) again sum to zero,
\begin{equation*}
\sum_{j=1}^4 I_{a_1a_j'}^{\Z_2^{(k)}}=
\sum_{j=1}^4 I_{a_1b_j}^{\Z_2^{(k)}}=\sum_{j=1}^4 I_{a_1c_j}^{\Z_2^{(k)}}=0 \qquad \text{ for } \quad  k=1,2,3
,
\end{equation*}
and the constant factor due to the intersections with O6-planes is twice the one from example~1,
\begin{equation*}
{\rm const}= \frac{1}{\pi^2} \, \ln 2
,
\end{equation*}
as can be seen by comparing the individual intersection numbers for the present model in table~\ref{tab:IntersectionNumbers_Ex3}
with those of example~1 in table~\ref{tab:IntersectionNumbers_Ex1+2}.
This completes the discussion of the open string K\"ahler metrics with charge under the D$6_{a_1}$-brane
and perturbatively exact holomorphic gauge kinetic function for the same D6-brane.

\section{Example on $T^6/\Z_6'$: the Standard Model on fractional D6-branes}\label{S:Z6p-Example}

On the $T^6/\Z_6'$ orbifold background, each torus three-cycle has three orbifold images.
This leads to a sum over these images when computing the holomorphic gauge kinetic function,
and matter states with identical charges can be localised at intersections of different orbifold 
images and thereby have distinct K\"ahler metrics. This happens e.g. for the various quark
families in the Standard Model on $T^6/\Z_6'$ presented in~\cite{Gmeiner:2007zz,Gmeiner:2008xq},
for which the gauge thresholds have been computed in~\cite{Gmeiner:2009fb}. More model building on the same orbifold background can be found e.g. in~\cite{Bailin:2006zf,Bailin:2007va,Bailin:2008xx}
In this section, we will give some examples for the decomposition into K\"ahler metrics for matter charged under the QCD stack $SU(3)_a$
or the `leptonic' stack $U(1)_d$ as well as for  the pertubatively exact
holomorphic gauge kinetic function ${\rm f}_{SU(3)_a}$ along the generic prescription in section~\ref{Ss:loop-KaehlerAdj}
with modifications for $h_{21}^{\rm bulk}=1$ discussed in section~\ref{Ss:Modifications-less-complex}. 
We will furthermore discuss two examples of orientifold invariant D6-branes, one parallel and the other 
perpendicular to the $\OR$ plane as classified in section~\ref{Ss:SO+Sp_Kaehler+GaugeKinetic},
and finally we will use the framework of section~\ref{Ss:U1_Kaehler+GaugeKinetic} and 
present the complete perturbatively exact holomorphic gauge kinetic functions for the single $U(1)_a$
and $U(1)_d$ factors as well as for the anomaly-free combination $U(1)_{B-L}=\frac{1}{3}U(1)_a + U(1)_d$. 

The D6-brane configuration is displayed in table~\ref{tab:smmodelz6p+hidden}.
\mathtab{
\begin{array}{|c|c||c|c|c|c|c|c|c|c|c|}\hline
\muc{8}{|c|}{\text{\bf D6-brane configuration for the SM with hidden sector $Sp(6)_{h_3}$ on } T^6/\Z_6'}
\\\hline\hline
\rotatebox{90}{\text{\!\!\!\!\!\!\!\!\!\!\!\!\!D$6_x$-brane}}
 & (\vec{\phi}_x) &\!\!\! \begin{array}{c} \text{Torus}\\ \text{wrappings} \\ (n^i_x,m^i_x)
  \end{array}\!\!\! 
&\!\!\! \begin{array}{c} (\tilde{\vec{X}}_x); (\tilde{\vec{Y}}_x) = \\\!\!
\left(\!\!\!\begin{array}{c} \frac{P_x+Q_x}{2} \\ \frac{V_x-U_x}{2} \end{array}\!\!\!\right)\!;\!
\left(\!\!\!\begin{array}{c} -\frac{U_x+V_x}{2} \\ \frac{Q_x-P_x}{2} \end{array}\!\!\!\right)\!\!
\end{array}\!\!\!   & (\sigma^1_x,\sigma^3_x) & (-1)^{\tau^{\Z_2}_x} & (\tau^1_x,\tau^3_x)
& (\vec{V}_{xx})
\\\hline\hline
a & \!\!(-\frac{1}{3},-\frac{1}{6},\frac{1}{2}) & \!\!(1,-1;1,0;0,1)\!\! & \left(\!\!\!\begin{array}{c} 0 \\ -1 \end{array}\!\!\!\right);
 \left(\!\!\!\begin{array}{c} 0 \\ 0\end{array}\!\!\!\right)
& (1;1) & (+) & (1,1)
&\!\!\!\left(\!\!\begin{array}{c} \frac{2}{\sqrt{3}} \\  \frac{2}{\sqrt{3}} \\ r \end{array}\!\!\right)\!\!\!
\\
h_3 & &
& & (0,0) &  (+)& (1,0) &
\\\hline
b & (\frac{1}{6},-\frac{1}{3},\frac{1}{6}) & \!\! (1,1;2,-1;1,1)\!\!  &  \left(\!\!\!\begin{array}{c} \frac{3}{2} \\ -\frac{3}{2} \end{array}\!\!\!\right);
 \left(\!\!\!\begin{array}{c} -\frac{3}{2} \\ -\frac{3}{2} \end{array}\!\!\!\right)
& (1,0) & (-) & (1,0)
&\!\!\!\left(\!\!\begin{array}{c} 2\sqrt{3} \\ 2\sqrt{3} \\ \frac{1}{r} + r \end{array}\!\!\right)\!\!\!
\\\hline
c & (-\frac{1}{3},\frac{1}{3},0) & \!\! (1,-1;-1,2;1,0)\!\!  &\left(\!\!\!\begin{array}{c} 1 \\ 0 \end{array}\!\!\!\right);
 \left(\!\!\!\begin{array}{c} 0 \\ 0 \end{array}\!\!\!\right)
& (1,1) & (+) & (1,1)
&\!\!\!\left(\!\!\begin{array}{c} \frac{2}{\sqrt{3}} \\ 2\sqrt{3} \\ \frac{1}{r} \end{array}\!\!\right)\!\!\!
\\\hline
d & (\frac{1}{6},\frac{1}{3},-\frac{1}{2}) & \!\! (1,1;1,-2;0,1)\!\!  & \left(\!\!\!\begin{array}{c} 0 \\ -3\end{array}\!\!\!\right);
 \left(\!\!\!\begin{array}{c}0 \\ 0\end{array}\!\!\!\right)
& (0,1) & (+) & (1,1)
&\!\!\!\left(\!\!\begin{array}{c} 2\sqrt{3} \\2\sqrt{3} \\  r \end{array}\!\!\right)\!\!\!
\\\hline
\end{array}
}{smmodelz6p+hidden}{Geometrical setup of the supersymmetric Standard Model example
with hidden sector $Sp(6)_{h_3}$ on the {\bf ABa} lattice in the $T^6/\mathbb{Z}_6'$
 background~\protect\cite{Gmeiner:2007zz,Gmeiner:2008xq,Gmeiner:2009fb}. Brane $b$ preserves
supersymmetry for the ratio of radii on the {\bf a}-type $T^2_{(3)}$ torus $r=1/\sqrt{3}$.
Details on the exceptional cycles and intersection numbers are given in~\protect\cite{Gmeiner:2007zz,Gmeiner:2008xq,Gmeiner:2009fb}.}
In order to compute the K\"ahler metrics for D6-branes intersecting at three non-vanishing angles, it is 
useful to note the following values of ratios of Gamma functions (cf. table~\ref{Tab:Gamma-values}),
\begin{equation*}
\prod_{i=1}^3 \left(
\frac{\Gamma(|\phi_{xy}^{(i)}|)}{\Gamma(1-|\phi_{xy}^{(i)}|)}\right)^{-\frac{\sgn(\phi^{(i)}_{xy})}{\sgn(I_{xy})}}
=\left\{\begin{array}{cr} \frac{25}{2} & (\vec{\phi}_{xy}) = \pm (\frac{1}{6},\frac{1}{6},-\frac{1}{3})\\
8 & \pm (\frac{1}{3},\frac{1}{3},-\frac{2}{3}) \\
10 & \pm (\frac{1}{6},\frac{1}{3},-\frac{1}{2})
\end{array}\right.
,
\end{equation*}
which occur in the model. Since the K\"ahler metrics are independent of the chirality, each value holds for both
orientations of angles. This is assured by the factor $\sgn(I_{xy})$ in the exponent.
The relative angles, beta function coefficients  and K\"ahler metrics related to branes $a$ and $d$, which include 
those for all quarks and leptons, are listed in table~\ref{tab:modelz6p+hidden-sectors_ax-Kaehler} as typical examples.
\mathtabfix{
\begin{array}{|c||c|c|c||c|c|c||c|c|c|}\hline
\muc{10}{|c|}{\text{\bf Angles, beta function coefficients and K\"ahler metrics for D6-branes $a$ and $d$ of the SM on } T^6/\Z_6'}
\\\hline\hline
\muc{10}{|c|}{\text{\bf Brane } a}
\\\hline\hline
y & (\vec{\phi}_{ay}) & b_{ay}^{\cal A} & K_{(\3_a,\ov{\N}_y)} 
 & (\vec{\phi}_{a(\theta y)}) & b_{a(\theta y)}^{\cal A} & K_{(\3_a,\ov{\N}_y)} 
 & (\vec{\phi}_{a(\theta^2 y)}) & b_{a(\theta^2 y)}^{\cal A} & K_{(\3_a,\ov{\N}_y)} 
\\\hline\hline
a & (0,0,0) & -6  & \!\!\!\begin{array}{c}  K_{\Adj_a}= \\ \sqrt{\frac{\pi \, r}{2}} \, \frac{f(S,U)}{v_2}  \end{array}\!\!\!
& \pm (\frac{1}{3},-\frac{1}{3},0) & 3 
& \muc{4}{|c|}{ K_{\Adj_a}= f(S,U) \sqrt{\frac{2\pi \, r}{v_1v_2}} }
\\\hline
a' &(-\frac{1}{3},\frac{1}{3},0)&\!\!\!\begin{array}{c} b_{aa'}^{\cal A} + b_{aa'}^{\cal M} \\ =\frac{3}{2} -1 \end{array}\!\!\! 
& \!\!\!\begin{array}{c} K_{\Anti_a} = \\ f(S,U) \sqrt{\frac{2\pi \, r}{v_1v_2}} \end{array}\!\!\!
& (0,0,0) &\!\!\!\begin{array}{c} b_{a(\theta a')}^{\cal A} + b_{a(\theta a')}^{\cal M} \\ = 6-4=2 \end{array}\!\!\! 
&  \!\!\!\begin{array}{c} K_{\Anti_a} = \\  f(S,U) \sqrt{\frac{ 4\pi}{\sqrt{3} \, v_1v_3}} \end{array}\!\!\!
& (\frac{1}{3},-\frac{1}{3},0) &\!\!\!\begin{array}{c} b_{a(\theta^2 a')}^{\cal A} + b_{a(\theta^2 a')}^{\cal M} \\ =\frac{3}{2} -1 \end{array}\!\!\! 
& \!\!\!\begin{array}{c} K_{\Anti_a} = \\  f(S,U) \sqrt{\frac{2\pi \, r}{v_1v_2}}  \end{array}\!\!\!
\\\hline
b & (\frac{1}{2},-\frac{1}{6},-\frac{1}{3}) & 0 & - 
& (-\frac{1}{6},\frac{1}{2},-\frac{1}{3}) & 0 & - 
&(\frac{1}{6},\frac{1}{6},-\frac{1}{3}) & 0 & - 
\\\hline
b' & (\frac{1}{6},-\frac{1}{2},\frac{1}{3}) & 2 &f(S,U) \, \sqrt{\frac{10}{v_1v_2v_3}}
& (-\frac{1}{2},\frac{1}{6},\frac{1}{3}) & 2 &  f(S,U) \, \sqrt{\frac{10}{v_1v_2v_3}}
& (-\frac{1}{6},-\frac{1}{6},\frac{1}{3}) & 1 &  f(S,U) \, \sqrt{\frac{25}{2\,v_1v_2v_3}}
\\\hline
c & (0,\frac{1}{2},-\frac{1}{2}) & 2  &  f(S,U) \, \sqrt{\frac{4\pi}{\sqrt{3} \,v_2v_3}}
& (\frac{1}{3},\frac{1}{6},-\frac{1}{2}) & 0 & -
& (-\frac{1}{3},-\frac{1}{6},\frac{1}{2}) & 1 &   f(S,U) \, \sqrt{\frac{10}{v_1v_2v_3}}
\\\hline
d & (\frac{1}{2},-\frac{1}{2},0) & 2 &  f(S,U) \, \sqrt{\frac{2\pi \, r}{v_1v_2}}
& (-\frac{1}{6},\frac{1}{6},0) & \frac{1}{2} &   f(S,U) \, \sqrt{\frac{2\pi \, r}{v_1v_2}}
& (\frac{1}{6},-\frac{1}{6},0) & \frac{1}{2} &   f(S,U) \, \sqrt{\frac{2\pi \, r}{v_1v_2}}
\\\hline
d' & (\frac{1}{6},-\frac{1}{6},0) & \frac{1}{2} &   f(S,U) \, \sqrt{\frac{2\pi \, r}{v_1v_2}}
& (\frac{1}{2},-\frac{1}{2},0) & 2 &  f(S,U) \, \sqrt{\frac{2\pi \, r}{v_1v_2}}
& (-\frac{1}{6},\frac{1}{6},0) & \frac{1}{2} &   f(S,U) \, \sqrt{\frac{2\pi \, r}{v_1v_2}}
\\\hline
h_3  & (0,0,0) & 0 & -   
& (\frac{1}{3},-\frac{1}{3},0) & 3 \cdot 0_3 & -  
&  (-\frac{1}{3},\frac{1}{3},0) & 3 \cdot 0_3 & - 
\\\hline\hline
\muc{10}{|c|}{\text{\bf Brane } d}
\\\hline\hline
x & (\vec{\phi}_{xd}) & b_{dx}^{\cal A} & K_{(\N_x,\ov{\1}_d)} 
 & (\vec{\phi}_{x(\theta d)}) & b_{(\theta d)x}^{\cal A} & K_{(\N_x,\ov{\1}_d)} 
 & (\vec{\phi}_{x(\theta^2 d)}) & b_{(\theta^2 d)x}^{\cal A} & K_{(\N_x,\ov{\1}_d)} 
\\\hline\hline
b & (0,-\frac{1}{3},\frac{1}{3}) & 3 \cdot 0_1 & -
& (\frac{1}{3},-\frac{2}{3},\frac{1}{3}) & 6 &  f(S,U) \, \sqrt{\frac{8}{v_1v_2v_3}}
& (-\frac{1}{3},0,\frac{1}{3}) & 4 & f(S,U) \, \sqrt{\frac{4\pi \sqrt{3}}{v_1v_3}}
\\\hline
c & (\frac{1}{2},0,-\frac{1}{2}) & 2 &  f(S,U) \, \sqrt{\frac{4\pi \sqrt{3}}{v_1v_3}}
& (-\frac{1}{6},-\frac{1}{3},\frac{1}{2}) & 3 &   f(S,U) \, \sqrt{\frac{10}{v_1v_2v_3}}
& (\frac{1}{6},\frac{1}{3},-\frac{1}{2}) & 0 & -
\\\hline
h_3  & (\frac{1}{2},-\frac{1}{2},0) & 24 \cdot 0_3 & - 
& (-\frac{1}{6},\frac{1}{6},0) & 6 \cdot 0_3 & -
& (\frac{1}{6},-\frac{1}{6},0) & 6 \cdot 0_3 & -
\\\hline\hline
x & (\vec{\phi}_{xd'}) & b_{d'x}^{\cal A} & K_{(\N_x,\1_d)} 
 & (\vec{\phi}_{x(\theta d')}) & b_{(\theta d')x}^{\cal A} & K_{(\N_x,\1_d)} 
 & (\vec{\phi}_{x(\theta^2 d')}) & b_{(\theta^2 d')x}^{\cal A} & K_{(\N_x,\1_d)} 
\\\hline\hline
b & (-\frac{1}{3},0,\frac{1}{3}) & 2 &  f(S,U) \, \sqrt{\frac{4\pi \sqrt{3}}{v_1v_3}}
& (0,-\frac{1}{3},\frac{1}{3}) & 3 \cdot 0_1 & - 
& (\frac{1}{3},-\frac{2}{3},\frac{1}{3}) & 3 &  f(S,U) \, \sqrt{\frac{8}{v_1v_2v_3}}
\\\hline
d' &  (-\frac{1}{3},\frac{1}{3},0) &\!\!\!\begin{array}{c} 2b_{dd'}^{\cal A} + b_{dd'}^{\cal M} \\ =9-9=0 \end{array}\!\!\!  & -
& (0,0,0) &\!\!\!\begin{array}{c} 2b_{d(\theta d')}^{\cal A} + b_{d(\theta d')}^{\cal M} \\ = 4-4=0\end{array}\!\!\! & -
& (\frac{1}{3},-\frac{1}{3},0) &\!\!\!\begin{array}{c} 2b_{d(\theta^2 d')}^{\cal A} + b_{d(\theta^2 d')}^{\cal M} \\ =9-9=0 \end{array}\!\!\! 
& -    
\\\hline
\end{array}
}{modelz6p+hidden-sectors_ax-Kaehler}{Relative angles, beta function coefficients and K\"ahler metrics 
for matter charged under $U(3)_a$ or $U(1)_d$ of the Standard Model example with hidden $Sp(6)_{h_3}$ on $T^6/\Z_6'$~\cite{Gmeiner:2008xq}. This includes all left- and right-handed quarks at $a(\theta^k b')$ and $a(\theta^k c)$  intersections, respectively, and leptons at $b(\theta^k d^{(\prime)})$ and $c(\theta^k d^{(\prime)})$  intersections. 
The beta function coefficients $b^{\cal A}_{(\theta^k d)a}$ and $b^{\cal A}_{a(\theta^k d)}$
with reversed lower indices differ by the global factor $N_a/N_d=3$, i.e.
 $(b^{\cal A}_{(\theta^k d)a})_{k=1,2,3}=(6,\frac{3}{2},\frac{3}{2})$ and
$(b^{\cal A}_{(\theta^k d')a})_{k=1,2,3}=(\frac{3}{2},6,\frac{3}{2})$. The corresponding $v_i$ dependent one-loop contributions to the 
holomorphic gauge kinetic functions $\delta_y \, {\rm f}^{\text{1-loop}}_{SU(3)_a}(v_i)$ and $\delta_x \, {\rm f}^{\text{1-loop}}_{SU(1)_d}(v_i)$
are given in table~\protect\ref{tab:modelz6p+hidden-sectors_ax-holomorphic}.
}

In contrast to the Standard Model on the six-torus, e.g.~\cite{Ibanez:2001nd}, or the $T^6/\Z_2 \times \Z_2$ orbifold,
e.g.~\cite{Cvetic:2001nr,Cvetic:2001tj,Gmeiner:2005vz}, particle generations
can arise at intersections of various orbifold image D6-branes. 
As an example, the $ac$ sector provides two generations of right-handed quarks with 
K\"ahler metric $K_{(\ov{\3},\1)}=f(S,U) \, \sqrt{\frac{4\pi}{\sqrt{3} \,v_2v_3}}$ at vanishing angle on $T^2_{(1)}$, 
while the $a(\theta^2 c)$ sector provides the third right-handed quark generation with K\"ahler metric 
$K_{(\ov{\3},\1)}= f(S,U) \, \sqrt{\frac{10}{v_1v_2v_3}}$ at three non-trivial angles.
The size of the physical Yukawa couplings is thus not only governed by the triangular worldsheets contributing to the holomorphic
factor~\cite{Cremades:2003qj,Cremades:2004wa}, but also by the values of the K\"ahler metrics. For matter localised at some intersection with one vanishing angle, e.g. $\phi^{(1)}_{ac}=0$, the
K\"ahler metric depends only on the volume of the remaining four-torus, e.g. $v_2v_3$ for the $ac$ sector.
 This makes it possible to obtain
Yukawa hierarchies by choosing unisotropic two-tori. More details for the interplay of these various effects will be
given in~\cite{Honecker:2011tbd}.

The holomorphic tree-level gauge couplings for $SU(3)_a$, the anomalous $U(1)_a$ and $U(1)_d$ and the massless linear
combination $U(1)_{B-L}$ read 
\begin{equation}
{\rm f}^{\text{tree}}_{SU(3)_a} = {\rm const.} \cdot U,
\quad
{\rm f}^{\text{tree}}_{U(1)_a} =  {\rm const.} \cdot 6 \,  U,
\quad
{\rm f}^{\text{tree}}_{U(1)_d} =  {\rm const.} \cdot 6 \, U,
\quad
{\rm f}^{\text{tree}}_{U(1)_{B-L}} = {\rm const.} \cdot \frac{20}{3} \, U,
\end{equation}
with the definition of the bulk complex structure $U$ on $T^6/\Z_6'$ given in~(\ref{Eq:Def-S+Ui=1}) 
and the constant identical for all four gauge groups. 
These gauge factors do at tree level not depend on the dilaton $S$, as can be seen from the 
corresponding bulk wrapping numbers in table~\ref{tab:smmodelz6p+hidden}.

The basic building blocks for the one-loop contributions to the holomorphic gauge kinetic functions 
of the strong interactions, ${\rm f}_{SU(3)_a}^{\text{1-loop}}$, as well as the (anomalous and unphysical) single $U(1)$ charges,
 ${\rm f}_{U(1)_a}^{\text{1-loop}}$ and ${\rm f}_{U(1)_d}^{\text{1-loop}}$, and their physical linear combination
 $U(1)_{B-L}=\frac{1}{3} U(1)_a + U(1)_d$, ${\rm f}_{U(1)_{B-L}}^{\text{1-loop}}$, 
are listed in table~\ref{tab:modelz6p+hidden-sectors_ax-holomorphic}. 
\mathtabfix{
\begin{array}{|c||c|c|c|}\hline
\muc{4}{|c|}{\text{\bf 1-loop contributions to the holomorphic gauge kinetic function } \delta_y \, {\rm f}_{SU(3)_a}^{\text{1-loop}}(v_i) \text{ \bf  for the SM on } T^6/\Z_6'}
\\\hline\hline
y & \delta_y {\rm f}_{SU(3)_a}^{\text{1-loop}}(v_i)
& \delta_{(\theta y)} {\rm f}_{SU(3)_a}^{\text{1-loop}}(v_i)
& \delta_{(\theta^2 y)} {\rm f}_{SU(3)_a}^{\text{1-loop}}(v_i)
\\\hline\hline
a &  \frac{3}{2\pi^2} \ln \eta(iv_2) & \muc{2}{|c|}{-\frac{3}{4\pi^2} \ln \eta(iv_3)
}
\\\hline
a' & -\frac{1}{8\pi^2} \ln \eta(iv_3) 
& \begin{array}{c}
-\frac{3}{2\pi^2} \ln \eta(iv_2)\\
 +\frac{1}{2\pi^2} \ln \eta(i \tilde{v}_1) +\frac{1}{2\pi^2} \ln \eta(i v_3) \\
\end{array}
& -\frac{1}{8\pi^2} \ln \eta(iv_3)  
\\\hline
b & - & -  & -
\\\hline
b' &  - & - & -
\\\hline
c & -\frac{1}{2\pi^2} \ln \eta(iv_1)  & -  & -
\\\hline
d & -\frac{1}{2\pi^2} \ln \eta(iv_3) &-\frac{1}{8\pi^2} \ln \eta(iv_3) &-\frac{1}{8\pi^2} \ln \eta(iv_3)
\\\hline
d' &-\frac{1}{8\pi^2} \ln \eta(iv_3)  &-\frac{1}{2\pi^2} \ln \eta(iv_3) &-\frac{1}{8\pi^2} \ln \eta(iv_3)
\\\hline
h_3 & - &- \frac{3}{8\pi^2} \ln \left(e^{-\pi v_3/4}\frac{\vartheta_1 (\frac{1 - i v_3}{2},i v_3)}{\eta (i v_3)} \right)  
&- \frac{3}{8\pi^2} \ln \left(e^{-\pi v_3/4}\frac{\vartheta_1 (\frac{1 - i v_3}{2},i v_3)}{\eta (i v_3)} \right)
\\\hline\hline
\muc{4}{|c|}{\text{\bf 1-loop contributions to the auxiliary holomorphic gauge kinetic function } \delta_x \, {\rm f}_{SU(1)_d}^{\text{1-loop}}}
\\\hline\hline
x & \delta_x {\rm f}_{SU(1)_d}^{\text{1-loop}}(v_i)
& \delta_{(\theta^{-1} x)} {\rm f}_{SU(1)_d}^{\text{1-loop}}(v_i)
& \delta_{(\theta^{-2} x)} {\rm f}_{SU(1)_d}^{\text{1-loop}}(v_i)
\\\hline\hline
b &  - \frac{3}{8\pi^2}  \ln \left(e^{-\pi v_1/4}\frac{\vartheta_1 (\frac{ - i v_1}{2},i v_1)}{\eta (i v_1)} \right) & - 
&  \begin{array}{c}
 -\delta_{\sigma^2_{bd},0}\delta_{\tau^2_{bd},0} \times  \frac{1}{\pi^2} \ln \eta(iv_2) \\
 - \frac{1}{2\pi^2} (1-\delta_{\sigma^2_{bd},0}\delta_{\tau^2_{bd},0}) \times \\
\times \ln \left(e^{-\pi (\sigma^2_{bd})^2 v_2/4}\frac{\vartheta_1 (\frac{\tau^2_{bd} - i \sigma^2_{bd} v_2}{2},i v_2)}{\eta (i v_2)} \right) 
\end{array}
\\\hline
b' & \begin{array}{c}
 -\delta_{\sigma^2_{bd'},0}\delta_{\tau^2_{bd'},0} \times   \frac{1}{2\pi^2} \ln \eta(iv_2) \\
 - \frac{1}{4\pi^2} (1-\delta_{\sigma^2_{bd'},0}\delta_{\tau^2_{bd'},0}) \times \\
\times \ln \left(e^{-\pi (\sigma^2_{bd'})^2 v_2/4}\frac{\vartheta_1 (\frac{\tau^2_{bd'} - i \sigma^2_{bd'} v_2}{2},i v_2)}{\eta (i v_2)} \right) 
\end{array}
&   - \frac{3}{8\pi^2}
 \ln \left(e^{-\pi v_1/4}\frac{\vartheta_1 (\frac{- i v_1}{2},i v_1)}{\eta (i v_1)} \right) & -
\\\hline
c & \begin{array}{c}
 - \delta_{\sigma^2_{cd},0}\delta_{\tau^2_{cd},0} \times  \frac{1}{2\pi^2} \ln \eta(iv_2) \\
 - \frac{1}{4\pi^2} (1-\delta_{\sigma^2_{cd},0}\delta_{\tau^2_{cd},0}) \times \\
\times \ln \left(e^{-\pi (\sigma^2_{cd})^2 v_2/4}\frac{\vartheta_1 (\frac{\tau^2_{cd} - i \sigma^2_{cd} v_2}{2},i v_2)}{\eta (i v_2)} \right) 
\end{array}  
& - & - 
\\\hline
h_3 &  - \frac{3}{\pi^2}  \ln \left(e^{-\pi v_3/4}\frac{\vartheta_1 (\frac{1 - i v_3}{2},i v_3)}{\eta (i v_3)} \right)
&  - \frac{3}{4\pi^2} \ln \left(e^{-\pi v_3/4}\frac{\vartheta_1 (\frac{1 - i v_3}{2},i v_3)}{\eta (i v_3)} \right)
&  - \frac{3}{4\pi^2} \ln \left(e^{-\pi v_3/4}\frac{\vartheta_1 (\frac{1 - i v_3}{2},i v_3)}{\eta (i v_3)} \right)
\\\hline
d' &   \frac{9}{8\pi^2} \ln \eta(iv_3) 
& \begin{array}{c} \delta_{\sigma^2_{dd'},0}\delta_{\tau^2_{dd'},0} \times \left(- \frac{1}{2\pi^2} \ln \eta(iv_2) + \frac{1}{\pi^2} \ln \eta(i \tilde{v}_2) \right)
\\ - (1-\delta_{\sigma^2_{dd'},0}\delta_{\tau^2_{dd'},0}) \times \\ \times \Biggl[
\frac{1}{4\pi^2}
 \ln \left(e^{-\pi (\sigma^2_{dd'})^2 v_2/4}\frac{\vartheta_1 (\frac{\tau^2_{dd'} - i \sigma^2_{dd'} v_2}{2},i v_2)}{\eta (i v_2)} \right) \\
- \frac{1}{2\pi^2}
\ln \left(e^{-\pi (\sigma^2_{dd'})^2 \tilde{v}_2/4}\frac{\vartheta_1 (\frac{\tau^2_{dd'} - i \sigma^2_{dd'} \tilde{v}_2}{2},i \tilde{v}_2)}{\eta (i \tilde{v}_2)} \right) \Biggr]
\end{array} 
&  \frac{9}{8\pi^2} \ln \eta(iv_3)
\\\hline
\end{array}
}{modelz6p+hidden-sectors_ax-holomorphic}{Two-torus volume $v_i$ dependent one-loop contributions to the holomorphic gauge kinetic functions 
involving branes $a$ and $d$. The total gauge kinetic function ${\rm f}^{\text{1-loop}}_{SU(3)_a}$ is given in the 
text. The auxiliary formal expressions $ \delta_x \, {\rm f}_{SU(1)_d}^{\text{1-loop}}(v_i)$ are basic building blocks for 
the holomorphic gauge kinetic functions of the single $U(1)_d$ gauge factor ${\rm f}_{U(1)_d}^{\text{1-loop}}$ 
and the massless $U(1)_{B-L}$ group ${\rm f}_{U(1)_{B-L}}^{\text{1-loop}}$.}
For $SU(3)_a$, summing up all contributions we obtain the two-torus volume $v_i$ dependent one-loop correction,
\begin{equation}
\begin{aligned}\label{Eq:Z6p-Ex-SU3-hol}
\delta_{\text{total}} {\rm f}^{\text{1-loop}}_{SU(3)_a}(v_i) =& \sum_{k=0}^2 \sum_{y= a,a',b,b',c,d,d',h_3} \delta_{(\theta^k y)} {\rm f}^{\text{1-loop}}_{SU(3)_a}(v_i)
\\
=& 
\frac{1}{2\pi^2} \Bigl(\ln \eta(i \tilde{v}_1)- \ln \eta(iv_1) \Bigr)
 - \frac{2}{\pi^2} \ln \eta(i v_3)- \frac{3}{4\pi^2} \ln \left(e^{-\pi v_3/4}\frac{\vartheta_1 (\frac{1 - i v_3}{2},i v_3)}{\eta (i v_3)} \right)
,
\end{aligned}
\end{equation}
and the `constant' contribution is given by
\begin{equation}
\begin{aligned}
\delta_{\text{total}} {\rm f}^{\text{1-loop}}_{SU(3)_a}(c) =
\frac{1}{2\pi^2} \ln \left[ 2^{15/8} \left(\frac{\sqrt{3} \, r \, v_1 v_3}{v_2^2}  \right)^{1/4} \right]
.
\end{aligned}
\end{equation}
The dependence on two-torus volumes in the `constant' factor arises because the stack $a$ is of the special type 
discussed in section~\ref{Ss:loop-KaehlerAdj} perpendicular to the $\OR$ invariant O6-plane orbit along $T^2_{(2)} \times T^2_{(3)}$.
The one-loop correction~(\ref{Eq:Z6p-Ex-SU3-hol}) is a rather short expression compared to the ones for $U(1)_d$ and $U(1)_{B-L}$ 
below since the third two-torus is of {\bf a}-type, which allows to combine the lattice sums from annulus and M\"obius strip.

For the  single (anomalous) $U(1)_a$ and $U(1)_d$ gauge factors summing up the generic  
one-loop contributions~(\ref{Eq:Delta_U1i_total}) gives
\begin{equation}
\begin{aligned}
\delta_{\text{total}}{\rm f}^{\text{1-loop}}_{U(1)_a}(v_i) &= 6 \times \left(
 \delta_{\text{total}}{\rm f}^{\text{1-loop}}_{SU(3)_a}(v_i) + \sum_{k=0}^2 \delta_{(\theta^k a')}{\rm f}^{\text{1-loop},{\cal A}}_{SU(3)_a} (v_i)
-  \sum_{k=0}^2 \delta_{(\theta^k a)}{\rm f}^{\text{1-loop}}_{SU(3)_a} (v_i)
 \right)
\\
&=  \frac{3}{\pi^2} \Bigl(\ln \eta(i \tilde{v}_1)- \ln \eta(iv_1) \Bigr)
-\frac{18}{\pi^2}\, \ln \eta(iv_2)
 \\
& - \frac{12}{\pi^2} \ln \eta(i v_3)- \frac{9}{2\pi^2} \ln \left(e^{-\pi v_3/4}\frac{\vartheta_1 (\frac{1 - i v_3}{2},i v_3)}{\eta (i v_3)} \right)
,
\end{aligned}
\end{equation}
and
\begin{equation}
\begin{aligned}
\delta_{\text{total}} {\rm f}^{\text{1-loop}}_{U(1)_d}(v_i) &= 2 \times \left( \sum_{k=0}^2 \sum_{y \neq d} \delta_{(\theta^{-k} y)} {\rm f}^{\text{1-loop}}_{SU(1)_d}(v_i) 
+ \sum_{k=0}^2 \delta_{(\theta^{-k} d')}{\rm f}^{\text{1-loop},{\cal A}}_{SU(1)_d} (v_i)
\right)
\\
&=-\frac{3}{2\pi^2}\ln \left(e^{-\pi v_1/4}\frac{\vartheta_1 (\frac{ - i v_1}{2},i v_1)}{\eta (i v_1)} \right) - \delta_{\sigma^2_{cd},0}\delta_{\tau^2_{cd},0} \times  \frac{1}{\pi^2} \ln \eta(iv_2) 
\\
&  -\frac{1}{\pi^2} \left( \delta_{\sigma^2_{bd},0}\delta_{\tau^2_{bd},0}  + \delta_{\sigma^2_{bd'},0}\delta_{\tau^2_{bd'},0}  \right) \ln \eta(i v_2) 
\\
& - \frac{1}{\pi^2} (1-\delta_{\sigma^2_{bd},0}\delta_{\tau^2_{bd},0})  \ln \left(e^{-\pi (\sigma^2_{bd})^2 v_2/4}\frac{\vartheta_1 (\frac{\tau^2_{bd} - i \sigma^2_{bd} v_2}{2},i v_2)}{\eta (i v_2)} \right) \\
& - \frac{1}{2\pi^2} (1-\delta_{\sigma^2_{bd'},0}\delta_{\tau^2_{bd'},0}) \ln \left(e^{-\pi (\sigma^2_{bd'})^2 v_2/4}\frac{\vartheta_1 (\frac{\tau^2_{bd'} - i \sigma^2_{bd'} v_2}{2},i v_2)}{\eta (i v_2)} \right) 
\\
& 
 - \frac{1}{2\pi^2} (1-\delta_{\sigma^2_{cd},0}\delta_{\tau^2_{cd},0}) 
 \ln \left(e^{-\pi (\sigma^2_{cd})^2 v_2/4}\frac{\vartheta_1 (\frac{\tau^2_{cd} - i \sigma^2_{cd} v_2}{2},i v_2)}{\eta (i v_2)} \right) 
 \\
 & +\delta_{\sigma^2_{dd'},0}\delta_{\tau^2_{dd'},0} \times \frac{2}{\pi} \left(- \ln \eta(iv_2) +  \ln \eta(i \tilde{v}_2) \right)
\\ 
& - (1-\delta_{\sigma^2_{dd'},0}\delta_{\tau^2_{dd'},0}) \frac{1}{\pi^2} \times  \Biggl[
 \ln \left(e^{-\pi (\sigma^2_{dd'})^2 v_2/4}\frac{\vartheta_1 (\frac{\tau^2_{dd'} - i \sigma^2_{dd'} v_2}{2},i v_2)}{\eta (i v_2)} \right) 
\\
& \hspace{40mm}
- \ln \left(e^{-\pi (\sigma^2_{dd'})^2 \tilde{v}_2/4}\frac{\vartheta_1 (\frac{\tau^2_{dd'} - i \sigma^2_{dd'} \tilde{v}_2}{2},i \tilde{v}_2)}{\eta (i \tilde{v}_2)} \right) \Biggr]
\\
&- \frac{9}{\pi^2} \ln \eta(iv_3)- \frac{9}{\pi^2}  \ln \left(e^{-\pi v_3/4}\frac{\vartheta_1 (\frac{1 - i v_3}{2},i v_3)}{\eta (i v_3)} \right)
.
\end{aligned}
\end{equation}
The lengthy expression for the latter is due to the continuous relative displacements and Wilson lines for all D6-branes
along $T^2_{(2)}$, which are crucial for making non-chiral matter massive and for breaking 
\mbox{$Sp(2)_c \to U(1)_c$}. The K\"ahler metrics for massless matter
depend only on the relative intersection angles and are thus not changed under this gauge symmetry breaking.
This means in particular that right-handed up- and down-type quarks as well as Higgses in the present example 
will have pairwise identical K\"ahler metrics.

The $U(1)$ anomaly matrix~(\ref{Eq:Def-AnomalyMatrix-Entries}) has the form
\begin{equation*}
\left(\begin{array}{ccc}  
 C_{aa} & C_{ab}  & C_{ad} \\  C_{ba} & C_{bb} & C_{bd}\\  C_{da} & C_{db}  & C_{dd} 
\end{array}\right)
= \frac{1}{2 \pi^2} \left(\begin{array}{ccc} 
0 & 9 & 0 \\ 9 & 0 & -3 \\ 0 & 9 & 0
\end{array}\right)
\end{equation*}
before the breaking of $Sp(2)_c \to U(1)_c$. The fact that this matrix has rank two is in agreement with the existence of a massless 
$U(1)_{B-L}=\frac{1}{3}U(1)_a + U(1)_d$ gauge group, for which the $v_i$ dependent one-loop corrections to the holomorphic
gauge kinetic function are given by
\begin{equation}
\begin{aligned}
\delta_{\text{total}} {\rm f}^{\text{1-loop}}_{U(1)_{B-L}} (v_i) &= \frac{1}{9} \, \delta_{\text{total}} {\rm f}^{\text{1-loop}}_{U(1)_a} (v_i)+ \delta_{\text{total}} {\rm f}^{\text{1-loop}}_{U(1)_d}
(v_i)+ 4 \, \left( -\delta_d \, {\rm f}^{\text{1-loop}}_{SU(3)_a}(v_i) + \delta_{d'} \, {\rm f}^{\text{1-loop}}_{SU(3)_a} (v_i)\right)
\\
&=\frac{1}{3\pi^2} \Bigl(\ln \eta(i \tilde{v}_1)- \ln \eta(iv_1) \Bigr)-\frac{3}{2\pi^2}\ln \left(e^{-\pi v_1/4}\frac{\vartheta_1 (\frac{ - i v_1}{2},i v_1)}{\eta (i v_1)} \right) 
\\
&  -\frac{1}{\pi^2} \left( \delta_{\sigma^2_{bd},0}\delta_{\tau^2_{bd},0}  + \delta_{\sigma^2_{bd'},0}\delta_{\tau^2_{bd'},0} +2  \right) \ln \eta(i v_2) 
\\
& - \frac{1}{\pi^2} (1-\delta_{\sigma^2_{bd},0}\delta_{\tau^2_{bd},0})  \ln \left(e^{-\pi (\sigma^2_{bd})^2 v_2/4}\frac{\vartheta_1 (\frac{\tau^2_{bd} - i \sigma^2_{bd} v_2}{2},i v_2)}{\eta (i v_2)} \right) \\
& - \frac{1}{2\pi^2} (1-\delta_{\sigma^2_{bd'},0}\delta_{\tau^2_{bd'},0}) \ln \left(e^{-\pi (\sigma^2_{bd'})^2 v_2/4}\frac{\vartheta_1 (\frac{\tau^2_{bd'} - i \sigma^2_{bd'} v_2}{2},i v_2)}{\eta (i v_2)} \right) 
\\
& - \delta_{\sigma^2_{cd},0}\delta_{\tau^2_{cd},0} \times  \frac{1}{\pi^2} \ln \eta(iv_2) 
 - \frac{1}{2\pi^2} (1-\delta_{\sigma^2_{cd},0}\delta_{\tau^2_{cd},0}) 
 \ln \left(e^{-\pi (\sigma^2_{cd})^2 v_2/4}\frac{\vartheta_1 (\frac{\tau^2_{cd} - i \sigma^2_{cd} v_2}{2},i v_2)}{\eta (i v_2)} \right) 
 \\
 & +\delta_{\sigma^2_{dd'},0}\delta_{\tau^2_{dd'},0} \times \frac{2}{\pi} \left(- \ln \eta(iv_2) +  \ln \eta(i \tilde{v}_2) \right)
\\ 
& - (1-\delta_{\sigma^2_{dd'},0}\delta_{\tau^2_{dd'},0}) \frac{1}{\pi^2} \times  \Biggl[
 \ln \left(e^{-\pi (\sigma^2_{dd'})^2 v_2/4}\frac{\vartheta_1 (\frac{\tau^2_{dd'} - i \sigma^2_{dd'} v_2}{2},i v_2)}{\eta (i v_2)} \right) 
\\
& \hspace{40mm}
- \ln \left(e^{-\pi (\sigma^2_{dd'})^2 \tilde{v}_2/4}\frac{\vartheta_1 (\frac{\tau^2_{dd'} - i \sigma^2_{dd'} \tilde{v}_2}{2},i \tilde{v}_2)}{\eta (i \tilde{v}_2)} \right) \Biggr]
\\
& - \frac{31}{3\pi^2} \ln \eta(i v_3)- \frac{19}{2\pi^2} \ln \left(e^{-\pi v_3/4}\frac{\vartheta_1 (\frac{1 - i v_3}{2},i v_3)}{\eta (i v_3)} \right)
,
\end{aligned}
\end{equation}
following the prescription~(\ref{Eq:linear_comb_U1_loop}) for the one-loop contributions.
The `constant' one-loop contributions are given by
\begin{equation}
\begin{aligned}
\delta_{\text{total}} {\rm f}^{\text{1-loop}}_{U(1)_a}(c)&=
\frac{3}{4\pi^2} \ln \left[2^{11/2} \; \sqrt{3} \; \frac{v_1v_3}{v_2^2} \, r\right],\\
\delta_{\text{total}} {\rm f}^{\text{1-loop}}_{U(1)_d}(c) &= \frac{3}{2 \pi^2} \ln 2,\\
\delta_{\text{total}} {\rm f}^{\text{1-loop}}_{U(1)_{B-L}} (c)&=
\frac{1}{12\pi^2} \ln \left[2^{47/2} \; \sqrt{3} \; \frac{v_1v_3}{v_2^2} \, r\right]
,
\end{aligned}
\end{equation}
where again the explicit $v_i$ dependence arises from the special $\OR$ invariance of the $a(\theta a')$ sector.

The K\"ahler metrics for the symmetric representation in the $cc$ sector of $Sp(2)_c$
and the antisymmetric representation in the $h_3h_3$ sector of $Sp(6)_{h_3}$ can be read off from
 table~\ref{tab:SO-Sp-GaugeKin+KaehlerMetric} with the (length)${}^2$ values of the three-cycles per two-torus
listed in table~\ref{tab:smmodelz6p+hidden},
\begin{equation}
K_{\Sym_c}^{cc} = f(S,U_l) \, \frac{\sqrt{2\pi}}{2^{4/3} v_2} \frac{1}{\sqrt{3r}}
,
\qquad
K_{\Anti_{h_3}}^{h_3h_3} = f(S,U_l) \, \frac{\sqrt{2\pi}}{2^{4/3} v_2} \sqrt{r}
.
\end{equation}
There exist three more antisymmetric representations of $Sp(2)_c$ at the intersections of orbifold images 
$c(\theta^k c)_{k=1,2}$ and one antisymmetric representation of $Sp(6)_{h_3}$ at the intersection 
$h_3(\theta^k h_3)_{k=1,2}$ with K\"ahler metrics
\begin{equation}
K_{\Anti_c}^{c(\theta^k c)} = f(S,U_l) \, \sqrt{\frac{2\pi}{r \, v_1v_2}}
,
\qquad
K_{\Anti_{h_3}}^{h_3(\theta^k h_3)} =f(S,U_l) \, \sqrt{\frac{2\pi \, r}{v_1v_2}}
.
\end{equation}
It is again obvious that the K\"ahler metrics of states in the same representation $\Anti_{h_3}$ but at different
intersections of orbifold images D6-branes, $h_3h_3$ and $h_3(\theta^k h_3)_{k=1,2}$, differ in their two-torus volume $v_i$
dependence.
This completes our discussion of examples for the non-trivial structure of D6-brane configurations
and the corresponding K\"ahler metrics and holomorphic gauge kinetic functions.

\section{Conclusions and Outlook}\label{S:Conclusions}

In this article, we completed the derivation of the perturbatively exact holomorphic gauge kinetic function and the 
K\"ahler metrics in the ${\cal N}=1$ supergravity formulation
via open string gauge threshold one-loop computations for all factorisable toroidal orbifolds 
taking into account all possible configurations of vanishing or non-vanishing intersection angles
and continuous or discrete Wilson line and displacement moduli per two-torus.
The at first complicated and lengthy formulas are considerably simplified by rewriting them in terms of 
dependences on beta function coefficients whenever possible. 
The K\"ahler metrics for adjoint matter on identical D6-branes in~(\ref{Eq:Def-Kaehler_adjoints}) depend on the 
kind of wrapped bulk, fractional or rigid three-cycle, whereas all other K\"ahler metrics in 
table~\ref{tab:Comparison-Kaehler-bifund}
only depend on the intersection angles and have an universal shape for all orbifold backgrounds on factorisable
tori considered here. 
The perturbatively exact holomorphic gauge kinetic functions on the other hand depend on both the number of $\Z_2$ symmetries 
of the orbifold background and the relative position of the D6-branes with respect to the O6-planes as compared in tables~\ref{tab:Comparison-gaugekin-bifund} 
and~\ref{tab:Comparison-gaugekin-antisym} for the part depending on the K\"ahler moduli. In addition, we found a one-loop correction 
to the gauge kinetic function for fractional and rigid D6-branes in~(\ref{Eq:3angle-f-angle}) depending on the complex structure moduli  and a constant
 contribution in~(\ref{Eq:ConstContr}) and the second line of~(\ref{Eq:aap-f-3angles}) due to the intersection with O6-planes, both of which 
have to our knowledge not been properly appreciated before in the transformation to the supergravity basis.
All results are given in terms of real geometric bulk moduli
with their complexifications by axions explained in section~\ref{Ss:loop-redef-S+Ui}, where we also briefly discuss some 
contradicting statements in the literature concerning one-loop field redefinitions of the bulk moduli under gauge transformations of 
anomalous massive Abelian groups.
The field theory of chiral matter at one vanishing angle has to our knowledge not been analysed before, neither in the 
intersecting D6-brane nor its T-dual magnetised D-brane language, but it is of crucial importance for model building,
as we have discussed in an example with Standard Model spectrum on $T^6/\Z_6'$, due to the possible hierarchical structure of 
the associated K\"ahler metrics for unisotropic choices of the two-torus volumes.
Moreover, we have corrected here the formula for bifundamental matter on parallel D6-branes compared to the literature~\cite{Blumenhagen:2007ip}, 
which assigned the one of the adjoints also to bifundamentals. 
In the context of symplectic gauge groups, we found a subtle ambiguity related to the assignment of constants to the K\"ahler metrics or 
gauge kinetic functions, which might be resolved by a complementary computation of scattering amplitudes along the lines 
derived for the six-torus in~\cite{Lust:2004cx,Blumenhagen:2006ci}.
It also remains to be seen how the field theory formulas generalise to non-factorisable orbifolds.

We proceeded to present some examples, first on $T^6/\Z_2 \times \Z_2$ with discrete torsion, where a detailed 
analysis of the rigid D6-brane geometry underlying the T-dual of a magnetised D9/D5-brane model in~\cite{Angelantonj:2009yj} was performed
and the corresponding K\"ahler metrics presented for the first time. 
As a second class of examples, we computed the K\"ahler metrics for matter charged under the QCD or the `leptonic' stack
in the Standard Model with hidden sector on $T^6/\Z_6'$, providing an explicit example for a different K\"ahler metric of the last
right-handed quark generation compared to the first two with a possible hierarchy if the volume of the first two-torus differs considerably from the other two. In the course of the computation we also found that in contrast to our previous statements~\cite{Gmeiner:2009fb}
the non-chiral antisymmetric pair on parallel orientifold image D6-branes
charged under $SU(3)_a$ cannot a acquire a mass by a parallel displacement.

Further directions of research will on the one hand include the application of the complete field theory results to the existing Standard
Model vacua on $T^6/\Z_6'$ while combining with worldsheet instanton contributions to the Yukawa couplings~\cite{Honecker:2011tbd}
along the lines derived for the six-torus in~\cite{Cremades:2003qj,Cremades:2004wa}, and with 
non-perturbative D-instanton contributions~\cite{Blumenhagen:2009qh}, and the search for new models on rigid D6-branes in $T^6/\Z_2 \times \Z_{2M}$ backgrounds with 
discrete torsion~\cite{Forste:2010gw} which is currently under way. In this context it will be interesting to analyse if 
there exist explicit models which are compatible with a low string scale,
\mbox{$M_{\rm string}\sim$ TeV}, as proposed in~\cite{Lust:2008qc,Lust:2009pz,Feng:2010yx,Anchordoqui:2011eg}.

On the other hand, the explicit formulas for one-loop mixing of a single $U(1)$ in (\ref{Eq:Delta_U1i_total}) and linearly combined massless 
$U(1)$s in (\ref{Eq:linear_comb_U1_loop}) open up new possibilities for studying the kinetic mixing of the observable 
Standard Model $U(1)$ with some dark $Z'$ photon, possibly including also the recently investigated RR photons~\cite{Camara:2011jg}; the classification of the one-loop holomorphic gauge kinetic functions in tables~\ref{tab:Comparison-gaugekin-bifund} 
and~\ref{tab:Comparison-gaugekin-antisym} with beta function coefficients as prefactors might also be useful for extrapolating to 
compactifications on smooth Calabi-Yau spaces, and last but not least it will be interesting to investigate if phenomenologically appealing 
spectra on D-branes are destabilised by instantons as studied e.g. in~\cite{Angelantonj:2011hs} for a simpler toy model.

\subsection*{Acknowledgements}

The work of G.\ H. \ is partially supported by the ``Research Center Elementary Forces and Mathematical Foundations'' (EMG)
at the Johannes Gutenberg-Universit\"at Mainz.

\noindent The author thanks Carlo Angelantonj and especially Emilian Dudas for valuable correspondences 
concerning the correct matching with the T-dual to the
D9/D5-brane model~\cite{Angelantonj:2009yj} discussed in section~\ref{Ss:Ex3_Angelantonj} in this article.


\clearpage
\begin{appendix}

\section{Reformulations of the M\"obius strip  contributions to the gauge thresholds}\label{App:A}

In~\cite{Gmeiner:2009fb}, we had derived the shape of the finite part of the r.h.s. of equation~(\ref{Eq:Integ-MS-vartheta}) as
\begin{equation*}
 -\frac{1}{4} \, \ln \left\{ \frac{\Gamma(|\nu|)^{{\rm sgn}(\nu)} }{\Gamma(1-|\nu|)^{{\rm sgn}(\nu)} }  \cdot \frac{\Gamma(\nu+\frac{1}{2} -\sgn(\nu)  H(|\nu|-\frac{1}{2}))}{\Gamma(-\nu+\frac{1}{2} +\sgn(\nu)  H(|\nu|-\frac{1}{2}))}\right\} 
 - \; \sgn(\nu) \; \frac{ 1 - 2 \, H(|2\nu|-1) }{4} \; \ln(2)
 ,
\end{equation*}
which becomes zero for $\nu = \pm \frac{1}{2}$.

Using the Gamma function identity
\begin{equation*}
\Gamma(z) \Gamma(z+\frac{1}{2}) = \sqrt{2\pi} \; 2^{-2z+\frac{1}{2}} \, \Gamma (2z)
,
\end{equation*}
the finite terms can be evaluated for the four ranges $-1 < \nu < -\frac{1}{2}$, $-\frac{1}{2} < \nu < 0$, $0< \nu < \frac{1}{2}$ and $\frac{1}{2} < \nu <1$ separately.
The result is as given in~(\ref{Eq:Integ-MS-vartheta}), i.e.
\begin{equation}
 -\frac{1}{4} \, \ln  \left( \frac{\Gamma(|2\nu|-H(|2\nu|-1))}{\Gamma(1 -|2\nu|+H( |2\nu|-1))}\right)^{\sgn(\nu)}
+  \left[ \nu - \frac{ \sgn(\nu)}{2} \right] \,  \ln (2)
,
\end{equation}
see also~\cite{Blumenhagen:2007ip} for $|\nu|< \frac{1}{2}$.
Defining 
\begin{equation*}
\begin{aligned}
X^{(n)}_{aa'} &\equiv \ln \left(\frac{\Gamma(|\phi^{(n)}_{aa'}|)}{\Gamma(1-|\phi^{(n)}_{aa'}|)} \right)^{\sgn(\phi^{(n)}_{aa'})}
,
\\
X^{(n)}_{a,\OR\Z_2^{(l)}} &\equiv \ln  \left( \frac{\Gamma(\nu_{a,\OR\Z_2^{(l)}}^{(n)})}{\Gamma(1 - \nu_{a,\OR\Z_2^{(l)}}^{(n)})}\right)^{\sgn(\phi_{a,\OR\Z_2^{(l)}}^{(n)})}
\quad
\text{with}
\qquad 
\nu_{a,\OR\Z_2^{(l)}}^{(n)} \equiv |2 \, \phi_{a,\OR\Z_2^{(l)}}^{(n)}|-H \big(|2\phi_{a,\OR\Z_2^{(l)}}^{(n)}|- 1 \bigr)
,
\end{aligned}
\end{equation*}
with $n \in \{1,2,3\}$ the two-torus index and $l \in\{0 \ldots 3\}$, where $l=0$ corresponds to the $\OR$ invariant O6-plane, and the Heaviside step function $H(x)$
as defined in equation~(\ref{EqApp:Heavyside}), we show that 
\begin{equation}\label{Eq:Identification-MS+Annulus-angles}
X^{(n)}_{a,\OR\Z_2^{(l)}} = X^{(n)}_{aa'}  
\qquad \text{ for all } \qquad 
l \in \{0 \ldots 3\}
\end{equation}
for all ${\cal N}=1$ supersymmetric D6-brane configurations at non-trivial angles.
Equation~(\ref{Eq:Annulus-Moebius-Gamma-Rewritten}) furthermore uses the relation among the different sign factors, 
\begin{equation}\label{Eq:cIORIaap-Relation}
c_a^{\OR\Z_2^{(l)}} \cdot \sgn(I_a^{\OR\Z_2^{(l)}})  \cdot \sgn(I_{aa'}) =-1,
\end{equation}
which we also show here on a case-by-case basis.

Without loss of generality, the angles $\phi_a^{(n)} \equiv - \phi_{a,\OR}^{(n)}$ w.r.t. the $\OR$ invariant O6-plane can be
chosen in the range 
\begin{equation}\label{Eq:Range-Angle-OR}
\begin{aligned}
0< | \phi_{a,\OR}^{(i)}|,  | \phi_{a,\OR}^{(j)}| <  | \phi_{a,\OR}^{(k)}| < 1 
\qquad \text{ and } \qquad 
0< | \phi_{a,\OR}^{(i)}|,  | \phi_{a,\OR}^{(j)}| < \frac{1}{2}
\\
\text{with } \qquad  \sgn(\phi_{a,\OR}^{(i)})= \sgn(\phi_{a,\OR}^{(j)})= - \sgn(\phi_{a,\OR}^{(k)})
,
\end{aligned}
\end{equation}
with $(i,j,k)$ a cyclic permutation of (1,2,3).
For the largest (absolute value of an) angle, three different ranges have to be considered:
$0< |\phi_{a,\OR}^{(k)}| < \frac{1}{2}$ or $ |\phi_{a,\OR}^{(k)}| = \frac{1}{2}$ or $ \frac{1}{2} < |\phi_{a,\OR}^{(k)}| < 1$. 

We first show that $X^{(n)}_{a,\OR} = X^{(n)}_{aa'}$: The angles $\phi^{(n)}_{aa'}$ between orientifold image branes D$6_a$ and D$6_a'$ 
in the range $ 0 < |\phi^{(n)}_{aa'}| < 1$, where we do not further constrain the two smaller angles as in~(\ref{Eq:Range-Angle-OR}),
are given as follows:
\begin{enumerate}
\item
for $0< |\phi_{a,\OR}^{(k)}| < \frac{1}{2}$: the angles between orientifold image D6-branes and those 
between D6-brane and $\OR$ invariant O6-plane are related by $\phi^{(n)}_{aa'} = 2 \, \phi_{a,\OR}^{(n)} $ for all $n \in \{1,2,3\}$.
The equality~(\ref{Eq:Identification-MS+Annulus-angles}) is obviously fulfilled.\\
The sign factor~(\ref{Eq:Def-sign_c}) of the M\"obius strip contribution to the beta function coefficient takes the value $c_a^{\OR}=-1$, and
since $\sgn(I_a^{\OR}) = \sgn(I_{aa'})$  equation~(\ref{Eq:cIORIaap-Relation}) is fulfilled.
\item
for $ |\phi_{a,\OR}^{(k)}| = \frac{1}{2}$: there are two different possibilities of shifting the angles $(\phi_{aa'}^{(n)})$
such that they are in the denoted range:
\begin{equation*}
\begin{aligned}
(\phi_{aa'}^{(i)},\phi_{aa'}^{(j)},\phi_{aa'}^{(k)}) &= (2 \, \phi_{a,\OR}^{(i)},2 \, \phi_{a,\OR}^{(j)},0) \, - 
\left\{\begin{array}{c}
 (\sgn(\phi_{a,\OR}^{(i)}) ,0,0) \\
 (0,\sgn(\phi_{a,\OR}^{(j)}) ,0) 
\end{array}\right.
\\
\text{and} &
\quad
\left( \sgn(\phi_{aa'}^{(i)}),\sgn(\phi_{aa'}^{(j)}) \right) = 
\left\{\begin{array}{c}
\left( -\sgn(\phi_{a,\OR}^{(i)}),\sgn(\phi_{a,\OR}^{(j)}) \right)
\\
\left(\sgn(\phi_{a,\OR}^{(i)}), -\sgn(\phi_{a,\OR}^{(j)}) \right)
\end{array}\right.
,
\end{aligned}
\end{equation*}
where the shift $\phi_{aa'}^{(k)} = 2 \,\phi_{a,\OR}^{(k)} - \sgn(\phi_{a,\OR}^{(k)})=0$ has already been performed.
It is now easy to see that the M\"obius strip contributions to the logarithm of Gamma functions from all three angles cancel out, and the annulus 
amplitude contributes a  Kaluza-Klein and winding
sum along the two-torus $T^2_{(k)}$, where  D$6_a$ and D$6_a'$ are parallel to each other and perpendicular to the $\OR$ invariant O6-plane.
\\
The constant~(\ref{Eq:Def-sign_c}) for the M\"obius strip contribution to the beta function coefficient vanishes, $c_a^{\OR}=0$.
\item
for $\frac{1}{2} < |\phi_{a,\OR}^{(k)}| < 1$, there exist the three different shifts of the angles $(\phi_{aa'}^{(n)})$ displayed in table~\ref{tab:Shifts-Angles-OR},
%
\mathtabfix{
\begin{array}{|c|c|}\hline
\multicolumn{2}{|c|}{\text{\bf Angles of a D$6_a$-brane with the $\OR$-invariant O6-plane and its image D$6_{a'}$-brane}}
\\\hline\hline
(\phi_{aa'}^{(i)},\phi_{aa'}^{(j)},\phi_{aa'}^{(k)})  & \left( \sgn(\phi_{aa'}^{(i)}), \sgn(\phi_{aa'}^{(j)}), \sgn(\phi_{aa'}^{(k)})   \right)
\\\hline\hline
(2 \, \phi_{a,\OR}^{(i)},2 \, \phi_{a,\OR}^{(j)},2 \, \phi_{a,\OR}^{(k)}) + \sgn(\phi_{a,\OR}^{(k)}) \cdot (1,1,-2) 
&\left(-\sgn(\phi_{a,\OR}^{(i)}),-\sgn(\phi_{a,\OR}^{(j)}),-\sgn(\phi_{a,\OR}^{(k)})    \right)
\\\hline
(2 \, \phi_{a,\OR}^{(i)},2 \, \phi_{a,\OR}^{(j)},2 \, \phi_{a,\OR}^{(k)}) + \sgn(\phi_{a,\OR}^{(k)}) \cdot (1,0,-1) 
& \left(- \sgn(\phi_{a,\OR}^{(i)}),\sgn(\phi_{a,\OR}^{(j)}),\sgn(\phi_{a,\OR}^{(k)})   \right) 
\\\hline
(2 \, \phi_{a,\OR}^{(i)},2 \, \phi_{a,\OR}^{(j)},2 \, \phi_{a,\OR}^{(k)}) + \sgn(\phi_{a,\OR}^{(k)}) \cdot (0,1,-1) 
& \left(\sgn(\phi_{a,\OR}^{(i)}),-\sgn(\phi_{a,\OR}^{(j)}),\sgn(\phi_{a,\OR}^{(k)})   \right) 
\\\hline
\end{array}
}{Shifts-Angles-OR}{Relation among the angles $(\phi_{aa'}^{(n)})$ of orientifold image D$6_a$ and D$6_a'$ branes and the angles $(\phi_{a,\OR}^{(n)})$
w.r.t. the $\OR$ invariant O6-plane for the maximal (absolute value of the) angle $|\phi_{a,\OR}^{(k)}|> \frac{1}{2}$. 
The shifts, e.g. $\phi_{aa'}^{(i)} = 2 \, \phi_{a,\OR}^{(i)} - \sgn(\phi_{a,\OR}^{(i)})$ on the first and second line, ensure that $|\phi_{aa'}^{(n)}|< 1$ for all $n$.
}
%
which lead to identical expressions for $X^{(n)}_{a,\OR}$ and coincide with $X^{(n)}_{aa'}$, as can be checked on a case-by-case basis.
\\
The sign~(\ref{Eq:Def-sign_c}) of the M\"obius strip contribution to the beta function coefficient is $c_a^{\OR}=1$, and since in table~\ref{tab:Shifts-Angles-OR}
we show that $\sgn(I_a^{\OR}) = - \sgn(I_{aa'})$  the signs satisfy the relation~(\ref{Eq:cIORIaap-Relation}).
\end{enumerate}
Now we show for definiteness that the relation~(\ref{Eq:Identification-MS+Annulus-angles}) is fulfilled for the $\OR\Z_2^{(2)}$ invariant O6-plane
and the maximal angle (more precisely the maximal absolute value) on any of the three two-tori. 
The relation between the angles of a given D$6_a$-brane w.r.t. to the  $\OR$ and $\OR\Z_2^{(2)}$ invariant O6-planes are displayed in 
table~\ref{tab:Shifts-Angles-ORZ2}.
\mathtabfix{
\begin{array}{|c|c|c||c|}\hline
\multicolumn{4}{|c|}{\text{\bf Angles of a D$6_a$-brane with the $\OR$- and $\OR\Z_2^{(2)}$-invariant O6-planes}}
\\\hline\hline
 & (\vec{\phi}_{a,\OR\Z_2^{(2)}}) & \sgn (\vec{\phi}_{a,\OR\Z_2^{(2)}}) & c_a^{\OR\Z_2^{(2)}}
\\\hline\hline
\sgn(\phi_{a,\OR}^{(1)}) = \sgn(\phi_{a,\OR}^{(3)}) & (\vec{\phi}_{a,\OR}) + \sgn(\phi_{a,\OR}^{(2)}) \cdot (\frac{1}{2},-1,\frac{1}{2}) 
& -  \sgn (\vec{\phi}_{a,\OR})
& 
\\
 |\phi_{a,\OR}^{(n)}| < \frac{1}{2} \text{ for all } n & |\phi_{a,\OR\Z_2^{(2)}}^{(2)}| > \frac{1}{2} &
& 1
\\
|\phi_{a,\OR}^{(2)}| > \frac{1}{2} &  |\phi_{a,\OR\Z_2^{(2)}}^{(n)}| < \frac{1}{2} \text{ for all } n & 
& -1
\\\hline
\!\!\!\begin{array}{c}
\sgn(\phi_{a,\OR}^{(1)}) \neq \sgn(\phi_{a,\OR}^{(3)}) \\
|\phi_{a,\OR}^{(n)}| < \frac{1}{2} \text{ for all } n
\\\hline
|\phi_{a,\OR}^{(3)}| >  \frac{1}{2}
\\\hline
|\phi_{a,\OR}^{(1)}| > \frac{1}{2}
\end{array}\!\!\!
& \begin{array}{c} (\vec{\phi}_{a,\OR}) + \sgn(\phi_{a,\OR}^{(3)}) \cdot  (\frac{1}{2},0,-\frac{1}{2})\\
|\phi_{a,\OR\Z_2^{(2)}}^{(n)}| < \frac{1}{2} \text{ for all } n 
\end{array}
&\!\!\begin{array}{c}
\\
 \left( -\sgn(\phi_{a,\OR}^{(1)}) , \sgn(\phi_{a,\OR}^{(2)}), - \sgn(\phi_{a,\OR}^{(3)})\right)\\\hline
 \left( -\sgn(\phi_{a,\OR}^{(1)}) , \sgn(\phi_{a,\OR}^{(2)}), \sgn(\phi_{a,\OR}^{(3)}) \right)\\\hline
\left( \sgn(\phi_{a,\OR}^{(1)}) , \sgn(\phi_{a,\OR}^{(2)}), - \sgn(\phi_{a,\OR}^{(3)})\right)
\end{array}\!\!
& -1
\\\hline
\end{array}
}{Shifts-Angles-ORZ2}{Relation between angles $(\phi_{a,\OR\Z_2^{(2)}}^{(n)})$ of a D$6_a$-brane  with the $\OR\Z_2^{(2)}$ invariant O6-plane 
and the angles $(\phi_{a,\OR}^{(n)})$ of the same D$6_a$-brane with the $\OR$ invariant O6-plane. The first column lists the five inequivalent 
shapes of the angles $(\phi_{a,\OR}^{(n)})$, the explicit expressions for $(\phi_{a,\OR\Z_2^{(2)}}^{(n)})$ in terms of shifts of the former angles are given
in the second column with classification of the maximal angle, and their sign relative to the one of $(\phi_{a,\OR}^{(n)})$ is given in the third column.
The last column lists the sign factor $c_a^{\OR\Z_2^{(2)}}$ that appears in the beta function coefficient.
}
For all five distinct configurations of angles, one can read off that $X^{(n)}_{a,\OR\Z_2^{(2)}} = X^{(n)}_{a,\OR}$, which combined with the already demonstrated
equality $X^{(n)}_{a,\OR} = X^{(n)}_{aa'}$ gives the desired result.
From table~\ref{tab:Shifts-Angles-ORZ2}, one can also read off that $c_a^{\OR\Z_2^{(2)}} \cdot \sgn(\tilde{I}_a^{\OR\Z_2^{(2)}}) = c_a^{\OR} \cdot \sgn(\tilde{I}_a^{\OR})$
for all ${\cal N}=1$ supersymmetric configuration of three non-vanishing angles, which implies that the relation~(\ref{Eq:cIORIaap-Relation}) of signs is fulfilled for the $\OR\Z_2^{(2)}$-invariant O6-plane. The case $|\phi_{a,\OR}^{(k)}| =\frac{1}{2}$ does not contribute to the logarithms of Gamma functions since this angle implies that two of the $\OR\Z_2^{(l)}$-invariant O6-planes
are parallel and the other two perpendicular to the D$6_a$-brane along $T^6_{(k)}$.
This completes the proof for all $T^6/\Z_{2N}$ orientifolds. 

The remaining two cases $l \in \{1,3\}$ for $T^6/\Z_2 \times \Z_{2M}$ orientifolds without and with discrete torsion are obtained from the $\OR\Z_2^{(2)}$ case
by permutation of the two-torus indices.

\section{Tree level gauge couplings for various orbifolds}\label{App:V-for-Z6+Z6p}

\subsection{The $T^6/\Z_4$ and $T^6/\Z_2 \times \Z_4$ orientifolds}\label{App:V-Z4}

The geometry of the $T^6/(\Z_4 \times \OR)$ orientifold has been discussed in detail in~\cite{Blumenhagen:2002gw},
and the $T^6/(\Z_2 \times \Z_4 \times \OR)$ orientifolds with and without discrete torsion in~\cite{Forste:2010gw}, see also~\cite{Honecker:2004np,Cvetic:2006by}.
A generic bulk three-cycle takes the form
\begin{equation}\label{Eq:Z2Z4-bulkWrappings}
\begin{aligned}
\Pi^{\rm bulk} =& P \, \rho_1 + Q \, \rho_2 + U \, \rho_3 + V \, \rho_4 ,
\\
\text{with} \qquad P & \equiv n^1 \, X 
, \qquad
Q \equiv m^1 \, Y 
, \qquad
U \equiv -m^1 \, X 
, \qquad
V \equiv n^1 \, Y 
,
\\
\text{and} \qquad X &\equiv n^2 n^3 - m^2 m^3
, \qquad
Y \equiv n^2  m^3 + m^2 n^3
,
\end{aligned}
\end{equation}
where the expansion in terms of one-cycle wrapping numbers $(n^i,m^i)$ applies to the \mbox{$T^6/\Z_2 \times \Z_4$} 
case and $T^6/\Z_4$ is obtained by permutation of two-torus indices, cf. table~\ref{Tab:T6ZN+T6Z2Z2M-shifts}.
The non-vanishing intersection numbers of the basic cycles $\rho_i$ are given by
\begin{equation}
\rho_1 \circ \rho_3 = \rho_2 \circ \rho_4 = \left\{\begin{array}{cc} -2 & T^6/\Z_4
\\ -4 & T^6/\Z_2 \times \Z_4 
\end{array}\right.
.
\end{equation}
To shorten the notation for the untilted or tilted shape of the first two-torus, it is useful to introduce
\begin{equation}
\begin{aligned}
\tilde{Q} & \equiv Q+ b \, V
, \qquad
\tilde{U} \equiv U - b \, P
\\
\tilde{\rho}_1 &\equiv \rho_1 + b \, \rho_3
, \qquad
\tilde{\rho}_4 \equiv \rho_4 - b \, \rho_2
.
\end{aligned}
\end{equation}
Using the identity $P \, \tilde{Q} = - V \, \tilde{U}$, the (length)${}^2$ of a three-cycle 
can then be rewritten as 
\begin{equation*}
\begin{aligned}
\prod_{i=1}^3 V_{aa}^{(i)} =& \frac{1}{r} \left( P^2 + V^2 \right) + r \left( \tilde{U}^2 + \tilde{Q}^2 \right)
\\
&=c_{\rm lattice} \times \left[\frac{1}{r} \left(\tilde{X}^0-r \, \tilde{X}^1 \right)^2 
+ r \left(\tilde{Y}^0-\frac{1}{r} \tilde{Y}^1 \right)^2 \right]
\\
&\stackrel{\text{SUSY}}{=} \left(\frac{\sqrt{c_{\rm lattice}}}{\sqrt{r}}  \, \tilde{X}^0-\sqrt{c_{\rm lattice}} \, \sqrt{r} \, \tilde{X}^1 \right)^2
,
\end{aligned}
\end{equation*}
where we introduced the lattice dependent constant factor
\begin{equation*}
c_{\rm lattice} = \left\{\begin{array}{cc}
1& {\bf a/bAA},{\bf a/bBB}\\
2 & {\bf a/bAB}
\end{array}\right.
,
\end{equation*}
and $r$ is the ratio of radii on the first two-torus. The decomposition of a bulk three-cycle
in terms of orientifold even and odd components,
\begin{equation*}
\Pi^{\rm bulk}_a = \sum_{i=0}^1 \left( \tilde{X}^i_a \, \Pi^{\rm even}_i + \tilde{Y}^i_a \, \Pi^{\rm odd}_i \right)
,
\end{equation*}
depends on the choice of the compactification background as detailed in table~\ref{Tab:T6Z4+T6Z2Z4-OR_even+odd}.
\begin{table}[ht]
\renewcommand{\arraystretch}{1.3}
\begin{equation*}
\begin{array}{|c||c|c|c|}\hline
\multicolumn{4}{|c|}{\text{\bf $\OR$ even \& odd 3-cycles on } T^6/\Z_4 \text{ \bf \& } T^6/\Z_2 \times \Z_4}
\\\hline\hline
\text{lattice} & {\bf a/bAA} & {\bf a/bAB} & {\bf a/bBB}
\\\hline\hline
\Pi^{\rm even}_0 & \tilde{\rho}_1 &  \tilde{\rho}_1 + \tilde{\rho}_4 & \tilde{\rho}_4
\\
\Pi^{\rm even}_1 & \rho_2 & \rho_2 + \rho_3 & \rho_3
\\\hline
\Pi^{\rm odd}_0 & \rho_3 & \rho_3 - \rho_2 & -\rho_2
\\
\Pi^{\rm odd}_1 & \tilde{\rho}_4 & \tilde{\rho}_4 - \tilde{\rho}_1 & -\tilde{\rho}_1
\\\hline\hline
\tilde{X}^0_a & P_a & \frac{P_a+V_a}{2} &  V_a
\\
\tilde{X}^1_a & \tilde{Q}_a & \frac{\tilde{Q}_a+\tilde{U}_a}{2} & \tilde{U}_a
\\\hline
\tilde{Y}^0_a & \tilde{U}_a & \frac{\tilde{U}_a - \tilde{Q}_a}{2} & -\tilde{Q}_a
\\
\tilde{Y}^1_a & V_a & \frac{V_a-P_a}{2} & -P_a
\\\hline
\end{array}
\end{equation*}
\caption{Orientifold even and odd bulk three-cycles and bulk wrapping numbers for the 
six inequivalent lattices of the 
$T^6/(\Z_4 \times \OR)$ and $T^6/(\Z_2 \times \Z_4 \times \OR)$ orientifolds.
}
\label{Tab:T6Z4+T6Z2Z4-OR_even+odd}
\end{table}
The on-trivial intersection numbers of the $\OR$ even and odd cycles in table~\ref{Tab:T6Z4+T6Z2Z4-OR_even+odd}
are 
\begin{equation*}
\Pi^{\rm even}_i \circ \Pi^{\rm odd}_j = -\delta_{ij}  \; c_{\rm lattice} \times \left\{\begin{array}{cc}
2 & T^6/\Z_4
\\ 4 & T^6/\Z_2 \times \Z_4
\end{array}\right.
,
\end{equation*}
and the supersymmetry constraints read  
\begin{equation*}
\tilde{X}_a^0 - r \, \tilde{X}_a^1 >0,
\qquad\quad
\tilde{Y}^0_a - \frac{1}{r} \, \tilde{Y}_a^1=0
.
\end{equation*}
The exceptional three-cycles do not contribute to the tree level value of the gauge coupling. Details on their orientifold 
projections can be found in~\cite{Blumenhagen:2002gw,Forste:2010gw}.

\subsection{The $T^6/\Z_6'$ and $T^6/\Z_2 \times \Z_6$ orientifolds}\label{App:V-Z6p}

The geometry of the $T^6/\Z_6'$ and $T^6/\Z_2 \times \Z_6$ orientifolds is explained in detail in~\cite{Gmeiner:2007zz}
and~\cite{Forste:2010gw}, respectively. We review here the basic steps for rewriting the tree level gauge coupling 
as linear function of the dilaton and complex structure modulus.

A bulk three-cycle can be written as 
\begin{equation*}
\Pi^{\rm bulk}= P\; \rho_1 + Q \; \rho_2 + U \; \rho_3 + V \; \rho_4
\end{equation*}
with the bulk wrapping numbers in the notation of $T^6/\Z_2 \times \Z_6$
(the abbreviations $X,Y$ in this equation should not be mixed with 
the $\OR$ even and odd bulk cycle wrapping numbers $\tilde{X}_a^i$ and $\tilde{Y}_a^i$)
\begin{equation*}
\begin{aligned}
 P \equiv&  n^1 \, X ,
\quad
Q\equiv n^1 \, Y ,
\quad
U \equiv m^1 \, X ,
\quad
V \equiv m^1 \, Y ,
\\
\text{with }  & \quad X \equiv n^2 n^3 - m^2 m^3 ,
\quad
Y \equiv n^2 m^3  + m^2 n^3 + m^2 m^3 
.
\end{aligned}
\end{equation*}
In order to shorten the notation, we define
\begin{equation*}
\begin{aligned}
\tilde{U} &\equiv U + b \, P
,
\qquad
\tilde{V} \equiv V + b \, Q
,
\\
\tilde{\rho}_1 &\equiv \rho_1 - b \, \rho_3
,
\qquad
\tilde{\rho}_2 \equiv \rho_2 - b \, \rho_4
,
\end{aligned}
\end{equation*}
and decompose a bulk three-cycle into its $\OR$ even and odd parts,
\begin{equation*}
\Pi^{\rm bulk}_a = \sum_{i=0}^1 \left( \tilde{X}^i_a \, \Pi^{\rm even}_i + \tilde{Y}^i_a \, \Pi^{\rm odd}_i \right)
,
\end{equation*}
with the explicit form of the three-cycles and bulk wrapping numbers given in table~\ref{Tab:T6Z6p+T6Z2Z6-OR_even+odd}.
The non-trivial intersection numbers read 
\begin{equation*}
\begin{aligned}
\Pi^{\rm even}_0 \circ \Pi^{\rm odd}_0 = -  c_{\rm lattice} \times \left\{\begin{array}{c} 4 \\ 8 \end{array}\right.
\qquad\quad
\Pi^{\rm even}_1 \circ \Pi^{\rm odd}_1 = - \frac{3}{c_{\rm lattice}} \times \left\{\begin{array}{cc}
4 & T^6/\Z_6'\\
8 & T^6/\Z_2 \times \Z_6
 \end{array}\right.
,
\end{aligned}
\end{equation*}
\begin{table}[ht]
\renewcommand{\arraystretch}{1.3}
\begin{equation*}
\begin{array}{|c||c|c|c|}\hline
\multicolumn{4}{|c|}{\text{\bf $\OR$ even \& odd 3-cycles on } T^6/\Z_6' \text{ \bf \& } T^6/\Z_2 \times \Z_6}
\\\hline\hline
\text{lattice} & {\bf a/bAA} & {\bf a/bAB} & {\bf a/bBB}
\\\hline\hline
\Pi^{\rm even}_0 & \tilde{\rho}_1 &  \tilde{\rho}_1 + \tilde{\rho}_2 & \tilde{\rho}_2
\\
\Pi^{\rm even}_1 & -\rho_3 + 2 \rho_4 & -\rho_3 + \rho_4 & -2\rho_3 + \rho_4
\\\hline
\Pi^{\rm odd}_0 & -\rho_3 & -\rho_3 - \rho_4 & -\rho_4
\\
\Pi^{\rm odd}_1 & - \tilde{\rho}_1 + 2 \, \tilde{\rho}_2 &  - \tilde{\rho}_1 + \tilde{\rho}_2 & -2\tilde{\rho}_1 + \tilde{\rho}_2
\\\hline\hline
\tilde{X}_a^0 & \frac{2P_a+Q_a}{2} & \frac{P_a+Q_a}{2} & \frac{P_a + 2Q_a}{2}
\\
\tilde{X}_a^1 & \frac{\tilde{V}_a}{2} & \frac{\tilde{V}_a-\tilde{U}_a}{2} & -\frac{\tilde{U}_a}{2}
\\\hline
\tilde{Y}_a^0 & -\frac{2\tilde{U}_a+\tilde{V}_a}{2} & -\frac{\tilde{U}_a + \tilde{V}_a}{2} & - \frac{\tilde{U}_a + 2 \tilde{V}_a}{2}
\\
\tilde{Y}_a^1 & \frac{Q_a}{2} & \frac{Q_a-P_a}{2} & - \frac{P_a}{2}
\\\hline
\end{array}
\end{equation*}
\caption{$\OR$ even and odd bulk three-cycles and bulk wrapping numbers for the 
six inequivalent lattices of the $T^6/(\Z_6' \times \OR)$ and $T^6/(\Z_2 \times \Z_6 \times \OR)$ orientifolds.
}
\label{Tab:T6Z6p+T6Z2Z6-OR_even+odd}
\end{table}
and supersymmetry conditions on the bulk cycles take the form
\begin{equation*}
\tilde{Y}_a^0 - \frac{\sqrt{3}}{c_{\rm lattice}} \; \frac{1}{r} \; \tilde{Y}_a^1=0
,
\qquad
 \tilde{X}_a^0 - \frac{\sqrt{3}}{c_{\rm lattice}} \; r \; \tilde{X}_a^1 >0
 \qquad
 \text{ with }
 \qquad
 c_{\rm lattice} = \left\{\begin{array}{cc}
1  & {\bf a/bAA}, {\bf a/bBB}\\
 3 & {\bf a/bAB}
 \end{array}\right.
,
\end{equation*}
with the complex structure $r \equiv \frac{R_2}{R_1}$ on the first two-torus.

Using the relation $\tilde{Y}_a^0\tilde{Y}_a^1=-\tilde{X}_a^0\tilde{X}_a^1$, the (length)${}^2$ 
of a supersymmetric cycle can be brought to the form
\begin{equation*}
\begin{aligned}
\prod_{i=1}^3 V_{aa}^{(i)} &=  \frac{4}{  3\; r} \left( \left[ P^2 + P Q + Q^2 \right] + r^2 \,
\left[ \tilde{U}^2 + \tilde{U} \, \tilde{V} + \tilde{V}^2 \right] \right)
\\
&=\frac{4}{3 \; r} \, c_{\rm lattice} \times \left[ \left(\tilde{X}_a^0 -\frac{\sqrt{3}}{c_{\rm lattice}} \; r \;   \tilde{X}_a^1 \right)^2  
+r^2 \, \left(\tilde{Y}_a^0 -\frac{\sqrt{3}}{c_{\rm lattice}} \; \frac{1}{r} \; \tilde{Y}_a^1\right)^2  \right]
\\
& \stackrel{\text{SUSY}}{=} \; \left( \frac{2 \, \sqrt{c_{\rm lattice}}}{\sqrt{3} \; \sqrt{r}}  \;\tilde{X}_a^0 -\frac{2}{\sqrt{c_{\rm lattice}}} \; \sqrt{r} \;  \tilde{X}_a^1 \right)^2
.
\end{aligned}
\end{equation*}
Details on the orientifold even and odd exceptional three-cycles can be found in~\cite{Gmeiner:2007zz,Forste:2010gw}.

\subsection{The $T^6/\Z_6$ and $T^6/\Z_2 \times \Z_6'$ orientifolds}\label{App:V-Z6}

While the other types of orbifolds have one or three complex structure moduli inherited from the torus, 
the tree level gauge coupling on the $T^6/\Z_6$ and $T^6/\Z_2 \times \Z_6'$ orientifolds depends only on the dilaton
since all complex structures are frozen by the underlying $\Z_3$ symmetry.

A generic bulk cycle can be written as~\cite{Honecker:2004kb,Forste:2010gw}
\begin{equation*}
\Pi^{\rm bulk} = X \; \rho_1 + Y \; \rho_2.
\end{equation*}
with the bulk three-cycles
\begin{equation*}
\begin{aligned}
X \equiv  n^1 n^2 n^3  - m^1 m^2 m^3 -\!\!\sum_{i\neq j\neq k\neq i}\!\! n^i m^j m^k
,
\qquad
Y \equiv \sum_{i\neq j\neq k\neq i} \!\!\left(n^i n^j m^k + n^i m^j m^k  \right)
.
\end{aligned}
\end{equation*}
The decomposition into $\OR$ even and odd components,
\begin{equation*}
\Pi^{\rm bulk} = \tilde{X}_a^0 \; \Pi_0^{\rm even} + \tilde{Y}_a^0 \; \Pi_0^{\rm odd}
, 
\end{equation*}
is detailed in table~\ref{Tab:T6Z6+T6Z2Z6p-OR_even+odd}
\begin{table}[ht]
\renewcommand{\arraystretch}{1.3}
\begin{equation*}
\begin{array}{|c||c|c|c|c|}\hline
\multicolumn{5}{|c|}{\text{\bf $\OR$ even \& odd 3-cycles on } T^6/\Z_6 \text{ \bf \& } T^6/\Z_2 \times \Z_6' }
\\\hline\hline
\text{lattice} & {\bf AAA} & {\bf AAB} & {\bf ABB} & {\bf BBB}
\\\hline\hline
\Pi^{\rm even}_0 & \rho_1 & \rho_1 + \rho_2 & \rho_2 & -\rho_1 + 2 \rho_2
\\
\Pi^{\rm odd}_0 & \rho_1 - 2 \rho_2 & \rho_1 - \rho_2 &  2\rho_1 - \rho_2 & \rho_1
\\\hline
\tilde{X}^0_a & X_a + \frac{Y_a}{2} & \frac{X_a+Y_a}{2} & \frac{X_a}{2}+Y_a & \frac{Y_a}{2}
\\
\tilde{Y}^0_a & -\frac{Y_a}{2} & \frac{X_a-Y_a}{2} &  \frac{X_a}{2} & X_a + \frac{Y_a}{2}
\\\hline
\end{array}
\end{equation*}
\caption{$\OR$ even and odd bulk three-cycles and bulk wrapping numbers for the 
four inequivalent lattices of the $T^6/(\Z_6 \times \OR)$ and $T^6/(\Z_2 \times \Z_6' \times \OR)$ orientifolds.
}
\label{Tab:T6Z6+T6Z2Z6p-OR_even+odd}
\end{table}
with intersection number
\begin{equation*}
\Pi^{\rm even}_0 \circ \Pi^{\rm odd}_0 =\left\{\begin{array}{cc} -4 & T^6/\Z_6\\ -8 & T^6/\Z_2 \times \Z_6' \end{array}\right.
\end{equation*}
for all four lattice orientations.
The bulk supersymmetry conditions are simply given by 
\begin{equation*}
\tilde{X}^0_a >0,
\qquad\qquad
\tilde{Y}^0_a=0
,
\end{equation*}
and the (length)${}^2$ of a supersymmetric bulk cycle can be rewritten as follows
\begin{equation}
\begin{aligned}
\prod_{i=1}^3 V_{aa}^{(i)} &= \left(\frac{2}{\sqrt{3}}\right)^3 \left(X^2 + XY + Y^2 \right)
\\
&=\frac{8 \, c_{\rm lattice}}{3^{3/2}}\; \left[ (\tilde{X}^0_a)^2 + \frac{3}{c_{\rm lattice}} (\tilde{Y}^0_a)^2 \right]
\\
& \stackrel{\text{SUSY}}{=} \; \left(\frac{2^{3/2} \, \sqrt{c_{\rm lattice}}}{3^{3/4}} \; \tilde{X}^0_a\right)^2
,
\end{aligned}
\end{equation}
where we introduced the factor
\begin{equation}
c_{\rm lattice} =
\left\{\begin{array}{cc}
1 & {\bf AAA}, {\bf  ABB} \\
3 & {\bf AAB}, {\bf BBB}
\end{array}\right.
\end{equation}
for the different lattice orientations.
Details on the orientifold projection on exceptional three-cycles can be found in~\cite{Honecker:2004kb,Forste:2010gw}.

\section{Details of the $T^6/\Z_2 \times \Z_2$ models with magnetised T-duals}\label{App:Magnetised}

In this appendix, we collect some technical facts, which are required for matching the magnetised D9- and D5-brane
models on $T^6/(\Z_2 \times \Z_2 \times \Omega)$ without torsion in~\cite{Angelantonj:2009yj} with 
the intersecting D6-brane models on $T^6/(\Z_2 \times \Z_2 \times \OR)$ with discrete torsion in section~\ref{S:Compare-Example}.
For all these examples, the background lattice is {\bf aaa}, and discrete Wilson lines are not taken into account, i.e. $\tau^i_x=0$
for all D$6_x$-branes.

The $\Z_2^{(k)}$ fixed point sets $F_k^x$ along the four-torus $T^2_i \times T^2_j$ depend only on the wrapping numbers 
$(n^i_x,m^i_x;n^j_x,m^j_x)$ being odd or even combined with the discrete displacements $(\sigma^i_x,\sigma^j_x)$
as displayed in table~\ref{tab:FixedPoints_OddEven}, for more details see appendix A.1 of~\cite{Gmeiner:2009fb}.
\begin{table}
\begin{equation*}
\begin{array}{|c|c||c|c|c|c|}\hline
\multicolumn{5}{|c|}{\text{\bf $\Z_2$ fixed points and wrapping numbers}}
\\\hline\hline
(n^i_x,m^i_x) & \sigma^i_x & \text{(odd,odd)} & \text{(odd,even)} & \text{(even,odd)}
\\\hline\hline
\alpha_i \in T^2_i & 0 & \left(\begin{array}{c} 1 \\ 3\end{array}\right)  
& \left(\begin{array}{c} 1 \\ 2\end{array}\right)  
& \left(\begin{array}{c} 1 \\ 4 \end{array}\right) 
\\\hline
& 1 & \left(\begin{array}{c} 2 \\ 4\end{array}\right) 
& \left(\begin{array}{c} 3 \\ 4\end{array}\right) 
& \left(\begin{array}{c} 2 \\ 3\end{array}\right) 
\\\hline
\end{array}
\end{equation*}
\caption{The fixed point sets $F_k^x$ depend on the properties of the torus wrapping numbers
and discrete displacements. The fixed point sets $F_k^x \in T^2_i \times T^2_j$ are obtained by
tensoring $\alpha_i\beta_j$.}
\label{tab:FixedPoints_OddEven}
\end{table}

\subsection{Example 1}\label{App:Mag-Ex1+2}

The rigid three-cycles of the first example in section~\ref{Ss:Ex1+2_Angelantonj} with wrapping numbers listed in 
table~\ref{tab:T-dual-Angelant-1+2} have the form
\begin{equation}\label{Eq:-Pi3-Ex1+2}
\begin{aligned}
\Pi_{a_i} &= \frac{1}{4} \left(\Pi_{135}^{\rm bulk} -\Pi_{146}^{\rm bulk} -\Pi_{236}^{\rm bulk} -\Pi_{245}^{\rm bulk}\right)
+ \frac{1}{4} \left( -\Pi_{246}^{\rm bulk} + \Pi_{235}^{\rm bulk} + \Pi_{145}^{\rm bulk} - \Pi_{136}^{\rm bulk}\right)
\\
& + \frac{(-1)^{\tau_{a_i}^{\Z_2^{(1)}}}}{4} \!\! \sum_{\beta\gamma \in F_1^i}\left( \varepsilon_{\beta\gamma}^{(1)} + \tilde{\varepsilon}_{\beta\gamma}^{(1)}\right)
 + \frac{(-1)^{\tau_{a_i}^{\Z_2^{(2)}}}}{4} \!\! \sum_{\alpha\gamma \in F_2^i}\left(\varepsilon_{\alpha\gamma}^{(2)} + \tilde{\varepsilon}_{\alpha\gamma}^{(2)}\right)
 + \frac{(-1)^{\tau_{a_i}^{\Z_2^{(3)}}}}{4} \!\! \sum_{\alpha\beta \in F_3^i}\left(\varepsilon_{\alpha\beta}^{(3)} - \tilde{\varepsilon}_{\alpha\beta}^{(2)}\right)
,
\\
\\
\Pi_{a_i}' &=\frac{1}{4} \left(\Pi_{135}^{\rm bulk} -\Pi_{146}^{\rm bulk} -\Pi_{236}^{\rm bulk} -\Pi_{245}^{\rm bulk}\right)
- \frac{1}{4} \left( -\Pi_{246} + \Pi_{235} + \Pi_{145} - \Pi_{136}\right)
\\
& + \frac{(-1)^{\tau_{a_i}^{\Z_2^{(1)}}}}{4} \!\! \sum_{\beta\gamma \in F_1^i}\left( -\varepsilon_{\beta\gamma}^{(1)} + \tilde{\varepsilon}_{\beta\gamma}^{(1)}\right)
 + \frac{(-1)^{\tau_{a_i}^{\Z_2^{(2)}}}}{4} \!\! \sum_{\alpha\gamma \in F_2^i}\left(-\varepsilon_{\alpha\gamma}^{(2)} + \tilde{\varepsilon}_{\alpha\gamma}^{(2)}\right)
 + \frac{(-1)^{\tau_{a_i}^{\Z_2^{(3)}}}}{4} \!\! \sum_{\alpha\beta \in F_3^i}\left(\varepsilon_{\alpha\beta}^{(3)} + \tilde{\varepsilon}_{\alpha\beta}^{(2)}\right)
,
\\
\\
\Pi_{a_i} + \Pi_{a_i}' &= \frac{1}{2} \left(\Pi_{135}^{\rm bulk} -\Pi_{146}^{\rm bulk} -\Pi_{236}^{\rm bulk} -\Pi_{245}^{\rm bulk}\right)
\\
&+\frac{(-1)^{\tau_{a_i}^{\Z_2^{(1)}}}}{2}  \sum_{\beta\gamma \in F_1^i} \tilde{\varepsilon}_{\beta\gamma}^{(1)}
\; + \frac{(-1)^{\tau_{a_i}^{\Z_2^{(2)}}}}{2} \sum_{\alpha\gamma \in F_2^i}\tilde{\varepsilon}_{\alpha\gamma}^{(2)}
\; + \frac{(-1)^{\tau_{a_i}^{\Z_2^{(3)}}}}{2}  \sum_{\alpha\beta \in F_3^i} \varepsilon_{\alpha\beta}^{(3)} 
.
\end{aligned}
\end{equation}
The bulk RR tadpoles cancel for $N_{a_1}= \ldots = N_{a_4}=4$, and the exceptional RR tadpoles cancel among the four different kinds of 
D$6_{a_i}$-branes if they have identical displacements and Wilson lines $(\vec{\sigma}_{a_ia_j}) = (\vec{0})=(\vec{\tau}_{a_ia_j})$.
For $(\vec{\sigma})_{a_i}=0$, each fixed point set is of the form $F_k^{a_i} = \{(11),(31),(13),(33)\}$.

\mathtabfix{
\begin{array}{|c||c|c||c|c|c|c|c|c|c|}\hline
\multicolumn{8}{|c|}{\text{\bf Relative $\Z_2^{(i)}$ eigenvalues in examples 1 \& 3}}
\\\hline\hline
(-1)^{\tau_{xy}^{\Z_2^{(k)}}} & a_ia_j & a_ia_j' & a_ib_j & a_i c_j & b_ib_j & c_ic_j & b_ic_j
\\\hline\hline
(+++) & \!\!\!\begin{array}{c} a_ia_i \\ {\tiny i=1\ldots 4} \end{array}\!\!\! &  \begin{array}{c} a_1a_4', \\ a_2a_3' \end{array}
&   \begin{array}{c} a_1b_1, \, a_2b_2, \\ a_3b_3 \equiv a_3 b_2', \, a_4b_4\equiv a_4b_1' \end{array} 
&   \begin{array}{c} a_1 c_1, \,  a_2c_2 \\ a_3c_3 \equiv a_3 c_1' , \, a_4c_4 \equiv a_4 c_2' \end{array} & \!\!\!\begin{array}{c} b_ib_i  \\ {\tiny i=1,2} \end{array}\!\!\!
&  \!\!\!\begin{array}{c}  c_ic_i  \\ {\tiny i=1,2} \end{array}\!\!\!  &  \!\!\!\begin{array}{c}  b_ic_i \\ {\tiny i=1,2} \end{array}\!\!\!
\\\hline
(+--) & \begin{array}{c} a_1a_2, \\ a_3a_4 \end{array} &  \begin{array}{c} a_1a_3' , \\ a_2a_4' \end{array} &  \begin{array}{c} a_1b_2, \, a_2b_1 \\
a_3b_4 \equiv a_3b_1', \, a_4b_3\equiv a_4 b_2'  \end{array} 
&  \begin{array}{c} a_1 c_2, \, a_2c_1 \\ a_3 c_4 \equiv a_3c_2' , \, a_4 c_3 \equiv a_4 c_1'  \end{array} 
 &b_1b_2 & c_1c_2 & b_1c_2, \, b_2c_1
\\\hline
(-+-) & \begin{array}{c} a_1a_3, \\ a_2 a_4  \end{array}  & \begin{array}{c}a_1a_2', \\ a_3a_4'  \end{array} 
&  \begin{array}{c} a_1b_3 \equiv a_1b_2', \, a_2b_4\equiv a_2 b_1' \\ a_3b_1, \, a_4b_2
 \end{array} 
&  \begin{array}{c} a_1 c_3 \equiv a_1c_1', \, a_2c_4 \equiv a_2 c_2' \\ a_3 c_1 , \, a_4 c_2  \end{array} 
& b_1b_2' & c_i c_i' & b_1c_1', \, b_2c_2'
\\\hline
(--+) & \begin{array}{c} a_1a_4,\\ a_2a_3  \end{array}  & a_ia_i'  &   \begin{array}{c} a_1b_4\equiv a_1b_1', \, a_2b_3 \equiv a_2 b_2' \\ a_3 b_2, \, a_4b_1
 \end{array} 
&  \begin{array}{c} a_1 c_4 \equiv a_1 c_2' , \, a_2c_3 \equiv a_2c_1' \\ a_3c_2 , \, a_4 c_1  \end{array} 
& b_ib_i' & c_1c_2' & b_1c_2', \, b_2 c_1'
\\\hline
\end{array}
}{Ex_1-3_Relative-Z2}{The first three columns contain the relative $\Z_2^{(i)}$ eigenvalues of the first example in~\protect\cite{Angelantonj:2009yj}.
These are the same signs as for the D$6_{a_i}$-branes in the third example, for which also the D$6_{b_i}$- and D$6_{c_i}$-branes are listed in the 
remaining columns.
}

\begin{table}
\begin{equation*}
\begin{array}{|c||c|c|c|c|}\hline
\multicolumn{5}{|c|}{\text{\bf Intersection numbers for example 1}}
\\\hline\hline
xy & I_{xy} & I_{xy}^{\Z_2^{(1)}} & I_{xy}^{\Z_2^{(2)}} & I_{xy}^{\Z_2^{(3)}}
\\\hline\hline
a_ia_j' & (-2) \cdot (-2) \cdot 2 
& (-1)^{\tau^{\Z_2^{(1)}}_{a_ia_j'}}\cdot (-2) \cdot 4 
& (-1)^{\tau^{\Z_2^{(2)}}_{a_ia_j'}} \cdot (-2)  \cdot 4
& (-1)^{\tau^{\Z_2^{(3)}}_{a_ia_j'}}\cdot 2 \cdot 4  
\\\hline\hline\hline
x & \eta_{\OR} \; \tilde{I}_{x}^{\OR} 
& \eta_{\OR\Z_2^{(1)}} \; \tilde{I}_{x}^{\OR\Z_2^{(1)}}
& \eta_{\OR\Z_2^{(2)}} \; \tilde{I}_{x}^{\OR\Z_2^{(2)}}
& \eta_{\OR\Z_2^{(3)}} \; \tilde{I}_{x}^{\OR\Z_2^{(3)}}
\\\hline\hline
a_i & 8 \times (-1) \cdot (-1) \cdot 1
& 8 \times (-1) \cdot 1 \cdot (-1)
& 8 \times 1 \cdot (-1) \cdot  (-1)
& - 8 \times 1 \cdot  (-1)   \cdot 1 
\\\hline
\end{array}
\end{equation*}
\caption{The torus and $\Z_2^{(i)}$ invariant intersection numbers of example 1 in section~\protect\ref{Ss:Ex1+2_Angelantonj}.
The net-chiralities can be seen to match with those given in the main text by remembering that the orientifold projection changes the $\Z_2^{(1)}$ and 
$\Z_2^{(2)}$ eigenvalues while leaving $\Z_2^{(3)}$ invariant,
$(\tau^{\Z_2^{(1)}}_{a_j'},\tau^{\Z_2^{(2)}}_{a_j'},\tau^{\Z_2^{(3)}}_{a_j'})  =  (\tau^{\Z_2^{(1)}}_{a_j}+1,\tau^{\Z_2^{(2)}}_{a_j}+1,\tau^{\Z_2^{(3)}}_{a_j})$.}
\label{tab:IntersectionNumbers_Ex1+2}
\end{table}
%

\subsection{Example 3}\label{App:Mag-Ex3}

The rigid D6-branes in section~\ref{Ss:Ex3_Angelantonj} 
with wrapping numbers listed in table~\ref{tab:T-dual-Angelant-3}
wrap the following three-cycles,
\begin{equation}\label{Eq:3-cycles-Ex3}
\begin{aligned}
\Pi_{a_i} =& \frac{1}{4} \left(\Pi_{135}^{\rm bulk} -\Pi_{146}^{\rm bulk} - 2 \, \Pi_{236}^{\rm bulk} +2 \, \Pi_{245}^{\rm bulk} \right)
+ \frac{1}{4} \left( -2 \, \Pi_{246}^{\rm bulk} + 2 \, \Pi_{235}^{\rm bulk} + \Pi_{145}^{\rm bulk} - \Pi_{136}^{\rm bulk}\right)
\\
& +\frac{(-1)^{\tau_{a_i}^{\Z_2^{(1)}}}}{4} \!\! \sum_{\beta\gamma \in F_1^{a_i}}\left( \varepsilon_{\beta\gamma}^{(1)} + 2 \, \tilde{\varepsilon}_{\beta\gamma}^{(1)}\right)
 + \frac{(-1)^{\tau_{a_i}^{\Z_2^{(2)}}}}{4} \!\! \sum_{\alpha\gamma \in F_2^{a_i}}\left(\varepsilon_{\alpha\gamma}^{(2)} + \tilde{\varepsilon}_{\alpha\gamma}^{(2)}\right)
 + \frac{(-1)^{\tau_{a_i}^{\Z_2^{(3)}}}}{4}\!\! \sum_{\alpha\beta \in F_3^{a_i}}\left(\varepsilon_{\alpha\beta}^{(3)} - \tilde{\varepsilon}_{\alpha\beta}^{(3)}\right)
,
\\
\\
\Pi_{b_i}=& \frac{1}{4}\, \Pi_{135}^{\rm bulk}
 +\frac{(-1)^{\tau_{b_i}^{\Z_2^{(1)}}}}{4} \!\!\sum_{\beta\gamma \in F_1^{b_i}} \varepsilon_{\beta\gamma}^{(1)}
 + \frac{(-1)^{\tau_{b_i}^{\Z_2^{(2)}}}}{4}\!\! \sum_{\alpha\gamma \in F_2^{b_i}} \varepsilon_{\alpha\gamma}^{(2)}
 + \frac{(-1)^{\tau_{b_i}^{\Z_2^{(3)}}}}{4} \!\! \sum_{\alpha\beta \in F_3^{b_i}} \varepsilon_{\alpha\beta}^{(3)}
,
\\
\\
\Pi_{c_i}=& -\frac{1}{4}\, \Pi_{146}^{\rm bulk}
 +\frac{(-1)^{\tau_{c_i}^{\Z_2^{(1)}}}}{4} \!\! \sum_{\beta\gamma \in F_1^{c_i}} \varepsilon_{\beta\gamma}^{(1)}
 + \frac{(-1)^{\tau_{c_i}^{\Z_2^{(2)}}}}{4} \!\! \sum_{\alpha\gamma \in F_2^{c_i}} \tilde{\varepsilon}_{\alpha\gamma}^{(2)}
 - \frac{(-1)^{\tau_{c_i}^{\Z_2^{(3)}}}}{4} \!\! \sum_{\alpha\beta \in F_3^{c_i}} \tilde{\varepsilon}_{\alpha\beta}^{(3)}
,
\end{aligned}
\end{equation}
and their orientifold image three-cycles are given by
\begin{equation}\label{Eq:OR-3-cycles-Ex3}
\begin{aligned}
\Pi_{a_i'} =& \frac{1}{4} \left(\Pi_{135}^{\rm bulk} -\Pi_{146}^{\rm bulk} - 2 \, \Pi_{236}^{\rm bulk} +2 \, \Pi_{245}^{\rm bulk} \right)
- \frac{1}{4} \left( -2 \, \Pi_{246}^{\rm bulk} + 2 \, \Pi_{235}^{\rm bulk} + \Pi_{145}^{\rm bulk} - \Pi_{136}^{\rm bulk}\right)
\\
& +\frac{(-1)^{\tau_{a_i}^{\Z_2^{(1)}}}}{4} \!\! \sum_{\beta\gamma \in F_1^{a_i}}\left(- \varepsilon_{\beta\gamma}^{(1)} + 2 \, \tilde{\varepsilon}_{\beta\gamma}^{(1)}\right)
 + \frac{(-1)^{\tau_{a_i}^{\Z_2^{(2)}}}}{4} \!\! \sum_{\alpha\gamma \in F_2^{a_i}}\left(-\varepsilon_{\alpha\gamma}^{(2)} + \tilde{\varepsilon}_{\alpha\gamma}^{(2)}\right)
 + \frac{(-1)^{\tau_{a_i}^{\Z_2^{(3)}}}}{4}\!\! \sum_{\alpha\beta \in F_3^{a_i}}\left(\varepsilon_{\alpha\beta}^{(3)} + \tilde{\varepsilon}_{\alpha\beta}^{(3)}\right)
,
\\
\\
\Pi_{b_i'}=& \frac{1}{4}\, \Pi_{135}^{\rm bulk}
 -\frac{(-1)^{\tau_{b_i}^{\Z_2^{(1)}}}}{4} \!\!\sum_{\beta\gamma \in F_1^{b_i}} \varepsilon_{\beta\gamma}^{(1)}
 - \frac{(-1)^{\tau_{b_i}^{\Z_2^{(2)}}}}{4}\!\! \sum_{\alpha\gamma \in F_2^{b_i}} \varepsilon_{\alpha\gamma}^{(2)}
 + \frac{(-1)^{\tau_{b_i}^{\Z_2^{(3)}}}}{4} \!\! \sum_{\alpha\beta \in F_3^{b_i}} \varepsilon_{\alpha\beta}^{(3)}
,
\\
\\
\Pi_{c_i'}=& -\frac{1}{4}\, \Pi_{146}^{\rm bulk}
 -\frac{(-1)^{\tau_{c_i}^{\Z_2^{(1)}}}}{4} \!\! \sum_{\beta\gamma \in F_1^{c_i}} \varepsilon_{\beta\gamma}^{(1)}
 + \frac{(-1)^{\tau_{c_i}^{\Z_2^{(2)}}}}{4} \!\! \sum_{\alpha\gamma \in F_2^{c_i}} \tilde{\varepsilon}_{\alpha\gamma}^{(2)}
 + \frac{(-1)^{\tau_{c_i}^{\Z_2^{(3)}}}}{4} \!\! \sum_{\alpha\beta \in F_3^{c_i}} \tilde{\varepsilon}_{\alpha\beta}^{(3)}
,
\end{aligned}
\end{equation}
where the orientifold transformation of exceptional cycles~(\ref{Eq:OR-images-Ex-cycles-Examples}) was used.

Constraints on the consistent sets of discrete choices of displacements $(\vec{\sigma})$
can be derived from the RR tadpole contributions of the individual three-cycles,
\begin{equation}
\begin{aligned}
\Pi_{a_i} + \Pi_{a_i}' =&  \frac{1}{2} \left(\Pi_{135}^{\rm bulk} -\Pi_{146}^{\rm bulk} - 2 \, \Pi_{236}^{\rm bulk} +2 \, \Pi_{245}^{\rm bulk} \right)
\\
& + (-1)^{\tau_{a_i}^{\Z_2^{(1)}}} \!\! \sum_{\beta\gamma \in F_1^{a_i}}\tilde{\varepsilon}_{\beta\gamma}^{(1)}
+ \frac{(-1)^{\tau_{a_i}^{\Z_2^{(2)}}}}{2} \sum_{\beta\gamma \in F_2^{a_i}}\tilde{\varepsilon}_{\beta\gamma}^{(2)}
+ \frac{(-1)^{\tau_{a_i}^{\Z_2^{(3)}}}}{2} \sum_{\alpha\beta \in F_3^{a_i}}\varepsilon_{\alpha\beta}^{(3)}
,
\\
\\
\Pi_{b_i} +\Pi_{b_i}' =& \frac{1}{2}\, \Pi_{135}^{\rm bulk}
 +  \frac{(-1)^{\tau_{b_i}^{\Z_2^{(3)}}}}{2} \sum_{\alpha\beta \in F_3^{b_i}} \varepsilon_{\alpha\beta}^{(3)}
,
\\
\\
\Pi_{c_i} + \Pi_{c_i}'=& -\frac{1}{4}\, \Pi_{146}^{\rm bulk}
  + \frac{(-1)^{\tau_{c_i}^{\Z_2^{(2)}}}}{2}\sum_{\alpha\gamma \in F_2^{c_i}} \tilde{\varepsilon}_{\alpha\gamma}^{(2)}
.
\end{aligned}
\end{equation}
The consistent match to the third example in~\cite{Angelantonj:2009yj} is obtained by combining considerations on the cancellation of 
exceptional RR tadpoles and matching of the $a_ia_j$ and $a_ia_j'$ sectors of the matter spectrum as follows:
\begin{itemize}
\item
Only the $a_i$ branes contribute to the $\Z_2^{(1)}$ twisted RR tadpoles. This does not constrain the displacements $\sigma^1_{a_i}$ along the 
first two-torus $T^2_{(1)}$, but the fixed point sets  $F_1^{a_i}$ and thus the displacements $(\sigma^2_{a_i},\sigma^3_{a_i})$ have to be pairwise
 identical among  two different $a_i$-branes with opposite $\Z_2^{(1)}$ eigenvalue in order to achieve cancellation of the exceptional RR tadpole at each
$\Z_2^{(1)}$ fixed point.
Setting $(\vec{\sigma}_{a_i})=(\vec{0})$ for all $a_i$ branes leads to a matching of all $a_ia_j$ and $a_ia_j'$ sectors with~\cite{Angelantonj:2009yj}
when including the corrections to the latter in table 2 of~\cite{Camara:2010zm}. 
\item
The $a_i$ branes have now been arranged in such a way that their exceptional $\Z_2^{(2)}$ and $\Z_2^{(3)}$ twisted RR tadpole contributions cancel.
Therefore, also the $\Z_2^{(2)}$ twisted RR tadpole has to cancel among the $c_i$-branes and the $\Z_2^{(3)}$ twisted RR tadpole among the $b_i$-branes.
This can only be achieved if the corresponding relative $\Z_2^{(k)}$ eigenvalue among each pair of branes is $(-)$ and the relative displacements are 
constrained to $(0,\sigma^2_{c_1c_2},0)$ and $(0,0,\sigma^3_{b_1b_2})$.
\item
The multiplicities in the $a_ib_j$,$a_ib_j'$, $a_ic_j$ and $a_ic_j'$ sectors are matched with~\cite{Angelantonj:2009yj} 
(up to the fact that the $\4$ and $\ov{\4}$ representations of $U(4)_{b_j}$ and $U(4)_{c_j}$ have to  be distinguished) by setting $\sigma^1_{b_j}=0=\sigma^1_{c_j}$.
The above choice of all $\sigma^i_x$ values implies that the overall amount of states with some $a_i$ charge fits, and all $U(2)^3_{a_i}$ anomalies cancel.
\item
The multiplicities in the $b_ic_j$ and $b_ic_j'$ sectors match with~\cite{Angelantonj:2009yj} for $\sigma^1_{b_i}=\sigma^1_{c_j}$
up to the issue of complex notations $\4$ and $\ov{\4}$ such that all $U(4)_{b_i}^3$ and $U(4)_{c_j}^3$ anomalies cancel.
\item
 The $b_ib_i'$ and $c_ic_i'$ sectors provide each a non-chiral pair of antisymmetric representations of $U(4)_{x_i}$ which reduces to
 the one antisymmetric of $Sp(4)_{x_i}$ listed in~\cite{Angelantonj:2009yj}.
 Finally, for vanishing relative displacements $\sigma^3_{b_1}=\sigma^3_{b_2}=0$ and $\sigma^2_{c_1}=\sigma^2_{c_2}=0$
   the $b_1b_2$, $b_1b_2'$, $c_1c_2$ and $c_1c_2'$ sectors provide also twice the amount of non-chiral bifundamental matter transforming under the unitary gauge factors
 compared to the matter in~\cite{Angelantonj:2009yj} transforming under the symplectic subgroups. This confirms the breaking pattern of the gauge symmetry discussed in 
 section~\ref{Ss:Ex3_Angelantonj}.
\end{itemize}
In summary, the assignments in table~\ref{tab:T-dual-Angelant-3} provide the best match with~\cite{Angelantonj:2009yj},
which gives full agreement in the $a_ia_j$ and $a_ia_j'$ sectors, and the differences involving branes $b_i$ and $c_i$ arise from the 
partial brane recombination discussed in section~\ref{Ss:Ex3_Angelantonj} and the associated breaking of unitary to symplectic subgroups.

The fixed point sets for the displacements $(\vec{\sigma})$ given in table~\ref{tab:T-dual-Angelant-3} 
can be read off from the assignment per two-torus in table~\ref{tab:FixedPoints_OddEven} to be
$F_1^{a_i}=\{(11),(31),(13),(33)\}$ for all D$6_{a_i}$-branes with 
$i=1 \ldots 4$ and $F_2^{a_i}=\{(11),(21),(13),(23)\}=F_3^{a_i}$. 
 For the D$6_{b_i}$-branes, all three fixed point sets are of the form
 $F_k^{b_i}=\{(11),(21),(12),(22)\}$ with $k=1,2,3$. Finally, for the D$6_{c_i}$-branes, the fixed point sets 
 are given by $F_1^{c_i}=\{(11),(41),(14),(44)\}$ and $F_2^{c_i}=\{(11),(21),(14),(24)\}=F_3^{c_i}$.

The torus and $\Z_2^{(k)}$ invariant intersection numbers for all D6-branes in example 3 are listed in table~\ref{tab:IntersectionNumbers_Ex3}.

\mathtabfix{
\begin{array}{|c||c|c|c|c|}\hline
\multicolumn{5}{|c|}{\text{\bf Intersection numbers for example 3}}
\\\hline\hline
xy & I_{xy} & I_{xy}^{\Z_2^{(1)}} & I_{xy}^{\Z_2^{(2)}} & I_{xy}^{\Z_2^{(3)}}
\\\hline\hline
a_ia_j' & (-4) \cdot (-2) \cdot 2 & (-1)^{\tau_{a_i a_j'}^{\Z_2^{(1)}}} \cdot   (-4)  \cdot (4 \, \delta_{(\sigma^2+\sigma^3)_{a_ia_j},0} )
& (-1)^{\tau_{a_i a_j'}^{\Z_2^{(2)}}} \cdot (-2) \cdot  (4 \, \delta_{(\sigma^1+\sigma^3)_{a_ia_j},0} )
& (-1)^{\tau_{a_i a_j'}^{\Z_2^{(3)}}} \cdot 2 \cdot  (4 \, \delta_{(\sigma^1+\sigma^2)_{a_ia_j},0} )
\\\hline
a_ib_j & (-2) \cdot (-1) \cdot 1 & (-1)^{\tau_{a_i b_j}^{\Z_2^{(1)}}} \cdot (-2) \cdot 1
& (-1)^{\tau_{a_i b_j}^{\Z_2^{(2)}}} \cdot (-1) \cdot  (2 \, \delta_{\sigma^1_{a_ib_j},0} )
& (-1)^{\tau_{a_i b_j}^{\Z_2^{(3)}}} \cdot 1 \cdot (2 \, \delta_{\sigma^1_{a_ib_j},0} ) 
\\\hline
a_ic_j & (-2) \cdot 1 \cdot (-1) & (-1)^{\tau_{a_i c_j}^{\Z_2^{(1)}}} \cdot (-2) \cdot 1
& (-1)^{\tau_{a_i c_j}^{\Z_2^{(2)}}} \cdot 1 \cdot  (2 \, \delta_{\sigma^1_{a_ic_j},0} )
& (-1)^{\tau_{a_i c_j}^{\Z_2^{(3)}}} \cdot (-1) \cdot  (2 \, \delta_{\sigma^1_{a_ic_j},0} )
\\\hline
b_ic_j & 0_{\pp} \cdot 1 \cdot (-1) & (-1)^{\tau_{b_i c_j}^{\Z_2^{(1)}}} \cdot 0_{\pp} \cdot \delta_{(\sigma^2+\sigma^3)_{b_ic_j},0} 
& (-1)^{\tau_{b_i c_j}^{\Z_2^{(2)}}} \cdot 1 \cdot  (2 \, \delta_{(\sigma^1+\sigma^3)_{b_ic_j},0} )
& (-1)^{\tau_{b_i c_j}^{\Z_2^{(3)}}} \cdot (-1) \cdot (2 \, \delta_{(\sigma^1+\sigma^2)_{b_ic_j},0} )
\\\hline\hline\hline
x & \eta_{\OR} \; \tilde{I}_{x}^{\OR} 
& \eta_{\OR\Z_2^{(1)}} \; \tilde{I}_{x}^{\OR\Z_2^{(1)}}
& \eta_{\OR\Z_2^{(2)}} \; \tilde{I}_{x}^{\OR\Z_2^{(2)}}
& \eta_{\OR\Z_2^{(3)}} \; \tilde{I}_{x}^{\OR\Z_2^{(3)}}
\\\hline\hline
a_i & 8 \times (-2) \cdot (-1) \cdot 1 
& 8 \times (-2) \cdot 1 \cdot (-1)
& 8 \times 1 \cdot (-1) \cdot (-1)
& - 8 \times 1 \cdot (-1) \cdot 1
\\\hline
b_i & 8 \times 0^{123}
& 8 \times 0_{\pp}  \cdot 1 \cdot (-1)
& 8 \times 1 \cdot 0_{\pp} \cdot (-1)
&  - 8 \times 1 \cdot (-1) \cdot 0_{\pp}
\\\hline
c_i & 8 \times 0_{\pp} \cdot (-1) \cdot 1
&  8 \times 0^{123}
& 8 \times 1 \cdot (-1) \cdot 0_{\pp}
& - 8 \times  1 \cdot 0_{\ap} \cdot 1
\\\hline
\end{array}
}{IntersectionNumbers_Ex3}{The torus and $\Z_2^{(i)}$ invariant intersection numbers for example 3 in section~\protect\ref{Ss:Ex3_Angelantonj}
match with the net-chiralities given in the main text when taking into account that
\mbox{$(\tau^{\Z_2^{(1)}}_{a_j'},\tau^{\Z_2^{(2)}}_{a_j'},\tau^{\Z_2^{(3)}}_{a_j'})  =  (\tau^{\Z_2^{(1)}}_{a_j}+1,\tau^{\Z_2^{(2)}}_{a_j}+1,\tau^{\Z_2^{(3)}}_{a_j})$}
for the orientifold image D6-branes $a_j$ and $a_j'$.}
%

\end{appendix}

\clearpage
\addcontentsline{toc}{section}{References}
\bibliographystyle{ieeetr}
\bibliography{refs_kaehler}

\end{document}